\documentclass[12pt,a4paper]{book}

\newif\ifpdf                % verifica se compilamos con TeX ou PDFTeX
\ifx\pdfoutput\undefined
    \pdffalse
\else
    \pdfoutput=1
    \pdftrue
\fi

\ifpdf       % En caso de PDFTeX...

    \usepackage[pdftex]{graphicx}
    \usepackage[pdftex,plainpages=false,bookmarks=true,bookmarksopen=false,linktocpage=true,pdfhighlight=/O,colorlinks=false,citecolor=blue]{hyperref}

\else       % En caso DVI...

    \usepackage[dvips]{graphicx}
    \usepackage[hypertex,colorlinks=false]{hyperref}

\fi

\usepackage{amsmath,amssymb}
\usepackage{afterpage,float}
\usepackage{fancyheadings}
\usepackage{calc,pifont}
\usepackage{enumerate}
\usepackage{array}
\usepackage{thesis}
\usepackage{PartInTOC}   % incluye la palabra "Parte" en la tabla de contenidos
\usepackage{mycommands}
\usepackage{makeidx}
\usepackage[english,galician,spanish]{babel}
\usepackage[numbers,sort&compress]{natbib}
\usepackage{hypernat}

\usepackage{mydefs}

\makeindex    % Des-comentar en el \'ultimo momento para crear el \'{\i}ndice.
\def\decimalpoint{\spanishdecimal{.}}
\decimalpoint

\begin{document}

% \hyphenation{pro-pa-ga-tion o-ccu-pied mo-du-la-ted}

% \input{auxil/mydefs}

\deactivatetilden  % Para que no transforme ~N en \~N, y evitar conflictos en la bibliograf\'{\i}a.
                    % Definido en el Babel spanish.ldf

\title{Contribuci\'on al estudio del transporte el\'ectrico en capas delgadas de cupratos superconductores: \protect \\ corrientes supercr\'{\i}ticas y paraconductividad}

\author{Jos\'e Mar\'{\i}a Vi\~na Rebolledo \protect \\ LBTS - Universidade de Santiago de Compostela}

\addtocounter{page}{-2} \thispagestyle{empty}
\selectlanguage{english}
\includegraphics[width=\textwidth]{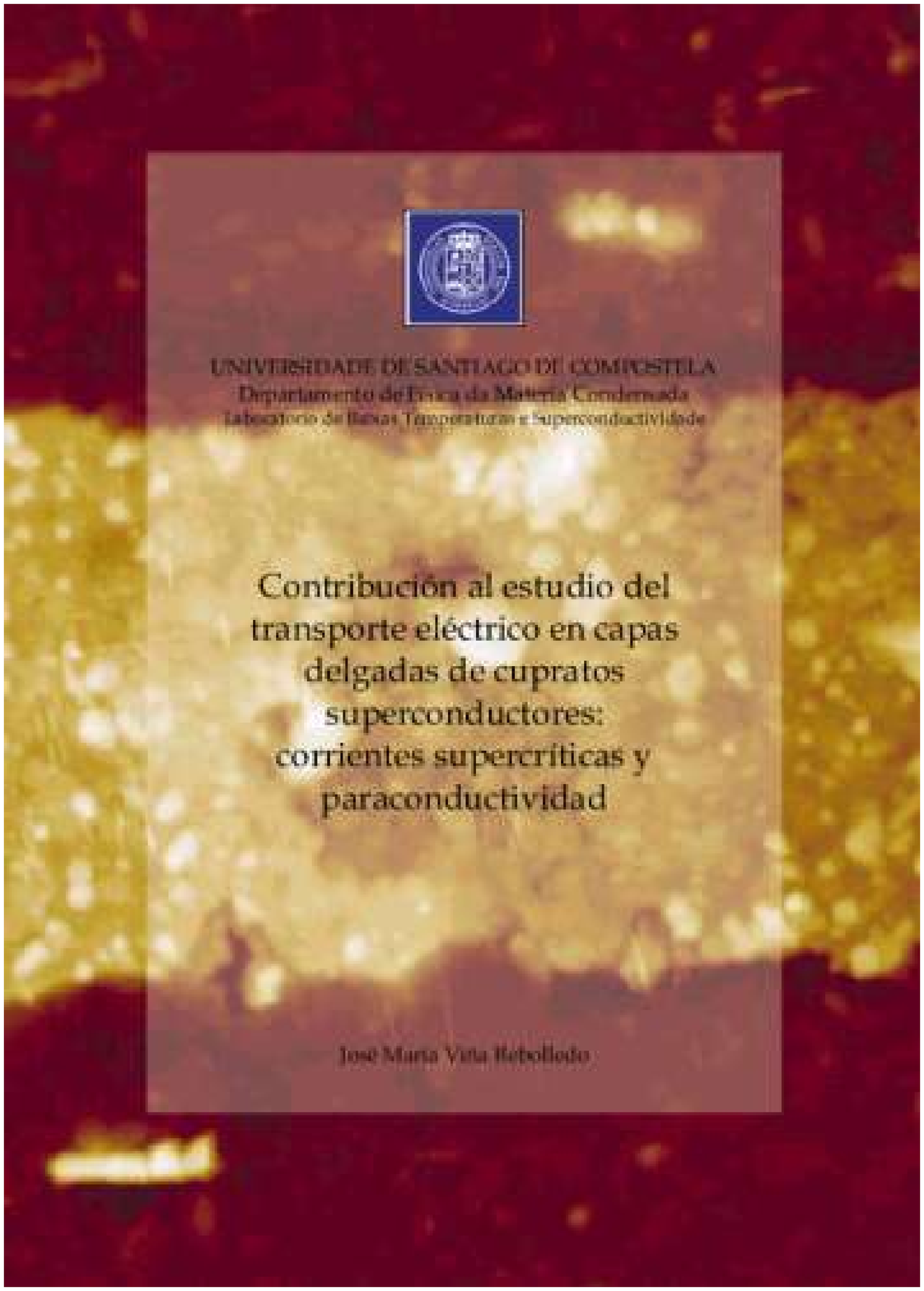}

\newpage
\thispagestyle{empty}

\vspace*{\stretch{1}} {\footnotesize

\noindent This is a special version of the original document, \\
adapted for the {\bf \href{http://arxiv.org}{arXiv.org} e-Print
archive}. \\ Ph.D. Thesis. 302 pages, 121 EPS figures. \\ In
Spanish, with summary and outline in English.
\\ Posted on 30th May 2003, cond-mat/0305712

\vspace*{0.2cm}

{\em \noindent Contribution to the study of the electric transport
\\ in cuprate superconducting thin films:  \\ supercritical
currents and paraconductivity.} \vspace*{0.2cm}

\noindent Jos\'e Mar\'{\i}a Vi\~na Rebolledo, \,\,
\href{mailto:fmjose@usc.es}{\texttt{fmjose@usc.es}}
\vspace*{0.2cm}

\noindent {\bf Abstract:} In type-II superconductors, in the
regime of very high applied current density, an abrupt transition
is experimentally observed from the mixed state to a highly
dissipative regime, probably the normal state. This provokes a
high voltage jump in the Current-Voltage Characteristics (CVC) of
the samples, at a current density $J^*$ larger than the critical
current $J_c$. In this work we study Y-123 (YBCO) microbridges
sputtered over SrTiO$_3$ substrates. We measure their CVCs for
different temperatures close to $T_c$, and with low applied
magnetic field, from 0 to 1~T. We compare the dependances of $J_c$
and $J^*$, and find that some of the typical experimental results
in this regime could be explained by only considering the effect
of the current self-field. This suggests that the depairing of the
supercarriers could be involved in the abrupt transition. We also
study the influence of thermal effects, by using a simplified
model to describe the thermal evolution of our samples. We find
that the uniform self-heating can produce an abrupt thermal
runaway, at high current density, which could explain the
occurrence of $J^*$, as well as its dependence on temperature and
applied magnetic field. These results do not totally exclude the
implication of other more intrinsic mechanisms (like Cooper pairs
depairing, or a change in the regime of the vortex dynamics), but
suggest that the uniform self-heating is more relevant than what
was considered up to now. In a shorter second part of the thesis,
the contribution of the author to the study of the
paraconductivity in his laboratory is reflected.

\vspace*{0.2cm}

\noindent Dissertation submitted to the \\ University of Santiago
de Compostela, Spain, \\ for the degree of Ph.D. \\ Faculty of
Physics, Santiago de Compostela (2003). \\
\href{http://www.usc.es}{\texttt{http://www.usc.es}}}

\newpage

% \selectlanguage{galician}   use spanish for arXiv compatibility
\selectlanguage{spanish}  \thispagestyle{empty}

\vspace*{-2cm}

\bmini[c]{\textwidth}{\normalsize
\begin{center}
\includegraphics[width=2.75cm]{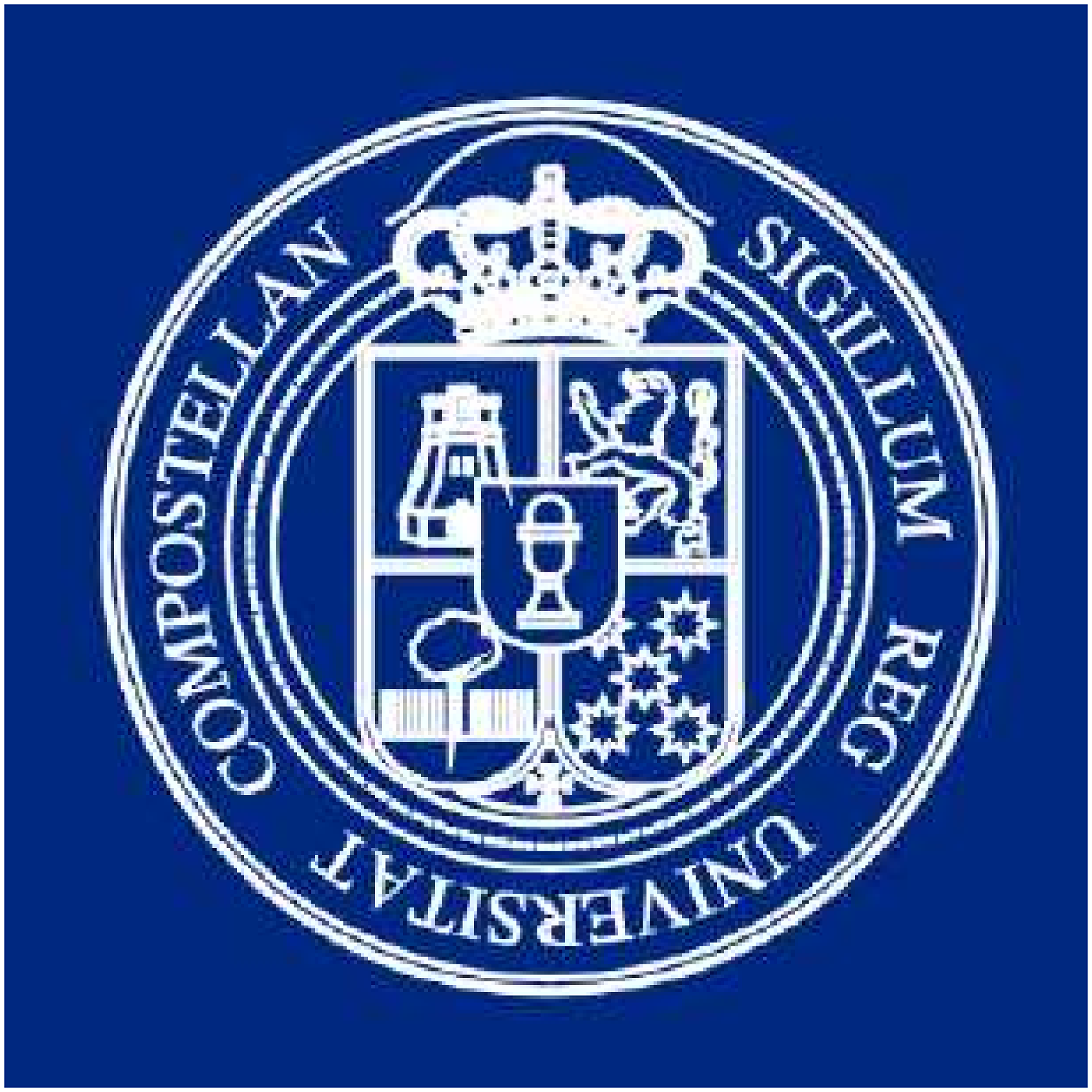} \\ %\centerline{}  % sello_usc, USC
\vspace{1cm}
 UNIVERSIDADE DE SANTIAGO DE COMPOSTELA \medskip\\
FACULTADE DE F\'{I}SICA \medskip\\ {\footnotesize DEPARTAMENTO DE
F\'{I}SICA DA MATERIA CONDENSADA \medskip \\ Laboratorio de Baixas
Temperaturas e Superconductividade \\ Unidad Asociada al
ICMM-CSIC} \vspace{1cm} \rule[-.5cm]{13cm}{.5pt}
\end{center}} \emini
%\vspace{.5cm}
%\hrule
%\vspace{\stretch{1.0}}
%\hrule
%\parbox{15cm}
%{\bcen \huge \textbf{Periodical and Fluctuating Fields\\
%in Active Spatially Extended Systems or...\\} \ecen}
%\hrule

%\hrule
\bmini[c]{\textwidth} {\bcen \LARGE {\textsc{ Contribuci\'on al
estudio del transporte el\'ectrico en capas delgadas de cupratos
superconductores:
\bigskip \\ corrientes supercr\'{\i}ticas y paraconductividad
\bigskip}} \ecen} \emini
%\hrule

%\vspace{3cm}
\vspace*{\stretch{4}}

\begin{flushright}
\bmini{7cm} \emph{Memoria presentada por Jos\'e Mar\'{\i}a Vi\~na
Rebolledo para optar ao grado de Doutor en F\'{\i}sica pola
Universidade de Santiago de Compostela.}\emini
\end{flushright}
\begin{flushright}
\bmini[r]{7cm}\emph{26 de decembro de 2002} \emini   % fecha de lectura
\end{flushright}

\newpage
\thispagestyle{empty} \selectlanguage{english}
\vspace*{\stretch{1}}
%\begin{minipage} %{1.3\textwidth}
{\footnotesize

{\em \noindent Contribuci\'on al estudio del transporte
el\'ectrico en
\\ capas delgadas de cupratos superconductores:
\\ corrientes supercr\'{\i}ticas y paraconductividad.
\vspace*{0.2cm}

\noindent (Contribution to the study of the electric transport
\\ in cuprate superconducting thin films:  \\ supercritical
currents and paraconductivity).} \vspace*{0.2cm}

\noindent Jos\'e Mar\'{\i}a Vi\~na Rebolledo, \,\,
\href{mailto:fmjose@usc.es}{\texttt{fmjose@usc.es}}
\vspace*{0.2cm}

\vspace*{0.2cm}

\noindent Dissertation submitted to the \\ University of Santiago
de Compostela, Spain, \\ for the degree of Ph.D. \\ Faculty of
Physics, Santiago de Compostela (2003). \\
\href{http://www.usc.es}{\texttt{http://www.usc.es}}
\vspace*{0.2cm}
\\In CD-ROM: ISBN 84-9750-180-2

\vspace*{0.2cm}

\noindent cond-mat/0305712. Typeset on \today.

\vspace*{0.2cm}

\noindent Ilustraci\'on de la cubierta: micropuente de YBCO
\\ superconductor barrido con microscopio de fuerza at\'omica.

%\vspace*{0.4cm}
%\noindent Ilustraci\'on de la cubierta: micropuente
%\\ de YBCO superconductor barrido mediante \\ microscop\'{\i}a de fuerza
%at\'omica. \\
}
%\end{minipage}

 \clearemptydoublepage
%%%%%%%%%%%%%%%%%%%%%%%%%%%%%%%%%%%%%%%%%%%%%%%%%%%%%%%%%%%%%%%%%%%%%%%%%%
% \selectlanguage{galician}    use spanish for arXiv compatibility
\selectlanguage{spanish}  \thispagestyle{empty}
%%%%%%%%%%%%%%%%%%%%%%%%%%%%%%%%%%%%%%%%%%%%%%%%%%%%%%%%%%%%%%%%%%%%%%%%%%
\vspace*{3.5cm}

D. F\'elix Vidal Costa e D. Antonio Veira Su\'arez, catedr\'atico e
profesor titular da Universidade de Santiago de Compostela
respectivamente,

\vspace{1cm}
\noindent{\bfseries\sffamily\large CERTIFICAN}

\vspace{1cm}

\noindent que a presente memoria, titulada \emph{``Contribuci\'on
al estudio del transporte el\'ectrico en capas delgadas de
cupratos superconductores: corrientes supercr\'{\i}ticas y
paraconductividad''}, foi realizada por \textbf{D. Jos\'e
Mar\'{\i}a Vi\~na Rebolledo} baixo a s\'ua direcci\'on, e que
constit\'ue a Tese que presenta para optar ao grado de Doutor en
F\'{\i}sica.

\vspace{1cm}

E, para que as\'{\i} conste, asinan a presente en Santiago de
Compostela a 26 de decembro de 2002.

\vspace{3cm}

%\hspace{\stretch{0.5}} {\bf V$^{\circ}$. e prace}
%\hspace{\stretch{3}} {\bf V$^{\circ}$. e prace}
%\hspace{\stretch{1}}\mbox{} \\ \noindent \hspace{\stretch{1}}
%\mbox{}\hspace{.2cm} {F\'elix Vidal Costa} \hspace{\stretch{4}}
%{Antonio Veira Su\'arez}
%\hspace{\stretch{1}}\mbox{}
%\vspace{\stretch{2}}

\bcen
\begin{tabular}{ccc}
  {V$^{\circ}$. e prace} & \hspace{3cm}& {V$^{\circ}$. e prace} \\
  F\'elix Vidal Costa &\hspace{3cm} & Antonio Veira Su\'arez \\
\end{tabular}
\ecen

%\bmini{\textwidth} \centering Jos\'e Mar\'{\i}a Vi\~na Rebolledo \emini
%\vspace{\stretch{1}}\mbox{}

\clearemptydoublepage \thispagestyle{empty} \mbox{}
 \vspace{1cm}

\bmini[c]{\textwidth}{\normalsize
\begin{center}
Facultade de F\'{\i}sica \\ Universidade de Santiago de Compostela
\\ 17 de marzo de 2003
\vspace{2cm}

{ \large Directores de la tesis} \vspace{.5cm}\\ Prof. Dr. D.
F\'elix Vidal Costa\\{\footnotesize Universidade de Santiago de
Compostela}
\medskip
\\Prof. Dr. D. Antonio Veira Su\'arez\\{\footnotesize Universidade de Santiago de
Compostela} \vspace{2cm}\\

{ \large Miembros del tribunal} \vspace{.5cm} \\ Prof. Dr. D.
Maurice-Xavier Fran\c{c}ois\\{\footnotesize Universit\'e Pierre et
Marie Curie (Paris VI)}
\medskip
\\ Prof. Dr. D. Jean-Paul Maneval\\{\footnotesize \'Ecole normale sup\'erieure, Paris}
\medskip
\\ Prof. Dr. D. Rafael Navarro Linares\\{\footnotesize Universidad de Zaragoza}
\medskip
\\ Prof. Dr. D. Jos\'e L. Vicent L\'opez\\{\footnotesize Universidad Complutense de Madrid}
\medskip
\\ Prof. Dr. D. Jes\'us Maza Frech\'{\i}n\\{\footnotesize Universidade de Santiago de Compostela}
\medskip
\end{center}} \emini

 \clearemptydoublepage
\selectlanguage{spanish}
% En M. Vargas Llosa, La org\'{\i}a perpetua, Taurus 1975, p.152:
% Pour qu'une chose soit int\'eressante, il suffit de la regader longtemps.
% G. Flaubert, carta a Le Poittevin. Carta s.f., sept. 1845.
% Correspondance, vol.I p.192. Ed. Conard.

%%%%%%%%%%%%%%%%%%%%%%%%%%%%%%%%%%%%%%%%%%%%%%%%%%%%%%%%%%%%%%%%%%%%%%%%%%
\clearemptydoublepage \thispagestyle{empty}
%%%%%%%%%%%%%%%%%%%%%%%%%%%%%%%%%%%%%%%%%%%%%%%%%%%%%%%%%%%%%%%%%%%%%%%%%%
\mbox{}\vspace{\stretch{1}}

\hspace{\stretch{1}}
\hspace{\stretch{3}}

\begin{flushright}
\begin{it}
Pour qu'une chose soit int\'eressante, \\ il suffit de la regarder
longtemps.

\hspace{\stretch{1}}

G. Flaubert, \emph{Corresp}. sept. 1845

\vspace{2cm} Quand on lit trop vite ou trop \\ doucement on
n'entend rien.

\hspace{\stretch{1}}

B. Pascal, \emph{Pens\'ees}, 41 (Papiers class\'es)
% 'Pens\'ees' nº41 (Sec. I, Papiers class\'es, II. Vanit\'e).
%  (j-) \'o 38  en vez de 41 ?

\end{it}

\end{flushright}

\vspace{\stretch{3}}\mbox{}

\clearemptydoublepage

\setlength{\parskip}{0ex plus 0.5ex}

\tableofcontents

\clearemptydoublepage

%\phantomsection  \addcontentsline{toc}{part}{Pr\'ologo}
%\part*{Pr\'ologo}

\setlength{\parskip}{1ex plus 0.5ex minus 0.2ex}
% summary and outline in English

\setlength{\parskip} {1ex plus 0.5ex minus 0.2ex}

%%%%%%%%%%%%%%%%%%%%%%%%%%%%%%%%%%%%%%%%%%%%%%%%%%%%%%%%%%%%%%%%%%

\selectlanguage{english} \chapter*{Summary in English}
\chaptermark{Summary in English}
%\chaptermark{Introducci\'on}
%\addtocontents{toc}{\protect\vspace{0.2cm}}
\addcontentsline{toc}{chapter}{Summary in English}

%-%-%-%-%-%-%-%-%-%-%-%-%-%-%-%-%-%-%-%-%-%-%-%-  figura -%-%-%-%-%-%-%-%

\newcommand{\figesqEJabst}{     % alias

\bfig
  \centering
  \includegraphics[width=\stfigw\textwidth]{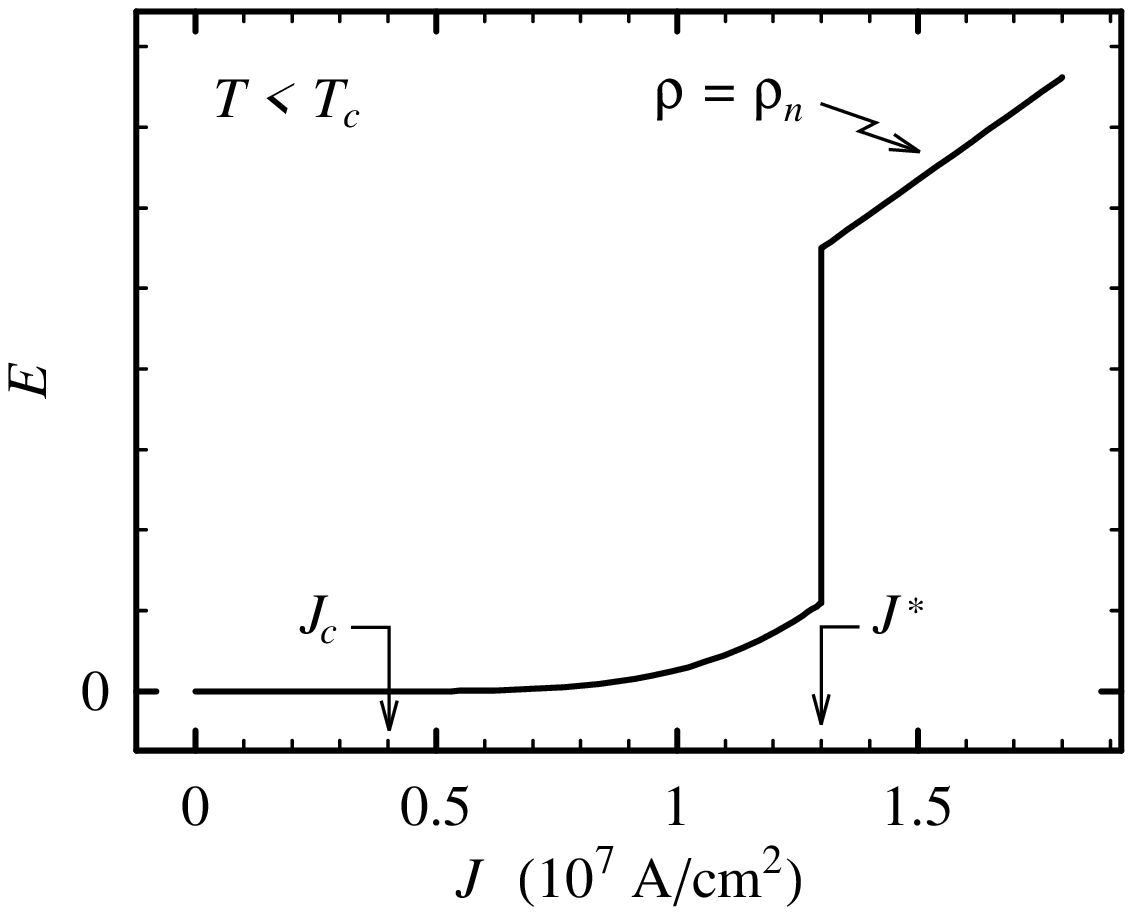}
  \caption[Schematic diagram of a typical \EJ\ curve for a \YBCOf\
superconductor at $T<\Tc$.]{Schematic diagram of a typical \EJ\
curve for a \YBCOf\ superconductor at $T<\Tc$. Below the
\emph{critical current} \Jc\ the vortices are pinned or do not
exist, there is no dissipation, and resistivity is zero. Above
\Jc\ the vortices move and dissipation appears. At \Jx\ the sample
abruptly transits to the normal state: the jump in resistivity can
be larger than that shown in this diagram.}
  \label{fig:abs:esqEJ}

\efig %%%%%%%%%%%%%%%%%%%%%%
}
% - - - - - - - - - - - - - - - - - - - - - - - - - - - - - - - - - - - %

\subsection*{Part I}

Some of the most interesting properties of high temperature
superconductors (HTSC), both from the fundamental and applied
points of view, are those associated with the electrical
transport. Since the discovery, in 1986, of these materials
\cite{Bednorz86}, such properties have been intensively studied,
and nowadays many of the mechanisms and phenomena implied in
electrical transport are known and understood
\cite{Ginsberg89,Narlikar}.

\figesqEJabst

Nevertheless, there still remain some important unanswered
questions about these properties. Among them are the phenomena
that occur in the mixed state at high current density, above the
critical current $\Jc(T,H)$ ---where dissipation appears without
destroying superconductivity--- up to the current density
$\Jx(T,H)$ where the material transits to the normal state,
usually in an abrupt jump. In figure \ref{fig:abs:esqEJ} a
schematic diagram of a typical \EJ\ characteristic curve of a
superconducting \YBCOf\ (\YBCO) is shown. To give some orders of
magnitude, let us say that in a good quality \YBCO\ thin film,
while \Jc\ is about $10^6 \Acms$ at zero applied magnetic field,
\Jx\ can be one order of magnitude larger, depending on
temperature. In good samples \mbox{---without} stoichiometrical or
structural defects that could generate weak links between
different parts of the \mbox{sample---} the physical origin of
\Jc\ is now clearly related to the depinning and movement of
magnetic vortices. (At zero applied field, these vortices are
those created by the self-field of the injected current
\cite[chap. 9]{Poole95}). However, on the contrary, the mechanisms
that originate \Jx\ are still not so clear.

This abrupt destruction of superconductivity at high current
density was already observed in metallic, low critical temperature
superconductors \cite{Klein85}, but the interest of its study
considerably grew since the discovery of HTSC \cite[and references
therein]{Kunchur02}. This is due, primarily, to the relevance of
the physical mechanisms possibly implied: high speed vortex flow,
breakage and electrical saturation of domain junctions, hot-spot
propagation, avalanches and thermal interchange, and other
possible processes of superconductivity destruction. Furthermore,
the studies around \Jx\ are also interesting from the practical
point of view for the design of devices exploiting the
superconducting transition, like current limiters, bolometric
detectors or ultra-fast switches.

During the last years, various laboratories ---including our
group--- have studied the transition at \Jx\ in HTSC
\cite{Klein85,Doettinger94,Xiao99,Chiaverini00,Curras01}. In our
laboratory, these studies begun with the Ph.D. Thesis of
Dr.~Severiano~R. Curr\'as, submitted in 2000 \cite{Curras00b}, which
may be considered as the \emph{local} precedent of our present
work, as well as of the studies also carried on by M.T.~Gonz\'alez
in the case of bulk textured HTSC
\cite{Gonzalez02b,Gonzalez02,Gonzalez03}. However, in spite of
these important affords, the behaviour and the physical origin of
$\Jx(T,H)$ in HTSC are still not well settled. As a continuation
of these studies, the central motivations of the present work
were:

i) To study in depth the causes of the jump, using faster devices
to permit detailed measurements of the phenomena that occur around
\Jx. In particular, in a typical transition in a \YBCO\ thin film,
at $77\und{K}$ and zero applied magnetic field, the resistivity
has been observed to increase by a factor of several hundreds in
less than $10\und{\mu s}$, which is the time resolution of our
experimental setup. The electronic devices used in the present
work are more that 1000 times faster than in the studies of S.~R.
Curr\'as, where the time resolution was about $30\und{ms}$. Although
our measurement time does not allow a complete discrimination of
the full jump, it is fast enough to observe the footpoint of the
transition close to \Jx, which, in particular, will let us study
the influence of the length of current pulses.

ii) Improve understanding of the role of the magnetic vortices
around \Jx. For that, we made studies with low applied magnetic
field, from the order of the self-field (typically
$\mH=\e{-2}\und{T}$) up to $\mH=1\und{T}$. This is a region of
fields still little explored in the bibliography
\cite{Xiao96,Doettinger94,Doettinger95,Lefloch99}, but that can
give relevant information about the causes of the transition.

iii) Study the influence of thermal effects. The superconducting
transition is strongly affected by apparently small changes in
temperature, which can happen even under excellent refrigeration
conditions. In the instant just before the jump, the dissipated
power density $\Wx=\Ex \times \Jx$ in our samples is typically
about $10^7\und{W/cm^3}$. We will analyze if this generated heat
can affect the transition observed in \Jx.

In the experimental work we have made use of microbridges
patterned on \YBCOf\ thin films, made in our laboratory. These
films were sputtered over crystalline substrates of \STOf. The
main advantage of these samples is, apart from their good quality,
their small size ---typically $150\und{nm} \times 10\und{\mu m}$
in cross section and $50\und{\mu m}$ long. On the one hand, due to
the reduced cross section, it is easy to reach high current
density applying low absolute currents. On the other hand, as a
result of the large surface/volume ratio and the fact that they
have a good thermal contact with the substrates, the heat removal
is appreciably better than in other kinds of sample.

The analysis of the thermal interchanges between microbridges and
their environment, essentially the substrate, constitute a
relevant and original part of this dissertation. In principle, to
do this thermal analysis one could think of solving the heat
conduction equation. Due to the complexity of the system, however,
the detailed solution becomes difficult even by numerical methods.
In this work we propose a simplified model to describe the thermal
evolution of our samples, based on elementary concepts. This model
will allow us to show that the thermal processes that take place
at high current density could explain, even quantitatively, some
of the essential aspects of the behavior of the \EJ\ curves close
to \Jx. In particular, our model, despite its simplicity,
addresses two of the main and more frequent arguments against
considering thermal effects as the fundamental cause of the
transition to the normal state in the regime of high current
density.

The first argument is the universal dependance of $\Jx(\eps)
\propto \eps^{3/2}$ \cite{Xiao96}, being $\eps \simeq 1-T/\Tc$ the
denominated \emph{reduced temperature}. Until now there were no
models essentially based in heating around \Jx\ that could account
for this dependance. We will show that our thermal model can
explain it.

A second argument used to discard the thermal origin of \Jx\ is
that the dissipated power right before the jump $\Wx=\Ex \times
\Jx$ augments when the applied magnetic field is increased
\cite{Xiao98,Xiao99,Pauly00}. This behavior seems to be
paradoxical if one attributes the transition to heat and supposes
that the temperature of the sample is increased up to above \Tc.
It is well known that applying a magnetic field has the effect of
decreasing the critical temperature \Tc: being the sample at a
given temperature, if we increase this field, we also
\emph{thermally} approximate the sample to the transition. Why
would we need then, to raise the temperature above \Tc, more
dissipation \Wx\ the closer to \Tc ? In principle, if the cause of
the jump were the heat, one would expect the necessary power to
provoke the transition to decrease with the applied field or, in
any case, to be independent of it \cite{Xiao99}. Although less
frequently, in some samples this increase of the power \Wx\ was
also observed when directly raising the temperature
\cite{Kamm00,Curras01}, not only the applied field. The thermal
model that we introduce solves this apparent paradox; we will see
that not only the value of the power $W$ is relevant to provoke a
thermal avalanche, but also that we need to include its derivative
with temperature $\indiff{W}{T}$. At equal values of dissipated
power, this derivative can be smaller close to $\Tc(H)$.

Our model of avalanche heating will explain the existence of \Jx,
and its dependance on $T$ and $H$, appealing only to a purely
thermal (extrinsic) mechanism. As will be shown, our proposal does
not exclude the implication, in the transition to the normal state
induced by a current, of other intrinsic mechanism (like Cooper
pair depairing, or a change in the regime of the vortex flow).
Nevertheless, even when these other intrinsic mechanisms are
present, our results suggest that the \emph{cascade heating} must
play an important role both in the measured values of \Jx\ and in
its dependance on $T$ and $H$.

\subsection*{Part II}

The second part of this dissertation summarizes the studies about
the influence of thermal fluctuations above \Tc\ on the electrical
conductivity (paraconductivity) of the superconducting cuprates.
This summary is made gathering the publications of our lab (LBTS)
in which the author of this dissertation has participated.

In the two first publications, the relaxation time of the
fluctuating Cooper pairs will be studied. In the three following
papers, we study the behaviour of the paraconductivity at
\emph{high} reduced temperatures, for $\eps \equiv \ln(T/\Tc)
\gtrsim 0.1$. The main result of this last study is the suggestion
that at high reduced temperatures the superconducting fluctuations
are dominated by the Heisenberg uncertainty principle, that limits
the \emph{shrinkage} of the superconducting wave function when $T$
increases above \Tc.

The contributions of the author of this dissertation to the
mentioned studies are, mainly, the measurements of resistivity as
a function of temperature, the extraction of the paraconductivity
from these measurements, and the analysis of the experimental
errors and uncertainties associated with the estimations of the
background and the critical temperature.

\comenta{Besides their relatively high critical temperature \Tc,
another important characteristic of the HTSC is the smallness, in
all directions, of the denominated \emph{correlation} or
\emph{coherence length} $\xi(T)$. This coherence length
approximately is the spatial dimension of the Cooper pairs, and
therefore the distance between the two paired electrons that
constitute it. In terms of the superconducting quantum wave
function $\Psi(r)$, $\xi(T)$ is the characteristic length where
$\Psi(r)$ can notably vary without expending too much energy.
($\xi(T)$ is the minimum dimension of the wave packet).

Applying the Heisenberg uncertainty principle, one can
qualitatively see that, around the superconducting transition,
$\xi$ and \Tc\ are linked \cite[cap. 1.3]{Tinkham96}:
\[ \xi \; \Delta p \geq \hbar,  \]
where $\Delta p \simeq \kB \; \Tc / v_F$ is the carriers momentum,
$v_F$ their Fermi velocity, and \kB\ the Boltzmann constant.
Therefore, crudely, \[ \xi \simeq \hbar \; v_f / (\kB \; \Tc).  \]
This correlation, of course very approximate \cite{Tinkham96},
shows that, since $v_F$ is expected to be similar in HTSC and in
the conventional, low \Tc\ superconductors, $\xi$ is expected to
be much smaller in HTSC that in conventional metallic
superconductors.

An important consequence of \Tc\ being high and, simultaneously,
$\xi(T)$ being small in HTSC, is that the thermodynamical
fluctuations are, around \Tc, very considerable in these
materials: on the one hand, the available thermal energy $\kB \;
\Tc$ is high and, on the other hand, the necessary energy to
create or destroy Cooper pairs is small. This last energy is,
approximately, the necessary to create one Cooper pair, multiplied
by the number of Cooper pairs that can be found in the coherence
volume. (This multiplication is because the superconductivity
coherence does not permit to create or destroy \emph{only one}
Cooper pair, what would be equivalent to vary $\Psi(r)$ in spatial
ranges shorter than $\xi$). This number of pairs is proportional
to the coherence volume $\xi(T)^D$, being $D = 3,2,1$ the
dimensionality of the sample. Since $\xi(T)$ is small in HTSC,
whatever the dimensionality is, the coherence volume is small too.
Furthermore, these materials are laminar, and in general $2 \leq D
\leq 3$, what contributes to make the coherence volume even
smaller. All this leads to that the creation and destruction of
Cooper pairs caused by thermal fluctuations is a relevant and
interesting phenomenon in HTSC.

One of the most appropriate magnitudes to experimentally
appreciate thermal fluctuations, because the ease and precision of
its measurement, but also because it is very affected by this
mechanism, is the electrical resistivity $\rho(T)$. The creation
of Cooper pairs by thermal fluctuations makes \rt\ noticeably
decrease, even for temperatures some degrees above \Tc. As a
matter of fact, the possible importance of these effects in HTSC
was already mentioned by Bednorz and M\"uller in their pioneer
article \cite{Bednorz86}. The conventional magnitude that takes
account of these effects is the denominated
\emph{paraconductivity}, the excess in electrical conductivity due
to fluctuations, that is defined as
\[ \Delta \sigma(T) = \frac{1}{\rho(T)} - \frac{1}{\rho_{bg}(T)}.
\]
In this expression, \rt\ is the experimental resistivity at a
given temperature $T$, while \rbgt\ is the named \emph{background
resistivity}, the resistivity that would exist at $T$ in the
absence of superconducting transition. For this, the only known
experimental access, commonly used, is the extrapolation of the
behaviour of \rt\ from the high temperature region, supposedly
lacking in fluctuations, to the region of lower temperature around
\Tc.+

This dissertation will treat, in this second part, some aspects of
the paraconductivity in the high temperature region of HTSC.}

\clearemptydoublepage

%%%%%%%%%%%%%%%%%%%%%%%%%%%%%%%%%%%%%%%%%%%%%%%%%%%%%%%%%%%%%%%%%%

\selectlanguage{english} \chapter*{General outline}
\chaptermark{General Outline}
%\addtocontents{toc}{\protect\vspace{0.2cm}}
\addcontentsline{toc}{chapter}{General outline}

The present work is divided in two parts. The
{\bf first part}, from chapter \ref{cap:intro} to \ref{cap:concl},
studies the transition from superconducting to normal state that
takes place in the regime of high current density, in type-II
superconductors.

\begin{description}

\item[{\bf Chapter \ref{cap:intro}}] gives a detailed introduction to the
phenomenon of the transition to the normal state induced by high
current density, in high temperature superconductors. It is an
extension of what was exposed in the previous Summary.

\item[{\bf Chapter \ref{cap:muestras}}] describes how we synthesized and
characterized the samples used in the experiments of this work. We
used $10\und{\mu m}$ wide and 20, 50 and $100\und{\mu m}$ long
microbridges, patterned by chemical wet photolithography on 150~nm
\YBCOf\ thin films. These films were synthesized by DC sputtering
over $1\und{cm^2}$ \STOf\ substrates. For characterization we used
x-ray diffractometry (XRD), profile meter, and atomic force
microscopy (AFM). We applied low resistance electrical contacts by
ultrasonic microsoldering.

\item[{\bf Chapter \ref{cap:disps}}] describes the experimental setup
used for the measurements, and comments its advantages over former
setups in our laboratory. We used a cryostat refrigerated with
liquid nitrogen with samples in helium atmosphere. Magnetic field
up to $\mH=1\und{T}$ was applied by surrounding coils. Current was
injected in the sample by a programmable current source, and
voltage was measured with a nanovoltmeter ---for low signals---
and a data acquisition card (DAQ) ---for fast measurements in the
high dissipation regime---.

\item[{\bf Chapter \ref{cap:resul}}] shows the different experimental data
acquired, and studies some of the thermal properties of the
system.

\item[{\bf Chapter \ref{cap:datos}}] analyses in detail the previous data,
and makes a review of some of the theoretical models that have
been proposed to explain the results. The effect of the self-field
in the regime of low applied magnetic fields is considered.

\item[{\bf Chapter \ref{cap:cal}}] proposes a simple model to describe the
thermal behavior of our samples, based in the feedback that the
increase of temperature causes in the dissipation, and in the
temperature distribution that the heat removal produces in the
substrate. Both elements can generate, under certain
circumstances, a thermal avalanche that increases the temperature
of the sample above the critical temperature in short periods of
time, causing the sample to transite to the normal state.

\item[{\bf Chapter \ref{cap:calpr}}] compares the predictions of the
model proposed in chapter \ref{cap:cal} with the experimental
data, and suggests an avalanche condition derived from our thermal
model. This chapter shows how, despite the simplicity of the
model, it can explain, quite accurately, the main phenomena
observed around the transition to the normal state at high current
density. This will lead to the conclusion that the thermal effects
can have special relevance in the mechanisms that break
superconductivity in this current regime.

\item[{\bf Chapter \ref{cap:concl}}] summarizes the main conclusions of this
first part.

\end{description}

The {\bf second part} gathers the main results of the study of the
paraconductivity (excess of electrical conductivity due to thermal
fluctuations) in the regime of high temperature, in which the
author of this dissertation was participant during his doctorate
period.

\begin{description}

\item[{\bf Chapter \ref{cap:introds}}] makes a brief introduction to this second
part explaining the importance of fluctuation effects in HTSC, and
considers some experimental details related to the measurement of
paraconductivity.

\item[{\bf Chapter \ref{cap:pubds}}] collects the publications of
our laboratory about this topic where the author has participated.
The main contributions were the experimental measurements of
paraconductivity in different thin films, and data analysis.

\end{description}

At the end,

\begin{description}

\item[{\bf Appendix \ref{cap:appMath}}] explains how some of the
calculations and graphics of this dissertation where done, what
could be of interest for those who want to continue or revise
these works.

\end{description}

\clearemptydoublepage

%%%%%%%%%%%%%%%%%%%%%%%%%%%%%%%%%%%%%%%%%%%%%%%%%%%%%%%%%%%%%%%%%%

\selectlanguage{spanish}

\chapter*{T\'erminos clave - Keywords}
\chaptermark{T\'erminos clave - Keywords}
%\addtocontents{toc}{\protect\vspace{0.2cm}}
\addcontentsline{toc}{chapter}{T\'erminos clave - Keywords}

%\begin{minipage}
\begin{flushleft}
\nohyphens

Superconductividad; \YBCOf\ (YBCO, \YBCO); capas delgadas, filmes
o pel\'{\i}culas superconductoras; alta densidad de corriente;
transici\'on superconductor-normal; avalancha o cascada;
realimentaci\'on; din\'amica de v\'ortices magn\'eticos; efectos
t\'ermicos; autocalentamiento; inestabilidad Larkin-Ovchinnikov
del movimiento de v\'ortices.

Superconductivity; \YBCOf\ (YBCO, \YBCO); superconducting thin
films; high current density; superconducting-normal (SN)
transition; avalanche, quench; feed-back; magnetic vortex
dynamics; thermal effects; self-heating; Larkin-Ovchinnikov (LO)
flux flow instability.

\normalhyphens \end{flushleft}
% \end{minipage}

\clearemptydoublepage

\clearemptydoublepage

\setlength{\parskip}{0ex plus 0.5ex}
 %% List of symbols.

\chapter*{Notaci\'on}
\chaptermark{NOTACI\'ON}
%\addtocontents{toc}{\protect\vspace{0.2cm}}
\addcontentsline{toc}{chapter}{Notaci\'on}

\begin{description}

\item[$a$] Ancho del micropuente.
\item[\alp] Factor geom\'etrico para corregir la distribuci\'on de temperatura dentro de \vs.

\item[$B$] $=\mH$ Densidad de flujo magn\'etico aplicado.
\item[\Bsf] Autocampo, densidad de flujo magn\'etico generado por la corriente aplicada.

\item[\cp] Capacidad calor\'{\i}fica por unidad de volumen, a presi\'on constante.

\item[$D$] $=\kappa/\cp$, difusividad t\'ermica.
\item[$d$] Espesor del filme.
\item[\DTci] Anchura de la transici\'on superconductora en $\rho(T)$, determinada como la anchura a mitad de altura de la derivada $\indiff{\rho}{T}$.
\item[\ds] $=1/\ro \;-\; 1/\rbg$, paraconductividad: exceso de conductividad el\'ectrica en
el estado normal debido a fluctuaciones termodin\'amicas.

\item[$E$] Campo el\'ectrico, determinado experimentalmente como
$E=V/l$.

\item[\Ex] Valor del campo el\'ectrico en el momento del salto, en \Jx.
\item[\EJ] Curva caracter\'{\i}stica campo el\'ectrico -- densidad de corriente.
\item[$\eps$] $\equiv\ln(\Tc/T)\simeq 1-T/\Tc$, temperatura
reducida. En el estado normal, donde $T>\Tc$, la definici\'on cambia
de signo.

\item[$H$] Campo magn\'etico.
\item[$h$] Coeficiente de transferencia de calor entre filme y substrato.
\item[\Hsf] Autocampo magn\'etico generado por la corriente aplicada.

\item[$I$] Intensidad de corriente.
\item[\Ix] Intensidad de corriente en que se produce el salto al estado normal.

\item[$J$] Densidad de corriente, establecida experimentalmente
como $J=I/(a\,d)$.
\item[\Jc] Densidad de corriente cr\'{\i}tica, en la que aparece disipaci\'on medible dentro del estado mixto (umbral en campo $E$).
\item[\Jcpin] Densidad de corriente cr\'{\i}tica en la que los v\'ortices salen de sus anclajes (umbral en velocidad $v$).
\item[\Jd] Densidad de corriente de desapareamiento de los pares de Cooper.
\item[\Jx] Densidad de corriente en que se produce el salto al estado normal.

\item[$\kappa$] Conductividad t\'ermica.

\item[$l$] Longitud del micropuente, distancia entre contactos de voltaje.

\item[$P$] $=W\;\vf$, potencia total disipada en el filme.
\item[\Px] $=\Wx\;\vf$, potencia total disipada en el momento del salto.

\item[$Q$] Calor total disipado en el filme.
\item[\Qp] Calor evacuado por unidad de tiempo desde la muestra a su entorno.

\item[$r(t)$] $\propto \sqrt{D\;t}$, distancia recorrida por el calor en un tiempo $t$.
\item[$\rho$] $\equiv E/J$, resistividad el\'ectrica.
\item[\rhoab] Resistividad el\'ectrica en los planos $ab$ del filme.
\item[\rbg] Resistividad el\'ectrica de fondo (\emph{background}), en ausencia de
transici\'on superconductora.
\item[\rhon] Resistividad el\'ectrica en el estado normal.

\item[$T$] Temperatura.
\item[$t$] Tiempo.
\item[\Tb] Temperatura del ba\~no.
\item[\Tc] Temperatura cr\'{\i}tica.
\item[\Tcz] Temperatura m\'as alta a la que $\rho(T)=0$, a campo $H$ aplicado nulo.
\item[\Tci] Temperatura en la que $\rho(T)$ tiene su punto de inflexi\'on en la transici\'on superconductora, determinada como el m\'aximo de la derivada $\indiff{\rho}{T}$.
\item[\Tf] Temperatura del filme.
\item[\Ts] Temperatura promedio del volumen $\vs(t)$.
\item[\tauB] $=\frac{\cp\;d}{h}$, tiempo bolom\'etrico de
enfriamiento.
\item[$\taue$] $=
(\tau_{ee}^{-1}+\tau_{ef}^{-1})^{-1}$, tiempo de relajaci\'on
energ\'etica de los portadores superconductores.
\item[$\tau_{ee}$] Tiempo de dispersi\'on electr\'on-electr\'on.
\item[$\tau_{ef}$] Tiempo de dispersi\'on electr\'on-fon\'on.
\item[\taui] Tiempo caracter\'{\i}stico de inercia t\'ermica de la
muestra.

\item[$V$] Diferencia de potencial, voltaje.
\item[$v$] $=E/\mH$ velocidad de los v\'ortices.
\item[\VI] Curva caracter\'{\i}stica voltaje -- corriente.
\item[\vf] $=l \times a \times d$, volumen del filme (micropuente).
\item[$\vs(t)$] Volumen total que se ha calentado en un tiempo $t$, de micropuente m\'as substrato.

\item[$W$] $=E \times J$, potencia disipada por unidad de volumen.
\item[\Wx] $=\Ex \times \Jx$, potencia disipada por unidad de volumen en el momento del salto.

\end{description}

\clearemptydoublepage

\setlength{\parskip}{1ex plus 0.5ex minus 0.2ex}
% reajusta el espacio entre p\'arrafos

%\renewcommand{\baselinestretch}{1.9} {\small}
% reajusta el interlineado. Est\'a en Thesis.sty, no aqu\'{\i}.

% \linespread{5}

%\include{esquema}
%\clearemptydoublepage

%\pagenumbering{arabic}

\part{Corrientes supercr\'{\i}ticas en filmes de \YBCOf}

%%%%%%%%%%%%%%%%%%%%%%%%%%%%%%%%%%%%%%%%%%%%%%%%%%%%%%%%%%%%%%%%%%
\chapter{Introducci\'on a la parte I}
%\chaptermark{Introducci\'on}
%\addtocontents{toc}{\protect\vspace{0.2cm}}
\label{cap:intro}
%%%%%%%%%%%%%%%%%%%%%%%%%%%%%%%%%%%%%%%%%%%%%%%%%%%%%%%%%%%%%%%%%%

%-%-%-%-%-%-%-%-%-%-%-%-%-%-%-%-%-%-%-%-%-%-%-%-  figura -%-%-%-%-%-%-%-%

\newcommand{\figesqEJ}{     % alias

\figura {fig:intro:esqEJ}   % label
{fig/intro/esqEJ2}           % file
{Representaci\'on esquem\'atica de una curva \EJ\ t\'{\i}pica de
un \YBCOf\ superconductor a $T<\Tc$. El eje vertical (el campo
el\'ectrico) no est\'a proporcionado con el horizontal, para que
se puedan apreciar los distintos reg\'{\i}menes de transporte. Por
debajo de la \emph{corriente cr\'{\i}tica} \Jc\ los v\'ortices
permanecen anclados o no existen, no hay disipaci\'on, y la
resistividad es cero. Por encima de \Jc\ los v\'ortices se mueven,
y aparece disipaci\'on: la resistividad aparente es mayor que
cero. Por encima de \Jx\ la muestra transita al estado normal. La
transici\'on en \Jx\ es abrupta, y el salto en resistividad puede
ser mucho m\'as grande de lo que se representa en
este esquema (ver texto).}  % caption
{Representaci\'on esquem\'atica de una curva \EJ\ t\'{\i}pica de
un cuprato
superconductor a $T<\Tc$.}  % toc
{\stfigw}                  % width \textwidth
}
% - - - - - - - - - - - - - - - - - - - - - - - - - - - - - - - - - - - %

%\subsection*{Contextualizaci\'on del problema \\ y objetivos del
%trabajo}

\section{Contextualizaci\'on del problema}

\subsection{Superconductividad}

\emph{Superconductividad} es un t\'ermino que se refiere a un
conjunto de propiedades especiales, el\'ectricas y magn\'eticas,
que presentan algunos materiales cuando se enfr\'{\i}an hasta
bajas temperaturas. Estas propiedades no fueron descubiertas hasta
que los refrigerantes criog\'enicos permitieron, a principios del
siglo XX, disminuir la temperatura en los experimentos hasta los
valores requeridos para su observaci\'on. En 1908, el f\'{\i}sico
holand\'es Heike Kamerlingh Onnes, en la universidad de Leiden,
logr\'o la licuefacci\'on del helio, y obtuvo de este modo un
refrigerante que le permiti\'o estudiar diversas propiedades
f\'{\i}sicas de la materia hasta temperaturas tan bajas como 1~K
($-272\und{\grado C}$). Tres a\~nos m\'as tarde, estudiando el
comportamiento del mercurio en esta regi\'on de temperaturas,
descubri\'o que su \Index{resistividad} el\'ectrica en corriente
continua ca\'{\i}a bruscamente a cero por debajo de 4.15~K
\cite{Onnes11}. \'Este fue el primer experimento en que se
apreciaban los sorprendentes efectos de la superconductividad. En
los a\~nos siguientes fue descubriendo diversos mecanismos que
destru\'{\i}an este estado de resistencia nula, como la
aplicaci\'on de corrientes el\'ectricas o de campos magn\'eticos
excesivamente elevados. Por sus investigaciones en el r\'egimen de
bajas temperaturas, Kamerlingh Onnes recibi\'o el Premio Nobel de
F\'{\i}sica en 1913.

Durante mucho tiempo se crey\'o que, a parte de esta ausencia de
resistencia el\'ectrica a temperaturas muy bajas, las restantes
propiedades f\'{\i}sicas de estos materiales eran las comunes a
los materiales normales. Pero en 1933 se descubri\'o que en
realidad los superconductores presenta no s\'olo una conducci\'on
el\'ectrica perfecta, sino tambi\'en un diamagnetismo perfecto. Un
simple conductor ideal lograr\'{\i}a mantener la densidad de flujo
magn\'etico \emph{constante} en su interior, debido a las
corrientes que se inducir\'{\i}an en su superficie sin coste
energ\'etico alguno: las variaciones de campo externo se
ver\'{\i}an compensadas por las corrientes inducidas. Pero un
superconductor hace m\'as que eso: no s\'olo mantiene la densidad
de flujo constante en su interior, sino que adem\'as este valor
constante es siempre cero. Si aplicamos campo magn\'etico a una
muestra de este tipo cuando est\'a en el estado normal ---a
temperatura elevada, por encima de la temperatura cr\'{\i}tica
\Tc---, y posteriormente la enfriamos hasta $T<\Tc$ para hacer que
transite al estado superconductor, todo el flujo magn\'etico
ser\'a expulsado de su interior. El campo magn\'etico queda
restringido a una delgada capa superficial, donde es apantallado
por las corrientes inducidas en una longitud de penetraci\'on
$\lambda$, caracter\'{\i}stica del material \cite{London35}. Este
fen\'omeno se conoce como efecto Meissner-Ochsenfeld
\cite{Meissner33}.

Con el paso de los a\~nos, y en paralelo a un trabajo de
an\'alisis de materiales que aumentaba sucesivamente el list\'on
de temperatura cr\'{\i}tica \Tc\ de la transici\'on
superconductora ---como ejemplos, 4.1~K en 1911 (Hg), 9.2~K en
1930 (Nb), 18.1~K en 1954 (Nb$_3$Sn), 23.2~K en 1973 (Nb$_3$Ge)
\cite[cap. 2]{Poole95}---, se avanzaba tambi\'en parcialmente en
la comprensi\'on de los mecanismos f\'{\i}sicos responsables de
estas novedosas propiedades de la materia
\cite{London35,Ginzburg50}. Por el momento, la mejor
interpretaci\'on de los mecanismos microsc\'opicos que originan la
superconductividad se la debemos a J. Bardeen, L. Cooper y J. R.
Schrieffer, que en 1957 publicaron la teor\'{\i}a hoy conocida
como BCS \cite{Bardeen57}, que dio a sus autores el Nobel de
F\'{\i}sica en 1972. En el contexto de esta teor\'{\i}a, las
part\'{\i}culas portadoras de \emph{supercorriente} son pares de
Cooper \cite{Cooper56}, parejas de electrones normales ligados
entre s\'{\i}, que al ser bosones (\emph{spin} entero) conforman
estados de alta coherencia. En la interacci\'on que mantiene
ligados a los electrones median los \'atomos de la red cristalina:
esquem\'aticamente, el paso de un electr\'on cerca de un \'atomo
desplaza a \'este ligeramente de su posici\'on de equilibrio, y
esta distorsi\'on de la red influye en un segundo electr\'on, cuyo
momento lineal se ve alterado, lig\'andose de este modo al
primero. La distribuci\'on de energ\'{\i}as de estos electrones no
es uniforme, existe una separaci\'on (\emph{gap}) en energ\'{\i}a
entre los estados normal y superconductor.

En 1986, J.~G. Bednorz y K.~A. M\"uller publicaron un
art\'{\i}culo titulado \emph{Possible High \Tc\ Superconductivity
in the Ba-La-Cu-O System} \cite{Bednorz86}, que dio pie a la
s\'{\i}ntesis de nuevos materiales superconductores, basados en
\'oxidos de cobre, con temperaturas cr\'{\i}ticas cada vez m\'as
elevadas: son los superconductores de alta temperatura, conocidos
por sus siglas anglosajonas como HTSC. Pronto, en 1987, se
super\'o el umbral de la temperatura del nitr\'ogeno l\'{\i}quido
a presi\'on atmosf\'erica (unos 77~K, $-196\und{\grado C}$), que
es un refrigerante mucho m\'as barato, eficiente y c\'omodo de
manejar que el helio l\'{\i}quido, con lo que el campo de
investigaci\'on y las aplicaciones pr\'acticas de la
superconductividad se hicieron mucho m\'as amplios. La
progresi\'on de los valores de \Tc\ a partir de ese momento tiene
como ejemplos los siguientes hitos: 30~K en 1986
(Ba$_x$La$_{5-x}$Cu$_5$O$_y$), 92~K en 1987 (\YBCOf), 110~K en
1988 (Bi$_2$Sr$_2$Ca$_2$Cu$_3$O$_{10}$), 133~K en 1993
(HgBa$_2$Ca$_2$Cu$_3$O$_{8+x}$) \cite[cap. 2]{Poole95}. El \YBCOf\
(abreviado YBCO o Y-123) sigue siendo uno de los HTSC m\'as
estudiados en la actualidad.

\subsection{Estado mixto}

Conforme se avanzaba en el conocimiento de estos materiales, se
descubri\'o que existen dos tipos de superconductores con
propiedades el\'ectricas y magn\'eticas diferentes. En los
superconductores de tipo I, el efecto Meissner-Ochsenfeld impide
que el flujo magn\'etico $B$ penetre en el interior de la muestra
mientras \'esta permanezca en estado superconductor. En
superconductores de tipo II ---de los que los HTSC forman parte
\mbox{destacada---,} por el contrario, resulta energ\'eticamente
favorable, a partir de que aplicamos un cierto campo o una cierta
corriente (la cual genera siempre un \emph{\Index{autocampo}}
magn\'etico \Bsf), la aparici\'on de \Index{v\'ortices} magn\'eticos
\cite{Abrikosov57} en el interior de la muestra antes de que se
rompa del todo la superconductividad: es el estado mixto
\cite[cap. 12]{Rose-Innes78}. Esto permite que el flujo magn\'etico
(ya sea el externo aplicado $B$ o el autocampo \Bsf) penetre m\'as
all\'a de la superficie, alcanzando mayor proporci\'on de volumen,
aunque siempre localizado dentro de esos v\'ortices. En una imagen
sencilla, la mayor parte del volumen de la muestra permanece en
estado superconductor, y aparecen muchos \emph{tubos} delgados que
la atraviesan, con n\'ucleos en estado normal, que permiten el paso
del campo magn\'etico a su trav\'es. En cada v\'ortice hay exactamente
un cuanto de flujo magn\'etico,
\be
\Phi _0 = \frac{h c}{2 e} \simeq 2 \times 10^{-11} {\rm~T~cm^2}.
\label{ec:fluxon} \ee Cuanto m\'as \index{campo magn\'etico} campo
aplicado se tenga o mayor sea la corriente inyectada en la
muestra, m\'as v\'ortices se crean. Mientras los v\'ortices puedan
permanecer anclados (ya sea en defectos de la red cristalina, en
puntos de dopaje, o en general en cualquier pozo de potencial) no
se producir\'a disipaci\'on, y el estado mixto seguir\'a
implicando un voltaje nulo al paso de corriente. Pero en el
momento en que las fuerzas de Lorentz, a que est\'an sometidos los
v\'ortices por la aplicaci\'on de la corriente el\'ectrica, venzan
esos pozos de potencial, los v\'ortices comenzar\'an a moverse,
generando un campo el\'ectrico paralelo a la corriente aplicada, y
en su mismo sentido. Aparece as\'{\i} disipaci\'on, y mantener
dicha corriente exige mantener una diferencia de potencial a lo
largo de la muestra. Para una profusa revisi\'on del
comportamiento de los \Index{v\'ortices} en el estado mixto,
remitimos a la referencia \cite{Blatter94}.

La densidad de corriente a la que comienza a verse
experimentalmente esta disipaci\'on suele denominarse
\emph{densidad de corriente cr\'{\i}tica}, y se representa por
\Jc. En su determinaci\'on experimental, normalmente se establece
que es aquella en la que se supera un cierto \Index{umbral} en el
campo el\'ectrico $E$ medido. Este umbral es, por supuesto, todo
lo peque\~no que la resoluci\'on permita. \Jc\ no es, por tanto,
la densidad de corriente que destruye la superconductividad, sino
la que empieza a generar disipaci\'on mensurable, manteni\'endose
la muestra en el estado mixto.

En estas muestras de tipo II podemos seguir aumentando la
corriente por encima de \Jc\ y detectar distintas tendencias del
movimiento de los \Index{v\'ortices}, siempre dentro del estado
mixto. El tipo de movimiento m\'as elemental de los v\'ortices es,
por ejemplo, el \emph{free flux flow} \cite{Kunchur93}, en que la
velocidad $v$ de los v\'ortices es proporcional a la corriente
aplicada $J$, y por lo tanto tambi\'en lo es el campo $E = v\;B$
que se mide a $B$ aplicado constante. Las curvas
caracter\'{\i}sticas \EJ\ (curvas \VI\ escaladas a las dimensiones
de las muestras), a grandes rasgos, tienen valor cero hasta \Jc\ y
son una l\'{\i}nea recta por encima de esta corriente \cite[cap.
13]{Rose-Innes78}.

Estas curvas caracter\'{\i}sticas pueden ser m\'as complejas
debido a diversos mecanismos. Por ejemplo, en los superconductores
de alta temperatura cr\'{\i}tica \Tc\ pueden estar involucrados
efectos t\'ermicos que hagan que se aprecie disipaci\'on incluso
antes de que las fuerzas de Lorentz desplacen a \emph{todos} los
v\'ortices de sus centros de anclaje. La energ\'{\i}a t\'ermica
crear\'{\i}a fluctuaciones locales que arrancar\'{\i}an a uno o
m\'as \Index{v\'ortices} de su lugar y, dependiendo de las
interacciones entre v\'ortices, se tendr\'{\i}an as\'{\i} diversas
posibles se\~nales de voltaje experimental \cite[cap.
9]{Tinkham96}. Esto es lo que ocurre por ejemplo en el \emph{flux
creep}, donde la variaci\'on de $E$ con $J$ ya no es lineal, sino
curva. Pero en estas y otras variantes de la linealidad la muestra
sigue en el estado mixto.

\subsection{Transici\'on al estado normal inducida por corrientes altas}

En estos superconductores de tipo II, si seguimos aumentando la
corriente aplicada muy por encima de \Jc\ llegamos a observar
experimentalmente una ruptura brusca del estado mixto. Si $J$
supera un cierto umbral de corriente \Jx\ (siendo \'esta hasta un
orden de magnitud mayor que \Jc) se induce una transici\'on
abrupta de toda la muestra a un estado de alta disipaci\'on, que
suele identificarse con el estado normal. El correspondiente salto
en \Index{resistividad} aparente $\rho = E/J$ es t\'{\i}picamente
de dos \'ordenes de magnitud. Esta transici\'on se observa en
todos los superconductores de tipo II, tanto en los de alta como
en los de baja \Tc\
\cite{Klein85,Samoilov95,Ruck97,Lefloch99,Doettinger94,Xiao99}. En
este trabajo nos referiremos frecuentemente a esta transici\'on
abrupta al estado normal llam\'andola simplemente \emph{el salto
en \Jx}.

\figesqEJ

En la figura \ref{fig:intro:esqEJ} se muestra un esquema de las
curvas \EJ\ de un \YBCOf\ (\YBCO) superconductor. Por encima de
\Jc\ la variaci\'on del campo $E$ con la corriente $J$ no es
lineal sino curva, t\'{\i}pica del \emph{flux creep}. Para dar
\'ordenes de magnitud digamos que, en una capa delgada de \YBCO,
si \Jc\ ronda los $10^6 \Acms$ en ausencia de campo magn\'etico
externo, \Jx\ puede llegar a ser un orden de magnitud mayor,
dependiendo de la temperatura. As\'{\i} como en buenas muestras
---sin defectos estequiom\'etricos o estructurales, que podr\'{\i}an generar uniones d\'ebiles entre diferentes partes de la
muestra--- el origen f\'{\i}sico de \Jc\ est\'a claramente
vinculado al  \Index{desanclaje} \index{pinning} y movimiento de
los v\'ortices magn\'eticos, no est\'a tan claro qu\'e mecanismo
induce el fen\'omeno observado en \Jx. Las causas de esta
transici\'on abrupta en \Jx\ podr\'{\i}an ser diversas, incluso
variar de una muestra a otra o de un tipo de material a otro. En
los \'ultimos veinticinco a\~nos se han considerado varios modelos
que explicar\'{\i}an este salto al estado normal. En general se
enmarcan en dos grandes conjuntos.

En un primer grupo de modelos, la causa de la transici\'on en \Jx\
se busca tambi\'en en el movimiento de los v\'ortices.
An\'alogamente a como en \Jc\ dicho movimiento sufre un cambio de
r\'egimen (ya sea porque se pasa del reposo al movimiento, o
porque se produce un movimiento global m\'as intenso que los
previos locales), en \Jx\ se producir\'{\i}a un cambio de
comportamiento de los v\'ortices al moverse.

En el otro gran grupo de modelos est\'an los que explican la
transici\'on exclusivamente con calentamiento. Se basan en la idea
de que la muestra ve elevada su temperatura por encima de \Tc\
destruy\'endose as\'{\i} la superconductividad.

Los distintos modelos, dentro de uno u otro grupo, describen
diversos posibles mecanismos para explicar c\'omo es y qu\'e causa
la transici\'on. Nos referiremos a varios de ellos en el
cap\'{\i}tulo \ref{cap:datos}. Sin que encaje claramente en alguno
de estos dos grupos previos, tampoco se debe olvidar la
posibilidad de que \Jx\ coincida con la \Index{corriente de
desapareamiento} \Jd\ de los pares de Cooper. Esta corriente \Jd\
es la que imprime tanta energ\'{\i}a cin\'etica a los pares de
Cooper que la existencia de estos se vuelve desfavorable
energ\'eticamente respecto a la de los electrones normales
disociados (\cite[sec. 4.4]{Tinkham96} y \cite[sec.
10.XIX]{Poole95}). De ser esta la causa, el fen\'omeno ser\'{\i}a
mucho m\'as intr\'{\i}nseco \index{intr\'{\i}nseco} que cualquiera
de los anteriores, e impondr\'{\i}a una cota superior
infranqueable a la corriente m\'axima que la muestra puede
transportar.

\section{Objetivos de este trabajo}

Este comportamiento de las curvas \EJ\ que se ha descrito se
observ\'o ya en los superconductores met\'alicos de baja
temperatura cr\'{\i}tica \cite{Klein85}, pero el inter\'es de su
estudio se ha acrecentado desde el descubrimiento de los
HTSC\footnote{A modo de ejemplo, la referencia \cite{Kunchur02} es
un art\'{\i}culo muy reciente en el que se relaciona la
bibliograf\'{\i}a m\'as destacada.}. Esto se debe en primer lugar
a la relevancia de los posibles mecanismos f\'{\i}sicos
implicados: movimiento de v\'ortices a grandes velocidades,
ruptura y saturaci\'on el\'ectrica de uniones entre dominios,
propagaci\'on de puntos calientes, avalanchas e intercambios
t\'ermicos, y otros posibles mecanismos de destrucci\'on de la
superconductividad. Adem\'as, los estudios en torno a \Jx\ son
tambi\'en interesantes desde el punto de vista aplicado, para el
dise\~no de dispositivos basados en la transici\'on
superconductora, como limitadores de corriente, detectores
bolom\'etricos o conmutadores ultrarr\'apidos.

Durante los \'ultimos a\~nos, varios laboratorios ---incluido
nuestro grupo--- han estudiado la transici\'on en \Jx\ en HTSC
\cite{Klein85,Doettinger94,Xiao99,Chiaverini00,Curras01}. En
nuestro laboratorio estos estudios se iniciaron con la tesis
doctoral de Severiano R. Curr\'as, defendida en el 2000
\cite{Curras00b}, que se puede considerar como el precedente
\emph{local} del presente trabajo, as\'{\i} como del que est\'a
llevando a cabo M.T.~Gonz\'alez en el caso de superconductores
texturados masivos \cite{Gonzalez02b,Gonzalez02,Gonzalez03}. En
cualquier caso, a pesar de estos esfuerzos, el comportamiento y el
origen f\'{\i}sico de $\Jx(T,H)$ en HTSC no est\'an todav\'{\i}a
bien establecidos. Como continuaci\'on de estos estudios, la
motivaciones principales del presente trabajo fueron:

i) Profundizar en las causas del salto haciendo medidas con
dispositivos electr\'onicos m\'as r\'apidos, que permitiesen
apreciar con m\'as detalle los fen\'omenos que ocurren en torno a
\Jx. Apuntemos que en una transici\'on t\'{\i}pica de una
pel\'{\i}cula de \YBCO, a $77\und{K}$ y sin campo magn\'etico
aplicado, la resistividad aparente puede aumentar en un factor de
varios cientos en menos de $10\und{\mu s}$, que es la resoluci\'on
temporal de nuestros dispositivos. La electr\'onica empleada en el
presente trabajo es m\'as de mil veces m\'as r\'apida que la
empleada en los trabajos de S.~R. Curr\'as, en los que la
resoluci\'on temporal era de unos $30\und{ms}$. Si bien nuestros
tiempos de medida tampoco permiten una discriminaci\'on completa
de todo el salto, son lo suficientemente r\'apidos como para
observar el arranque de la transici\'on cerca de \Jx, lo que en
particular nos permitir\'a estudiar la influencia de la duraci\'on
de los pulsos de corriente.

ii) Intentar conocer mejor el papel que desempe\~nan los
v\'ortices magn\'eticos en torno a \Jx. Para ello nos propusimos
estudiar el salto con campos magn\'eticos aplicados bajos, desde
del orden del \Index{autocampo} (t\'{\i}picamente
$\mH=\e{-2}\und{T}$) hasta $\mH=1\und{T}$. Esta es una regi\'on de
campos todav\'{\i}a poco explorada en la bibliograf\'{\i}a sobre
el tema \cite{Xiao96,Doettinger94,Doettinger95,Lefloch99}, y que
sin embargo puede aportar informaci\'on relevante acerca de las
causas de la transici\'on.

iii) Estudiar la influencia de los efectos t\'ermicos. La
transici\'on superconductora se ve fuertemente afectada por
aparentemente peque\~nas variaciones de temperatura, que se pueden
originar incluso bajo excelentes condiciones de refrigeraci\'on.
En el instante previo al salto la potencia disipada por unidad de
volumen $\Wx=\Ex \times \Jx$ en nuestras muestras ronda
t\'{\i}picamente los $10^7\und{W/cm^3}$. Analizaremos si este
calor generado puede influir en la transici\'on que se observa en
\Jx.

En nuestro estudio hemos empleado micropuentes grabados sobre
capas delgadas de \YBCOf\ depositadas sobre substratos cristalinos
de \STOf. Las principales ventajas de nuestras muestras, crecidas
en nuestro laboratorio, se deben, adem\'as de a su buena calidad,
a su reducido tama\~no, t\'{\i}picamente de $150\und{nm} \times
10\und{\mu m}$ de secci\'on por $50\und{\mu m}$ de largo. Por un
lado, en virtud de su secci\'on tan peque\~na, es f\'acil alcanzar
altas densidades de corriente aplicando corrientes absolutas muy
bajas. Pero adem\'as, gracias a su elevada relaci\'on
superficie/volumen y a que tienen buen contacto t\'ermico con el
substrato cristalino, la evacuaci\'on del calor es notablemente
mayor que en otros tipos de muestras.

El an\'alisis de los intercambios t\'ermicos entre los
micropuentes y su entorno, esencialmente el substrato, constituye
una parte muy relevante y original de esta tesis. Para hacer este
an\'alisis t\'ermico en principio se podr\'{\i}a emplear
directamente la ecuaci\'on de conducci\'on del calor, pero debido
a la complejidad del sistema su resoluci\'on detallada se hace
dif\'{\i}cil incluso por m\'etodos num\'ericos. En este trabajo
propondremos un modelo simplificado para describir la evoluci\'on
t\'ermica de nuestras muestras basado en conceptos muy
elementales. Este modelo nos permitir\'a mostrar que los procesos
t\'ermicos que tienen lugar a altas densidades de corriente
podr\'{\i}an explicar, incluso cuantitativamente, algunos de los
aspectos esenciales del comportamiento de las curvas \EJ\ en las
cercan\'{\i}as de \Jx. En particular, podemos mencionar ya
aqu\'{\i}que nuestro modelo t\'ermico, a pesar de su sencillez, da
respuesta a dos de los principales y m\'as habituales argumentos
empleados en contra de que los efectos t\'ermicos sean la causa
esencial de la transici\'on al estado normal en el r\'egimen de
altas densidades de corriente.

El primer argumento es el comportamiento universal de $\Jx(\eps)
\propto \eps^{3/2}$ \cite{Xiao96,Curras01}, siendo $\eps \simeq
1-T/\Tc$ la denominada temperatura reducida. Hasta el momento no
hab\'{\i}a un modelo basado esencialmente en el calentamiento
alrededor de \Jx\ que diese cuenta de esta dependencia. Veremos
que nuestro modelo t\'ermico s\'{\i} puede explicarla.

Un segundo argumento empleado para descartar el origen
esencialmente t\'ermico de \Jx\ es que la potencia disipada justo
antes del salto $\Wx=\Ex \times \Jx$ aumenta cuando se incrementa
el campo magn\'etico aplicado \cite{Xiao98,Xiao99,Pauly00}. Este
comportamiento resulta parad\'ojico si se piensa en el calor como
causa de la transici\'on, y se supone por tanto que la temperatura
de la muestra aumenta hasta por encima de \Tc. Es conocido que
aplicar campo magn\'etico tiene el efecto de disminuir la
temperatura cr\'{\i}tica \Tc: con la muestra a una temperatura
dada, si aumentamos el campo aproximamos \emph{t\'ermicamente} la
muestra a la transici\'on. ?`Por qu\'e, entonces, para elevar la
temperatura por encima de \Tc\ habr\'{\i}amos de necesitar m\'as
calor \Wx\ cuanto m\'as cerca de \Tc ? En principio, si lo que
induce el salto fuese el calor, se podr\'{\i}a esperar que la
potencia necesaria para causar la transici\'on disminuyese con el
\index{campo magn\'etico} campo aplicado o, en todo caso, fuese
independiente de \'el \cite{Xiao99}. Aunque con menos frecuencia,
en algunas muestras este aumento de la potencia \Wx\ tambi\'en se
ha observado al aumentar directamente la temperatura
\cite{Kamm00,Curras01}, no s\'olo el campo aplicado. El modelo
t\'ermico que proponemos resuelve esta aparente paradoja: veremos
que no s\'olo el valor de la potencia $W$ es relevante para causar
una \Index{avalancha}, sino que tenemos que implicar tambi\'en a
su derivada con la temperatura $\indiff{W}{T}$. A igualdad de
potencia disipada, esta derivada puede ser menor m\'as cerca de
$\Tc(H)$.

Nuestro modelo de \Index{calentamiento} permitir\'a explicar la
existencia de \Jx, y su dependencia respecto de $T$ y $H$,
invocando s\'olo un mecanismo puramente t\'ermico
\index{extr\'{\i}nseco} (extr\'{\i}nseco). Como veremos, nuestra
propuesta no excluye que en la transici\'on al estado normal
inducida por una corriente \emph{supercr\'{\i}tica} est\'en
tambi\'en involucrados mecanismos m\'as \index{intr\'{\i}nseco}
intr\'{\i}nsecos (desapareamiento de pares de Cooper, cambio de
r\'egimen del movimiento de v\'ortices\ldots). Sin embargo,
incluso en el caso de que dichos mecanismos intr\'{\i}nsecos
est\'en presentes, nuestros resultados sugieren que el
\emph{calentamiento en cascada} debe de desempe\~nar un papel
importante tanto en los valores de \Jx\ medidos como en su
dependencia respecto de $T$ y $H$.

\comenta{
 A pesar de su sencillez, veremos que este modelo
permite explicar los aspectos esenciales del comportamiento de las
curvas \EJ\ en las cercan\'{\i}as de \Jx. Mencionamos dos ejemplos
de este comportamiento que consideramos significativos: a) Por un
lado est\'a la dependencia que presenta \Jx\ con la temperatura,
t\'{\i}pica de un mecanismo intr\'{\i}nseco.\Jx\ var\'{\i}a con la
temperatura reducida $\eps \simeq 1-T/\Tc$ con una ley de
potencias $\Jx(\eps) \propto \eps^{3/2}$ \cite{Xiao96}. No ha
habido hasta el momento un modelo basado exclusivamente en el
calentamiento que haya podido dar cuenta de dicha dependencia. b)
Por otro lado se observa que la potencia disipada }

\comenta{Estos dos argumentos, tomados como ejemplo entre otros,
se han empleado para descartar que la causa de la transici\'on en
\Jx, en pel\'{\i}culas delgadas, tenga que ver con el
calentamiento. En este trabajo, sin embargo, analizaremos
detallados resultados experimentales en el contexto del modelo
t\'ermico que proponemos, y veremos que los efectos del
calentamiento no se pueden descartar ni siquiera con las buenas
condiciones de refrigeraci\'on que presentan estas muestras.
Concretamente, empleando s\'olo nuestro modelo de calentamiento
seremos capaces de explicar tanto la dependencia de $\Jx(\eps)$
como la de $\Wx(H)$, aclarando la aparente paradoja: ... para
entender c\'omo peque\~nos incrementos de temperatura son m\'as
cr\'{\i}ticos en algunas regiones que en otras, y c\'omo pueden
llegar a desestabilizar el sistema.}
  % cap 1
\clearemptydoublepage

%%%%%%%%%%%%%%%%%%%%%%%%%%%%%%%%%%%%%%%%%%%%%%%%%%%%%%%%%%%%%%%%%%
\chapter{S\'{\i}ntesis, caracterizaci\'on y preparaci\'on de las muestras}
%\chaptermark{S\'{\i}ntesis, caracterizaci\'on y preparaci\'on de las muestras}
%\addtocontents{toc}{\protect\vspace{0.2cm}}
\label{cap:muestras}
%%%%%%%%%%%%%%%%%%%%%%%%%%%%%%%%%%%%%%%%%%%%%%%%%%%%%%%%%%%%%%%%%%

%-%-%-%-%-%-%-%-%-%-%-%-%-%-%-%-%-%-%-%-%-%-%-%-  figura -%-%-%-%-%-%-%-%

\newcommand{\figMuestSputtering}{  % alias

\figura {fig:muest:sputtering}      % label
{fig/muest/sputtering}              % file
{Banco de pulverizaci\'on (\emph{sputtering}) Plassys MP400
modificado. En primer plano, elevada sobre cuatro patas, la
campana cil\'{\i}ndrica de deposici\'on, con peque\~nas mirillas
laterales. Sobre ella, una gran campana prism\'atica de
metacrilato, que se dispuso para unas pruebas de aislamiento. En
segundo plano, el ordenador de control y los dispositivos de
alimentaci\'on y monitorizaci\'on. Al fondo de todo, un compresor
y un sistema de refrigeraci\'on. Hacia la derecha salen los tubos
que llevan a la bomba de vac\'{\i}o primario y a las balas de
gases.}                  % caption
{Banco de pulverizaci\'on (\emph{sputtering}) Plassys MP400 modificado.}                  % toc
{\stfigw}            % width \textwidth
}
% - - - - - - - - - - - - - - - - - - - - - - - - - - - - - - - - - - - %

%-%-%-%-%-%-%-%-%-%-%-%-%-%-%-%-%-%-%-%-%-%-%-%-  figura -%-%-%-%-%-%-%-%

\newcommand{\figMuestDifractom}{  % alias

\figura {fig:muest:difractom}      % label
{fig/muest/difractom}              % file
{Geometr\'{\i}a Bragg-Brentano de difracci\'on de rayos-x del
equipo Siemens D-5005 empleado para la caracterizaci\'on
estructural de nuestras muestras. Se se\~nalan los \'angulos
$\theta$ y $2\,\theta$ que determinan los brazos de la
fuente y del detector respectivamente.}                  % caption
{Esquema del difract\'ometro de rayos-x empleado para la caracterizaci\'on estructural.}                  % toc
{.7}            % width \textwidth
}
% - - - - - - - - - - - - - - - - - - - - - - - - - - - - - - - - - - - %

%-%-%-%-%-%-%-%-%-%-%-%-%-%-%-%-%-%-%-%-%-%-%-%-  figura -%-%-%-%-%-%-%-%

\newcommand{\figMuestRocking}{  % alias

\figura {fig:muest:rocking}      % label
{fig/muest/rocking}              % file
{Curva de basculamiento (\emph{rocking curve}) t\'{\i}pica para el
pico (005) de una de nuestras capas delgadas de \YBCOf. La anchura
a mitad de altura (FWHM) nos informa de la dispersi\'on de
orientaciones de los microgranos que conforma la muestra, en torno
a la orientaci\'on preferente. En este caso, la mayor\'{\i}a de
los granos est\'an alineados en una misma direcci\'on con una
dispersi\'on menor de $0.3\grado$.}
% caption
{Curva de basculamiento (\emph{rocking curve}) para el pico (005) del \YBCO.}                  % toc
{\stfigw}            % width \textwidth
}
% - - - - - - - - - - - - - - - - - - - - - - - - - - - - - - - - - - - %

%-%-%-%-%-%-%-%-%-%-%-%-%-%-%-%-%-%-%-%-%-%-%-%-  figura -%-%-%-%-%-%-%-%

\newcommand{\figMuestRayosx}{  % alias

\figura {fig:muest:rayosx}      % label
{fig/muest/rayosx}              % file
{Difractograma $\theta$--$2\,\theta$ t\'{\i}pico para una de
nuestras capas delgadas, en representaci\'on logar\'{\i}tmica. La
orientaci\'on preferente de la muestra es con la direcci\'on $c$
perpendicular al plano del substrato: s\'olo se detectan picos de
la familia (001). Los picos (003) y (006), se\~nalados con un
asterisco, coinciden con los picos (100) y (200) del substrato de
\STOf, por lo que la
intensidad es mayor.}                  % caption
{Difractograma $\theta$--$2\,\theta$.}                  % toc
{\stfigw}            % width \textwidth
}
% - - - - - - - - - - - - - - - - - - - - - - - - - - - - - - - - - - - %

%-%-%-%-%-%-%-%-%-%-%-%-%-%-%-%-%-%-%-%-%-%-%-%-  figura -%-%-%-%-%-%-%-%

\newcommand{\figMuestEsqmascara}{  % alias

\figura {fig:muest:esqmascara}      % label
{fig/muest/esqmascara}              % file
{Esquema de la m\'ascara para el fotolitografiado de los
micropuentes. Sus distintos elementos no est\'an representados en
proporci\'on relativa, para poder apreciar los m\'as peque\~nos. Hay
tres micropuentes de 100, 50 y $20\und{\mu m}$ de largo y
$10\und{\mu m}$ de ancho, unidos por un puente grande de
$1\und{cm}$ de largo en total, por $200\und{\mu m}$ de ancho.
Hacia arriba salen los contactos para las sondas de voltaje. Los
ancheamientos del puente se reservan para los
contactos de corriente.}                  % caption
{Esquema de la m\'ascara para el fotolitografiado de los micropuentes.}                  % toc
{\stfigw}            % width \textwidth
}
% - - - - - - - - - - - - - - - - - - - - - - - - - - - - - - - - - - - %

%-%-%-%-%-%-%-%-%-%-%-%-%-%-%-%-%-%-%-%-%-%-%-%-  figura -%-%-%-%-%-%-%-%

\newcommand{\figMuestInsoladora}{  % alias

\figura {fig:muest:insoladora}      % label
{fig/muest/insoladora}              % file
{Banco de insolaci\'on ultravioleta que empleamos para
fotolitografiar los micropuentes en las pel\'{\i}culas de \YBCOf.
Bajo el ocular, en el centro de la mesa, se alinean la m\'ascara y
la muestra. El ocular se aparta luego lateralmente hacia la
derecha y en su lugar, sobre la muestra, se dispone l\'ampara de
xen\'on que se ve en el extremo izquierdo de la mesa, con un
protector negro (KarlSuss IN470). }
% caption
{Insoladora ultravioleta que empleamos para el fotolitografiado.}                  % toc
{\stfigw}            % width \textwidth
}
% - - - - - - - - - - - - - - - - - - - - - - - - - - - - - - - - - - - %

%-%-%-%-%-%-%-%-%-%-%-%-%-%-%-%-%-%-%-%-%-%-%-%-  figura -%-%-%-%-%-%-%-%

\newcommand{\figMuestPerfilom}{  % alias

\figura {fig:muest:perfilom}      % label
{fig/muest/perfilom}              % file
{Resultados t\'{\i}picos de las medidas de perfilometr\'{\i}a en un micropuente, en tres diferentes puntos de su longitud. El resultado es, en cada medida, un corte transversal de la muestra: el espesor en funci\'on de la posici\'on $x$ a lo ancho. El espesor promedio ronda los 150~nm.}                  % caption
{Resultados t\'{\i}picos de las medidas de perfilometr\'{\i}a en un micropuente.}                  % toc
{\stfigw}            % width \textwidth
}
% - - - - - - - - - - - - - - - - - - - - - - - - - - - - - - - - - - - %

%-%-%-%-%-%-%-%-%-%-%-%-%-%-%-%-%-%-%-%-%-%-%-%-  figura -%-%-%-%-%-%-%-%

\newcommand{\figMuestVideo }{  % alias

\figura {fig:muest:video}      % label
{fig/muest/video}              % file
{Im\'agenes por v\'{\i}deo digital de un micropuente, la de abajo con mayor aumento que la de arriba. Las \'areas blancas redondas que se aprecian parcialmente en los bordes de la imagen superior son capas de oro depositadas por evaporaci\'on, para la posterior aplicaci\'on de contactos el\'ectricos.}                  % caption
{Im\'agenes por v\'{\i}deo digital de un micropuente.}                  % toc
{0.7}            % width \textwidth
}
% - - - - - - - - - - - - - - - - - - - - - - - - - - - - - - - - - - - %

%-%-%-%-%-%-%-%-%-%-%-%-%-%-%-%-%-%-%-%-%-%-%-%-  figura -%-%-%-%-%-%-%-%

\newcommand{\figMuestAfmpc}{  % alias

\figura {fig:muest:afmpc}      % label
{fig/muest/afmpc}              % file
{Conjunto de dispositivos que conforman el microscopio de fuerza
at\'omica (AFM) de nuestro laboratorio, un Digital Instruments
NanoScope E. A la izquierda, bajo un visor \'optico para inspecci\'on
ocular, la unidad con el portamuestras y los sensores. \'Esta se
conecta a la unidad de control e interface (\emph{Scanning Probe
Microscope Controller}), que a su vez va conectada a un ordenador
PC que gestiona los datos adquiridos. Vemos las dos pantallas del
ordenador en las que se monitoriza la
medida.}                  % caption
{Microscopio de fuerza
at\'omica (AFM) de nuestro laboratorio}                  % toc
{\stfigw}            % width \textwidth
}
% - - - - - - - - - - - - - - - - - - - - - - - - - - - - - - - - - - - %

%-%-%-%-%-%-%-%-%-%-%-%-%-%-%-%-%-%-%-%-%-%-%-%-  figura -%-%-%-%-%-%-%-%

\newcommand{\figMuestAfm}{  % alias

\figura {fig:muest:afm}      % label
{fig/muest/afm}              % file
{Detalle de la unidad de medida del AFM (LFM-3/269). En la base
cil\'{\i}ndrica se ven dos contadores digitales, que muestran las
se\~nales de desviaci\'on de la punta (\emph{cantilever}) en
vertical y horizontal. Sobre esta base, otros cilindros
met\'alicos alojan los dispositivos piezoel\'ectricos del
portamuestras ---pues en este sistema es la muestra quien se mueve
para mantener constante la desviaci\'on de la punta de AFM---. Por
\'ultimo, un cabezal prism\'atico aloja el \emph{cantilever}, el
l\'aser, y los fotodiodos de detecci\'on. Sobre la unidad, sujeto
por un brazo articulado, tenemos un visor para inspecci\'on ocular
de la muestra. }
% caption
{Detalle de la unidad de medida del AFM.}                  % toc
{\stfigw}            % width \textwidth
}
% - - - - - - - - - - - - - - - - - - - - - - - - - - - - - - - - - - - %

%-%-%-%-%-%-%-%-%-%-%-%-%-%-%-%-%-%-%-%-%-%-%-%-  figura -%-%-%-%-%-%-%-%

\newcommand{\figMuestSysNovTres}{  % alias

\figura {fig:muest:sys93}      % label
{fig/muest/sys93}              % file
{Dos im\'agenes compuestas por diversos barridos de AFM solapados.
Corresponden al mismo micropuente (muestra \fI), pero la de la
izquierda fue tomada justo despu\'es de la fotolitograf\'{\i}a,
mientras que la de la derecha es tras un proceso posterior de
limpieza por ultrasonidos, en ba\~no de acetona durante 15~min.
Parte de los residuos del proceso de grabado se han desprendido,
estrechando el
micropuente.}                  % caption
{Dos im\'agenes de AFM de un mismo micropuente, antes y despu\'es de una limpieza por ultrasonidos.}                  % toc
{.95}            % width \textwidth
}
% - - - - - - - - - - - - - - - - - - - - - - - - - - - - - - - - - - - %

%-%-%-%-%-%-%-%-%-%-%-%-%-%-%-%-%-%-%-%-%-%-%-%-  figura -%-%-%-%-%-%-%-%

\newcommand{\figMuestSysDNueveVeinte}{  % alias

\figura {fig:muest:sys119y120}      % label
{fig/muest/sys119y120}              % file
{Dos im\'agenes compuestas por diversos barridos de AFM solapados.
Algunas \'areas no barridas se han completado simuladamente, para
facilitar la interpretaci\'on de las im\'agenes. Corresponde a dos
micropuentes distintos: a la izquierda, la muestra \fJ, y a la
derecha la \fC. Se aprecia que presentan m\'as rugosidad
superficial que el micropuente \fI\ de la figura
\ref{fig:muest:sys93}. Esto no es relevante para la calidad
superconductora: mientras que la muestra \fC\ es de mucha mejor
calidad que la \fI, la \fJ\ es, por el contrario, mucho peor,
tanto que no es superconductora. \'Esta no es, probablemente, una
rugosidad intr\'{\i}nseca a la muestra, sino causada por los
precipitados aislantes o los residuos del fotolitografiado
comentados en el texto. }
% caption
{Dos im\'agenes de AFM de sendos micropuentes.}                  % toc
{.95}            % width \textwidth
}
% - - - - - - - - - - - - - - - - - - - - - - - - - - - - - - - - - - - %

%-%-%-%-%-%-%-%-%-%-%-%-%-%-%-%-%-%-%-%-%-%-%-%-  figura -%-%-%-%-%-%-%-%

\newcommand{\figMuestMsoldador}{  % alias

\figura {fig:muest:msoldador}      % label
{fig/muest/msoldador}              % file
{Microsoldador Kulicke \& Soffa 4523 Manual Wire Bonder empleado
para la aplicaci\'on de contactos el\'ectricos sobre los
micropuentes.}                  % caption
{Microsoldador KS 4523 para la aplicaci\'on de contactos el\'ectricos sobre los micropuentes.}                  % toc
{\stfigw}            % width \textwidth
}
% - - - - - - - - - - - - - - - - - - - - - - - - - - - - - - - - - - - %

%-%-%-%-%-%-%-%-%-%-%-%-%-%-%-%-%-%-%-%-%-%-%-%-  figura -%-%-%-%-%-%-%-%

\newcommand{\figMuestPorta}{  % alias

\figura {fig:muest:porta}      % label
{fig/muest/porta}              % file
{Fotograf\'{\i}a de un filme sobre el portamuestras de cobre
intermedio. El ancho del substrato es de 1~cm. Sobre el
superconductor, con micropuentes grabados por fotolitograf\'{\i}a,
se ven las capas circulares de oro sobre las que se sueldan,
mediante ultrasonidos, hilos de aluminio. Estos hilos, en su otro
extremo, se llevan a las patillas del portamuestras intermedio, de
las que a su vez salen hilos trenzados de cobre, soldados con
aleaci\'on de esta\~no. Sobre el substrato del filme se ven los
cortes con bistur\'{\i} para abrir los caminos conductores
residuales del litografiado. Las marcas oscuras de los laterales
del portamuestras de cobre son
se\~nales identificativas.}                  % caption
{Fotograf\'{\i}a de un filme sobre el portamuestras de cobre
intermedio.}                  % toc
{\stfigw}            % width \textwidth
}
% - - - - - - - - - - - - - - - - - - - - - - - - - - - - - - - - - - - %

%-%-%-%-%-%-%-%-%-%-%-%-%-%-%-%-%-%-%-%-%-%-%-%-  figura -%-%-%-%-%-%-%-%

\newcommand{\figMuestContactUno}{  % alias

\figura {fig:muest:contact1}      % label
{fig/muest/contact1}              % file
{Detalle de una patilla del portamuestras de cobre intermedio. Por
un lado llegan del filme los hilos de aluminio microsoldados con
ultrasonidos. Por el otro, sale un hilo de cobre (en este caso
sencillo),
soldado previamente con aleaci\'on de esta\~no.}                  % caption
{Detalle de una patilla del portamuestras de cobre intermedio, con los hilos de contactos el\'ectricos.}                  % toc
{.7}            % width \textwidth
}
% - - - - - - - - - - - - - - - - - - - - - - - - - - - - - - - - - - - %

%-%-%-%-%-%-%-%-%-%-%-%-%-%-%-%-%-%-%-%-%-%-%-%-  figura -%-%-%-%-%-%-%-%

\newcommand{\figMuestContactDos}{  % alias

\figura {fig:muest:contact2}      % label
{fig/muest/contact2}              % file
{Detalle de las microsoldaduras de hilos de Al/Si-1\% de
$25\und{\mu m}$ de di\'ametro. Sobre estas soldaduras se aplica
luego una gota de pasta conductora para mejorar la resistencia
mec\'anica y aumentar el \'area de contacto.}                  % caption
{Detalle de las microsoldaduras de hilos de aluminio dopados con silicio.}                  % toc
{\stfigw}            % width \textwidth
}
% - - - - - - - - - - - - - - - - - - - - - - - - - - - - - - - - - - - %

Las muestras que se emplearon para los experimentos del presente
trabajo se elaboraron en este Laboratorio de Baixas Temperaturas e
Superconductividade. Son capas delgadas de \YBCOf\ crecidas por
pulverizaci\'on cat\'odica sobre substratos monocristalinos
\index{substrato} de \STOf. En estas pel\'{\i}culas
superconductoras se grabaron micropuentes con configuraci\'on de
cuatro contactos \index{contactos el\'ectricos} en l\'{\i}nea,
para medidas de voltaje a cuatro hilos. Las dimensiones
t\'{\i}picas de los micropuentes son de $150\und{nm} \times
10\und{\mu m}$ de secci\'on por $50\und{\mu m}$ de largo.

Adem\'as de tener buena calidad, ser muy homog\'eneas y con
temperaturas cr\'{\i}ticas rondando los $90\und{K}$, las
principales ventajas de este tipo de muestras se encuentran, a los
efectos de los experimentos que realizamos, en su reducido
tama\~no. Como ya mencionamos en el cap\'{\i}tulo anterior, debido
a la peque\~nez de su secci\'on es f\'acil alcanzar altas
densidades de corriente aplicando corrientes absolutas muy bajas.
Inyectando en el micropuente $50\und{mA}$, por ejemplo, se alcanza
una densidad de corriente superior a $3\Exund{6}{\Acms}$.
Adem\'as, debido a la elevada relaci\'on superficie/volumen que
tienen, de casi $7\Exund{5}{cm^{-1}}$, y a que adem\'as presentan
buen contacto t\'ermico con el substrato cristalino, la
evacuaci\'on del calor en estos filmes es notablemente mayor que
en otros tipos de muestras. Veremos que los efectos t\'ermicos
tienen un papel muy importante en la regi\'on de altas densidades
de corriente, donde la disipaci\'on es elevada, as\'{\i} que
optimizar la refrigeraci\'on es una de los requisitos esenciales
en estos experimentos.

\section{S\'{\i}ntesis de muestras mediante pulverizaci\'on cat\'odica}

La \Index{pulverizaci\'on cat\'odica} (\emph{\Index{sputtering}})
es una t\'ecnica de crecimiento de materiales en la cual se
extraen \'atomos de la superficie de un blanco, mediante el
bombardeo con iones de un plasma gaseoso. El blanco es normalmente
una pastilla policristalina con la misma composici\'on
estequiom\'etrica que el material que se pretende crecer. Los
\'atomos de este \Index{blanco}, arrancados poco a poco, se
condensan sobre un \Index{substrato} cristalino, y se va
construyendo as\'{\i} una pel\'{\i}cula delgada con alto grado de
ordenaci\'on. El substrato y el material que sobre \'el se crece
deben tener \Index{par\'ametros de red} cristalina lo m\'as
parecidos posible.

\figMuestSputtering

Nuestras muestras se crecieron por pulverizaci\'on cat\'odica DC
reactiva, con disposici\'on de blanco y substrato alineados en el
mismo eje vertical, y con oxigenado inmediato en el propio banco
(t\'ecnica \emph{in situ}). Explicamos qu\'e significa todo esto:

La pulverizaci\'on cat\'odica DC aplica un voltaje continuo entre
los electrodos que generan la descarga del plasma. En nuestro
caso, el blanco es el c\'atodo, y es bombardeado por iones
positivos de ox\'{\i}geno. Estos iones, tras bombardear la
superficie del blanco, se neutralizan y regresan al proceso como
\'atomos neutros.

En general, que el proceso sea reactivo implica la presencia, en
el gas de \emph{sputtering}, de un gas que reacciona con el
material que se crece, dop\'andolo o favoreciendo la formaci\'on
de otro compuesto distinto. En nuestro caso, el propio gas de
\emph{sputtering} que empleamos es el que reacciona con el
material, oxigen\'andolo. As\'{\i} que la atm\'osfera empleada, en
este caso, tiene doble funci\'on: por un lado genera el plasma que
bombardea el blanco, pero por otro tambi\'en reacciona con el
material que se crece, mejorando sus propiedades.

A pesar de esta \Index{oxigenaci\'on}, esta pulverizaci\'on hace
crecer el \YBCOf, con estructura \Index{tetragonal}, no
superconductora. Este crecimiento se hace a temperaturas entre 650
y $850\und{\grado C}$ y con presi\'on de O$_2$ entre $10^{-2}$ y 1
Torr. Tras el crecimiento, un proceso de oxigenaci\'on \emph{in
situ} permite alcanzar el dopaje \'optimo de O$_2$, y hace que la
muestra transite a una fase \Index{ortorr\'ombica}, ya
superconductora. Este proceso se realiza manteniendo a la muestra
a $600\und{\grado C}$ durante media hora, con una presi\'on de
O$_2$ de 500 Torr. Las \Index{rampa} de descenso desde la
temperatura de crecimiento hasta la de oxigenaci\'on es de
$70\und{\grado C/min}$. Mientras se disminuye la temperatura, la
presi\'on de ox\'{\i}geno se aumenta paulatinamente hasta los 500
Torr.

Hemos empleado el banco de pulverizaci\'on comercial Plassys MP400
de que disponemos en nuestro laboratorio  ---que se muestra en la
figura \ref{fig:muest:sputtering}---, modificado para el
crecimiento en alta presi\'on ($\simeq 1\und{mbar}$). Para una
detallada descripci\'on del banco y de las modificaciones que se
realizaron recomendamos consultar la tesis doctoral de S.~R.
Curr\'as \cite{Curras00b}: gran parte de ese trabajo est\'a dedicado a
la puesta a punto y aplicaci\'on del banco que empleamos en nuestro
laboratorio, y se recogen en esa memoria tanto una ilustrativa
introducci\'on a esta t\'ecnica de crecimiento como detalles pr\'acticos
de inter\'es.

Los par\'ametros de crecimiento que dieron las mejores muestras
(entre las que se encuentran las de esta memoria) son los que se
muestran resumidos en la tabla \ref{tab:muest:sputt}.

\begin{table}
  \centering
  \caption[Par\'ametros del crecimiento por pulverizaci\'on cat\'odica de
las pel\'{\i}culas empleadas en este trabajo]{Principales
par\'ametros del crecimiento por pulverizaci\'on cat\'odica de las
pel\'{\i}culas empleadas en este trabajo.} \label{tab:muest:sputt}
\begin{tabular}{ll}
  \\[-8pt]
  \hline
  Par\'ametro & Valor \\ \hline
  Potencia de pulverizaci\'on & $\sim 110\und{W}$ \\
  Corriente de pulverizaci\'on & $\sim 400\und{mA}$ \\
  Distancia blanco-substrato & 3 cm \\
  Presi\'on de O$_2$ en crecimiento  & 1.5--2.0 Torr \\
  Temperatura del substrato & $830\und{\grado C}$ \\
  Substrato  & \STOf\ orientado (100) \\
  Tama\~no substrato & $1 \times 1 \und{cm^2}$ $\times 1\und{mm}$ \\
  Tiempo de crecimiento & 2 h \\
  Tiempo de oxigenaci\'on & 30 min \\
  Presi\'on de oxigenaci\'on & 500 Torr \\
  Temperatura de oxigenaci\'on & $600\und{\grado C}$ \\
    \hline

\end{tabular}

\end{table}

%Para este trabajo se prepararon m\'as de veinte muestras distintas.
%No todas resultaron de calidad satisfactoria, y algunas no
%soportaron las primeras pruebas del montaje experimental, las
%altas corrientes o los reciclados t\'ermicos, y resultaron
%destruidas.

Los datos experimentales que se muestran en este trabajo
corresponden a las muestras cuyas caracter\'{\i}sticas se recogen
en la tabla \ref{tab:muest:muest}. Las de peor calidad, con
transiciones m\'as anchas y con temperaturas cr\'{\i}ticas m\'as
bajas (como las muestras \fF\ y \fH), se emplearon en los
experimentos en los que la muestra corr\'{\i}a m\'as riesgo de
destruirse debido al \Index{calentamiento}, como caracterizaciones
t\'ermicas con pulsos muy largos o en reg\'{\i}menes de muy alta
disipaci\'on (incluso en el estado normal). Otras muestras de
mejor calidad (como las muestras \fA, \fD\ y \fG) se reservaron
para las medidas m\'as importantes y delicadas, las relativas a
las variaciones de la corriente de salto con la temperatura y el
campo. La determinaci\'on de los par\'ametros que se muestran en
esta tabla \ref{tab:muest:muest} (extra\'{\i}dos de las medidas de
\Index{resistividad} en funci\'on de temperatura) se ver\'a con
detalle en el cap\'{\i}tulo \ref{cap:resul}.

\begin{table}
  \centering

    \caption[Relaci\'on de las muestras empleadas en los experimentos]
  {Relaci\'on de las muestras (M) empleadas en los experimentos de
  este trabajo, y sus alias en la nomenclatura de nuestro
  laboratorio. Sus par\'ametros caracter\'{\i}sticos: longitud $l$ del
  micropuente, presi\'on $P$ de crecimiento, temperatura de
  inflexi\'on en la transici\'on \Tci, ancho de la transici\'on \DTci,
  temperatura \Tcz\ m\'as alta con $\rho(T)=0$, y resistividades
  $\rho_{100}$ y $\rho_{300}$ medidas a 100~K y 300~K. La muestra
  \fJ\ no es superconductora debido a un percance sucedido durante el
  proceso de oxigenaci\'on. Los par\'ametros que se dejan en blanco no se midieron.
  Todos los micropuentes tiene un ancho $a=10\und{\mu m}$ y un
  espesor promedio $d \simeq
  150\und{nm}$.}
  \label{tab:muest:muest}

\begin{tabular}{clccccccc}
    \\[-8pt]
  \hline
   M   & \fns{Alias}  & $l$ & $P$ & \Tci & \DTci & \Tcz & $\rho_{100}$ & $\rho_{300}$ \\
      &   & \fns{${\rm \mu m}$} & \fns{Torr} & \fns{K} & \fns{K} & \fns{K} & \fns{\mOcm} & \fns{\mOcm}   \\ \hline
  \fA & \fns{\fAa} & 50 & 1.54 & 90.6 & 0.9 & 89.8 & 116.7 & 373.7 \\
  \fB & \fns{\fBa} & 100 & 1.59 & 90.0 & 2.2 & 87.0 & 236.0 & 707.6 \\
  \fC & \fns{\fCa} & 100 & 1.84 & 90.8 & 1.2 & 89.5 & 91.2 & 269.9 \\
  \fD & \fns{\fDa} & 50 & 1.54 & 90.6 & 0.9 & 89.7 & 147.3 & 464.5 \\
  \fE & \fns{\fEa} & 100 & 1.65 & 90.9 & 1.6 & 80.0 & 109.3 & 337.2 \\
  \fF & \fns{\fFa} & 50 & 1.69 & 89.7 & 1.4 & 86.3 & 220.1 & 641.3 \\
  \fG & \fns{\fGa} & 50 & 1.82 & 90.7 & 0.6 & 89.8 & 115.7 & $\sim$400 \\
  \fH & \fns{\fHa} & 20 & 1.82 & 89.4 & 1.6 & 85.0 & 170.7 & 536.9 \\
  \fI & \fns{\fIa} & 50 & 1.70 & 88.0 & 3 & 80.0 &  &   \\
  \fJ & \fns{\fJa} & 50 & 1.84 &  - & -  &  - &   &  \\ \hline
\end{tabular}

\end{table}

La primera caracterizaci\'on se hace en cuanto la muestra sale del
proceso de s\'{\i}ntesis: tras aplicar \index{contactos
el\'ectricos} contactos el\'ectricos con microsoldadura
\index{soldadura} directamente sobre el filme, en sus cuatro
esquinas (m\'etodo Montgomery \cite{Montgomery71}), se hace una
primera medida de la \Index{resistividad} de la muestra en
funci\'on de la temperatura, principalmente para apreciar si hay
transici\'on superconductora. Este m\'etodo de medida de la
resistividad, abarcando toda la muestra, hace que se manifiesten
las \Index{inhomogeneidades}\footnote{\label{notaInhomog}Empleamos
el t\'ermino \emph{\Index{inhomog\'eneo}}, habitual en ciencia de
materiales en contraposici\'on a \emph{\Index{homog\'eneo}}, para
indicar un leve grado de heterogeneidad. Las
\emph{inhomogeneidades} son, por tanto, peque\~nas deficiencias
m\'as o menos localizadas que alteran parcialmente la homogeneidad
general de la muestra.} (estequiom\'etricas, de espesor\ldots) que
\'esta pueda tener en toda su extensi\'on ($1\und{cm^2}$), sobre
todo en los bordes del substrato donde la \emph{pluma}, la nube de
pulverizaci\'on, llega m\'as deficientemente: as\'{\i} se miden
transiciones superconductoras anchas y a temperaturas inferiores a
la \'optima, que ronda los 90~K. Dichas \Index{inhomogeneidades}
no se ponen de manifiesto en los micropuentes que posteriormente
se grabar\'an, ya que \'estos abarcan un \'area mucho m\'as
peque\~na y se realizan en la parte central de la pel\'{\i}cula,
m\'as homog\'enea. La temperatura cr\'{\i}tica de los micropuentes
suele ser al menos dos grados m\'as elevada que la primera
estimaci\'on que se hace en todo el filme.

\section{Caracterizaci\'on estructural: difractometr\'{\i}a de rayos-x}

Mediante difractometr\'{\i}a de rayos-x (XRD) estudiamos la
calidad estructural de las pel\'{\i}culas, centrando nuestra
atenci\'on en tres aspectos: la presencia de fases secundarias, la
direcci\'on de crecimiento ---que nos da informaci\'on acerca de
la orientaci\'on global preferente de la pel\'{\i}cula--- y la
\Index{mosaicidad} ---que nos dice c\'omo es la distribuci\'on de
orientaciones, en torno a esa preferente, de los microgranos que
componen la muestra---. Empleamos un \Index{difract\'ometro}
Siemens D-5005 disponible en los servicios generales de nuestra
universidad.

\figMuestDifractom

En la figura \ref{fig:muest:difractom} vemos un esquema del
difract\'ometro en el que se indican los \'angulos $\theta$ y
$2\,\theta$ que determinan los brazos de la fuente y del detector
respectivamente. Destaquemos que, en esta convenci\'on, el \'angulo
$2\,\theta$ del detector es el que forma respecto a la
prolongaci\'on del brazo de la fuente. El emisor de rayos-x es un
\'anodo de Cu que hacemos funcionar en torno a 40~kV y 30~mA.
Empleamos la radiaci\'on K$_\alpha$ del Cu, eliminando la K$_\beta$
con un monocromador en el detector. El emisor tiene un colimador
(en la figura, etiquetado como AB) con 0.2~mm de apertura, y el
receptor otro (DB) con 0.1~mm. Una tercera rendija (SB), con
0.1~mm de apertura, elimina la radiaci\'on dispersada. La resoluci\'on
del goni\'ometro es de 0.01\grado.

Con este \Index{difract\'ometro} se pueden mover los brazos de la
fuente de rayos-x y del detector ---bien desacoplados, bien
acoplados de diferentes maneras---, mientras que la muestra
permanece est\'atica. Esta versatilidad permite realizar distintos
tipos de medidas. Para las que nos interesan, emplearemos dos
tipos de movimientos acoplados entre los brazos: difractogramas
$\theta$--$2\,\theta$ y curvas de basculamiento.

Esta independencia del movimiento de los brazos permite, adem\'as,
corregir cualquier leve \Index{desalineamiento} de la muestra:
\'esta debe ser tangente a la circunferencia de enfoque en la
geometr\'{\i}a Bragg-Brentano que empleamos (elemento P en fig.
\ref{fig:muest:difractom}) para obtener m\'aximas intensidades en
el detector, y al situarla en el \Index{portamuestras} no se
consigue tanta precisi\'on (siempre hay peque\~nas desviaciones de
hasta $1\grado$). Se pueden desplazar ambos brazos un cierto
\'angulo para hacerlos equidistantes de la verdadera recta normal
a la muestra. Para establecer cu\'al es este \'angulo de
correcci\'on tenemos que entender antes en qu\'e consisten las
curvas de basculamiento.

\subsection{Curvas de basculamiento}

\figMuestRocking

En las medidas de \Index{basculamiento} (\emph{\Index{rocking
curves}}) mantenemos fijo el \'angulo $2\,\theta$ del detector
respecto a la fuente, de manera que fijamos las posiciones
relativas de ambos brazos. Elegimos para esto el \'angulo
correspondiente a una reflexi\'on de Bragg del \YBCO\ conocida,
t\'{\i}picamente la del pico (005)
---veremos en la siguiente caracterizaci\'on por rayos-x que la
pel\'{\i}cula est\'a orientada con la direcci\'on $c$
perpendicular al substrato---. Movemos los dos brazos
as\'{\i}acoplados, variando el \'angulo $\theta$ de la fuente.
Vemos un resultado t\'{\i}pico de este tipo de medida en la figura
\ref{fig:muest:rocking}. Estas medidas dan informaci\'on acerca de
la dispersi\'on de orientaciones de los granos que la conforman en
torno a la orientaci\'on preferente (\Index{mosaicidad}). El
par\'ametro que caracteriza esta dispersi\'on es la anchura de la
curva de \Index{basculamiento} a la mitad de su altura (FWHM,
\emph{Full Width at Half Maximum}). En el ejemplo de la figura,
vemos que la mayor parte de los granos que conforman la
pel\'{\i}cula est\'an alineados en torno a la direcci\'on
preferente con menos de $0.3\grado$ de dispersi\'on. Este valor
pone de manifiesto el alto grado de ordenaci\'on de nuestras
muestras. En todas las pel\'{\i}culas medidas la FWHM es similar,
y su anchura est\'a limitada por la propia mosaicidad del
substrato ($\sim 0.2\grado$).

%Este par\'ametro probablemente mejora cuando se recortan los
%micropuentes, pero debido a su reducido tama\~no, que da una se\~nal
%de difracci\'on muy baja, no es f\'acil comprobarlo. De la misma
%manera que la \Tc\ del micropuente es mejor que la de toda la
%pel\'{\i}cula, pues no incluye zonas deficientes de los bordes del
%substrato, es muy probable que esta medida de la
%\Index{mosaicidad} hecha s\'olo en el micropuente diese una
%dispersi\'on de orientaciones mucho menor.

Para las curvas de basculamiento empleamos una velocidad de
barrido de $0.5\und{\grado/min}$, y variamos $\theta$ unos pocos
grados en torno a la reflexi\'on de Bragg que hemos elegido.

Esta medida tambi\'en nos dice cu\'anto tenemos que corregir el
\Index{desalineamiento} de la muestra respecto de los brazos, al
ver c\'omo se desv\'{\i}a el m\'aximo de la curva de basculamiento
respecto de la posici\'on esperada. Esto implica, claro,
presuponer que la orientaci\'on preferente (m\'aximo de
intensidad) es con el eje $c$ perpendicular al plano del
substrato, pero esta es una suposici\'on razonable: cualquier leve
desviaci\'on en conjunto de toda la muestra es energ\'eticamente
desfavorable ---a la temperatura de deposici\'on empleada de $\sim
830\grado$---, y es de esperar que los microgranos tengan una
dispersi\'on de sus orientaciones en torno a esta orientaci\'on
preferente. Una vez corregido el desalineamiento de la muestra,
repetimos esta medida de basculamiento, y realizamos el siguiente
difractograma.

\subsection{Difractogramas}

En la difractometr\'{\i}a $\theta$--$2\,\theta$ ambos brazos se
mueven acopladamente de manera que una variaci\'on del \'angulo de
la fuente implica una variaci\'on doble del \'angulo del detector
(recordemos en la figura \ref{fig:muest:difractom} que este
\'ultimo se define sobre la prolongaci\'on del brazo de la
fuente). Esto es equivalente a decir que ambos brazos mantienen el
mismo \'angulo $90\grado-\theta$ respecto de la normal al plano de
la muestra, uno a cada lado. Es la medida de difractometr\'{\i}a
m\'as convencional.

\figMuestRayosx

La velocidad de barrido que empleamos para esta medida es de
$2\und{\grado/min}$. En la figura \ref{fig:muest:rayosx} se
muestra el \Index{difractograma} t\'{\i}pico de nuestras
pel\'{\i}culas de \YBCO, crecidas seg\'un hemos descrito en la
secci\'on previa. Se puede apreciar que las muestras han crecido
orientadas con el eje cristalogr\'afico $c$ perpendicular al
substrato: s\'olo se detectan picos de la familia (001). Los picos
m\'as intensos, de la forma (003h) ---se\~nalados con un asterisco
en el difractograma--- coinciden con los picos (h00) de substrato:
la pel\'{\i}cula es muy delgada y el substrato muy masivo,
as\'{\i}que la intensidad de los picos de este \'ultimo es
notablemente mayor.

Tambi\'en se puede observar en este difractograma la total ausencia
de fases secundarias en la muestra, dentro de la resoluci\'on
experimental ($\sim 1\%$).

\section{Preparaci\'on de las muestras mediante fotolitograf\'{\i}a}

Sobre los filmes crecidos por deposici\'on cat\'odica sobre un
substrato con superficie de $1\und{cm^2}$ grabamos mediante
\Index{fotolitograf\'{\i}a} los micropuentes sobre los que
posteriormente realizaremos las medidas. Este proceso nos permite
crear \index{circuito} circuitos superconductores con la forma
apropiada para nuestros experimentos. En concreto, para este
trabajo hemos grabado varios micropuentes en cada pel\'{\i}cula,
de distintas longitudes, pero con anchos iguales de $10\und{\mu
m}$, con disposici\'on de cuatro contactos en l\'{\i}nea para
medir a cuatro hilos. \index{contactos el\'ectricos}

El fotolitografiado es un proceso similar al revelado
fotogr\'afico convencional. Se cubre la pel\'{\i}cula con una
resina fotosensible, y se estimula con radiaci\'on ultravioleta
que pase a trav\'es de una m\'ascara con el dise\~no del circuito
deseado. De este modo, la parte de la resina que estaba en sombra
por la m\'ascara queda inalterada, pero la resina iluminada se
vuelve atacable por agentes qu\'{\i}micos que \emph{revelan} la
imagen de la m\'ascara. Hay resinas positivas y negativas, en
funci\'on de que la iluminaci\'on ultravioleta las haga sensibles
o resistentes a los l\'{\i}quidos reveladores correspondientes.
Dependiendo de cu\'al empleemos, la m\'ascara con el circuito
deber\'a ir en negativo o en positivo.

Una vez revelada la imagen de la m\'ascara en la resina, tenemos
que parte del filme de \YBCOf\ queda al descubierto, mientras que
otra parte, con la forma del circuito que deseamos grabar,
permanece protegido por la resina. En este estado, atacamos
qu\'{\i}micamente la pel\'{\i}cula, de modo que eliminemos la
parte de \YBCOf\ desprotegida. Posteriormente eliminamos la resina
restante, y tenemos as\'{\i}\Index{grabado} el circuito
superconductor de la m\'ascara.

\figMuestEsqmascara

En la figura \ref{fig:muest:esqmascara} mostramos un esquema de la
m\'ascara que grabamos sobre nuestras pel\'{\i}culas. Las
proporciones de los distintos elementos no est\'an a escala, para
poder apreciarlos. Los micropuentes tiene 100, 50 y $20\und{\mu
m}$ de largo por $10\und{\mu m}$ de ancho, pero el puente grande
que los une tiene en total casi $1\und{cm}$ de largo por
$200\und{\mu m}$ de ancho. De este salen lateralmente contactos de
$100\und{\mu m}$ de ancho, para las sondas de voltaje.
\index{contactos el\'ectricos} En los ancheamientos del puente
grande, a los extremos y entre los micropuentes, se dispondr\'an
los contactos de corriente.

Los \index{contactos el\'ectricos} contactos de voltaje pueden
parecer muy anchos y muy alejados de los micropuentes, pero
recordemos que en las medidas \EJ\ en que centraremos nuestro
inter\'es toda esa parte de la pel\'{\i}cula permanece en estado
superconductor, y por lo tanto no es resistiva. S\'olo los
micropuentes son suficientemente estrechos como para que en ellos
la densidad de corriente sea superior a la cr\'{\i}tica \Jc\ en
que aparece disipaci\'on. En las medidas que hagamos por encima de
\Tc, sin embargo, toda la muestra se vuelve resistiva, as\'{\i}
que deberemos tener en cuenta convenientemente toda esa parte
a\~nadida de \YBCOf\ a mayores del micropuente cuando, por
ejemplo, queramos calcular la \Index{resistividad} a partir de
medidas de voltaje.

\figMuestInsoladora

El proceso de fotolitografiado incluye los siguientes pasos:

\begin{enumerate}

\item Aplicaci\'on de la \Index{fotorresina}. Se emplean un par de gotas de Shipley Microposit
S1400-37, y se homogeneiza el espesor de esta capa protectora
centrifugando la muestra a 4000~r.p.m durante unos treinta
segundos (centrifugadora Headway Research PWH 101).

\item Endurecimiento de la fotorresina. Se introduce la muestra en un
horno a $90\und{\grado C}$ durante 30~min para eliminar el
disolvente de la resina.

\item Impresi\'on del \Index{circuito}. Bajo una l\'ampara insoladora ultravioleta
(KarlSuss IN470 con l\'ampara de xen\'on, en la figura
\ref{fig:muest:insoladora}) se alinean con cuidado la m\'ascara y la
muestra, y se iluminan durante 10~s.

\item Segundo endurecimiento. Horneado igual que el previo, para
reforzar la adhesi\'on de la fotorresina a la pel\'{\i}cula, que
puede estar debilitada por la acci\'on del revelador. As\'{\i} se
aumenta la definici\'on en el revelado y se consigue mayor
resoluci\'on lateral. (\cite[p. 64]{Salgueiro96} y \cite[p.
545]{Moreau89}).

\item Revelado. Se introduce la muestra en l\'{\i}quido revelador
(Shipley Microposit Developer) durante unos 15~s, agit\'andola
leve pero constantemente. As\'{\i} se elimina la resina
sensibilizada. Posteriormente se aclara con agua destilada. Hasta
aqu\'{\i} el proceso de fotolitografiado es reversible, ya que el
\YBCOf\ todav\'{\i}a no se ha atacado. Podr\'{\i}amos volver a
iniciar el proceso, eliminando toda la resina con un
\Index{ba\~no} de acetona.

\item Grabado qu\'{\i}mico (o mordido). Se introduce la muestra en una disoluci\'on
acuosa de \'acido ortofosf\'orico al 1\% durante unos 5--10~s,
agitando ligeramente. El \'acido elimina las regiones de \YBCOf\
que no est\'an protegidas con la resina, grabando el circuito de
la m\'ascara. Volvemos a aclarar r\'apidamente con agua destilada.

\item Limpiado y secado. Eliminamos la resina restante, que todav\'{\i}a protege al
\YBCOf, con un ba\~no de acetona, y secamos con aire seco
comprimido. A partir de este momento el \YBCOf\ queda
desprotegido, por lo que se debe poner extremo cuidado en la
manipulaci\'on de la muestra: no debe entrar en contacto con el
\'acido o el agua destilada, y hay que conservarla en ambiente seco.

\item El grabado puede ser deficiente en los bordes del substrato,
en donde suelen quedar caminos superconductores que ponen en
contacto distintas regiones de los micropuentes. Como el borde es
una regi\'on alejada de los delicados micropuentes, podemos abrir
estos caminos con un leve corte de bistur\'{\i}.

\item Es conveniente realizar una limpieza en profundidad de la
muestra resultante, mediante \Index{ultrasonidos} en un ba\~no de
acetona durante al menos quince minutos. Los efectos de esta
limpieza de ver\'an en el apartado de caracterizaci\'on por AFM.

\end{enumerate}

\figMuestVideo

En la figura \ref{fig:muest:video} se ven dos im\'agenes de
v\'{\i}deo de uno de los micropuentes as\'{\i} grabados. En la
imagen superior, tomada con menos aumentos que la inferior, se ven
tambi\'en los contactos \index{contactos el\'ectricos} laterales
para las sondas de voltaje. Las \'areas de color claro, de forma
redondeada, son parte de unas capas circulares de oro que se
depositaron para la posterior aplicaci\'on de contactos
el\'ectricos. Veremos esto en detalle en una secci\'on posterior.
En la imagen inferior se aprecian con m\'as detalle algunos
defectos del \Index{substrato}, en forma de muescas sobre su
superficie, debidos probablemente a que eran substratos reciclados
de crecimientos anteriores.

Para m\'as informaci\'on acerca del proceso de fotolitografiado se
pueden consultar las siguientes referencias:
\cite{Moreau89,Salgueiro96,Curras95,Curras00b}.

\section{Caracterizaci\'on~geom\'etrica: perfilometr\'{\i}a y AFM}

S\'olo despu\'es de haber grabado los micropuentes se puede medir
el espesor de las pel\'{\i}culas: para hacerlo hace falta tener un
escal\'on entre el substrato y la pel\'{\i}cula cuya altura
podamos medir, y este escal\'on s\'olo se consigue eliminando
parte de la pel\'{\i}cula crecida sobre el substrato. Por otra
parte, lo que nos interesa es el espesor del micropuente que
estudiaremos, no el del conjunto de toda la pel\'{\i}cula.

\figMuestPerfilom

Las primeras caracterizaciones del espesor de nuestros puentes las
hicimos con un \Index{perfil\'ometro} de punta de contacto DekTak
3, disponible en la Escola de \'Optica e Optometr\'{\i}a de esta
universidad. En la figura \ref{fig:muest:perfilom} mostramos
algunos resultados de perfilometr\'{\i}as sobre uno de nuestros
micropuentes, en tres diferentes puntos de su longitud. El
resultado es, en cada una de las medidas, un corte transversal del
micropuente, pues se mide el espesor $d$ en funci\'on de la
posici\'on $x$ a lo ancho de la muestra. Vemos que el promedio de
espesor ronda los 150~nm, y que la superficie es, en el caso de
esta muestra, bastante irregular, con oscilaciones de hasta el
25\%. Veremos que esto no es igual en todas las muestras y que, en
todo caso, la mayor parte de estas irregularidades son
precipitados aislantes que no intervienen en la secci\'on eficaz
conductora del micropuente.

Parece que la anchura del micropuente var\'{\i}a a lo largo de su
longitud (las tres perfilometr\'{\i}as tienen anchuras ligeramente
diferentes), pero esto podr\'{\i}a deberse tambi\'en a un
inevitable \Index{desalineamiento} del micropuente con la punta:
si la direcci\'on del desplazamiento de la punta durante el
barrido no es exactamente perpendicular al micropuente podemos
estar leyendo una anchura falseadamente m\'as ancha.

\figMuestAfmpc \figMuestAfm

Esta caracterizaci\'on mediante perfilometr\'{\i}a es bastante
simple: s\'olo se pueden medir espesores, y adem\'as en regiones
muy localizadas. Para poder estudiar otros aspectos interesantes
de este tipo de muestras, en nuestro laboratorio decidimos
adquirir un \Index{microscopio} de fuerza at\'omica (\Index{AFM}),
lo que en concreto nos ha permitido medir con m\'as detalle la
geometr\'{\i}a de los micropuentes. El equipo de que disponemos
para medidas de AFM es un microscopio Digital Instruments
NanoScope E con unidad de medida LFM-3/269 (figuras
\ref{fig:muest:afmpc} y \ref{fig:muest:afm}), que en un futuro
permitir\'a un estudio sistem\'atico de la geometr\'{\i}a del
micropuente y su influencia en los par\'ametros de transporte. Por
el momento lo hemos empleado para tener una imagen de la forma de
los micropuentes m\'as detallada que la que nos aporta el
perfil\'ometro.

La microscop\'{\i}a de fuerza at\'omica que empleamos es de
contacto. La muestra se desplaza lateralmente bajo una punta
nanosc\'opica de nitruro de silicio\footnote{Referencia comercial:
Nanoprobe SPM tip type OTR8-35.}, que est\'a en el extremo de un
peque\~no brazo flexible (\emph{cantilever}). Las irregularidades
de la superficie de la muestra hacen que este brazo flexible se
curve m\'as o menos. Un sistema de \Index{realimentaci\'on} mide
la desviaci\'on del brazo empleando un l\'aser que se refleja
sobre \'el, y ajusta en la direcci\'on vertical un dispositivo
piezoel\'ectrico que est\'a en la base de la muestra, para
mantener dicha desviaci\'on en un valor constante. El sistema
devuelve el dato de la altura de la muestra en funci\'on de
cu\'anto haya tenido que alterar la posici\'on de este
piezoel\'ectrico vertical.

\figMuestSysNovTres \figMuestSysDNueveVeinte

Los piezoel\'ectricos horizontales de nuestro AFM permiten hacer
barridos en un \'area cuadrada m\'axima de unas $12\times 12 \und{\mu
m^2}$, suficiente para hacer medidas de la altura del escal\'on del
micropuente respecto al substrato en alguna regi\'on de su longitud.
Pero si queremos tomar una imagen completa de un micropuente
debemos solapar varios de estos barridos de AFM, para componer una
imagen mayor. Esto es lo que se hizo en las im\'agenes de las
figuras \ref{fig:muest:sys93} y \ref{fig:muest:sys119y120}.

En esas im\'agenes se puede apreciar como algunas pel\'{\i}culas
muestran superficies m\'as irregulares que otras: los peque\~nos
granos blancos, de distinto tama\~no dependiendo de la muestra,
son bultos que sobresalen en altura del resto del puente. Estos
bultos son, de lo que se deduce de medidas de microscop\'{\i}a
electr\'onica de barrido \index{microscopio} \cite{Curras00b},
precipitados el\'ectricamente aislantes, posiblemente compuestos
ricos en cobre
\cite{Bhatt94,Catana93,Loquet93,Kim92,Wuyts92,Gavaler91,Habermeier91},
quiz\'a causados por las caracter\'{\i}sticas del blanco empleado
en el crecimiento \cite{Li95}. Tambi\'en pueden ser restos de todo
el proceso de fotolitografiado. Por este motivo los excluimos de
la secci\'on eficaz conductora de nuestros
micropuentes\footnote{Si estos precipitados no se manifiestan en
las medidas de rayos-x es, probablemente, debido a que son
cristales con orientaciones aleatorias y con una contribuci\'on
vol\'umica peque\~na.}. Todas las muestras presentan peque\~nas
variaciones del espesor dependiendo del punto concreto de su
longitud que consideremos, pero en promedio rondan los 150~nm, si
descontamos estos precipitados aislantes. Esto es razonable, pues
todas las pel\'{\i}culas han sido crecidas con procesos de
\emph{\Index{sputtering}} de igual duraci\'on. Por todo esto hemos
considerado que todas las muestras empleadas en este trabajo
tienen un espesor promedio $d \simeq 150\und{nm}$, con un
\Index{error} que estimamos rondando $\pm 10\und{nm}$.

En la figura \ref{fig:muest:sys93} se ponen de manifiesto los
efectos de una limpieza con \Index{ultrasonidos}, en
\Index{ba\~no} de acetona, tras el
fotolitografiado\footnote{Empleamos una m\'aquina P-Selecta
Ultrasons, que mediante un transductor aplica ultrasonidos de
40~kHz a una cubeta de agua. En esta cubeta se introduce un vaso
con acetona de modo que el nivel de l\'{\i}quidos sea semejante, y
en el interior de la acetona se dispone la muestra para su
limpieza. Las ondas sonoras se propagan hasta las cercan\'{\i}as
de la muestra. Un efecto de cavitaci\'on origina burbujas
microsc\'opicas de baja presi\'on, que eliminan las impurezas
superficiales.}. La imagen de la izquierda, con restos de \YBCOf\
mal eliminados por el grabado, fue tomada antes de la limpieza. En
la imagen de la derecha, tomada despu\'es de dicha limpieza por
ultrasonidos durante quince minutos, se ven como algunos de los
restos (probablemente no superconductores, atacados a medias por
los agentes qu\'{\i}micos) se han desprendido, haciendo que el
micropuente sea m\'as estrecho. Esta limpieza tambi\'en mejora la
nitidez de los barridos AFM: se aprecian menos trazas horizontales
a lo largo del puente, que son claros defectos generados por
suciedad atrapada en la punta.

Estos restos, y en general el proceso de fotolitografiado, hacen
que el ancho del micropuente no sea exactamente igual en toda su
longitud, como ya ve\'{\i}amos en las medidas por
perfilometr\'{\i}a \label{par:errorgeometrico}. (Tampoco podemos
estar seguros de que todo el ancho de micropuente que vemos sea de
igual calidad superconductora, pues los bordes podr\'{\i}an estar
afectados por el ataque qu\'{\i}mico). Pero las variaciones no son
grandes en torno al valor promedio $a \simeq 10 \und{\mu m}$, cuyo
\Index{error} m\'aximo estimamos en un 20\%. Si los micropuentes
fuesen m\'as anchos, estas mismas deficiencias laterales se
traducir\'{\i}an en un error relativo mucho menor.

La longitud del micropuente est\'a relativamente mejor establecida,
con un \Index{error} que estimamos en $\pm 2\und{\mu m}$, lo que
para los puentes de $l=50\und{\mu m}$ supone un 4\%.

De este modo, el promedio geom\'etrico de los errores lleva a que el
\Index{error} total de la \Index{resistividad} medida en estos
micropuentes sea de aproximadamente un 21\%.

\section{Aplicaci\'on de contactos el\'ectricos}

Previa a la aplicaci\'on de hilos el\'ectricos se depositan sobre el
\YBCOf\ una capa de oro de $1\und{\mu m}$ de espesor y de
aproximadamente 1~mm de di\'ametro en los lugares en los que se
realizar\'an las \index{soldadura} soldaduras, mediante una m\'ascara
met\'alica y un evaporador Balzers BAE 250 o un pulverizador
cat\'odico \index{pulverizaci\'on cat\'odica} Edwards Scancoat Six. El
empleo de uno u otro sistema depende del n\'umero de muestras que se
quiera preparar simult\'aneamente, para reducir en todo lo posible
el gasto de oro: el evaporador gasta m\'as metal que el
pulverizador, pero cubre un \'area mayor de aplicaci\'on, por lo que
se puede poner mayor n\'umero de muestras.

Tras la aplicaci\'on de estas capas circulares de oro en cada
\index{contactos el\'ectricos} contacto (que se pueden apreciar en
las figuras \ref{fig:muest:video} y \ref{fig:muest:porta}),
sometemos la muestra a un proceso que mejore la
\Index{interdifusi\'on} de oro e \YBCOf. Se cubre la pel\'{\i}cula
con un trozo de \YBCO\ masivo de sus mismas dimensiones, y se
introduce as\'{\i} en un horno tubular, con un peque\~no flujo de
ox\'{\i}geno a presi\'on ligeramente superior a la atmosf\'erica,
a $600\und{\grado C}$ y durante 15~min.

\figMuestPorta \figMuestMsoldador \figMuestContactUno \figMuestContactDos

Luego se sit\'ua la muestra en un \Index{portamuestras} intermedio
de cobre, adherida con grasa termoconductora Apiezon N, tal y como
se muestra en la figura \ref{fig:muest:porta}. El portamuestras
tiene unas patillas met\'alicas para los hilos el\'ectricos que
provienen de la muestra, y a las que tambi\'en se sueldan los
hilos que llevan al resto del \Index{circuito} experimental. Estas
patillas se realizaron con cobre de una plancha para circuitos
impresos. (En algunos casos, como prueba, tambi\'en se emple\'o
lat\'on, pero resultaba m\'as inc\'omodo de manejar). Soldamos,
desde cada \index{contactos el\'ectricos} contacto de la
pel\'{\i}cula a una de las patillas, hilos de aluminio con un 1\%
de silicio, de $25\und{\mu m}$ de di\'ametro, empleando para esto
el \Index{microsoldador} por contacto con \Index{ultrasonidos}
Kulicke \& Soffa 4523 Manual Wire Bonder de la figura
\ref{fig:muest:msoldador}. Normalmente se disponen varios hilos
para una misma uni\'on el\'ectrica por seguridad, por si alguno se
suelta durante su manipulaci\'on. Se pueden ver im\'agenes
detalladas de estas microsoldaduras \index{soldadura} por
ultrasonidos en las figuras \ref{fig:muest:contact1} y
\ref{fig:muest:contact2}. Posteriormente, para mejorar su
resistencia mec\'anica y aumentar el \'area de \index{contactos
el\'ectricos} contacto el\'ectrico, se aplica una gota de pasta
conductora Dupont 4929 sobre las soldaduras de la muestra y de las
patillas, que se seca bajo una l\'ampara de infrarrojos durante
una media hora.

Las microsoldaduras del hilo de Al/Si-1\% se realizan por presi\'on
de una punta de carburo de wolframio pulida y
plana\footnote{Referencia comercial P/N 40427-0004-151.} mientras
se aplican \Index{ultrasonidos}. Cada \index{soldadura} soldadura
se caracteriza por tres par\'ametros en el microsoldador: la
potencia de ultrasonidos (power), el tiempo de soldadura (time) y
la fuerza de empuje de la punta sobre la soldadura (force). Estos
par\'ametros se regulan en una escala del 0 al 10. En la tabla
\ref{tab:muest:microsold} se recogen los m\'as adecuados para las
superficies que hemos empleado (oro sobre la muestra, y lat\'on o
cobre en las patillas del portamuestras intermedio). Las patillas
de cobre, de plantillas para circuitos impresos, son m\'as delgadas
que las de lat\'on. Para hacer soldaduras sobre ellas
apropiadamente, se necesita aumentar el tiempo y la potencia
mientras que se disminuye la fuerza de la aguja, en comparaci\'on
con los par\'ametros que se emplean para el lat\'on, m\'as masivo.

\begin{table}
  \centering
    \caption[Par\'ametros m\'as convenientes para las microsoldaduras.]{Par\'ametros m\'as convenientes para las microsoldaduras realizadas con el KS 4523 Wire Bonder. Se regulan dentro de una escala del 0 al 10.}\label{tab:muest:microsold}
  \begin{tabular}{cccc}
        \\[-8pt]
    \hline
    Superficie & Power & Time & Force \\
    \hline
    Oro & 2--3 & 2--3 & 2--3 \\
    Cobre & 3.5--5 & 3--4 & 1--2 \\
    Lat\'on & 3--3.5 & 3 & 2--2.5 \\ \hline
  \end{tabular}

\end{table}

Los contactos aplicados sobre el filme presentan una
\index{contactos el\'ectricos} resistencia inferior a
$0.1\und{\Omega}$, al menos diez veces menor que la propia
resistencia t\'{\i}pica del micropuente superconductor en las
regiones m\'as disipativas de la curva \VI, y quinientas veces
menor que la resistencia del micropuente a $100\und{K}$. Adem\'as,
los contactos aplicados est\'an alejados del micropuente una
distancia que es al menos diez veces la longitud de \'este: toda
esa distancia entre los contactos soldados y el micropuente objeto
de nuestro estudio est\'a cubierta por caminos el\'ectricos
superconductores. De este modo, es de esperar que la disipaci\'on
en los contactos el\'ectricos tenga poco efecto sobre el
micropuente: la propia disipaci\'on del micropuente ser\'a mucho
mayor.

Una vez preparada la muestra y sus contactos en el portamuestras
intermedio, ya est\'a dispuesta para introducirse en el
dispositivo de medida que explicamos en el siguiente
cap\'{\i}tulo.
  % cap 2
\clearemptydoublepage
%%%%%%%%%%%%%%%%%%%%%%%%%%%%%%%%%%%%%%%%%%%%%%%%%%%%%%%%%%%%%%%%%%
\chapter{Dispositivos de medida}
%\chaptermark{Dispositivos de medida}
%\addtocontents{toc}{\protect\vspace{0.2cm}}
\label{cap:disps}
%%%%%%%%%%%%%%%%%%%%%%%%%%%%%%%%%%%%%%%%%%%%%%%%%%%%%%%%%%%%%%%%%%

%-%-%-%-%-%-%-%-%-%-%-%-%-%-%-%-%-%-%-%-%-%-%-%-  figura -%-%-%-%-%-%-%-%

\newcommand{\figDispsCircuito}{  % alias

\figura {fig:disps:circuito}      % label
{fig/disps/circuito}              % file
{Esquema del circuito el\'ectrico de medida a cuatro hilos,
empleado en nuestros experimentos. La corriente se inyecta
mediante la fuente de corriente I, y la ca\'{\i}da de voltaje en
el micropuente se mide con el dispositivo V (un
nanovolt\'{\i}metro o una tarjeta de adquisici\'on, dependiendo
del r\'egimen de disipaci\'on). En serie con el micropuente se
dispone una resistencia conocida para medir, con la tarjeta de
adquisici\'on, un voltaje V$_i$ que nos sirva para saber qu\'e
corriente est\'a siendo aplicada. Esto es relevante s\'olo para
las medidas r\'apidas de toda la curva \VI, porque en general
tomamos como valor de la corriente el que programamos en la
fuente.}
% caption
{Esquema del circuito el\'ectrico de medida a cuatro hilos.}                  % toc
{0.7}            % width \textwidth
}
% - - - - - - - - - - - - - - - - - - - - - - - - - - - - - - - - - - - %

%-%-%-%-%-%-%-%-%-%-%-%-%-%-%-%-%-%-%-%-%-%-%-%-  figura -%-%-%-%-%-%-%-%

\newcommand{\figDispsCriostatorack}{  % alias

\figura {fig:disps:criostatorack}      % label
{fig/disps/criostatorack}              % file
{Fotograf\'{\i}a de los principales elementos experimentales. A la
izquierda, conjunto de dispositivos electr\'onicos y ordenador de
control. A la derecha, vista general del criostato Oxford, y de
las bobinas para la aplicaci\'on de campo magn\'etico hasta
$\mH=1\und{T}$. Los cables que llegan a la muestra bajan desde el
cabezal del portamuestras ---en la parte superior de la imagen,
sobre el criostato--- hasta un punto centrado entre las bobinas.
Al fondo, en la pared, el cuadro de llaves para la purga de las
c\'amaras y entrada/salida de gases.}
% caption
{Fotograf\'{\i}a de los principales elementos experimentales}                  % toc
{\stfigw}            % width \textwidth
}
% - - - - - - - - - - - - - - - - - - - - - - - - - - - - - - - - - - - %

%-%-%-%-%-%-%-%-%-%-%-%-%-%-%-%-%-%-%-%-%-%-%-%-  figura -%-%-%-%-%-%-%-%

\newcommand{\figDispsEsqoxford}{  % alias

\bfig
  \centering
  \includegraphics[width=.85\textwidth,height=0.75\textheight,keepaspectratio=true]{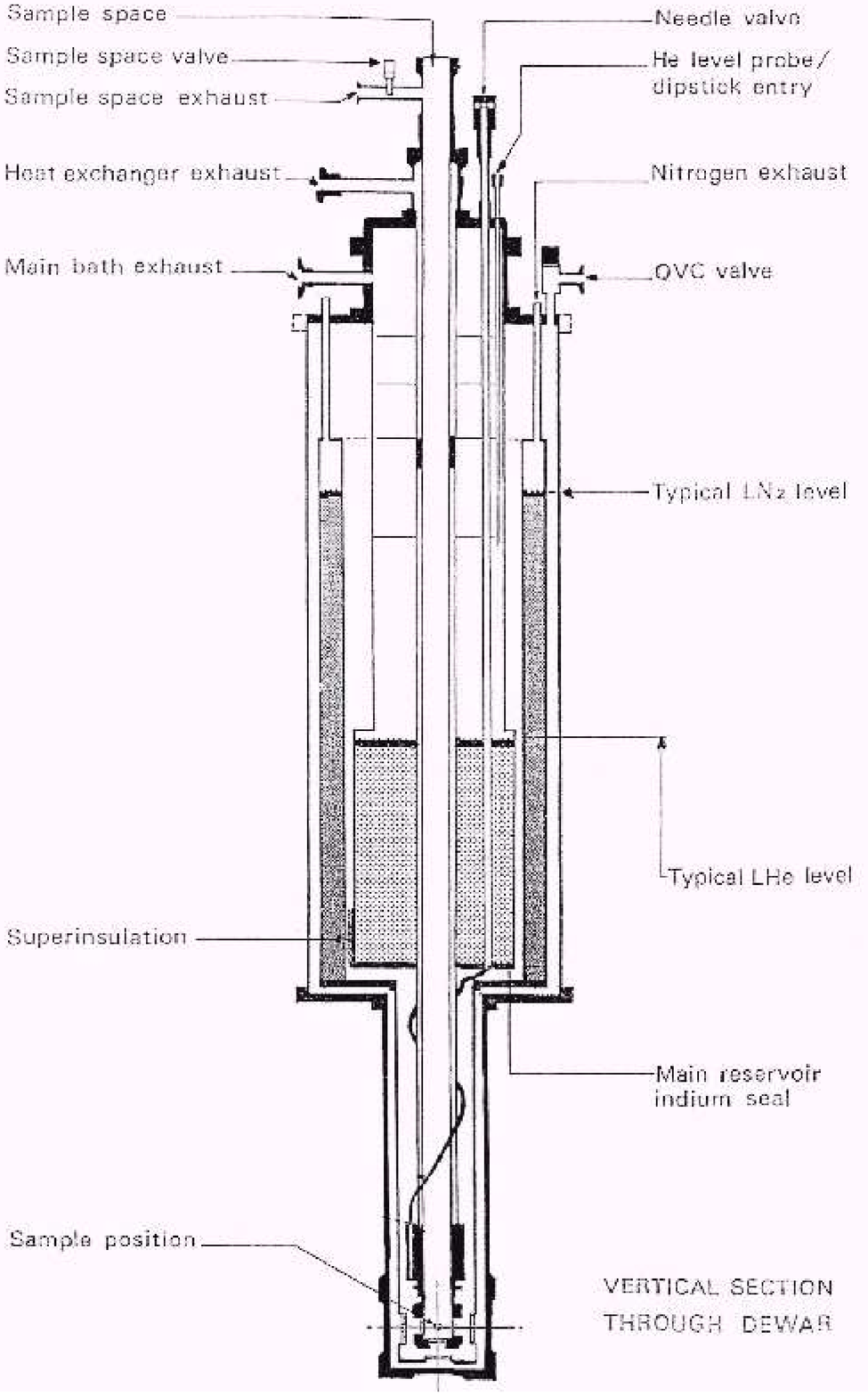}
  \caption[Corte longitudinal esquem\'atico del criostato.]{Corte longitudinal esquem\'atico del criostato empleado para las medidas experimentales. Aunque est\'a dise\~nado sobre todo para trabajar con helio l\'{\i}quido (LHe) como refrigerante, nosotros hemos empleado nitr\'ogeno l\'{\i}quido en su lugar. Otra c\'amara rellena de nitr\'ogeno l\'{\i}quido (LN$_2$) apantalla t\'ermicamente el dep\'osito principal. El refrigerante circula hasta la parte inferior de la c\'amara central, donde se aloja la muestra. El flujo se regula con una v\'alvula de aguja, y la temperatura se compensa con un calentador controlado electr\'onicamente.}                  % caption
  \label{fig:disps:esqoxford}

\efig %%%%%%%%%%%%%%%%%%%%%%

}
% - - - - - - - - - - - - - - - - - - - - - - - - - - - - - - - - - - - %

%-%-%-%-%-%-%-%-%-%-%-%-%-%-%-%-%-%-%-%-%-%-%-%-  figura -%-%-%-%-%-%-%-%

\newcommand{\figDispsBastonporta}{  % alias

\bfig
  \centering
  \includegraphics[width=.85\textwidth,height=\textwidth,keepaspectratio=true]{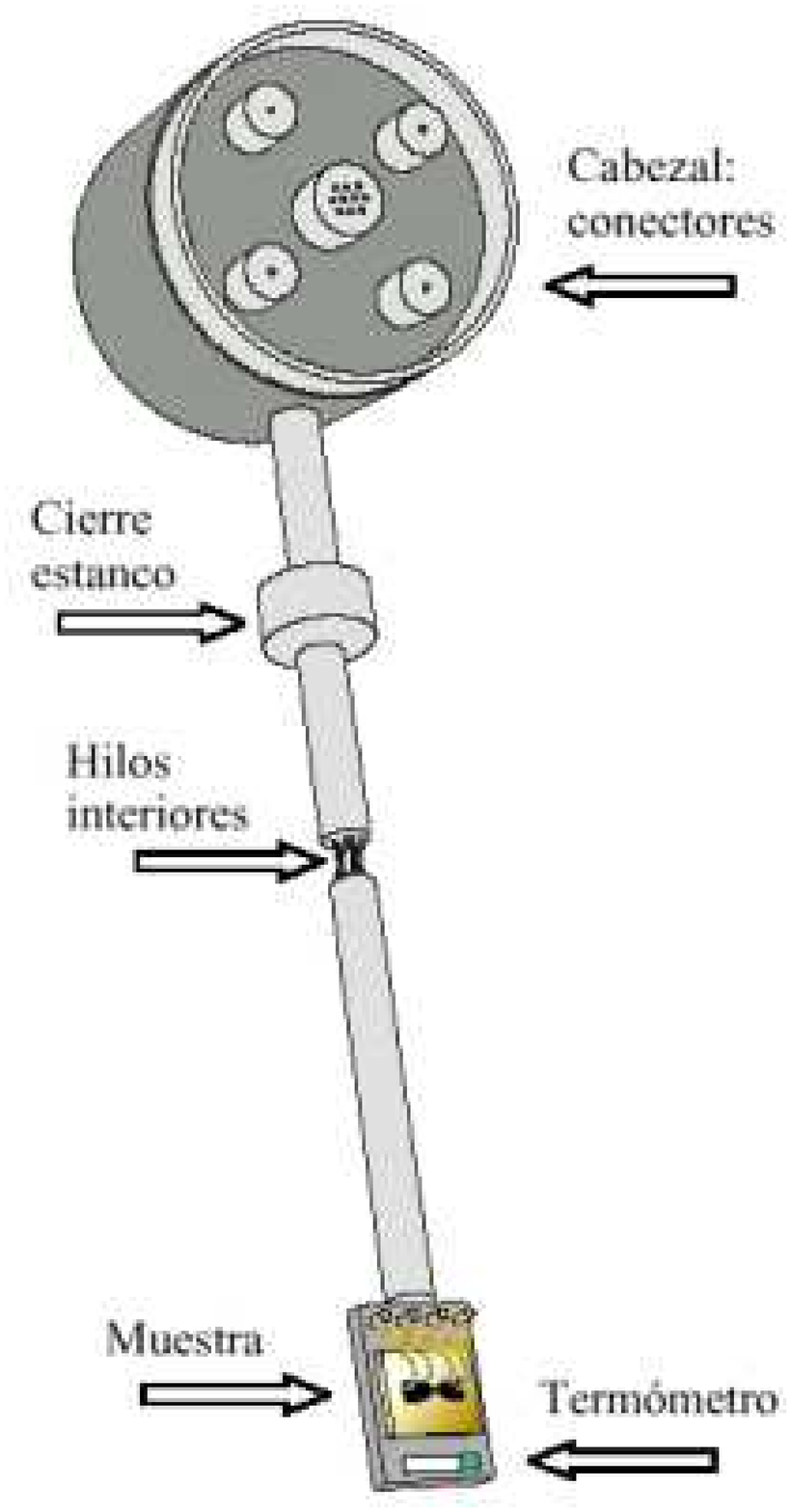}
  \caption[Esquema del bast\'on portamuestras.]{Esquema del bast\'on portamuestras que sit\'ua a la muestra y su term\'ometro en su lugar apropiado, al fondo del criostato. Por su interior circulan los hilos el\'ectricos que llegan desde los dispositivos de medida. El propio cabezal cierra herm\'eticamente la c\'amara. }                  % caption
  \label{fig:disps:bastonporta}

\efig %%%%%%%%%%%%%%%%%%%%%%

}
% - - - - - - - - - - - - - - - - - - - - - - - - - - - - - - - - - - - %

%-%-%-%-%-%-%-%-%-%-%-%-%-%-%-%-%-%-%-%-%-%-%-%-  figura -%-%-%-%-%-%-%-%

\newcommand{\figDispsOxfordUno}{  % alias

\bfig
  \centering
  \includegraphics[width=.95\textwidth,height=0.9\textheight,keepaspectratio=true,clip=true]{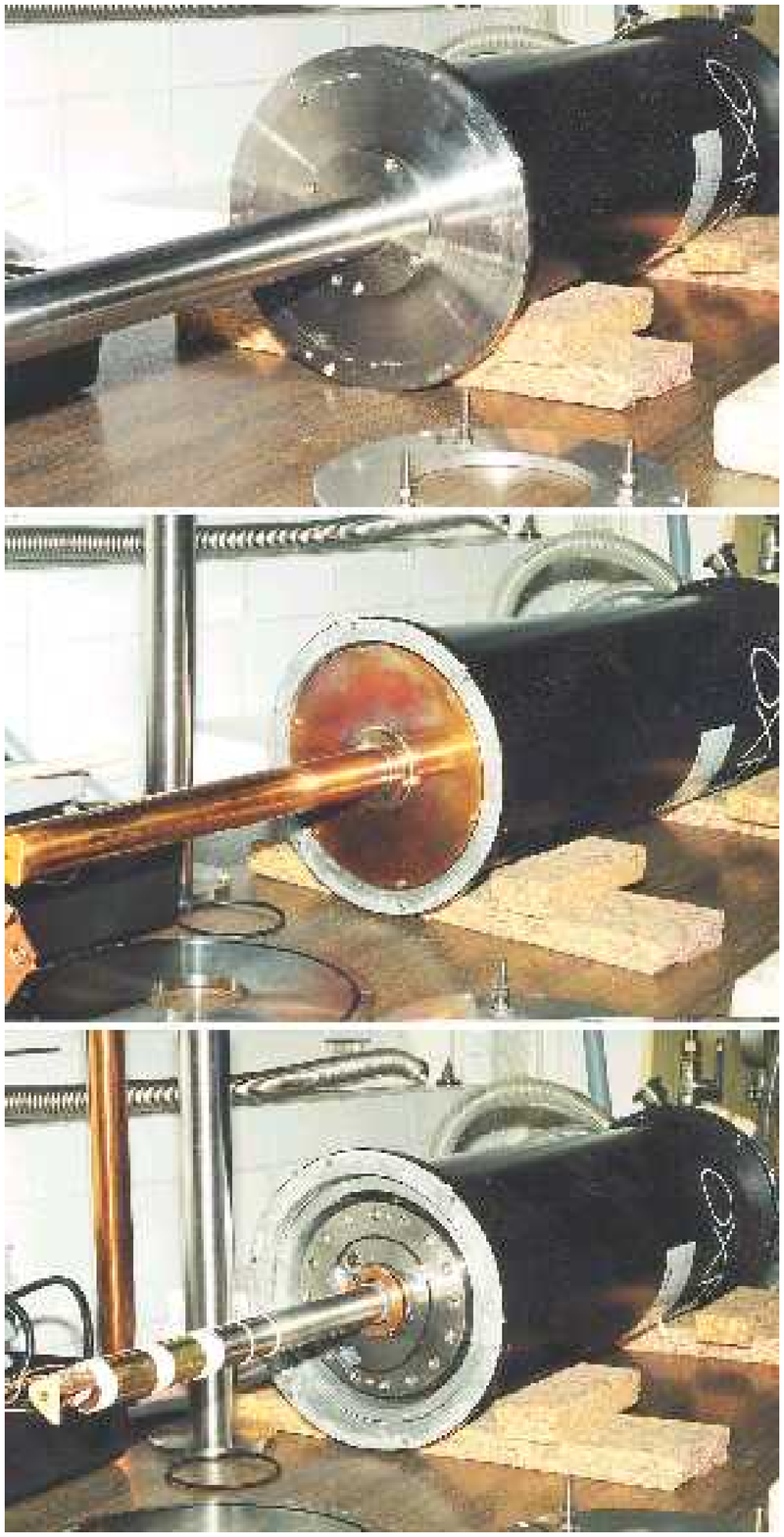}
  \caption[Fotograf\'{\i}as de las paredes del criostato Oxford.]{Fotograf\'{\i}as de las distintas paredes que constituyen el
  criostato Oxford. Fueron tomadas durante una reparaci\'on del
  criostato.}                  % caption
  \label{fig:disps:oxford1}

\efig %%%%%%%%%%%%%%%%%%%%%%

}
% - - - - - - - - - - - - - - - - - - - - - - - - - - - - - - - - - - - %

%-%-%-%-%-%-%-%-%-%-%-%-%-%-%-%-%-%-%-%-%-%-%-%-  figura -%-%-%-%-%-%-%-%

\newcommand{\figDispsOxfordDos}{  % alias

\bfig
  \centering
  \includegraphics[width=.95\textwidth,height=0.9\textheight,keepaspectratio=true]{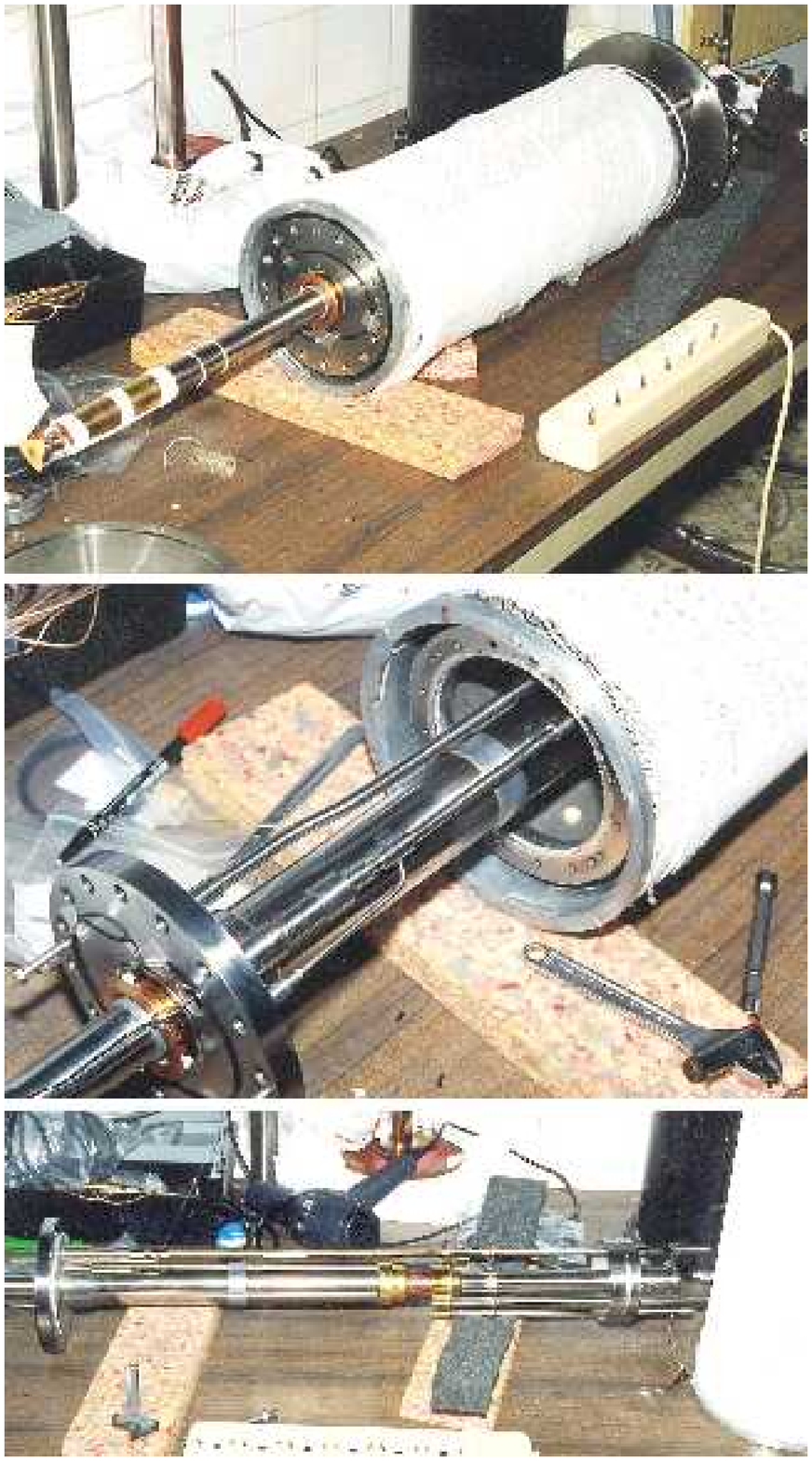}
  \caption[Fotograf\'{\i}as de las paredes y tubos del criostato.]{Fotograf\'{\i}as de las paredes y tubos interiores que
  constituyen el criostato Oxford. Fueron tomadas durante una
  reparaci\'on del criostato.}                  % caption
  \label{fig:disps:oxford2}

\efig %%%%%%%%%%%%%%%%%%%%%%

}
% - - - - - - - - - - - - - - - - - - - - - - - - - - - - - - - - - - - %

Los elementos b\'asicos que componen nuestro dispositivo
experimental son un criostato para la regulaci\'on de la
temperatura, unas bobinas para aplicar campo magn\'etico externo,
y diversas fuentes y volt\'{\i}metros para medir la ca\'{\i}da de
voltaje en el micropuente al paso de una corriente el\'ectrica.

\section{Criostato y campo magn\'etico}
\label{sec:criostato}

Al principio, en nuestro laboratorio estos experimentos se
realizaban con la muestra sumergida directamente en un
\Index{ba\~no} de nitr\'ogeno l\'{\i}quido, cuya temperatura se
regulaba presurizando el ba\~no con gas \cite{Curras00b}. Cuando
tuvimos la necesidad de aplicar campo magn\'etico externo,
introdujimos cerca de la muestra, dentro del ba\~no, unas
peque\~nas \Index{bobinas de Helmholtz} que apenas alcanzaban los
$150\und{Gs}$, lo que en principio estim\'abamos ser\'{\i}a el
\Index{autocampo} de la corriente inyectada en la muestra. Al no
observar efectos sobre las medidas con este campo, vimos la
necesidad de emplear las bobinas magn\'eticas de nuestro
\Index{criostato} Oxford, m\'as potentes, lo que no permit\'{\i}a
con facilidad usar un ba\~no l\'{\i}quido presurizado. De este
modo, los experimentos que se muestran en este trabajo se
realizaron con la muestra en atm\'osfera de helio gas, y la
temperatura regulada con los elementos de dicho criostato. No se
aprecian substanciales diferencias entre las medidas tomadas con
ba\~no l\'{\i}quido \cite{Curras00b,Curras02,Gonzalez02} o ba\~no
gaseoso, al menos mientras la muestra no transita completamente al
estado normal y se vuelve altamente disipativa. Esto es razonable,
pues en los tiempos tan cortos de medida que empleamos la
interacci\'on m\'as importante ocurre entre el micropuente y su
substrato y quiz\'a el \Index{portamuestras}, no con la
atm\'osfera. El coeficiente de transferencia de calor $h$ entre la
muestra y el \Index{substrato} es muy superior al que presentan la
muestra o el substrato con la atm\'osfera circundante, ya sea
\'esta l\'{\i}quida o gaseosa. \label{par:substratogas} Los
valores son de $1000~{\rm W/cm^2\,K}$ para filme-substrato
\cite{Nahum91}, y de $1~{\rm W/cm^2\,K}$ para filme-entorno cuando
este \'ultimo es nitr\'ogeno l\'{\i}quido
\cite{Mosqueira93,Mosqueira93b,Mosqueira94}, por lo que ser\'a
incluso menor si es gas. Los efectos de la atm\'osfera se
har\'{\i}an relevantes para mayores duraciones de medida.
Estimaremos estos tiempos en cap\'{\i}tulos posteriores.

La muestra, sobre su \Index{portamuestras} intermedio de cobre, se
dispone a su vez sobre otro portamuestras mayor, tambi\'en de
cobre, que se introduce en una c\'amara refrigerada de nuestro
criostato. Con los portamuestras de cobre aseguramos una buena
termalizaci\'on del substrato ---y por lo tanto tambi\'en de la
pel\'{\i}cula superconductora--- con el entorno de la muestra, en
nuestro caso helio gas. Ambos est\'an adheridos entre s\'{\i} con
grasa Apiezon N, igual que el substrato al portamuestras
intermedio.

Adherido igualmente con Apiezon N al portamuestras de cobre
situamos un \Index{term\'ometro} resistivo Pt-100. \'Este se
alimenta con una fuente DC de voltaje\footnote{Hewlett-Packard
(Agilent)~6181}, y ponemos una resistencia tamp\'on de
$100\und{k\Omega}$ en serie para mantener constante la corriente
en $100\und{\mu A}$. El voltaje del \Index{term\'ometro} lo
medimos con un volt\'{\i}metro\footnote{Hewlett-Packard
(Agilent)~3457A}. De esta manera determinamos la temperatura del
conjunto de la muestra y los portamuestras de cobre.

\figDispsCriostatorack \figDispsOxfordUno
\figDispsOxfordDos \figDispsEsqoxford \figDispsBastonporta

En la fotograf\'{\i}a de la figura \ref{fig:disps:criostatorack}
podemos ver el \Index{criostato} empleado para la refrigeraci\'on
de las muestras, y las bobinas para la aplicaci\'on de campos
magn\'eticos. En la serie de fotograf\'{\i}as de las figuras
\ref{fig:disps:oxford1} y \ref{fig:disps:oxford2}, tomadas durante
su puesta a punto, podemos ver la estructura interna del
criostato. En la figura \ref{fig:disps:esqoxford} se muestra un
esquema del corte longitudinal del criostato, con sus distintas
c\'amaras. La c\'amara que aloja la muestra es un tubo vertical
que recorre todo el criostato. En el fondo se sit\'ua la muestra,
en un punto centrado entre las bobinas que generan el campo
magn\'etico. Para ello se dispone de un largo bast\'on de
$1.30\und{m}$, esquematizado en la figura
\ref{fig:disps:bastonporta}, que lleva los cables desde el cabezal
superior hasta el fondo del criostato, donde est\'a el
portamuestras con la muestra y su \Index{term\'ometro}. Esta
c\'amara de la muestra es estanca: se cierra con el propio cabezal
del bast\'on \Index{portamuestras}, y se llena de helio gas. La
atm\'osfera de helio se refrigera mediante un serpent\'{\i}n, en
la base de la c\'amara, por el que circula gas evaporado de un
l\'{\i}quido criog\'enico, en nuestro caso nitr\'ogeno. El flujo
del gas se regula con una v\'alvula de aguja, y se genera por su
propia presi\'on, hacia la atm\'osfera del laboratorio, pero se
puede forzar ligeramente succionando con una bomba. La temperatura
de la c\'amara de la muestra se regula mediante un calentador
resistivo controlado electr\'onicamente\footnote{Controlador de
temperatura Oxford ITC-4. En nuestra configuraci\'on, el
term\'ometro del controlador no es el que est\'a adherido al
portamuestras sino otro, fijo al fondo de la c\'amara central del
criostato: esto hace que la temperatura que se programa y la que
realmente alcanza la muestra (medida con su propio term\'ometro)
sean levemente diferentes.}, que contrarresta esta refrigeraci\'on
criog\'enica. El dep\'osito de l\'{\i}quido refrigerante est\'a
apantallado del exterior por una c\'amara rellena de nitr\'ogeno
l\'{\i}quido. Esta c\'amara ser\'{\i}a especialmente relevante en
el caso de que el refrigerante fuese helio l\'{\i}quido, pero en
nuestro caso, en que empleamos nitr\'ogeno, tambi\'en es de
utilidad para hacer el sistema m\'as estable t\'ermicamente. Todas
estas c\'amaras se a\'{\i}slan del entorno con un vac\'{\i}o
intermedio que las envuelve. Para obtener este vac\'{\i}o
empleamos una bomba turbomolecular\footnote{Pfeiffer-Balzers TSH
332.} dentro de la cual, en buenas condiciones, se alcanzan
vac\'{\i}os de $7\Exund{-7}{mbar}$, lo que genera vac\'{\i}os
intermedios en el criostato mejores que $10^{-3}\und{mbar}$. En
los casos en que el criostato responde peor o se ponen de
manifiesto peque\~nas fugas, dejamos bombeando continuamente el
vac\'{\i}o intermedio para garantizar un buen aislamiento
t\'ermico. En la tapa superior del criostato se encuentran las
llaves y v\'alvulas de seguridad de las distintas c\'amaras.

La medida de una serie de curvas \VI\ a distintos campos y a
temperatura constante puede prolongarse hasta casi hora y media.
Durante este tiempo la temperatura debe permanecer todo lo estable
que sea posible. A temperaturas cercanas a la de ebullici\'on del
nitr\'ogeno l\'{\i}quido, el refrigerante que empleamos, la
estabilizaci\'on es ligeramente mejor que a temperaturas mayores.
Como ejemplos, mencionamos que en una medida concreta realizada a
76.2~K, la lectura de la temperatura oscilaba en la cent\'esima de
Kelvin, entre 76.22 y 76.24~K, durante 90 minutos. En otra medida
a 84.1~K, la lectura durante un tiempo semejante oscilaba entre
84.12 y 84.19~K.

El electroim\'an consiste en unas bobinas externas de cobre, con
polos c\'onicos truncados que dirigen el campo convenientemente a la
zona en que se sit\'ua la muestra. Se refrigera por flujo de agua, y
permite la aplicaci\'on de \index{campo magn\'etico} campos magn\'eticos
de hasta poco m\'as de $\mH=1\und{T}$, mediante corrientes de hasta
casi $20\und{A}$ con una fuente de potencia\footnote{KSM Brookmans
PK Herts Stabilised Power Supply, 20A-250V.}. El campo se mide con
una magnet\'ometro de efecto Hall\footnote{Oxford Hall Effect
Magnetometer 5200} cuya sonda se sit\'ua en la misma zona de la
muestra, exteriormente al criostato.

%%%%%%%%%%%%%%%%%%%%%%%%%%%%%%%%%%%%%%%%
\section{Circuito el\'ectrico}
%%%%%%%%%%%%%%%%%%%%%%%%%%%%%%%%%%%%%%%%

\figDispsCircuito

La muestra, una vez preparada como hasta aqu\'{\i} hemos descrito,
se conecta dentro del \Index{circuito} el\'ectrico que se
esquematiza en la figura \ref{fig:disps:circuito}. Mediremos la
ca\'{\i}da de voltaje en el micropuente con una configuraci\'on de
cuatro hilos. La corriente se inyecta con una fuente de corriente
programable y de precisi\'on\footnote{Keithley 2400 SourceMeter.},
por lo que tomamos como valor de la corriente el que programamos
en la propia fuente. De todos modos, disponemos de una resistencia
conocida en serie con la muestra, para que en las medidas de
curvas \VI\ r\'apidas, en las que programamos varios valores de
corriente seguidos en una misma medida, podamos identificar
c\'omodamente a qu\'e \Index{pulso} de corriente corresponde cada
dato adquirido.

El dispositivo que mide el voltaje de la muestra depender\'a de en
qu\'e r\'egimen de disipaci\'on nos encontremos. Para se\~nales
d\'ebiles emplearemos un nanovolt\'{\i}metro, y para el r\'egimen
de alta disipaci\'on, donde el \Index{calentamiento} es
notablemente influyente en los resultados, empleamos una tarjeta
de adquisici\'on mucho m\'as r\'apida \cite{NIan007}, con el fin
de minimizar los efectos de autocalentamiento.

Debido a las deficiencias de la \Index{toma de tierra} que tiene
la instalaci\'on el\'ectrica de nuestra Facultad, es habitual
apreciar en nuestros experimentos algunos problemas de ruido
par\'asito, sobre todo en las medidas de se\~nales m\'as bajas.
Tras diversas pruebas encontramos que la configuraci\'on m\'as
aislada se consegu\'{\i}a uniendo entre s\'{\i} las tierras de las
fuentes (la de la muestra y la del term\'ometro), y \'estas a la
carcasa del \Index{criostato}, pero dejando flotantes las tierras
de los dispositivos de medida. Conectar con las tierras
mencionadas tambi\'en la de la alimentaci\'on trif\'asica de la
fuente de las bobinas no mejoraba la calidad de la medida, pero
parec\'{\i}a proteger las muestras de algunas destrucciones
espont\'aneas que ten\'{\i}an lugar al manipular dicha fuente,
sobre todo al encenderla.

\subsection{Medidas de voltajes bajos con nanovolt\'{\i}metro}

Para determinar experimentalmente la densidad de corriente
cr\'{\i}tica \Jc\ emplearemos un criterio umbral en campo
el\'ectrico de $\Ec=10\und{\mu V/cm}$, muy habitual en la
bibliograf\'{\i}a. Para una longitud t\'{\i}pica del micropuente
$l=50\und{\mu m}$, nos vemos por lo tanto obligados a medir
voltajes tan peque\~nos como $5\Exund{-8}{V}$, para lo cual
empleamos un nanovolt\'{\i}metro Hewlett-Packard (Agilent)~34420A.
Las medidas t\'{\i}picas se realizan en 10~NPLC (\emph{number of
power line cycles}, 1~NPLC = $20\und{ms}$), diez \Index{ciclos de
red}, a corriente aplicada constante (superior a $1\und{mA}$).
Como la disipaci\'on es tan baja podemos despreciar los efectos
t\'ermicos, y hacer medidas de larga duraci\'on con el fin de
minimizar el ruido. Cada punto de la curva \EJ\ se adquiere
haciendo la semidiferencia de dos medidas con corrientes de igual
valor absoluto, pero de sentidos contrario: de este modo se
eliminan los efectos \Index{termoel\'ectricos}, que son bastante
notables cuando la se\~nal es tan baja. Midiendo a 10~NPLC tenemos
una resoluci\'on mejor que el nanovoltio, lo que para un puente
t\'{\i}pico de $50\und{\mu m}$ de largo implica una resoluci\'on
en el campo el\'ectrico mejor que $0.2\und{\mu V/cm}$. A pesar de
los promedios en varios ciclos de red, el ruido experimental
aumenta el error por encima de esta resoluci\'on, aunque se
mantiene siempre por debajo del umbral de campo $\Ec=10\und{\mu
V/cm}$ que elegimos para la determinaci\'on de \Jc.

\subsection{Medidas r\'apidas de voltajes elevados con tarjeta de
adquisici\'on}

Cuando, al aumentar la corriente inyectada, la disipaci\'on se
hace tan elevada que los efectos t\'ermicos se vuelven muy
relevantes, nos vemos obligados a hacer las medidas a gran
velocidad para tratar de minimizarlos. Con el fin de reducir los
tiempos de medida que en este tipo de experimentos se ven\'{\i}an
usando en nuestro laboratorio \cite{Curras01,Curras00b}, en los
que cada punto de la curva \EJ\ se tomaba en t\'{\i}picamente
$30\und{ms}$, parte de este trabajo consisti\'o en la
instalaci\'on, integraci\'on y programaci\'on de una tarjeta
adquisidora de datos (DAQ \cite{NIan007}) National Instruments PCI
6035E, que permite medidas de voltaje de hasta
$2\Exund{5}{muestras/s}$, o de un punto cada $5\und{\mu s}$. Tiene
una resoluci\'on de 16 bits en cada uno de sus rangos bipolares de
medida ($\pm 50$~mV, $\pm 500$~mV, $\pm 5$~V y $\pm 10$~V). Usando
dos canales, uno para la muestra y otro para la resistencia en que
medimos la corriente aplicada, la velocidad de medida por canal se
reduce a la mitad de ese valor: un punto cada $10\und{\mu s}$.

Esto nos permite estudiar la evoluci\'on temporal de la se\~nal de
voltaje, a corriente aplicada constante, de lo que podremos
obtener informaci\'on a cerca del comportamiento t\'ermico de la
muestra. Como esta variaci\'on es muy r\'apida (veremos que los
cambios m\'as dr\'asticos se producen en apenas $100\und{\mu s}$,
y se tienen estabilizaciones en $1\und{ms}$), cualquier
dispositivo que mida a velocidades m\'as lentas se perder\'a estos
detalles. La transici\'on al estado normal en \Jx\ ocurre tan
abruptamente que ni siquiera con nuestro sistema de adquisici\'on
r\'apida podemos apreciar todos los detalles del salto, pero
s\'{\i}que alcanza para ver el arranque de la transici\'on y
estudiar los efectos que la duraci\'on de la medida tiene en la
determinaci\'on de \Jx.

La velocidad de la tarjeta de adquisici\'on tambi\'en nos permite
medir curvas \EJ\ completas en breves periodos de tiempo, de
manera que garanticemos la estabilidad de la temperatura durante
toda la medida. La fuente de corriente que empleamos no permite
aplicar pulsos m\'as breves que de unos $0.7\und{ms}$, as\'{\i}
que estas medidas de curvas completas las realizamos con
\index{rampa} rampas escalonadas: cada \emph{pelda\~no} de
corriente constante tiene la duraci\'on m\'as corta que la fuente
permite, y se recorren varios valores de corriente entre \Jc\ y
\Jx\ en, t\'{\i}picamente, 20-30~ms, midiendo constantemente el
voltaje. El tener pelda\~nos de corriente constante nos permite
descartar efectos \Index{inductivos}, eliminando los datos tomados
en las transiciones entre pelda\~nos.

Al aplicar densidades de corriente elevadas, cercanas a \Jx, se
observan variaciones de voltaje en el tiempo que dura cada
escal\'on aplicado, signos claros de \Index{calentamiento} de la
muestra. En muchas de las curvas \EJ\ que mostraremos, el valor de
$E$ para cada $J$ se ha tomado como el promedio de los valores
medidos en un pelda\~no estable de la \Index{rampa}. (Salvo en los
puntos m\'as cercanos a la transici\'on, en que no siempre se
promedian, para mostrar su evoluci\'on). Haremos estudios de los
efectos que tienen en la determinaci\'on de la curva \EJ\ y del
valor de \Jx\ tanto la duraci\'on de la rampa como el hecho de
medir con rampa escalonada o con \index{pulso} pulsos aislados.

Las medidas de curvas \VI\ con la DAQ suelen hacerse empleando su
ganancia m\'axima de un factor 100, para medir en la escala
bipolar de $\pm 50\und{mV}$, al menos en la regi\'on por debajo de
la transici\'on. Estas medidas tienen una resoluci\'on de
$100\und{mV}/2^{16} = 1.5\und{\mu V}$, pero una dispersi\'on
debido al ruido experimental de unos $80\und{\mu V}$. Haciendo
medidas en escalones obtenemos entre 60 y 70 puntos \'utiles
(descontando las transiciones) con una desviaci\'on t\'{\i}pica de
$30\und{\mu V}$. Con 60 puntos, esto supone un intervalo de
confianza al 95\% de $~2 \times 30\und{\mu V}/\sqrt{60} =
8\und{\mu V}$. Para un puente de $50\und{\mu m}$ de largo esto
implica un error en el campo el\'ectrico menor que
$\pm2\und{mV/cm}$. Este error es menor del 0.1\% para los valores
t\'{\i}picos en torno al salto, pero muy grande para los rangos en
torno a \Jc, por eso en ese caso empleamos el nanovolt\'{\i}metro,
que mide promediando el ruido sobre varios ciclos de red.

Por comodidad en el an\'alisis, en estas medidas en rampa
consideramos que la corriente $I$ inyectada es la nominal de la
fuente en el cero y en un punto elevado de la rampa. Con estos dos
puntos fijos se escalan todas las lecturas de la resistencia
patr\'on, simult\'aneas con las del voltaje de la
muestra\footnote{Esta tarjeta DAQ no mide simult\'aneamente en
todos sus canales, sino que conmuta la lectura de uno a otro en
intervalos regulares de tiempo. En nuestras medidas, un canal
tiene un retraso respecto del otro de $10\und{\mu s}$, lo que
resulta despreciable dentro de la duraci\'on t\'{\i}pica de un
escal\'on de corriente constante, en torno a $0.75\und{ms}$.},
tambi\'en promediando en cada escal\'on de la rampa. La
desviaci\'on t\'{\i}pica de estas medidas (que se realizan con la
DAQ en la escala de $\pm5$~V) es unos $10\und{\mu A}$, lo que para
60 adquisiciones implicar\'{\i}a un intervalo de confianza al 95\%
de unos $3\und{\mu A}$ en la medida de la corriente. Para un
micropuente t\'{\i}pico, esto supone un error en la
determinaci\'on de $J$ de $200\und{\Acms}$, poco m\'as de 30
partes por mill\'on en torno a \Jx.

Para la medida de curvas de \Index{resistividad} en funci\'on de
la temperatura se emple\'o indistintamente el nanovolt\'{\i}metro
o la tarjeta de adquisici\'on, lo que resultase m\'as c\'omodo en
cada momento, a pesar de que estas medidas se hacen siempre en un
r\'egimen de no muy alta disipaci\'on ($J\sim 100\und{\Acms}$). La
tarjeta DAQ proporciona resultados m\'as ruidosos y con peor
resoluci\'on que el nanovolt\'{\i}metro, pero las curvas $\rho(T)$
que se mostrar\'an en esta primera parte de la memoria no son
m\'as que caracterizaciones generales de las muestras, y no
centramos nuestro inter\'es en ellas. Veremos un caso singular de
medida de resistividad en funci\'on de la temperatura con altas
corrientes, a prop\'osito para ver efectos t\'ermicos. En este
caso, por supuesto, se emple\'o la DAQ para hacer medidas
r\'apidas.

\section{Otros aspectos}

Gran parte del tiempo de este trabajo de doctorado se invirti\'o en
la preparaci\'on de este montaje experimental que hemos descrito.
Los aspectos que principalmente ocuparon el tiempo dedicado a la
preparaci\'on del experimento fueron dos:

Por un lado, la puesta a punto del \Index{criostato}. Durante la
b\'usqueda de una fuga en una de sus c\'amaras estancas, cuya
localizaci\'on\footnote{Con un detector de fugas de helio Balzers
HLT~150.} y reparaci\'on ocup\'o varias semanas, se tomaron las
fotograf\'{\i}as mostradas previamente.

Por otro lado, se dedic\'o largo tiempo a la elaboraci\'on de un
programa escrito inicialmente en \Index{HP-BASIC} y posteriormente
en \Index{Visual Basic}, que controla y sincroniza los diversos
dispositivos electr\'onicos y permite realizar medidas de curvas
\VI\ en breves intervalos de tiempo. El tiempo invertido en el
dise\~no de este programa ha dado buenos frutos, pues hemos podido
tomar con comodidad y rapidez las medidas que se muestran en
sucesivos cap\'{\i}tulos. Adem\'as, se facilita enormemente la
continuidad de trabajo, pues la modularidad y sencillez de uso del
programa permiten adaptarlo a nuevas necesidades con poco trabajo
extra.
  % cap 3
\clearemptydoublepage
%%%%%%%%%%%%%%%%%%%%%%%%%%%%%%%%%%%%%%%%%%%%%%%%%%%%%%%%%%%%%%%%%%
\chapter{Resultados experimentales}
%\chaptermark{Resultados experimentales}
%\addtocontents{toc}{\protect\vspace{0.2cm}}
\label{cap:resul}
%%%%%%%%%%%%%%%%%%%%%%%%%%%%%%%%%%%%%%%%%%%%%%%%%%%%%%%%%%%%%%%%%%

%-%-%-%-%-%-%-%-%-%-%-%-%-%-%-%-%-%-%-%-%-%-%-%-  figura -%-%-%-%-%-%-%-%

\newcommand{\figResulRhoAB}{  % alias
\bfig
  \centering
  \includegraphics[width=.8\textwidth,clip=true]{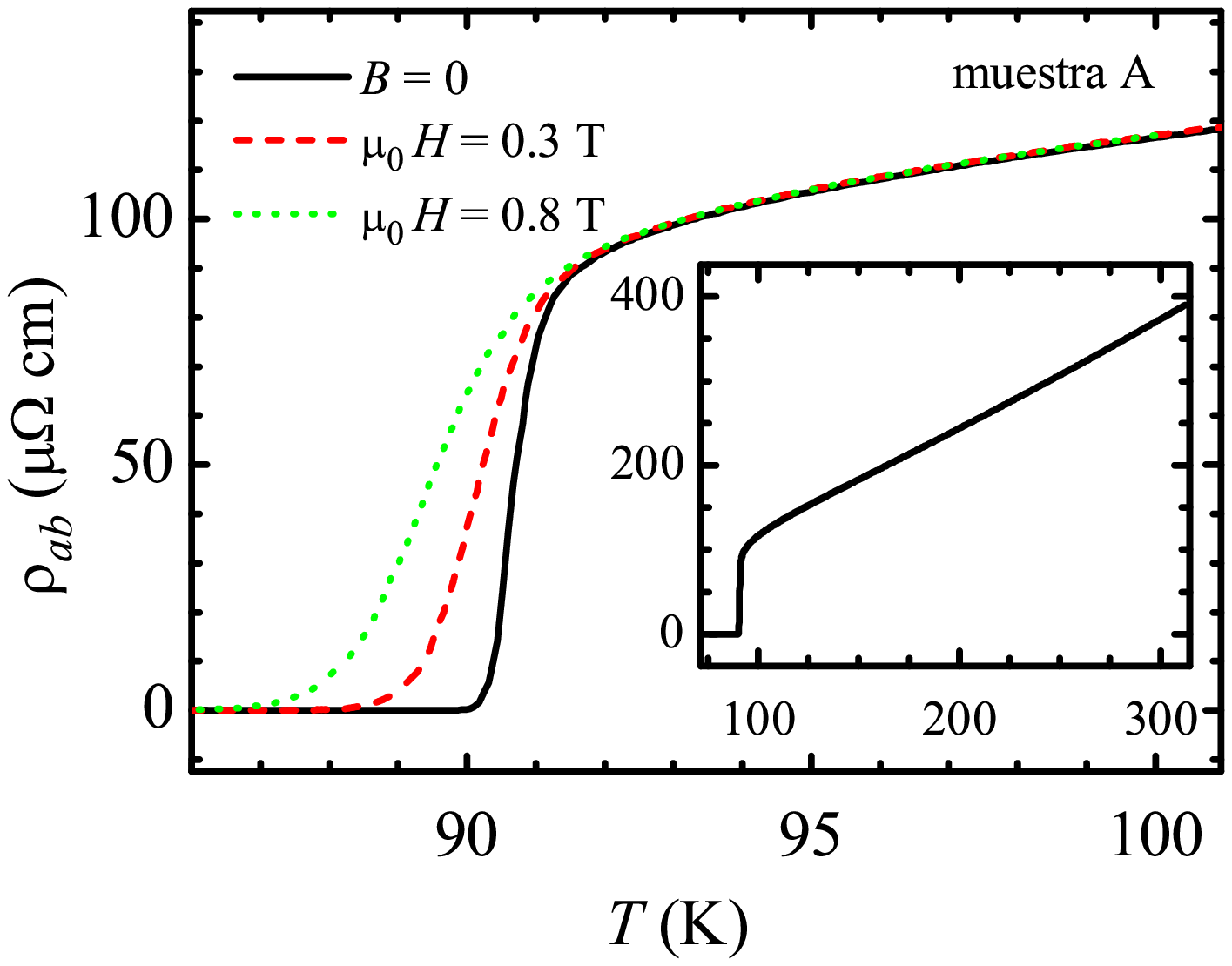}
  \includegraphics[width=.8\textwidth,clip=true]{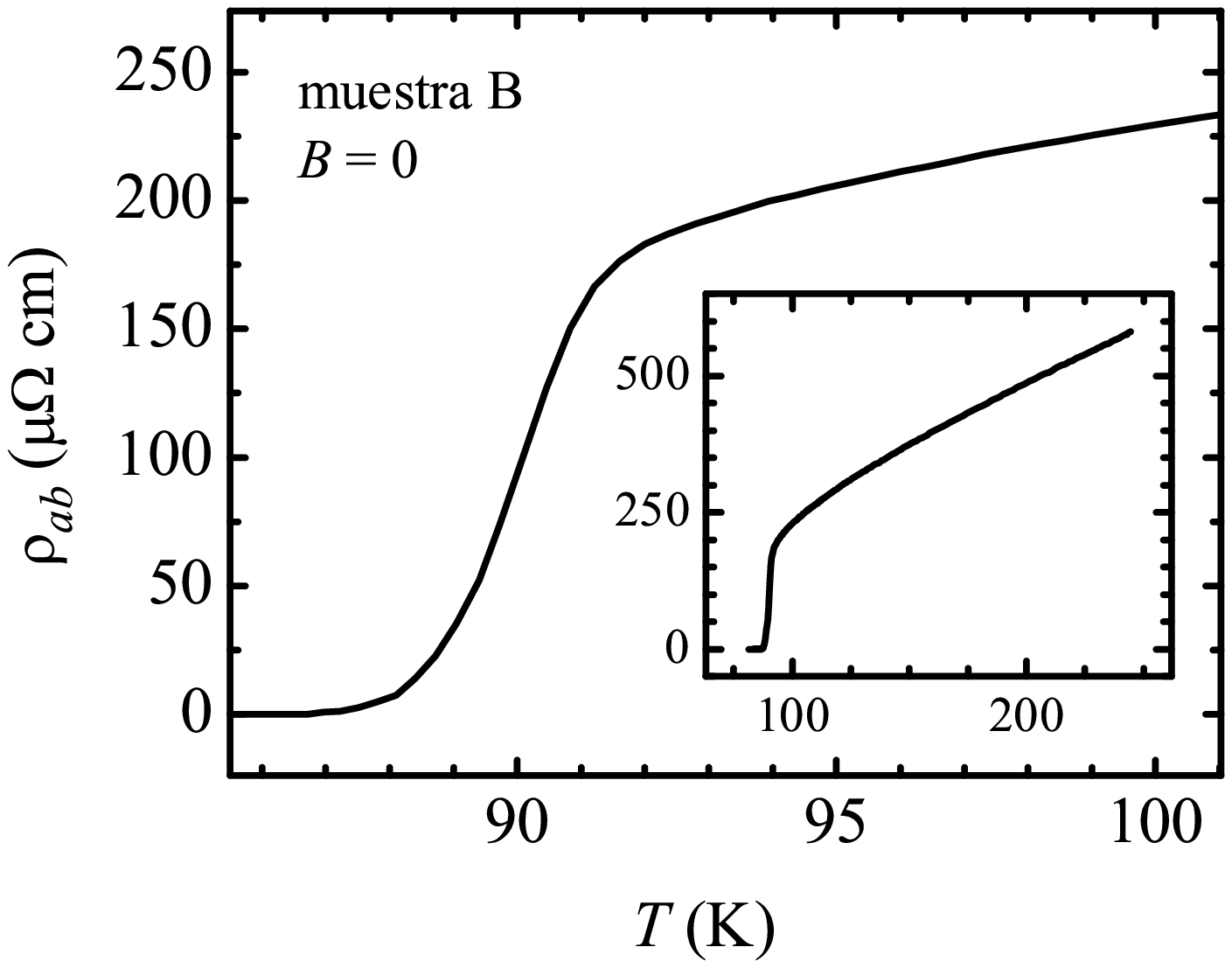}
  \caption[Resistividad en funci\'on de $T$ para las muestras \fA\ y \fB.]{Resistividad en funci\'on de la temperatura para las muestras \fA\ y \fB, para $B=0$ y con campo magn\'etico aplicado.}
  \label{fig:resul:rhoAB}
\efig %%%%%%%%%%%%%%%%%%%%%%
}
% - - - - - - - - - - - - - - - - - - - - - - - - - - - - - - - - - - - %

%-%-%-%-%-%-%-%-%-%-%-%-%-%-%-%-%-%-%-%-%-%-%-%-  figura -%-%-%-%-%-%-%-%

\newcommand{\figResulRhoDG}{  % alias
\bfig
  \centering
  \includegraphics[width=.8\textwidth,clip=true]{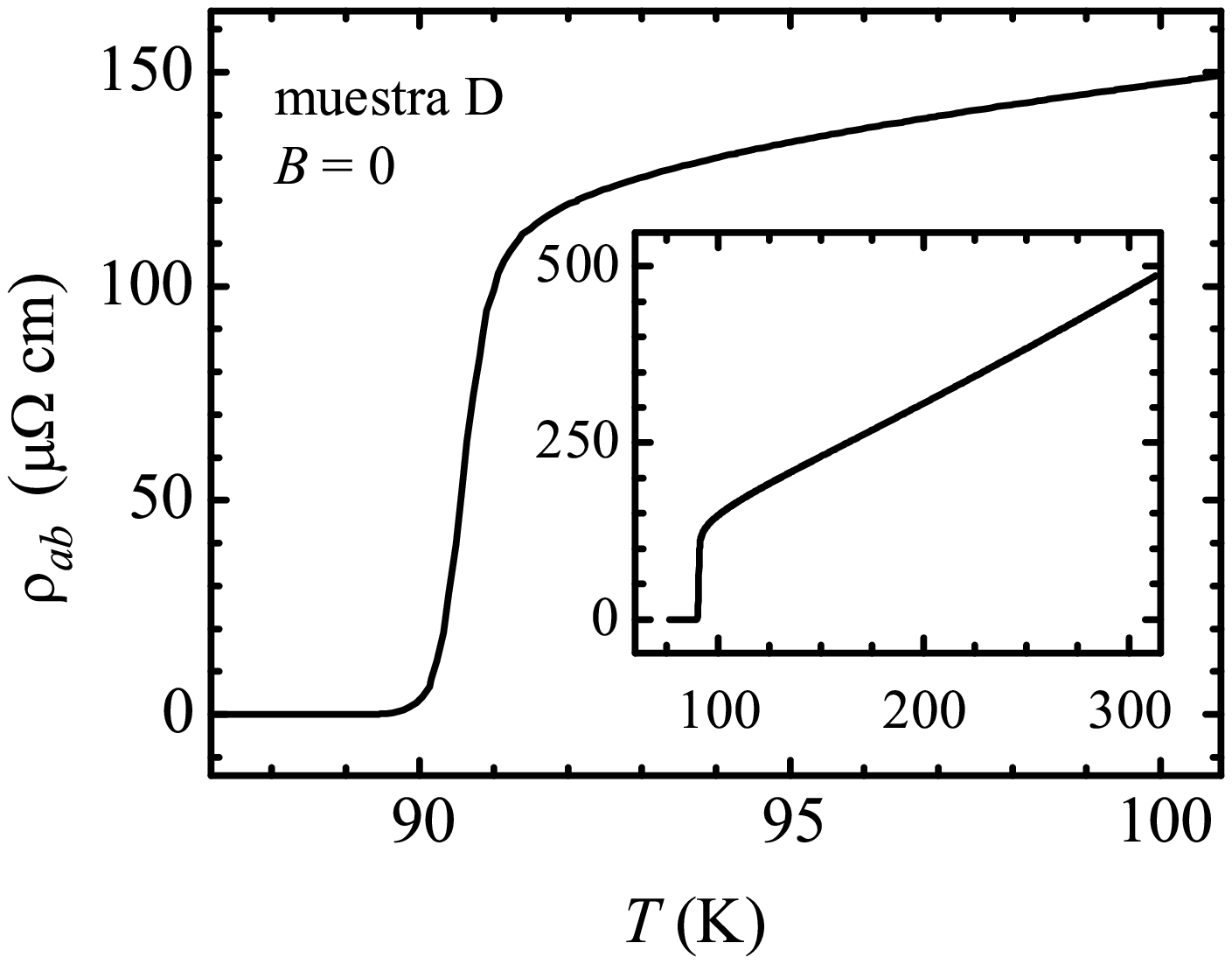}
  \includegraphics[width=.8\textwidth,clip=true]{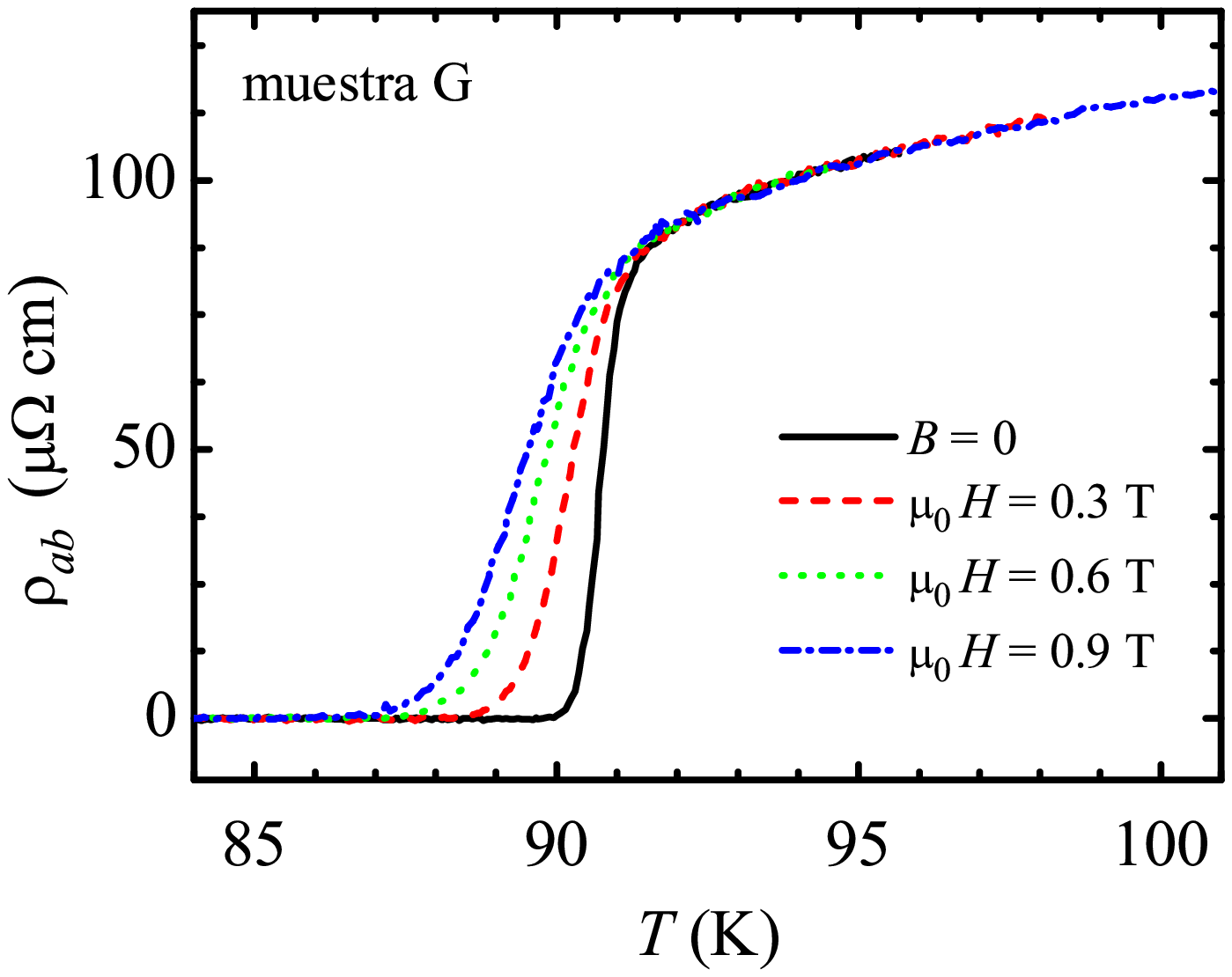}
  \caption[Resistividad en funci\'on de $T$ para las muestras \fD\ y \fG.]{Resistividad en funci\'on de la temperatura para las muestras \fD\ y \fG, para $B=0$ y con campo magn\'etico aplicado.}
  \label{fig:resul:rhoDG}
\efig %%%%%%%%%%%%%%%%%%%%%%
}
% - - - - - - - - - - - - - - - - - - - - - - - - - - - - - - - - - - - %

%-%-%-%-%-%-%-%-%-%-%-%-%-%-%-%-%-%-%-%-%-%-%-%-  figura -%-%-%-%-%-%-%-%

\newcommand{\figResulDerivA}{  % alias

\figura {fig:resul:derivA}      % label
{fig/resul/derivA}              % file
{Un ejemplo de derivada de la resistividad en funci\'on de la temperatura, empleada para la determinaci\'on de la temperatura cr\'{\i}tica. El punto de la transici\'on con derivada m\'axima (el punto de inflexi\'on de la transici\'on) es \Tci, y es la temperatura que tomaremos como la \Tc\ de la muestra. El ancho de esta curva a mitad de altura (FWHM) es el par\'ametro \DTci, indicativo de la estrechez de la transici\'on, y por lo tanto de la homogeneidad estequiom\'etrica de la muestra. La temperatura m\'axima \Tcz\ a la que la resistividad es cero tambi\'en se pone de manifiesto en esta representaci\'on.}                  % caption
{Ejemplo de derivada de la resistividad en funci\'on de la temperatura.}                  % toc
{\stfigw}            % width \textwidth
}
% - - - - - - - - - - - - - - - - - - - - - - - - - - - - - - - - - - - %

%-%-%-%-%-%-%-%-%-%-%-%-%-%-%-%-%-%-%-%-%-%-%-%-  figura -%-%-%-%-%-%-%-%

\newcommand{\figResulEJTA}{  % alias

\figura {fig:resul:EJTA}      % label
{fig/resul/EJTA}              % file EJTAdisc
{Curvas caracter\'{\i}sticas \EJ\ para la muestra \fA, a distintas
temperaturas por debajo de \Tc, a campo magn\'etico aplicado nulo.
En las distintas gr\'aficas se representan las mismas curvas, pero
disminuyendo sucesivamente el rango de representaci\'on en campo
$E$ para apreciar los detalles.}
% caption
{Curvas caracter\'{\i}sticas \EJ\ para la muestra \fA, a distintas
temperaturas por debajo de \Tc, a campo magn\'etico aplicado nulo.}                  % toc
{\stfigw}            % width \textwidth
}
% - - - - - - - - - - - - - - - - - - - - - - - - - - - - - - - - - - - %

%-%-%-%-%-%-%-%-%-%-%-%-%-%-%-%-%-%-%-%-%-%-%-%-  figura -%-%-%-%-%-%-%-%

\newcommand{\figResulEJTAup}{  % alias

\figura {fig:resul:EJTAup}      % label
{fig/resul/EJTAupc}              % file
{En (a), detalle de la figura \ref{fig:resul:EJTA}, para apreciar
la regi\'on normal por encima de \Jx. En l\'{\i}nea continua se
incluyen otras isotermas que, por claridad, no estaban en la
figura previa. En (b), ampliaci\'on de la regi\'on encuadrada
arriba, para apreciar las isotermas m\'as cercanas a \Tc. Ver
texto.}
% caption
{Curvas \EJ\ para la muestra \fA, a distintas temperaturas por
debajo de \Tc, a $B=0$:
detalle de las isotermas cercanas a \Tc.}                  % toc
{\stfigw}            % width \textwidth
}
% - - - - - - - - - - - - - - - - - - - - - - - - - - - - - - - - - - - %

%-%-%-%-%-%-%-%-%-%-%-%-%-%-%-%-%-%-%-%-%-%-%-%-  figura -%-%-%-%-%-%-%-%

\newcommand{\figResulEJTAlog}{  % alias

\figura {fig:resul:EJTAlog}      % label
{fig/resul/EJTAlogc}              % file
{Curvas caracter\'{\i}sticas \EJ\ para la muestra \fA, a distintas
temperaturas por debajo de \Tc, a campo magn\'etico aplicado nulo.
Son las mismas isotermas de las figuras \ref{fig:resul:EJTA} y
\ref{fig:resul:EJTAup}, pero ahora en representaci\'on
logar\'{\i}tmica para apreciar globalmente sus tendencias. El
l\'{\i}mite en voltaje impuesto a la fuente de alimentaci\'on
establece el campo $E$ m\'aximo en unos $4\Exund{3}{V/cm}$.}
% caption
{Curvas \EJ\ para la muestra \fA, a distintas
temperaturas por debajo de \Tc, a $B=0$: representaci\'on logar\'{\i}tmica.}                  % toc
{\stfigw}            % width \textwidth
}
% - - - - - - - - - - - - - - - - - - - - - - - - - - - - - - - - - - - %

%-%-%-%-%-%-%-%-%-%-%-%-%-%-%-%-%-%-%-%-%-%-%-%-  figura -%-%-%-%-%-%-%-%

\newcommand{\figResulEJHA}{  % alias

\figura {fig:resul:EJHA}      % label
{fig/resul/EJHAp}              % file EJHApsdis
{Curvas caracter\'{\i}sticas \EJ\ para la muestra \fA, a distintos
campos magn\'eticos aplicados $0<\mH<1\und{T}$, a temperatura
constante $T=76.2\und{K}$. En las distintas gr\'aficas se
representan las mismas curvas, pero disminuyendo sucesivamente el
rango de representaci\'on en campo $E$ para apreciar los detalles.
El l\'{\i}mite en voltaje impuesto a la fuente de alimentaci\'on,
impide ver la regi\'on normal.}
% caption
{Curvas caracter\'{\i}sticas \EJ\ para la muestra \fA, a distintos
campos magn\'eticos aplicados, a $T=76.2\und{K}$.}                  % toc
{\stfigw}            % width \textwidth
}
% - - - - - - - - - - - - - - - - - - - - - - - - - - - - - - - - - - - %

%-%-%-%-%-%-%-%-%-%-%-%-%-%-%-%-%-%-%-%-%-%-%-%-  figura -%-%-%-%-%-%-%-%

\newcommand{\figResulEJHAOchCuatro}{  % alias

\figura {fig:resul:EJHA84}      % label
{fig/resul/EJHA84}              % file
{Curvas caracter\'{\i}sticas \EJ\ para la muestra \fA, a distintos
campos magn\'eticos aplicados $0<\mH<1\und{T}$, a temperatura
constante $T=84.1\und{K}$. En contraste con la figura
\ref{fig:resul:EJHA}, ahora la regi\'on normal presenta un voltaje
inferior al umbral fijado en la fuente, por lo que se puede medir.
La recta de trazo discontinuo corresponde a la resistividad
extrapolada desde el estado normal para $T=84.1\und{K}$. Todas las
isotermas, en la regi\'on normal, se desv\'{\i}an de esta
tendencia recta, probablemente debido a la elevaci\'on de la
temperatura de la muestra por la disipaci\'on. La tendencia se
ajusta aproximadamente a una funci\'on proporcional al cuadrado de
la corriente aplicada. }
% caption
{Curvas caracter\'{\i}sticas \EJ\ para la muestra \fA, a distintos
campos magn\'eticos aplicados, a $T=84.1\und{K}$.}                  % toc
{\stfigw}            % width \textwidth
}
% - - - - - - - - - - - - - - - - - - - - - - - - - - - - - - - - - - - %

%-%-%-%-%-%-%-%-%-%-%-%-%-%-%-%-%-%-%-%-%-%-%-%-  figura -%-%-%-%-%-%-%-%

\newcommand{\figResulMedidaEJ}{  % alias

\figura {fig:resul:medidaEJ}      % label
{fig/resul/medidaEJ}              % file
{Experimento t\'{\i}pico para la determinaci\'on de curvas \EJ\ en
el r\'egimen de alta disipaci\'on, con tarjeta adquisidora (DAQ).
Arriba, la rampa escalonada de corriente que se aplica,
determinada sobre una resistencia conocida en serie con el
micropuente. Abajo, el voltaje le\'{\i}do en el micropuente. Se
muestra s\'olo la parte final de la rampa. En el momento en que se
produce la transici\'on al estado normal la impedancia del
circuito aumenta notablemente, y si la fuente tiene un l\'{\i}mite
m\'aximo en voltaje la corriente aplicada cae bruscamente.
Posteriormente termina la medida y se elimina la
corriente.}                  % caption
{Experimento t\'{\i}pico para la determinaci\'on de curvas \EJ\ en
el
r\'egimen de alta disipaci\'on.}                  % toc
{\stfigw}            % width \textwidth
}
% - - - - - - - - - - - - - - - - - - - - - - - - - - - - - - - - - - - %

%-%-%-%-%-%-%-%-%-%-%-%-%-%-%-%-%-%-%-%-%-%-%-%-  figura -%-%-%-%-%-%-%-%

\newcommand{\figResulMedidaEJz}{  % alias

\figura {fig:resul:medidaEJz}      % label
{fig/resul/medidaEJz}              % file
{Experimento t\'{\i}pico para la determinaci\'on de curvas \EJ\ en
el r\'egimen de alta disipaci\'on, con tarjeta adquisidora (DAQ):
detalle de la figura \ref{fig:resul:medidaEJ}. Arriba, la rampa
escalonada de corriente que se aplica. Abajo, el voltaje
le\'{\i}do en el micropuente. En la regi\'on de mayor
disipaci\'on, cerca de la transici\'on, el voltaje crece
suavemente mientras la corriente permanece en un escal\'on
constante:
esto se atribuye al aumento de temperatura de la muestra.}                  % caption
{Experimento t\'{\i}pico para la determinaci\'on de curvas \EJ\ en
el
r\'egimen de alta disipaci\'on: detalle.}                  % toc
{\stfigw}            % width \textwidth
}
% - - - - - - - - - - - - - - - - - - - - - - - - - - - - - - - - - - - %

%-%-%-%-%-%-%-%-%-%-%-%-%-%-%-%-%-%-%-%-%-%-%-%-  figura -%-%-%-%-%-%-%-%

\newcommand{\figResulMedidaEJzVI}{  % alias

\figura {fig:resul:medidaEJzVI}      % label
{fig/resul/medidaEJzVI}              % file
{De las medidas de la figura \ref{fig:resul:medidaEJz} se compone
una gr\'afica caracter\'{\i}stica \VI. Algunos puntos dispersos
correspondientes a los saltos de escalones de corriente son
posteriormente eliminados, pues est\'an muy afectados de
incertidumbre y efectos inductivos. La variaci\'on del voltaje por
calentamiento puesta de manifiesto en la figura anterior se
aprecia tambi\'en aqu\'{\i} cuando la disipaci\'on es mayor: los
puntos
experimentales est\'an m\'as dispersos a corrientes altas.}                  % caption
{Gr\'afica caracter\'{\i}stica \VI\ experimental: detalle.}                  % toc
{\stfigw}            % width \textwidth
}
% - - - - - - - - - - - - - - - - - - - - - - - - - - - - - - - - - - - %

%-%-%-%-%-%-%-%-%-%-%-%-%-%-%-%-%-%-%-%-%-%-%-%-  figura -%-%-%-%-%-%-%-%

\newcommand{\figResulRampas}{  % alias

\figura {fig:resul:rampas}      % label
{fig/resul/rampas}              % file
{Rampas de corriente escalonadas de distinta duraci\'on, variando
el n\'umero de escalones entre la corriente inicial y final.
Arriba, corriente aplicada. La corriente de salto se reconoce por
un peque\~no pico, causado por el aumento repentino de impedancia
en el circuito. Ahora la limitaci\'on de voltaje en la fuente no
impide seguir aumentando la corriente tras el salto. Abajo,
voltaje le\'{\i}do en el micropuente. Las rectas de trazo
discontinuo unen las corrientes y los voltajes de salto. Mientras
que la corriente apenas var\'{\i}a un 0.5\%, el voltaje aumenta
hasta un 14\% debido al
calentamiento. }                  % caption
{Rampas de corriente escalonadas de distinta duraci\'on.}                  % toc
{\stfigw}            % width \textwidth
}
% - - - - - - - - - - - - - - - - - - - - - - - - - - - - - - - - - - - %

%-%-%-%-%-%-%-%-%-%-%-%-%-%-%-%-%-%-%-%-%-%-%-%-  figura -%-%-%-%-%-%-%-%

\newcommand{\figResulCompdetalle}{  % alias

\figura {fig:resul:compdetalle}      % label
{fig/resul/compdetalle}              % file
{Comparaci\'on de los valores de \Jx\ determinados con rampas de
corriente de distinta resoluci\'on. La barra de error de cada punto
la determina la altura del escal\'on de su rampa correspondiente.
Las rampas con mayor resoluci\'on son tambi\'en las m\'as largas
temporalmente, por lo que acusan un desplazamiento sistem\'atico de
\Jx\ hacia
valores m\'as bajos debido al calentamiento. Las l\'{\i}neas son gu\'{\i}as para los ojos.}                  % caption
{Comparaci\'on de los valores de \Jx\ determinados con rampas de
corriente de distinta resoluci\'on.}                  % toc
{\stfigw}            % width \textwidth
}
% - - - - - - - - - - - - - - - - - - - - - - - - - - - - - - - - - - - %

%-%-%-%-%-%-%-%-%-%-%-%-%-%-%-%-%-%-%-%-%-%-%-%-  figura -%-%-%-%-%-%-%-%

\newcommand{\figResulPulsorhoI}{  % alias

\figura {fig:resul:pulsorhoI}      % label
{fig/resul/pulsorhoI}              % file
{Medida de resistividad (proporcional al voltaje) a corriente
constante para una muestra a temperatura mayor que \Tc. A estas
corrientes tan elevadas, la disipaci\'on causa un calentamiento de
la muestra que se traduce en una variaci\'on del voltaje
le\'{\i}do, an\'aloga a la que se observa en las curvas \EJ\ a
$T<\Tc$. La temperatura parece estabilizarse en
aproximadamente $1\und{ms}$.}                  % caption
{Medida de voltaje a corriente constante para una muestra a
temperatura mayor que \Tc.}                  % toc
{\stfigw}            % width \textwidth
}
% - - - - - - - - - - - - - - - - - - - - - - - - - - - - - - - - - - - %

%-%-%-%-%-%-%-%-%-%-%-%-%-%-%-%-%-%-%-%-%-%-%-%-  figura -%-%-%-%-%-%-%-%

\newcommand{\figResulRhoTJ}{  % alias

\figura {fig:resul:rhoTJ}      % label
{fig/resul/rhoTJ}              % file
{De varias medidas como la de la figura \ref{fig:resul:pulsorhoI}
a altas corrientes y a distintas temperaturas, se obtiene una
curva $\rhoab(T)$ desplazada respecto de la que se mide con
corrientes bajas. En las gr\'aficas de detalle vemos c\'omo el
desplazamiento horizontal en temperatura \DT\ de una respecto de
la otra es mayor cuanto m\'as es la disipaci\'on $W=J^2\;\rhoab$. De
esto se obtiene una correlaci\'on entre \DT\ y $W$.}
% caption
{Curva $\rhoab(T)$ desplazada en temperatura debido al calentamiento de la muestra.}                  % toc
{\stfigw}            % width \textwidth
}
% - - - - - - - - - - - - - - - - - - - - - - - - - - - - - - - - - - - %

%-%-%-%-%-%-%-%-%-%-%-%-%-%-%-%-%-%-%-%-%-%-%-%-  figura -%-%-%-%-%-%-%-%

\newcommand{\figResulDTW}{  % alias

\figura {fig:resul:DTW}      % label
{fig/resul/DTW}              % file
{Correlaci\'on entre la potencia disipada y el incremento de
temperatura de la muestra, en el estado normal, aunque no muy
lejos de la transici\'on. Ver texto para detalles.}
% caption
{Correlaci\'on entre la potencia disipada y el incremento de
temperatura de la muestra.}                  % toc
{\stfigw}            % width \textwidth
}
% - - - - - - - - - - - - - - - - - - - - - - - - - - - - - - - - - - - %

%-%-%-%-%-%-%-%-%-%-%-%-%-%-%-%-%-%-%-%-%-%-%-%-  figura -%-%-%-%-%-%-%-%

\newcommand{\figResulCalent}{  % alias

\figura {fig:resul:calent}      % label
{fig/resul/calent}              % file
{Muestra \fH: arriba, evoluciones temporales del voltaje
experimental al aplicar durante largo rato distintas intensidades
de corriente. La evoluci\'on acusa el calentamiento de la muestra:
al 95\% del valor m\'aximo se llega siempre pasado $\sim 1\und{ms}$.
Abajo, disminuci\'on brusca de la corriente, tras los largos pulsos
anteriores, hasta un valor m\'as bajo, de poca disipaci\'on.}                  % caption
{Evoluciones temporales del voltaje causadas por el calentamiento
y enfriamiento de la muestra.}                  % toc
{\stfigw}            % width \textwidth
}
% - - - - - - - - - - - - - - - - - - - - - - - - - - - - - - - - - - - %

%%%%%%%%%%%%%%%%%%%%%%%%%%%%%%%%%%%%%%%%
\section{Resistividad en funci\'on de la temperatura}
%%%%%%%%%%%%%%%%%%%%%%%%%%%%%%%%%%%%%%%%

\figResulRhoAB
\figResulRhoDG

En las figuras \ref{fig:resul:rhoAB} y \ref{fig:resul:rhoDG} se
muestran las resistividades el\'ectricas de algunos de los
micropuentes empleados en este trabajo. Algunas curvas fueron
medidas tambi\'en con \Index{campo magn\'etico} aplicado, como
caracterizaci\'on complementaria.

Estas medidas de \Index{resistividad} se realizaron aplicando
corrientes bajas, para disminuir todo lo posible la disipaci\'on y
la alteraci\'on de la medida con \Index{calentamiento}. Se
inyectan en el micropuente t\'{\i}picamente $10\und{\mu A}$, lo
que supone una densidad de corriente de $6.7\Exund{2}{\Acms}$.
Disminuir la corriente en un factor cinco no produce variaciones
de la medida, por lo que los efectos t\'ermicos pueden
descartarse. Como las pel\'{\i}culas han sido crecidas en la
direcci\'on cristalogr\'afica $c$, la resistividad que se mide
sobre los micropuentes es la correspondiente al promedio en las
direcciones $a$ y $b$: por este motivo denominamos a esta
resistividad \rhoab. De este tipo de medidas se extraen varios de
los par\'ametros descriptivos de las muestras. Concretamente,
empleamos estas caracterizaciones para determinar la temperatura
cr\'{\i}tica de cada muestra.

\section{Determinaci\'on de la temperatura cr\'{\i}tica \Tc}

\figResulDerivA

Del an\'alisis exclusivo de $\rho(T)$ s\'olo podemos estimar la
localizaci\'on de \Tc\ fij\'andonos en alguna discontinuidad fruto
de la transici\'on superconductora. Ya sea por la existencia de
\Index{inhomogeneidades}\footnote{Ver nota \ref{notaInhomog} de la
p\'agina \pageref{notaInhomog}.} estequiom\'etricas en la muestra
(varios dominios con temperaturas cr\'{\i}ticas diferentes) o,
a\'un en el mejor de los casos, por la inevitable presencia de las
fluctuaciones termodin\'amicas, la transici\'on aparece siempre
redondeada\footnote{La mejor manera de estimar \Tc\ en muestras de
elevada calidad, sin \Index{inhomogeneidades}, es hacer medidas de
un observable menos afectado por fluctuaciones, por ejemplo de la
susceptibilidad magn\'etica con el campo paralelo a los planos
superconductores. Pero ni siquiera los mejores dispositivos
mejoran la resoluci\'on en 0.1~K.}. De las medidas de $\rho(T)$
obtenemos unos cuantos par\'ametros que caracterizan esta
transici\'on. La temperatura m\'as alta \Tcz\ a la que la
\Index{resistividad} es nula, a corrientes bajas y sin campo
magn\'etico aplicado

%, es la que emplearemos como temperatura cr\'{\i}tica \Tc\ de cada muestra.

Otros par\'ametros caracter\'{\i}sticos relevantes son la
temperatura en que la transici\'on alcanza su punto medio, y la
anchura de dicha transici\'on. En la figura \ref{fig:resul:derivA}
se representa un ejemplo de derivada de la \Index{resistividad} en
funci\'on de la temperatura para uno de los micropuentes, y se
muestra c\'omo se determinan estos par\'ametros. El punto de
inflexi\'on de la transici\'on, donde la derivada es m\'axima,
establece la temperatura \Tci. La anchura a mitad de altura (FWHM)
de esta derivada caracteriza la anchura de la transici\'on, dando
el par\'ametro \DTci. Todos estos par\'ametros para las distintas
muestras fueron recopilados previamente en la tabla
\ref{tab:muest:muest} (p. \pageref{tab:muest:muest}).

En general, tomaremos como temperatura cr\'{\i}tica \Tc\ de cada
muestra el valor de \Tci, salvo cuando se indique lo contrario
espec\'{\i}ficamente.

%%%%%%%%%%%%%%%%%%%%%%%%%%%%%%%%%%%%%%%%%%%%%%%%%%%%
\section{Curvas caracter\'{\i}sticas campo el\'ectrico -- densidad de corriente}
%%%%%%%%%%%%%%%%%%%%%%%%%%%%%%%%%%%%%%%%%%%%%%%%%%%%

Las curvas caracter\'{\i}sticas campo el\'ectrico--densidad de
corriente \EJ\ son las curvas caracter\'{\i}sticas \VI\ de las
muestras escaladas a sus correspondientes dimensiones
geom\'etricas, para facilitar la comparaci\'on de los resultados.
De este modo, $E=V/l$ y $J=I/(a\;d)$, donde $l$ es la longitud del
micropuente, y $a$ y $d$ su ancho y espesor respectivamente.

\subsection{Curvas a diferentes temperaturas y campos magn\'eticos}

\figResulEJTA \figResulEJTAup \figResulEJTAlog \figResulEJHA
\figResulEJHAOchCuatro

En las figuras \ref{fig:resul:EJTA}--\ref{fig:resul:EJTAlog} vemos
varias representaciones de las isotermas \EJ\ a campo aplicado
nulo para una de nuestras muestras. Se var\'{\i}a el rango de
representaci\'on para apreciar los distintos detalles. Cada una de
estas curvas isotermas (curvas a constante temperatura \emph{del
\Index{ba\~no}}, que no necesariamente de la muestra) se midi\'o
en dos partes
---con un nanovolt\'{\i}metro (medidas largas en el r\'egimen de baja
disipaci\'on) y con una tarjeta de adquisici\'on (medidas r\'apidas con
alta disipaci\'on)--- que se empalmaron posteriormente en el
an\'alisis de datos, para reconstruir las curvas completas.

Se impone un l\'{\i}mite superior en voltaje a la fuente de
alimentaci\'on, para evitar el deterioro de la muestra por
disipaciones excesivas. Este l\'{\i}mite lo determina la
experiencia, tras haber destruido varios micropuentes por exceso
de calor. Puede variar de una muestra a otra, sobre todo en
funci\'on de la longitud del micropuente, lo que parece indicar
que esta longitud es un par\'ametro directamente relacionado con
la capacidad de refrigeraci\'on del sistema, y merecer\'{\i}a un
estudio detallado. En el caso que se muestra, se fija el campo $E$
m\'aximo en unos $4\Exund{3}{V/cm}$.

En las isotermas m\'as cercanas a \Tc\ se aprecia que la
transici\'on se suaviza, y se pierde el salto abrupto. En la
figura \ref{fig:resul:EJTAup}.b se compara la isoterma medida a
$T=92.1\und{K}$, por encima de \Tc, con la recta \EJ\
correspondiente al valor de la \Index{resistividad} a esa
temperatura, medida a corrientes bajas, de la figura
\ref{fig:resul:rhoAB}. La isoterma \EJ\ experimental se
desv\'{\i}a de esta tendencia recta a corrientes altas,
probablemente debido al aumento de la temperatura de la muestra
causada por la elevada disipaci\'on. Este efecto debe de
manifestarse tambi\'en en las otras isotermas por debajo de \Tc,
as\'{\i} que es probable que todas est\'en afectadas de
calentamiento en las regiones de alta disipaci\'on (principalmente
en el estado normal).

Las figuras \ref{fig:resul:EJHA} y \ref{fig:resul:EJHA84} recogen
otras curvas caracter\'{\i}sticas \EJ\ para esta misma muestra,
pero esta vez manteniendo constante la temperatura y variando el
campo magn\'etico aplicado externamente. La figura
\ref{fig:resul:EJHA} es a $T=76.2\und{K}$, y presenta una zona en
el estado normal, tras el salto en \Jx, con campo $E$ superior al
\Index{umbral} impuesto en nuestra fuente. En la figura
\ref{fig:resul:EJHA84} podemos apreciar esa regi\'on normal para
una temperatura superior $T=84.1\und{K}$, m\'as cerca de \Tc,
donde los voltajes correspondientes son asequibles a nuestro
experimento dentro del l\'{\i}mite fijado. N\'otese que esto
\emph{no} es porque la \Index{resistividad} sea menor cuando crece
la temperatura: para resistividades similares o incluso mayores,
el campo $E=J\;\rhon$ con $J \gtrsim \Jx$ correspondiente a la
regi\'on normal es menor cuando \Jx\ disminuye.

En la figura \ref{fig:resul:EJHA84} representamos tambi\'en la
recta \EJ\ que corresponde a la \Index{resistividad} extrapolada
desde la zona normal hasta $T=84.1\und{K}$, como gu\'{\i}a
comparativa. Los puntos en la regi\'on normal de las curvas \EJ\
se desv\'{\i}an nuevamente de esta tendencia recta, con una
dependencia aparentemente cuadr\'atica respecto de la corriente.

Las curvas \EJ\ de nuestras muestras son an\'alogas a las que en
la bibliograf\'{\i}a presentan muestras similares. Ver, por
ejemplo, las referencias \cite{Wordenweber91,Xiao98,Wen01}. Como
ya mencionamos en la sec. \ref{sec:criostato}, tampoco hay
diferencias entre estas curvas y las medidas con una
refrigeraci\'on l\'{\i}quida en vez de gaseosa
\cite{Curras00b,Curras02,Gonzalez02}, por lo menos mientras la
muestra no transita completamente al estado normal y se vuelve
altamente disipativa.

\subsection{Obtenci\'on de las curvas \EJ}
\label{ssec:medidaEJ}

\figResulMedidaEJ \figResulMedidaEJz \figResulMedidaEJzVI

Las figuras \ref{fig:resul:medidaEJ}--\ref{fig:resul:medidaEJzVI}
muestran un experimento t\'{\i}pico para la determinaci\'on de las
curvas \EJ\ mostradas previamente, en el rango en que la
disipaci\'on exige medidas r\'apidas con la tarjeta adquisidora.
Se aplica una \Index{rampa} de corriente escalonada, y se mide el
voltaje en el micropuente y en la resistencia patr\'on, a una
velocidad de $10^5\und{adquisiciones/s}$ por canal (se hace una
adquisici\'on cada $5\und{\mu s}$, alternando los dos canales: las
medidas no son simult\'aneas). Cuando se produce la transici\'on
en \Jx\ la resistencia de la muestra aumenta, de modo que la
impedancia del \Index{circuito} crece notablemente. La fuente de
corriente acusa este fen\'omeno: si el l\'{\i}mite m\'aximo de
voltaje impuesto es bajo, la corriente cae bruscamente y el estado
normal no es asequible al experimento.

En la regi\'on de mayor disipaci\'on, cerca de la transici\'on, el
voltaje crece suavemente mientras la corriente permanece en un
escal\'on constante: esto se atribuye a un aumento de temperatura de
la muestra. \index{calentamiento} Cuando de estas medidas se
compone la gr\'afica \VI\ (fig. \ref{fig:resul:medidaEJzVI}, que una
vez escalada a las dimensiones de la muestra dar\'a la \EJ), se
aprecia c\'omo los puntos experimentales aumentan su dispersi\'on
cuando estamos cerca de \Jx\ debido a este efecto t\'ermico. Para la
presentaci\'on final de las curvas \EJ\ se eliminan los puntos que
coinciden con un salto de escal\'on de corriente, pues en estos la
incertidumbre y los efectos inductivos son demasiado grandes.

\figResulRampas

La figura \ref{fig:resul:rampas} muestra los efectos de variar la
duraci\'on total de la \Index{rampa}. Como el ancho temporal de
los pelda\~nos que nuestra fuente puede aplicar tiene un
m\'{\i}nimo en algo menos de 1~ms, la duraci\'on de la rampa la
establecemos variando el n\'umero de escalones totales,
haci\'endolos de mayor o menor altura. En esta figura podemos ver
c\'omo el valor de la corriente de salto no se ve especialmente
afectado por el tiempo que dure la rampa, al menos en este rango
en que nos encuadramos. Las rectas de trazos discontinuos unen los
puntos en que se produce la transici\'on, tanto en la gr\'afica de
corriente como en la de voltaje. El valor del voltaje de salto,
sin embargo, s\'{\i} que acusa los efectos t\'ermicos: la
variaci\'on al aumentar el tiempo de la \Index{rampa} en un factor
ocho es del 14\%.

\figResulCompdetalle

Los valores de \Jx\ y de \Ex\ que emplearemos en los an\'alisis
posteriores los extraemos del \'ultimo punto de la curva \EJ\ que
medimos antes de la transici\'on. Cometemos por tanto un
\Index{error} en esta determinaci\'on relacionado con la resoluci\'on
de la \Index{rampa} de corriente que aplicamos: dependiendo de la
altura de los escalones que apliquemos, determinaremos con mayor o
menor exactitud estos valores. Pero dentro de los rangos
temporales en que nos movemos, aumentar la resoluci\'on de la rampa
de corriente no mejora mucho la determinaci\'on de \Jx. En la figura
\ref{fig:resul:compdetalle} comparamos los valores de \Jx\ que se
determinaron con rampas de distinta resoluci\'on, que se refleja en
la barra de \Index{error} de cada punto. Los correspondientes a
las rampas con m\'as detalle est\'an sistem\'aticamente desplazados
hacia valores m\'as bajos de \Jx. Esto se debe probablemente a
efectos de \Index{calentamiento} de la muestra: las rampas m\'as
detalladas son tambi\'en las m\'as largas temporalmente, y calientan
por m\'as tiempo la muestra, incluso si las iniciamos directamente
en escalones muy cercanos al valor esperado de \Jx.

En el l\'{\i}mite opuesto, empleando pulsos aislados, de
sucesivamente mayor corriente, se obtienen valores compatibles con
estos: como la lectura del voltaje muy cerca de \Jx\ se ve
inevitablemente afectada de calentamiento, se hace dif\'{\i}cil
saber cu\'al es exactamente el \'ultimo valor \emph{previo} a la
transici\'on. Como hemos visto, esta influencia del calentamiento
se manifiesta sobre todo en la incertidumbre de \Ex, pero es de
suponer que indirectamente afecta tambi\'en al valor de \Jx.

Tras todas estas consideraciones, hemos empleado rampas de
corriente de duraci\'on y resoluci\'on que compensan en lo posible
estos problemas (el calentamiento por un lado, y la falta de
resoluci\'on por el otro). De este modo, hemos procurado que los
valores de \Jx\ que mostramos tengan un \Index{error} inferior al
5\%, entendiendo \'este como la dispersi\'on de los resultados al
emplear diferentes par\'ametros de \Index{rampa} dentro de los
rangos que permiten nuestros dispositivos (sin considerar otros
errores sistem\'aticos, como los provenientes de la caracterizaci\'on
geom\'etrica de las muestras).

%%%%%%%%%%%%%%%%%%%%%%%%%%%%%%%%%%%%%%%%%%%%%%%%%%%%
\section{Consideraciones sobre el calentamiento en torno al salto}
%%%%%%%%%%%%%%%%%%%%%%%%%%%%%%%%%%%%%%%%%%%%%%%%%%%%

\subsection{Estimaci\'on del aumento de temperatura durante la medida}

Hemos realizado medidas de la correlaci\'on entre la potencia
disipada y el incremento de temperatura, en una muestra de igual
geometr\'{\i}a a las empleadas en el resto de las medidas, en el
estado normal, aunque no muy lejos de la transici\'on. En el
estado normal se puede suponer que la \Index{resistividad} no
depende de la corriente aplicada (lejos de la regi\'on m\'as
afectada de fluctuaciones t\'ermicas, y suponiendo que no hay
magnetorresistencia), y atribuir toda diferencia que se observe en
la resistividad, dependiendo de la corriente aplicada, a un
aumento de temperatura.

De varias medidas de la evoluci\'on de la \Index{resistividad} en
funci\'on del tiempo como la de la figura
\ref{fig:resul:pulsorhoI}, a altas corrientes aplicadas y a
distintas temperaturas, se obtiene una curva $\rhoab(T)$ como la
de la figura \ref{fig:resul:rhoTJ} (l\'{\i}nea de puntos),
desplazada respecto de las que se miden con corrientes bajas
(l\'{\i}nea continua) en las que se supone que no hay
\Index{calentamiento} apreciable. El desplazamiento horizontal en
temperatura \DT\ de una respecto de la otra es mayor cuanto m\'as
es la disipaci\'on $W=J^2\;\rhoab$. De esto se obtiene una
correlaci\'on de \DT\ con la potencia disipada por unidad de
volumen $W = E \times J = \rhoab \times J^2$. Si no involucramos
en esta correlaci\'on variables temporales, esto s\'olo tiene
sentido en estados estacionarios, cuando $\rho$ (que es
proporcional a $E$) est\'a bien definida; es decir, cuando pasado
cierto tiempo, se estabiliza. En estas medidas (fig.
\ref{fig:resul:pulsorhoI}), el voltaje parece estabilizarse en
menos de 1~ms, y es este valor estable el que tomamos para
determinar $E$ (del que extraemos $W$) y \DT.

Los resultados de la correlaci\'on cerca de la transici\'on se
muestran en la figura \ref{fig:resul:DTW}. Si extrapolamos este
comportamiento t\'ermico hasta $T<\Tc$, a regiones
superconductoras\footnote{Esta extrapolaci\'on parece en principio
una buena aproximaci\'on, pues los principales mecanismos
t\'ermicos no var\'{\i}an notablemente por volverse
superconductora la muestra: lo m\'as relevante es la interacci\'on
con el medio, principalmente el substrato. En todo caso la
superconductividad mejora la \Index{conductividad t\'ermica} del
micropuente, por lo que esta correlaci\'on dar\'{\i}a una cota
superior al aumento de temperatura.}, tenemos que para las
potencias t\'{\i}picas disipadas antes del salto, $\Wx \simeq
10^7\und{\Wcmc}$, el aumento de temperatura m\'aximo ser\'{\i}a de
$0.5\und{K}$. Esto dar\'{\i}a pie a descartar el calentamiento
homog\'eneo como causa de la transici\'on abrupta: no parece, de
estas medidas, que se est\'e produciendo un aumento de temperatura
de la muestra de decenas de grados, hasta elevarse por encima de
\Tc. Sin embargo, veremos en los cap\'{\i}tulos \ref{cap:cal} y
\ref{cap:calpr} que esta conclusi\'on es precipitada, y que hace
falta un an\'alisis m\'as detallado del comportamiento t\'ermico
del sistema para poder descartar por completo el calentamiento.

\figResulPulsorhoI \figResulRhoTJ \figResulDTW

\subsection{Tiempos t\'ermicos caracter\'{\i}sticos}

\figResulCalent

En la figura \ref{fig:resul:calent}.a ponemos de manifiesto los
estados casi estacionarios que se alcanzan por debajo de \Jx, a
$J$ constante, tras un calentamiento inicial de la muestra.
Aplicamos aqu\'{\i} pulsos m\'as largos que los hasta ahora
mostrados, y vemos que efectivamente el valor del voltaje ---y por
lo tanto el de la temperatura--- se estabiliza pasado
aproximadamente un milisegundo, con independencia de la
disipaci\'on que se genere: es un tiempo que caracteriza la
inercia t\'ermica del sistema.\index{tiempo de inercia t\'ermica}g

Esta muestra \fH\ presenta una transici\'on superconductora m\'as
ancha que otras muestras, lo que pone de manifiesto su menor
homogeneidad. Estas \Index{inhomogeneidades} podr\'{\i}an ser las
causantes de que los tiempos t\'ermicos se prolonguen, al aparecer
disipaci\'on extra, comparados con los de muestras de mejor
calidad. En la figura \ref{fig:resul:pulsorhoI}, por ejemplo,
ve\'{\i}amos c\'omo la mayor parte de la variaci\'on de
temperatura se produce en los primeros $0.3\und{ms}$, siendo
posteriormente casi constante.

Tras aplicar durante 200~ms esas corrientes elevadas, disminuimos
$J$ bruscamente hasta un valor m\'as bajo
$J_r=0.3\Exund{5}{\Acms}$, siempre el mismo, en la figura
\ref{fig:resul:calent}.b. (Esta densidad de corriente $J_r$ era,
antes del experimento, del orden de \Jc, de modo que no generaba
una disipaci\'on muy alta y no se apreciaba calentamiento al
aplicarla). Vemos, en contraste con la medida de calentamiento de
la figura \ref{fig:resul:calent}.a, que la velocidad de
enfriamiento es ahora dependiente del valor de la corriente
aplicada \emph{previamente}, y ocurre en tiempos notablemente
m\'as largos. La explicaci\'on de este resultado podr\'{\i}a ser
que esta corriente residual $J_r$ es ahora m\'as disipativa de lo
que era \emph{antes} del experimento, pues la \Index{resistividad}
de la muestra ha aumentado tras el calentamiento previo. Como el
aumento de temperatura ---y por lo tanto de resistividad--- est\'a
relacionado con cu\'anta corriente est\'abamos inyectando, $J_r$
es m\'as disipativa en unos casos que en otros, y la velocidad de
enfriamiento presenta tiempos caracter\'{\i}sticos diferentes. En
cualquier caso, estos tiempos son m\'as largos que en el caso del
calentamiento, donde las diferencias entre una disipaci\'on y otra
eran menores, relativamente hablando.

%%%(j-) esto no me convence, hay que pensarlo mejor. Si la disipaci\'on
%%%ralentiza el enfriamiento, tambi\'en deber\'{\i}a ralentizar, incluso
%%%m\'as, el calentamiento. Y sin embargo digo que el calentamiento se
%%%estabiliza siempre en torno a 0.3 ms...
%%% O es que las diferencias son m\'as peque\~nas en el caso de alta disipaci\'on,
%%% y se notan m\'as las peque\~nas diferencias aqu\'{\i} abajo...
  % cap 4
\clearemptydoublepage

%%%%%%%%%%%%%%%%%%%%%%%%%%%%%%%%%%%%%%%%%%%%%%%%%%%%%%%%%%%%%%%%%%
\chapter{An\'alisis de datos y comparaci\'on con los modelos te\'oricos}
%\chaptermark{An\'alisis de datos y comparaci\'on con los modelos te\'oricos}
%\addtocontents{toc}{\protect\vspace{0.2cm}}
\label{cap:datos}
%%%%%%%%%%%%%%%%%%%%%%%%%%%%%%%%%%%%%%%%%%%%%%%%%%%%%%%%%%%%%%%%%%

%-%-%-%-%-%-%-%-%-%-%-%-%-%-%-%-%-%-%-%-%-%-%-%-  figura -%-%-%-%-%-%-%-%

\newcommand{\figDatosEJmJc}{  % alias

\figura {fig:datos:EJmJc}      % label
{fig/datos/EJmJc}              % file
{Campo el\'ectrico de la muestra \fA\ a distintos campos magn\'eticos
aplicados y a temperatura constante. La representaci\'on es en
$J-\Jc(H)$, para hacer coincidir el arranque de todas las curvas
en el mismo punto. La dependencia no lineal respecto de la
corriente descarta que el r\'egimen din\'amico de v\'ortices sea de
\emph{free
flux flow}.}                  % caption
{Curvas de $E$ frente a $J-\Jc(H)$ de la muestra \fA, a distintos campos magn\'eticos y a $T=76\und{K}$.}                  % toc
{\stfigw}            % width \textwidth
}
% - - - - - - - - - - - - - - - - - - - - - - - - - - - - - - - - - - - %

%-%-%-%-%-%-%-%-%-%-%-%-%-%-%-%-%-%-%-%-%-%-%-%-  figura -%-%-%-%-%-%-%-%

\newcommand{\figDatosEJmodelo}{  % alias

\figura {fig:datos:EJmodelo}      % label
{fig/datos/EJmodelo}              % file
{Comparaci\'on del modelo que se propone en la ecuaci\'on \ref{ec:EJT}
con los datos experimentales de la muestra \fA, con $B=0$. La
temperatura inicial se indica al lado de cada curva. Los
par\'ametros empleados son $\Ez(1) = 1.04\Exund{-2}{V/cm}$, $\Jz(1)
= 1.26\Exund{7}{\Acms}$, $n = 5.86$, $g = 1.52$, $f = 1.26$ y $\Tc
= 89.8\und{K}$}                  % caption
{Comparaci\'on del modelo propuesto para las curvas \EJ\ con los datos experimentales, con $B=0$.}                  % toc
{\stfigw}            % width \textwidth
}
% - - - - - - - - - - - - - - - - - - - - - - - - - - - - - - - - - - - %

%-%-%-%-%-%-%-%-%-%-%-%-%-%-%-%-%-%-%-%-%-%-%-%-  figura -%-%-%-%-%-%-%-%

\newcommand{\figDatosEJmodeloB}{  % alias

\figura {fig:datos:EJmodeloB1}      % label
{fig/datos/EJmodeloB1}              % file
{Comparaci\'on del modelo que se propone en la ecuaci\'on
\ref{ec:EJTB1} con los datos experimentales de la muestra \fA, con
$\mH=1\und{T}$. Los par\'ametros son $\Jz(1) =
8.00\Exund{6}{\Acms}$,
$n = 2.85$, $f = 1.08$ y $\Tc = 89.8\und{K}$}                  %caption
{Comparaci\'on del modelo propuesto para las curvas \EJ\ con los datos experimentales, con $\mH=1\und{T}$.}                  % toc
{\stfigw}            % width \textwidth
}
% - - - - - - - - - - - - - - - - - - - - - - - - - - - - - - - - - - - %

%-%-%-%-%-%-%-%-%-%-%-%-%-%-%-%-%-%-%-%-%-%-%-%-  figura -%-%-%-%-%-%-%-%

\newcommand{\figDatosJcpin}{  % alias

\figura {fig:datos:Jcpin}      % label
{fig/datos/Jcpin}              % file
{Determinaci\'on de las corrientes cr\'{\i}ticas \Jcpin\ y \Jc.
Abajo, \Jc\ se determina estableciendo un umbral para el campo $E$
de $10\und{\mu V/cm}$ (l\'{\i}nea horizontal). Arriba, \Jcpin\ se
determina con un umbral en la velocidad de movimiento de
v\'ortices, fij\'andolo de tal modo que \Jcpin\ y \Jc\ coincidan
en la curva medida con m\'as campo. Para las restantes curvas a
campo menor hay que hacer una extrapolaci\'on ---pues los datos
son ruidosos en esos rangos--- de modo que \Jcpin\ es siempre
menor que \Jc.}
% caption
{Determinaci\'on de las corrientes cr\'{\i}ticas \Jcpin\ y \Jc}                  % toc
{\stfigw}            % width \textwidth
}
% - - - - - - - - - - - - - - - - - - - - - - - - - - - - - - - - - - - %

%-%-%-%-%-%-%-%-%-%-%-%-%-%-%-%-%-%-%-%-%-%-%-%-  figura -%-%-%-%-%-%-%-%

\newcommand{\figDatosJcJcpin}{  % alias

\figura {fig:datos:JcJcpin}      % label
{fig/datos/JcJcpin}              % file
{Comparaci\'on de la corriente cr\'{\i}tica \Jc, definida con un
umbral en el campo $E$ de $10\und{\mu V/cm}$, con \Jcpin, definida
con un umbral en la velocidad de los v\'ortices $v$ (fig.
\ref{fig:datos:Jcpin}). \Jcpin\ es, en general, ligeramente menor
que \Jc, aunque sus dependencias respecto de $T$ y $H$ son muy parecidas. Las curvas corresponden a $T=$ 76.2, 80.4 y 84.1~K.} % caption
{Comparaci\'on de la corriente cr\'{\i}tica \Jc\ con \Jcpin.}                  % toc
{\stfigw}            % width \textwidth
}
% - - - - - - - - - - - - - - - - - - - - - - - - - - - - - - - - - - - %

%-%-%-%-%-%-%-%-%-%-%-%-%-%-%-%-%-%-%-%-%-%-%-%-  figura -%-%-%-%-%-%-%-%

\newcommand{\figDatosJcJxTcompA}{  % alias

\figura {fig:datos:JcJxTcompA}      % label
{fig/datos/JcJxTcompA}              % file
{Corriente cr\'{\i}tica \Jc\ y de salto \Jx\ en funci\'on de la
temperatura, a campo magn\'etico aplicado nulo. Ambas dependencias
se ajustan con un modelo de exponente $3/2$ en temperatura
reducida (ec. \ref{ec:JT32}). Los par\'ametros para \Jx\ son
$J_0=80.7 \times 10^6$, $b=0.97$, y $J_0=41.2 \times 10^6$,
$b=0.99$ para \Jc.}
% caption
{Corriente cr\'{\i}tica \Jc\ y de salto \Jx\ en funci\'on de la temperatura, a campo magn\'etico aplicado nulo.}                  % toc
{\stfigw}            % width \textwidth
}
% - - - - - - - - - - - - - - - - - - - - - - - - - - - - - - - - - - - %

%-%-%-%-%-%-%-%-%-%-%-%-%-%-%-%-%-%-%-%-%-%-%-%-  figura -%-%-%-%-%-%-%-%

\newcommand{\figDatosJcJxTA}{  % alias

\figura {fig:datos:JcJxTA}      % label
{fig/datos/JcJxTA}              % file
{Cociente de la corriente cr\'{\i}tica entre la de salto
$\Jc/\Jx$, en funci\'on de la temperatura, a campo aplicado nulo.
Se incluyen tambi\'en unos cuantos puntos tomados a campos
distinto de cero, para apuntar la tendencia que presentar\'{\i}an
esas curvas. Las l\'{\i}neas son gu\'{\i}as para los ojos.}
% caption
{Cociente de la corriente cr\'{\i}tica entre la de salto $\Jc/\Jx$, a campo aplicado nulo.}                  % toc
{\stfigw}            % width \textwidth
}
% - - - - - - - - - - - - - - - - - - - - - - - - - - - - - - - - - - - %

%-%-%-%-%-%-%-%-%-%-%-%-%-%-%-%-%-%-%-%-%-%-%-%-  figura -%-%-%-%-%-%-%-%

\newcommand{\figDatosJxTAE}{  % alias

\figura {fig:datos:JxTAE}      % label
{fig/datos/JxTAE}              % file
{Corriente de salto \Jx\ en funci\'on de la temperatura para dos
muestras distintas, a $B=0$. Ambas dependencias se ajustan con un
modelo de exponente $3/2$ en temperatura reducida $\eps=1-T/\Tc$
(ec. \ref{ec:JT32}, par\'ametros $J_0=80.7\Exund{6}{\Acms}$,
$b=0.97$
---muestra \fA--- y $J_0=102.6\Exund{6}{\Acms}$, $b=0.98$ ---muestra \fE---).
Un exponente $1/2$ queda, sin embargo, descartado (ec.
\ref{ec:JT12}, $J_0=14.7\Exund{6}{\Acms}$, $b=1.04$).}
% caption
{Corriente de salto \Jx\ en funci\'on de la temperatura para dos muestras distintas.}                  % toc
{\stfigw}            % width \textwidth
}
% - - - - - - - - - - - - - - - - - - - - - - - - - - - - - - - - - - - %

%-%-%-%-%-%-%-%-%-%-%-%-%-%-%-%-%-%-%-%-%-%-%-%-  figura -%-%-%-%-%-%-%-%

\newcommand{\figDatosJcJxvarias}{  % alias

\figura {fig:datos:JcJxvarias}      % label
{fig/datos/JcJxvarias}              % file
{Corriente cr\'{\i}tica \Jcpin\ y de salto \Jx\ de varias
muestras, en funci\'on del campo magn\'etico aplicado y para
varias temperaturas. La corriente cr\'{\i}tica \Jcpin\ parece
acusar antes que \Jx\ la influencia del campo externo: var\'{\i}a
para campos peque\~nos, mientras que, hasta aproximadamente
$\mH=0.05\und{T}$, \Jx\ apenas se altera. Las l\'{\i}neas son
gu\'{\i}as para los ojos.}
% caption
{Corriente cr\'{\i}tica \Jcpin\ y de salto \Jx\ de varias muestras, en funci\'on del campo magn\'etico aplicado y para varias temperaturas.}                  % toc
{\stfigw}            % width \textwidth
}
% - - - - - - - - - - - - - - - - - - - - - - - - - - - - - - - - - - - %

%-%-%-%-%-%-%-%-%-%-%-%-%-%-%-%-%-%-%-%-%-%-%-%-  figura -%-%-%-%-%-%-%-%

\newcommand{\figDatosJcEJxH}{  % alias

\figura {fig:datos:JcEJxH}      % label
{fig/datos/JcEJxH}              % file
{En (a), cociente de la corriente cr\'{\i}tica y la de salto
\Jcpin/\Jc\ en funci\'on del campo magn\'etico, para distintas
muestras y temperaturas. (Datos de la figura
\ref{fig:datos:JcJxvarias}). Se pone de manifiesto que \Jcpin\
var\'{\i}a m\'as r\'apidamente que \Jx\ a campos bajos, y s\'olo a
partir de unos $\mH=0.1\und{T}$ var\'{\i}an a la par. Las
tendencias de todas las curvas para las distintas muestras y
temperaturas son similares: en (b) se representa la derivada de
las curvas de (a) respecto del campo aplicado, y se ve
como todas son coincidentes. Las l\'{\i}neas son gu\'{\i}as para los ojos.}                  % caption
{Cociente de la corriente cr\'{\i}tica y la de salto $\Jcpin/\Jc$ en funci\'on del campo magn\'etico.}                  % toc
{\stfigw}            % width \textwidth
}
% - - - - - - - - - - - - - - - - - - - - - - - - - - - - - - - - - - - %

%-%-%-%-%-%-%-%-%-%-%-%-%-%-%-%-%-%-%-%-%-%-%-%-  figura -%-%-%-%-%-%-%-%

\newcommand{\figDatoscampoHI}{  % alias

\figura {fig:datos:campoHI}      % label
{fig/datos/campoHI}              % file
{Intensidad del campo magn\'etico creado por una corriente que
circula por una muestra infinita de secci\'on rectangular.
(Blanco: m\'as intenso. Negro: intensidad nula). El perfil de la
secci\'on de la muestra son los rect\'angulos de l\'{\i}nea
delgada. Arriba, la corriente est\'a uniformemente distribuida en
la secci\'on. Abajo, la corriente se concentra s\'olo en los dos
filetes laterales de la muestra, de secci\'on cuadrada. El campo
es m\'aximo en los laterales en ambos casos, pero cuando la
corriente se concentra, la intensidad es mayor en esos puntos que
cuando se reparte.}                  % caption
{Campo magn\'etico creado por una corriente que circula por una muestra infinita de secci\'on rectangular.}                  % toc
{\stfigw}            % width \textwidth
}
% - - - - - - - - - - - - - - - - - - - - - - - - - - - - - - - - - - - %

%-%-%-%-%-%-%-%-%-%-%-%-%-%-%-%-%-%-%-%-%-%-%-%-  figura -%-%-%-%-%-%-%-%

\newcommand{\figDatosExJxHTA}{  % alias

\figura {fig:datos:ExJxHTA}      % label
{fig/datos/ExJxHTA}              % file
{Campo \Ex\ y corriente \Jx\ de salto en funci\'on del campo
aplicado de la muestra \fA, para varias temperaturas. Se indica la
incertidumbre del campo \Ex, que se vuelve notable a campos altos,
cuando la disipaci\'on en el momento del salto es mayor. En (a)
las l\'{\i}neas son gu\'{\i}as para la vista que unen los puntos
experimentales: son pr\'acticamente \emph{paralelas},
diferenciadas s\'olo por un factor multiplicativo. En (b), las
l\'{\i}neas son ajustes a la ecuaci\'on \ref{ec:BSJx}, teniendo en
cuenta el autocampo seg\'un se explica en el texto. \Ex\
var\'{\i}a un orden de magnitud
mientras que \Jx\ apenas lo hace un factor dos.}                  % caption
{Campo \Ex\ y corriente \Jx\ de salto en funci\'on del campo aplicado, para varias muestras y temperaturas.}                  % toc
{\stfigw}            % width \textwidth
}
% - - - - - - - - - - - - - - - - - - - - - - - - - - - - - - - - - - - %

%-%-%-%-%-%-%-%-%-%-%-%-%-%-%-%-%-%-%-%-%-%-%-%-  figura -%-%-%-%-%-%-%-%

\newcommand{\figDatosExHTAfit}{  % alias

\figura {fig:datos:ExHTAfit}      % label
{fig/datos/ExHTAfit}              % file
{Campo \Ex\ y corriente \Jx\ de salto en funci\'on del campo
aplicado de la muestra \fA, para $T=76.2\und{K}$. El ajuste de la
ecuaciones \ref{ec:BSEx} y \ref{ec:BSJx} en funci\'on del campo
aplicado (l\'{\i}nea discontinua) es v\'alido para \Jx\ pero no
para \Ex. Sin embargo, el modelo ajusta bien ambas curvas si
consideramos que el campo total es el externo aplicado m\'as el
autocampo de la corriente inyectada, dado por un
t\'ermino $c\;I^*$.}                  % caption
{Campo \Ex\ y corriente \Jx\ de salto en funci\'on del campo: ajuste con el modelo BS.}                  % toc
{\stfigw}            % width \textwidth
}
% - - - - - - - - - - - - - - - - - - - - - - - - - - - - - - - - - - - %

%-%-%-%-%-%-%-%-%-%-%-%-%-%-%-%-%-%-%-%-%-%-%-%-  figura -%-%-%-%-%-%-%-%

\newcommand{\figDatosExJxHvarias}{  % alias

\figura {fig:datos:ExJxHvarias}      % label
{fig/datos/ExJxHvarias}              % file
{Campo \Ex\ y corriente \Jx\ de salto en funci\'on del campo
aplicado, para varias muestras y temperaturas. Al aumentar la
temperatura, \Ex\ crece levemente y \Jx\ disminuye. La
dispersi\'on en valores de \Ex\ es la mitad que la de \Jx, pero
\Ex\ var\'{\i}a un orden de magnitud mientras que \Jx\ apenas un
factor dos. Las l\'{\i}neas que unen los puntos son s\'olo
gu\'{\i}as para la vista.}
% caption
{Campo \Ex\ y corriente \Jx\ de salto en funci\'on del campo aplicado, para varias muestras y temperaturas.}                  % toc
{\stfigw}            % width \textwidth
}
% - - - - - - - - - - - - - - - - - - - - - - - - - - - - - - - - - - - %

%-%-%-%-%-%-%-%-%-%-%-%-%-%-%-%-%-%-%-%-%-%-%-%-  figura -%-%-%-%-%-%-%-%

\newcommand{\figDatosJxJcNA}{  % alias

\figura {fig:datos:JxJcNA}      % label
{fig/datos/JxJcNAD}              % file
{Normalizaci\'on, al correspondiente valor con campo aplicado nulo,
de las corrientes \Jx\ y \Jcpin\ en funci\'on del campo y para
varias temperaturas. La dependencia de \Jx\ coincide para todas
las temperaturas, y se ajusta con la ecuaci\'on \ref{ec:JcXu} con
$H_0(T)=0.36$, $n=0.63$. Por el contrario, \Jcpin\ no escala tan
bien, y necesita valores de $H_0$ y $n$ diferentes para cada
temperatura (0.13 y 1.29 en 76.2~K; 0.077 y 1.20 en 80.4~K; 0.036
y 1.07 en 84.1~K), y ni siquiera de esa forma el ajuste es bueno.
Este comportamiento se repite en las otras muestras estudiadas.}
% caption
{Normalizaci\'on de las corrientes \Jx\ y \Jcpin\ en funci\'on del campo y para varias temperaturas.}                  % toc
{\stfigw}            % width \textwidth
}
% - - - - - - - - - - - - - - - - - - - - - - - - - - - - - - - - - - - %

%-%-%-%-%-%-%-%-%-%-%-%-%-%-%-%-%-%-%-%-%-%-%-%-  figura -%-%-%-%-%-%-%-%

\newcommand{\figDatosEtJ}{  % alias

\figura {fig:datos:EtJ}      % label
{fig/datos/EtJ}              % file
{Experimento realizado aplicando corrientes constantes, de valores
muy parecidos entre s\'{\i}, cercanas a la corriente de salto. Se
mide la evoluci\'on temporal del voltaje. La corriente de
$6.09\und{\Acms}$ lleva a un voltaje aparentemente estable en
tiempos mayores de 1~ms, pero la de 6.13 (un 0.6\% mayor) causa
una avalancha tan pronto se aplica (dentro de la resoluci\'on
temporal de la medida). Los valores intermedios de corriente
conllevan una transici\'on en momentos diferentes, m\'as pronto
cuanto
m\'as alta es $J$.}                  % caption
{Evoluci\'on temporal del voltaje aplicando distintas corrientes
cercanas a \Jx.}                  % toc
{\stfigw}            % width \textwidth
}
% - - - - - - - - - - - - - - - - - - - - - - - - - - - - - - - - - - - %

%-%-%-%-%-%-%-%-%-%-%-%-%-%-%-%-%-%-%-%-%-%-%-%-  figura -%-%-%-%-%-%-%-%

\newcommand{\figDatosWxT}{  % alias

\figura {fig:datos:WxT}      % label
{fig/datos/WxT}              % file
{Variaci\'on de la potencia disipada justo antes del salto \Wx\ con la temperatura de ba\~no. La l\'{\i}nea es una gu\'{\i}a para la vista que resalta la tendencia descendente.}                  % caption
{Variaci\'on de la potencia disipada \Wx\ con la temperatura de ba\~no.}                  % toc
{\stfigw}            % width \textwidth
}
% - - - - - - - - - - - - - - - - - - - - - - - - - - - - - - - - - - -%

%-%-%-%-%-%-%-%-%-%-%-%-%-%-%-%-%-%-%-%-%-%-%-%-  figura -%-%-%-%-%-%-%-%

\newcommand{\figDatosWxHA}{  % alias

\figura {fig:datos:WxHA}      % label
{fig/datos/WxHA}              % file
{Variaci\'on de la potencia disipada justo antes del salto \Wx\ con el campo magn\'etico externo aplicado.}                  % caption
{Variaci\'on de la potencia disipada \Wx\ con el campo magn\'etico externo aplicado.}                  % toc
{\stfigw}            % width \textwidth
}
% - - - - - - - - - - - - - - - - - - - - - - - - - - - - - - - - - - - %

En este cap\'{\i}tulo veremos en primer lugar la forma funcional
que presentan las curvas caracter\'{\i}sticas \EJ\ dependiendo de
la temperatura y el campo magn\'etico. Esto nos servir\'a
posteriormente para el estudio del salto en \Jx, viendo c\'omo se
desv\'{\i}an los datos respecto de la tendencia general al llegar
a la regi\'on de altas corrientes.

Haremos tambi\'en un repaso a los principales modelos te\'oricos
que se emplean para explicar esta ruptura abrupta del estado
superconductor. Veremos c\'omo var\'{\i}an los distintos
par\'ametros cr\'{\i}ticos (\Jc, \Jx\ y \Ex, principalmente) en
funci\'on de la temperatura y el campo magn\'etico aplicado, e
intentaremos encajar estas dependencias en algunos de los modelos
mencionados. Veremos que el efecto del autocampo magn\'etico (el
campo generado por la corriente inyectada) es determinante en la
regi\'on de bajos campos aplicados.

Por \'ultimo, estudiaremos la dependencia de la potencia disipada
antes del salto $\Wx=\Ex \times \Jx$ respecto de la temperatura y
el campo magn\'etico. Veremos c\'omo estas dependencias llevan a
ciertas conclusiones acerca de la naturaleza de la transici\'on en
\Jx, conclusiones que no tienen por qu\'e ser necesariamente ciertas
si no se hacen con cuidado.

%%%%%%%%%%%%%%%%%%%%%%%%%%%%%%%%%%%%%%%%
\section{Dependencia del voltaje respecto de la temperatura y la corriente}
\sectionmark{Dependencia del voltaje respecto de $T$ y $J$}
%%%%%%%%%%%%%%%%%%%%%%%%%%%%%%%%%%%%%%%%

\figDatosEJmJc

\figDatosEJmodelo

\figDatosEJmodeloB

Para el estudio de las corrientes de salto que posteriormente
haremos en los cap\'{\i}tulos \ref{cap:cal} y \ref{cap:calpr}
necesitaremos conocer c\'omo es la dependencia del campo $E$
respecto de la corriente aplicada $J$ y de la temperatura de la
muestra. Emplearemos para ajustar nuestros datos algunos modelos
habituales que describen las curvas \EJ, pero debemos se\~nalar
que \emph{no estamos interesados} aqu\'{\i} en el significado de
estos modelos: no queremos estudiar los mecanismos f\'{\i}sicos
que producen disipaci\'on en el estado mixto\footnote{Las curvas
\EJ\ del estado mixto en HTSC han sido estudiadas en profundidad
en m\'ultiples trabajos, y concretamente para pel\'{\i}culas de
\YBCO\
\cite{Hettinger89,Zeldov90,Bernstein90,Wordenweber90,Tinkham91,Gupta93,Kunchur93,Roberts94,Woltgens95,Kilic98,Prester98,Wang00,Wen01,Landau00,Landau01,Landau02}.
Para una revisi\'on de los mecanismos implicados remitimos
tambi\'en a la tesis doctoral de M.~T. Gonz\'alez
\cite{Gonzalez03}}, sino, dada una dependencia para el voltaje
experimental, ver c\'omo se producen desviaciones respecto de la
tendencia en el r\'egimen de alta disipaci\'on, debido a los
mecanismos que causan la transici\'on al estado normal.

En la figura \ref{fig:datos:EJmJc} se representan los valores de
$E$ en funci\'on de $J-\Jc(H)$, para hacer coincidir todas las
curvas \EJ\ en el punto en que comienza la disipaci\'on. La forma
curvada hacia arriba que presentan es caracter\'{\i}stica de
\emph{\Index{flux creep}}
\cite{Tinkham91,Tinkham96,Landau00,Landau01,Landau02}, no se
explica con una din\'amica de \Index{v\'ortices} simple, de
\emph{\Index{free flux flow}}: en este caso los datos mostrados
deber\'{\i}an seguir una tendencia rectil\'{\i}nea.

Al analizar los datos correspondientes a la muestra \fA\ con $B=0$
\mbox{---mostrados} previamente en la figura \ref{fig:resul:EJTA}
y ahora de nuevo, resumidamente, en la figura
\mbox{\ref{fig:datos:EJmodelo}---,} encontramos que un modelo
cr\'{\i}tico en ley de potencias (\emph{breakdown power law},
motivada por el \emph{flux creep} de los v\'ortices)
\cite{Prester98,Gonzalez02} unifica \emph{todas} las isotermas
\EJ\ en el rango que estamos estudiando, $J>\Jc$ y $77\und{K}
\lesssim T <\Tc$. Este modelo da el campo el\'ectrico, \emph{a
temperatura constante}, con una funci\'on de tres par\'ametros
libres: \be E(J)= \Ez \; \left(\frac{J}{\Jz}-1\right)^n.
\label{ec:EJT} \ee La dependencia respecto de la temperatura
est\'a impl\'{\i}cita en los par\'ametros \Ez\ y \Jz: de nuestros
ajustes a los datos experimentales vemos que var\'{\i}an con la
temperatura reducida $\eps \equiv \ln(\Tc/T)$ como \be \Ez(\eps) =
\Ez(1) \; \eps^g \label{ec:parEc}\ee \be \Jz(\eps) = \Jz(1) \;
\eps^f. \label{ec:parJz}\ee Insertando las ecuaciones
\ref{ec:parEc} y \ref{ec:parJz} en la ecuaci\'on \ref{ec:EJT}
obtenemos una \'unica funci\'on que nos da el valor de $E(J,T)$.
En estas ecuaciones, los ajustes son \'optimos cuando se toma como
valor de \Tc\ el par\'ametro \Tcz\ ---la temperatura m\'axima a la
que la \Index{resistividad} es cero, sin campo magn\'etico
aplicado y a corrientes bajas---, en vez de \Tci\ como hacemos en
general.

Tambi\'en necesitaremos una expresi\'on an\'aloga para los resultados
con campo magn\'etico aplicado. Para $\mH=1\und{T}$, por ejemplo,
(figura \ref{fig:datos:EJmodeloB1}) encontramos que cada curva
isoterma \EJ\ se describe mejor con una ley de potencias
convencional \cite{Prester98,Gonzalez02}, de la forma
\be E(J)= \left(\frac{J}{\Jz}\right)^n\label{ec:EJTB1} \ee
donde de nuevo la dependencia respecto de la temperatura la recoge
el par\'ametro \Jz, con la misma ecuaci\'on \ref{ec:parJz}.

El que las funciones que describen las curvas \EJ\ sean tan
diferentes dependiendo de que haya o no campo aplicado
\index{campo magn\'etico} (ecuaciones \ref{ec:EJT} y \ref{ec:EJTB1})
se entiende si pensamos en cu\'al es el origen de la disipaci\'on. El
campo el\'ectrico $E$ por encima de \Jc\ se origina por el
movimiento de los v\'ortices magn\'eticos del estado mixto en
presencia de una corriente aplicada \cite[cap. 13]{Rose-Innes78}.
En el caso $\mH=1\und{T}$, los \Index{v\'ortices} dentro de la
muestra son principalmente debidos al campo externo aplicado, y
est\'an presentes independientemente de cu\'al sea la $J$ inyectada.
Incluso a corrientes muy bajas existe \emph{flux creep} y se
observa disipaci\'on. Por el contrario, en el caso $B=0$ los
v\'ortices son s\'olo los que genera el \Index{autocampo} asociado a
$J$, de manera que habr\'a m\'as o menos v\'ortices dependiendo del
valor concreto de $J$. Cuando la corriente aplicada $J$ tiene
valores bajos no se tienen apenas v\'ortices que puedan causar
disipaci\'on, por lo que una funci\'on que imponga un \Index{umbral}
al campo $E$ medible, como hace el \emph{breakdown power law} de
la ecuaci\'on \ref{ec:EJT}, describe mejor los resultados
experimentales.

Se puede apreciar la bondad de estas funciones en la gr\'aficas
\ref{fig:datos:EJmodelo} y \ref{fig:datos:EJmodeloB1}, en donde se
contrastan los ajustes con los resultados experimentales. Los
par\'ametros empleados se indican en el pie de figura
correspondiente. En los dos casos se ha tomado $\Tc = \Tcz =
89.8\und{K}$ (muestra \fA). Ambas funciones subestiman el voltaje
experimental a corrientes y temperaturas altas, pero esto es lo
apropiado: para hacer c\'alculos es importante que no recojan un
posible \Index{calentamiento} entremezclado en las medidas, y lo
que hacen es seguir la tendencia de los datos con menos
disipaci\'on. Veremos en los cap\'{\i}tulos \ref{cap:cal} y
\ref{cap:calpr} que la desviaci\'on de los datos experimentales
respecto de la tendencia general, en esos rangos de alta
disipaci\'on, se puede explicar con el aumento de temperatura de
la muestra. Este mismo calentamiento puede explicar tambi\'en la
transici\'on al estado normal.

%%%%%%%%%%%%%%%%%%%%%%%%%%%%%%%%%%%%%%%%
\section{Modelos habituales para el salto}
%%%%%%%%%%%%%%%%%%%%%%%%%%%%%%%%%%%%%%%%

Como ya mencionamos previamente, el origen f\'{\i}sico de la
corriente cr\'{\i}tica \Jc, donde se mide por primera vez
disipaci\'on en el estado mixto, est\'a claramente vinculado al
\Index{desanclaje} \index{pinning} y movimiento de los
\Index{v\'ortices} magn\'eticos. No est\'a tan claro sin embargo
qu\'e mecanismo induce la transici\'on abrupta observada en \Jx.
Las causas podr\'{\i}an ser diversas, incluso variar de una
muestra a otra o de un tipo de material a otro. En los \'ultimos
veinticinco a\~nos se han considerado varios modelos que
explicar\'{\i}an este salto al estado normal. A modo de
contextualizaci\'on hacemos un breve resumen de estos modelos. En
general se enmarcan en dos grandes conjuntos.

\subsection{Transici\'on causada por calentamiento}

\index{calentamiento} Seg\'un este grupo de modelos, la
transici\'on al estado normal se producir\'{\i}a simplemente
porque la temperatura de la muestra aumenta hasta por encima de
\Tc, sin que medie ning\'un cambio de r\'egimen en el sistema de
\Index{v\'ortices}.

Este aumento de temperatura podr\'{\i}a generarse a partir de un
punto inicial localizado, pero que se propaga r\'apidamente. De
esta posibilidad dan cuenta las teor\'{\i}as de punto caliente
(\emph{\Index{hot-spot}}), un t\'ermino bastante gen\'erico que
engloba distintos tipos de mecanismos: \Index{inhomogeneidades}
\cite{Xiao98b}, distribuciones no uniformes de temperatura
mantenidas por la corriente \cite{Skocpol74,Skokov93},
\emph{\Index{phase slip centers}}
\cite{Jelila98,Maneval01,Reymond02}\ldots

Otra posibilidad es que el calentamiento sea m\'as o menos
uniforme en toda la muestra, debido a la disipaci\'on que producen
los \Index{v\'ortices} en su movimiento. En muestras escasamente
refrigeradas (como cualquier muestra masiva en un medio
l\'{\i}quido o gaseoso) en las que el calor disipado no se evacua
al entorno de manera eficiente, se suele considerar que \'esta es
la causa m\'as probable de transici\'on al estado normal
\cite{Yang99,Elschner99,Tournier00}. En otro trabajo desarrollado
en nuestro laboratorio paralelamente al presente se estudian este
tipo de muestras \cite{Gonzalez02b,Gonzalez02,Gonzalez03}.

En muestras peliculares depositadas sobre substratos cristalinos,
sin embargo, la evacuaci\'on del calor es mucho mejor, y suele
descartase que el calentamiento uniforme sea la causa de la
transici\'on \cite{Xiao96,Xiao99,Curras00b,Curras01}.

Nos referiremos a esta posibilidad del calentamiento como a un
\emph{efecto extr\'{\i}nseco}, \index{extr\'{\i}nseco} en el
sentido de que ser\'{\i}a evitable si las condiciones de
refrigeraci\'on fuesen \'optimas, si las medidas se hiciesen tan
r\'apidamente que no tuviese tiempo la muestra a calentarse, o si
las muestras fuesen de mejor calidad y no tuviesen puntos
calientes.

\subsection{Transici\'on causada por movimiento de v\'ortices}

A estos mecanismos, por contraposici\'on a los del grupo anterior,
los llamaremos \index{intr\'{\i}nseco} \emph{intr\'{\i}nsecos},
pues estar\'{\i}an presentes aun en condiciones ideales de
refrigeraci\'on y de s\'{\i}ntesis de las muestras, y s\'olo
podr\'{\i}an eliminarse con un anclaje de \Index{v\'ortices}
ideal.

\subsubsection{Avalancha de Larkin-Ovchinnikov (LO)}

En el modelo quiz\'a m\'as extendido de este otro grupo
\cite{Larkin76} se interpreta que los v\'ortices presentes en el
estado mixto se desplazan contra una \Index{viscosidad} del medio
que no es lineal con la velocidad. Esta viscosidad aumenta con la
velocidad de los \Index{v\'ortices} hasta llegar a un m\'aximo, pero a
partir de ese punto vuelve a disminuir, contrariamente a lo
habitual. Al superar ese m\'aximo los v\'ortices, sometidos en todo
momento a las \Index{fuerzas de Lorentz} que los empujan,
encuentran cada vez menos oposici\'on a su movimiento: incrementan
su velocidad, se reduce con ello cada vez m\'as la viscosidad, se
aceleran progresivamente y generan una \Index{avalancha}. De este
modo causan en muy poco tiempo altas disipaciones que hacen
transitar la muestra al estado
normal\comenta{(j-)\footnote{Algunos autores mencionan la
posibilidad de que la transici\'on no sea exactamente al estado
normal, sino a otro estado de v\'ortices con una resistividad
aparente cercana a la normal \cite{Xiao}}}.

Esta variaci\'on de la viscosidad que experimentan los
\Index{v\'ortices} se debe a que el \Index{tiempo de dispersi\'on}
interelectr\'onica $\tau_{ee}$ es finito, mayor o del orden del
tiempo de dispersi\'on electr\'on-fon\'on $\tau_{ef}$ que los
termaliza con el entorno. De este modo, cuando un conjunto de
electrones sufre alguna variaci\'on energ\'etica, esta
energ\'{\i}a no se distribuye con suficiente velocidad entre todos
los electrones, y ya no los podemos considerar termalizados entre
s\'{\i}. La descripci\'on de la funci\'on de distribuci\'on
electr\'onica requiere un tratamiento matem\'atico complicado
\cite{Larkin76}, cuyas interpretaciones f\'{\i}sicas son variadas
\cite{Klein85,Doettinger94,Xiao96}. Imaginemos qu\'e sucede en
estas condiciones cuando un \index{v\'ortices} v\'ortice se mueve
en el seno del superconductor. Un v\'ortice es una isla de
material en estado normal rodeada por un mar superconductor. Al
desplazarse, la regi\'on que previamente ocupaba vuelve al estado
superconductor, y la regi\'on que acaba de ocupar transita al
estado normal. Si el desplazamiento del v\'ortice es
suficientemente r\'apido, el \Index{tiempo de relajaci\'on}
energ\'etica puede entrar en juego con efectos relevantes: no
damos tiempo a que la nueva regi\'on ocupada transite
\emph{totalmente} al estado normal, mientras que la zona que se ha
dejado atr\'as todav\'{\i}a no ha vuelto \emph{del todo} al estado
superconductor. El efecto sobre la funci\'on de distribuci\'on
electr\'onica es que hay menos cuasipart\'{\i}culas (electrones
normales) \emph{dentro} del v\'ortice, y m\'as fuera de \'el, que
cuando el v\'ortice est\'a en reposo o se mueve lentamente. Esto
se interpreta como una disminuci\'on del di\'ametro del v\'ortice,
que al estrecharse reduce su rozamiento con el entorno:
as\'{\i}disminuye la \Index{viscosidad} que experimenta en su
movimiento.

Una variante de este modelo LO, publicada por Bezuglyj y
Shklovskij (BS) \cite{Bezuglyj92}, introduce correcciones que
consideran el inevitable \Index{calentamiento} que se tiene en
cualquier estado disipativo, como es el estado mixto. Este modelo
ser\'a el que empleemos para el an\'alisis de nuestros datos, pues ya
se ha comprobado que la variante sencilla, de LO, no explica
correctamente estos experimentos
\cite{Doettinger94,Xiao96,Doettinger97,Gonzalez02}. En todo caso,
en este modelo BS el origen del salto sigue siendo la
\Index{avalancha} de \Index{v\'ortices} que hemos descrito, y se
considera que el calentamiento es una correcci\'on de orden menor.
Los autores que emplean estos modelos para explicar sus
experimentos siempre estiman que el calor se evacua con mucha
eficiencia desde la muestra al entorno, y que se puede despreciar
salvo quiz\'a como peque\~na correcci\'on.

\subsubsection{Otros mecanismos de din\'amica de v\'ortices}

Hay otras posibilidades, dentro de los modelos que atribuyen a la
din\'amica de v\'ortices la causa del salto, que algunos autores
no descartan. As\'{\i} como en ciertos materiales (\BSCOf) la
transici\'on abrupta ocurre en un r\'egimen claramente de
\emph{\Index{free flux flow}}, en que los v\'ortices se mueven
libremente y presentan una curva caracter\'{\i}stica \EJ\ muy
lineal \cite{Xiao98}, en otros sin embargo (\YBCOf) las curvas
\EJ\ previas al salto muestran dependencias no lineales. Como ya
mencionamos, en estos casos se interpreta que los v\'ortices
siguen globalmente anclados en toda la curva \EJ, y que el voltaje
que se observa se debe a peque\~nas avalanchas locales o a saltos
finitos de un punto de anclaje a otro (flujos activados
t\'ermicamente (\emph{TAFF}), \emph{\Index{flux creep}}
\cite{Blatter94}, criticidad auto-organizada (\emph{SOC})
\cite{Pla91,Xiao97,Bassler98,Altshuler02}\ldots). De esta manera,
la corriente \Jx\ coincidir\'{\i}a con la de \index{pinning}
\Index{desanclaje}, en la que todos los v\'ortices se liberan de
los centros en que estaban clavados, generando una brusca
disipaci\'on que hace transitar a la muestra. Pero en principio no
parece muy probable que estos mecanismo den cuenta de la brusca y
reproducible transici\'on observada. En \emph{flux creep}, por
ejemplo, no hay un cambio brusco en el momento en que todos los
v\'ortices se sueltan de sus centros de anclaje: esta transici\'on
ocurre suavemente \cite{Blatter94,Tinkham96}.

Recientemente se ha propuesto un mecanismo esencialmente opuesto a
la avalancha LO, en el que los v\'ortices aumentan su tama\~no, en vez
de reducirlo, cuando se mueven a altas velocidades
\cite{Kunchur02}. Esto ocurre cuando el \Index{tiempo de
dispersi\'on} interelectr\'onica $\tau_{ee}$ es menor que el de
dispersi\'on electr\'on-fon\'on $\tau_{ef}$, lo que sucede a
temperaturas muy bajas y a campos magn\'eticos elevados. Al elevarse
la temperatura de los electrones debido a la disipaci\'on, aumenta
la regi\'on normal en torno a los v\'ortices, incrementando \'estos su
tama\~no, y la \Index{viscosidad} se reduce igualmente debido a que
los gradientes en torno a los \Index{v\'ortices} se suavizan con
este aumento de tama\~no. El efecto final es el mismo: la velocidad
de los v\'ortices aumenta, y se alcanza una inestabilidad.

\subsection{Otras posibilidades como causa de la transici\'on}

Otras interpretaciones, como las basadas en el efecto Josephson
intr\'{\i}nseco \cite{Kleiner92,Kleiner94,Kleiner94b,Tanabe96} o
la cristalizaci\'on de la red de v\'ortices \cite{Koshelev94}, se
tuvieron en cuenta en su momento, pero fueron descartadas por los
experimentos para explicar este fen\'omeno en concreto
\cite{Xiao96}. Hemos mencionado aqu\'{\i} las que siguen de
actualidad y se aplican con frecuencia a distintas muestras para
explicar los resultados experimentales.

Pero existen otras interpretaciones que no se pueden descartar
todav\'{\i}a, y que no encajan claramente en alguno de los dos
grupos previos. Por ejemplo, es posible que electrones normales,
que se inyectan a gran velocidad en el superconductor desde el
resto del \Index{circuito}, distorsionen el equilibrio din\'amico
de los pares de Cooper \cite{Sabouret02,Jelila98}. Tampoco se debe
olvidar la posibilidad de que \Jx\ coincida con la
\Index{corriente de desapareamiento} \Jd\ de los pares de Cooper.
Esta corriente \Jd\ es la que imprime tanta energ\'{\i}a
cin\'etica a los pares de Cooper que la existencia de estos se
vuelve desfavorable energ\'eticamente respecto a la de los
electrones normales disociados (\cite{Kunchur93}, \cite[sec.
4.4]{Tinkham96} y \cite[sec. 10.XIX]{Poole95}). Aunque esta
posibilidad se suele descartar porque la \Jx\ experimental
$=\Ix/d\;a$ dista notablemente de la \Jd\ te\'orica
\cite{Curras00b,Curras02}, bastar\'{\i}a con que la distribuci\'on
de corriente no fuese uniforme en toda la secci\'on de la muestra,
sino concentrada en una regi\'on menor (por ejemplo en filamentos
laterales, por efecto Meissner \cite{Fuchs98,Fuchs98B} o por
lentitud de los \Index{v\'ortices} en penetrar al interior de la
muestra \cite{Zhang01,Liu02}), para que la densidad de corriente
real sea mucho mayor que la aparente. Esta posibilidad est\'a
siendo estudiada con m\'as detalle en otro trabajo paralelo que se
desarrolla en nuestro laboratorio \cite{Gonzalez03}, y aqu\'{\i}
la consideraremos brevemente para el caso de los filmes.

\section{Dependencias de los par\'ametros cr\'{\i}ticos \Jx, \Ex\ y \Jc}

\subsection{Determinaci\'on de los par\'ametros cr\'{\i}ticos}

Los puntos \Jx\ y \Ex\ de cada curva son en general f\'aciles de
determinar: en ellos se manifiesta una discontinuidad clara de la
curva \EJ, y tomamos estas coordenadas como los \'ultimos puntos
experimentales que podemos medir antes de la transici\'on. La
incertidumbre de la medida, como hemos visto en la secci\'on
\ref{ssec:medidaEJ}, proviene de la resoluci\'on en corriente con
la que midamos la curva \EJ, pero tambi\'en del
\Index{calentamiento} que altera el valor de \Ex. Adem\'as, a
campos magn\'eticos aplicados altos, o a temperaturas cercanas a
\Tc, la transici\'on al estado normal deja de ser abrupta, y el
valor de \Jx\ y, sobre todo, el de \Ex, tienen m\'as
incertidumbre. En general no estudiaremos estos datos con mayor
\Index{error}, aunque en la referencia \cite{Gonzalez02} se
propone un criterio para determinar estos par\'ametros que
permitir\'{\i}a ampliar el estudio a estas regiones.

En medidas m\'as detalladas, cuando queramos medir \Jx\ con m\'as
resoluci\'on, \Jx\ ser\'a la corriente m\'as alta que podamos aplicar
durante el tiempo $t$ que dure el experimento sin causar una
transici\'on al estado normal.

\figDatosJcpin

Para la determinaci\'on experimental de la corriente cr\'{\i}tica
\Jc, lo habitual es ver a qu\'e corriente se alcanza un cierto
campo \Index{umbral} $E$, t\'{\i}picamente de $10\und{\mu V/cm}$.
En la figura \ref{fig:datos:Jcpin}.b vemos d\'onde queda ese
umbral, y las corrientes cr\'{\i}ticas \Jc\ que determina para
cada campo. De esta manera, como hemos venido explicando, \Jc\ es
la corriente en la que aparece por primera vez disipaci\'on
mensurable con nuestros dispositivos. Esta es la corriente que
hemos estudiado y comparado con \Jx\ en funci\'on de la
temperatura, a campo aplicado nulo. La disipaci\'on experimental
$V=E\;l$ es proporcional a la velocidad que adquieren los
v\'ortices de la muestra en su movimiento.

Sin embargo, para el estudio de estos efectos tan relacionados con
la din\'amica de los \Index{v\'ortices}, esta corriente
cr\'{\i}tica \Jc\ no es la m\'as significativa cuando inyectamos
v\'ortices extra, es decir, cuando aplicamos campo magn\'etico
externo. Como la disipaci\'on que medimos experimentalmente es
proporcional no s\'olo a la velocidad de los v\'ortices, sino
tambi\'en al n\'umero de ellos que est\'an presentes en la
muestra, al aplicar campo magn\'etico es conveniente establecer un
criterio \Index{umbral} diferente. Si fij\'asemos un umbral en el
voltaje, estar\'{\i}amos estudiando puntos diferentes de la
din\'amica de los v\'ortices: con campo aplicado bajo, ser\'{\i}a
un punto en que los v\'ortices se mueven r\'apido; con campo alto,
el mismo voltaje experimental se alcanzar\'{\i}a con velocidades
menores de los v\'ortices.

\figDatosJcJcpin

Por este motivo definimos otra corriente cr\'{\i}tica \Jcpin, como
aquella en que los v\'ortices presentes en la muestra alcanzan un
cierto \Index{umbral} \emph{de velocidad}, lo m\'as bajo posible
experimentalmente: se pretende reflejar en ella el momento en que
los v\'ortices se sueltan de sus centros de anclaje
(\emph{\Index{pinning}} \cite[cap. 9.V y 9.VI]{Poole95}). En la
figura \ref{fig:datos:Jcpin}.a vemos c\'omo se establece esta
\Jcpin. La velocidad de los v\'ortices es $v=E/B=E/\mH$. Fijamos
el \Index{umbral} de velocidades en la m\'as baja que hemos podido
medir: de esta forma, \Jc\ y \Jcpin\ coinciden en la curva \EJ\
medida con mayor campo magn\'etico aplicado, pero para las otras
curvas establecemos \Jcpin\ con una extrapolaci\'on a velocidades
menores
---pues los datos experimentales son ruidosos en esos rangos---,
de manera que \Jcpin\ es en general algo menor que \Jc\ (ver fig.
\ref{fig:datos:JcJcpin}), aunque con dependencias respecto de $T$
y $H$ muy semejantes. A campos tan bajos, no se aprecian
diferencias muy significativas entre los valores de \Jc\ y \Jcpin.

\subsection{Dependencia respecto de la temperatura de \Jx\ y \Jc}
\label{ssec:depT}

\figDatosJcJxTcompA \figDatosJxTAE \figDatosJcJxTA

En la figura \ref{fig:datos:JcJxTcompA} mostramos la variaci\'on
con la temperatura de \Jx\ y \Jc\ para una de las muestras. Ambas
series de puntos se ajustan a una funci\'on dependiente de la
temperatura reducida $\eps \simeq (1-T/\Tc)$ con exponente 3/2,
\be J(T) = J_0 \; (1-b \; T/\Tc)^{3/2}, \label{ec:JT32}\ee donde
$J_0$ es la corriente en $\eps=1$, $T=0\und{K}$. Esta dependencia
con un exponente 3/2 aparece con mucha frecuencia para corrientes
cr\'{\i}ticas \Jc\ o de \index{corriente de desapareamiento}
desapareamiento \Jd\ \cite{Tinkham96}. El par\'ametro $b$, del
orden de 1, se incluye aqu\'{\i} para dar cierta flexibilidad al
modelo: es equivalente a dejar libre la temperatura cr\'{\i}tica
\Tc, pues el modelo no tiene por qu\'e ajustar exactamente con el
valor de \Tc\ que hemos elegido. Los par\'ametros empleados para
ajustar los puntos se indican en el correspondiente pie de figura.
Los par\'ametros $b$ (o, equivalentemente, las \Tc\ que se extraen
del ajuste) no son exactamente iguales en las dos series de datos,
pero no podemos esperar mucho m\'as de estos modelos en
potencias, que son aproximaciones muy sencillas. Lo que
pretendemos aqu\'{\i} es simplemente discriminar entre un modelo
con exponente $3/2$ y otros exponentes posibles. En la figura
\ref{fig:datos:JxTAE} se compara la dependencia de \Jx\ respecto
de la temperatura de otra muestra con la que acabamos de mostrar
en la figura \ref{fig:datos:JcJxTcompA}: tambi\'en sigue la misma
tendencia. Esta dependencia $\Jx(T)$ con exponente 3/2 se pone de
manifiesto en todas las muestras \cite{Curras00b,Curras01,Xiao96},
quedando descartada una posible funcionalidad con exponente 1/2,
\be J(T) = J_0 \; (1-b \; T/\Tc)^{1/2}, \label{ec:JT12}\ee
representada en la figura \ref{fig:datos:JxTAE} con una l\'{\i}nea
de puntos. Esta \'ultima funcionalidad se encuentra en un modelo
que explica el origen de la corriente de salto \Jx\ por puntos
calientes \index{hot-spot} \cite{Skocpol74,Gurevich87,Xiao96}: no
es raro encontrar en la bibliograf\'{\i}a que los fen\'omenos
t\'ermicos se descartan para explicar el origen de \Jx\ s\'olo
porque los datos no se ajustan a la ecuaci\'on \ref{ec:JT12}
\cite{Curras00b,Curras01,Xiao96}. Veremos con detalle en los
cap\'{\i}tulos siguientes lo inapropiado de este razonamiento: el
\Index{calentamiento} puede manifestarse de otras formas distintas
de las de \emph{hot-spot}.

En la figura \ref{fig:datos:JcJxTA} se muestra el cociente
$\Jc/\Jx$ en funci\'on de la temperatura, poniendo de manifiesto
que lejos de \Tc\ ambas temperaturas var\'{\i}an muy a la par, y
m\'as cerca de \Tc\ $(T/\Tc \tiende 1)$ \Jc\ comienza a caer m\'as
r\'apidamente con la temperatura que \Jx. En la figura se incluyen
tambi\'en algunos puntos tomados con \Index{campo magn\'etico}
aplicado, para ver la tendencia que manifiestan, aunque este es un
estudio que no se ha hecho en detalle.

Esta dependencia t\'ermica con exponente 3/2 que presenta $\Jx(T)$
coincide con la de $\Jc(T)$, y tambi\'en con la \index{corriente de
desapareamiento} corriente de desapareamiento $\Jd(T)$
\cite{Kunchur93,Curras00b,Curras01}, por lo que resulta inevitable
atribuir a \Jx\ un origen relacionado con los mecanismos que
generan \Jcpin, \Jc\ o \Jd. Pero s\'olo estos datos no son
concluyentes, y no bastan para discernir unas causas de las otras
\cite{Xiao96}.

\subsection{Dependencia respecto del campo magn\'etico de \Jx, \Jcpin\ y \Ex}

\figDatosJcJxvarias

\figDatosJcEJxH

En la figura \ref{fig:datos:JcJxvarias} representamos la corriente
cr\'{\i}tica \Jcpin\ y la de salto \Jx\ de varias muestras, en
funci\'on del \Index{campo magn\'etico} aplicado y para varias
temperaturas distintas. Es notable la diferencia de comportamiento
a campos bajos: la corriente cr\'{\i}tica\footnote{Aunque nuestra
definici\'on de \Jcpin\ no coincida exactamente con la m\'as
frecuentemente empleada \Jc\ (recordemos que para la comparaci\'on
con \Jx\ que queremos realizar \Jcpin\ resulta m\'as conveniente),
sus valores son muy parecidos, y la variaci\'on con el campo
aplicado an\'aloga a la que en la bibliograf\'{\i}a se encuentra
para \Jc\ \cite{Hettinger89,Kahan93,Xu94,Cao97}.} \Jcpin\ parece
acusar antes que \Jx\ la influencia del campo externo: var\'{\i}a
para campos peque\~nos, mientras que \Jx\ apenas se altera hasta
que el campo aplicado supera la d\'ecima de Tesla. La figura
\ref{fig:datos:JcEJxH}.a resalta esta diferencia de
comportamiento: se representa el cociente $\Jcpin/\Jx(H)$ para las
mismas curvas de la figura \ref{fig:datos:JcJxvarias}, y se ve
c\'omo a campos bajos \Jcpin\ var\'{\i}a m\'as r\'apidamente que
\Jx. A partir de aproximadamente la d\'ecima de Tesla comienzan a
variar a la par, haci\'endose constante el cociente. Todas las
tendencias son muy similares, como pone de manifiesto la figura
\ref{fig:datos:JcEJxH}.b, en la que se representa la derivada
respecto al campo del cociente $\Jcpin/\Jx$: todas las derivadas
se colapsan en la misma curva.

La tendencia de $\Jx(H)$ parece apuntar a que existe un campo
magn\'etico de baja intensidad que se suma al externamente aplicado,
de manera que s\'olo cuando este \'ultimo supera un cierto valor
(cercano a la d\'ecima de Tesla) puede competir con aquel y hacer
notar sus efectos. El que esto se aprecie s\'olo en el
comportamiento de $\Jx(H)$ y no en el de $\Jcpin(H)$ hace pensar
que este otro campo magn\'etico extra es el \Index{autocampo}
generado por la corriente inyectada en la muestra: las corrientes
en torno a \Jcpin\ son menores que los valores de \Jx, por lo que
sus autocampos ser\'an tambi\'en menores, y los efectos de \'estos menos
notorios.

Conviene por tanto tener una estimaci\'on de los valores del
autocampo generado por la corriente inyectada, para analizar
correctamente los datos. El campo $\mathbf{H}\und{(Oe)}$ que
genera una corriente $I\und{(A)}$ al circular por un conductor de
longitud infinita y de secci\'on rectangular, teniendo \'esta un
v\'ertice inferior en $(x_1,y_1)$ y el superior diagonalmente
opuesto en $(x_2,y_2)$ (cm), se expresa como \cite{Duran68}
\be \mathbf{H}(x,y) = \frac{\mu_0 \; I}{2\; \pi (x_2-x_1) (y_2-y_1)} \; \left( h_x(x,y) \; \mathbf{\hat{x}} + h_y(x,y) \;
\mathbf{\hat{y}}\right), \label{ec:campoHI} \ee donde
\[\begin{split} h_x(x,y) = & \sum_{i,\,j=1}^2 (-1)^{i+j}
(y-y_i)\arctan\left(\frac{x_j-x}{y-y_i}\right)+ \\
 & + \sum_{i,\,j=1}^2 (-1)^{i+j} (x_j-x)\ln\left(\sqrt{(x-x_j)^2+(y-y_i)^2}\right),
\end{split}\]
\[\begin{split} h_y(x,y) = & \sum_{i,\,j=1}^2 (-1)^{i+j+1}
(x-x_i)\arctan\left(\frac{y_j-y}{x-x_i}\right)+ \\
 & + \sum_{i,\,j=1}^2 (-1)^{i+j+1}
 (y_j-y)\ln\left(\sqrt{(x-x_i)^2+(y-y_j)^2}\right).
\end{split}\]

\figDatoscampoHI

En la figura \ref{fig:datos:campoHI} representamos la intensidad
del campo magn\'etico calculada seg\'un estas ecuaciones para la
corriente que circula por una secci\'on rectangular de ancho $a$ y
de alto $d$, en dos casos diferentes. Por un lado, cuando la
corriente est\'a uniformemente distribuida en toda la secci\'on, de
modo que calculamos el campo con $x_1=-a/2$, $x_2=+a/2$,
$y_1=-d/2$, $y_2=+d/2$. Por otro lado, cuando la corriente se
concentra s\'olo en unos filetes laterales de secci\'on cuadrada y de
lado $d$, de modo que el campo total es la suma del campo
producido por la corriente en $x_1=a/2-d$, $x_2=a/2$, $y_1=-d/2$,
$y_2=+d/2$ m\'as el de la corriente sim\'etrica al otro lado de la
muestra. El campo es m\'aximo en los laterales $x=-a/2$ y $x=a/2$ en
ambos casos, pero cuando la corriente se concentra, la intensidad
es mayor en esos puntos que cuando se reparte por toda la secci\'on.

En el caso de que la corriente est\'e \emph{uniformemente}
distribuida, el campo magn\'etico en las coordenadas $(a/2,0)$ tiene
s\'olo componente $\mathbf{\hat{y}}$, y la expresi\'on de su
intensidad
---igual, pero de signo contrario, a la de la posici\'on sim\'etrica
$(-a/2,0)$--- queda reducida a
\be\begin{split} H_u(a/2,0) = & \frac{2\times 10^{-3} \; I}{a \; d} \times \\
& \times \left( 2\;a\; \arctan\left(\frac{d}{2\;a}\right) +
\frac{d}{2} \ln\left(\frac{d^2/ + 4\;a^2}{d^2}\right) \right),
\end{split}\label{ec:campoenw2} \ee
lo que en el caso l\'{\i}mite $a \gg d$ se simplifica a
\cite{Lefloch99} \be H_u(a/2,0) = \frac{I}{500\;a}
\;\ln\left(\frac{2\;a}{d}\right). \label{ec:campoUnif} \ee
Siguiendo en este caso l\'{\i}mite $a \gg d$, la correspondiente
expresi\'on para el campo en ese mismo punto cuando la corriente
se distribuye s\'olo por dos filetes \emph{laterales} de longitud
de penetraci\'on $\lambda$ ($a \gg \lambda$) resulta \be
H_l(a/2,0) = \frac{I}{2000\;d\;\lambda} \; \left(4 \lambda
\arctan\left(\frac{d}{2\lambda}\right) + d \ln
\left(1+\frac{4\lambda^2}{d^2}\right)\right).
\label{ec:campoLateral}\ee Para los valores t\'{\i}picos de
nuestros micropuentes ($d=150\und{nm}$, $a=10\und{\mu m}$), el
autocampo en el borde con corriente uniforme es $H_u(a/2,0) \simeq
1000 \;I$, mientras que si suponemos que la corriente se concentra
en una longitud de penetraci\'on $\lambda = d$ por cada lado, el
autocampo es $H_l(a/2,0) \simeq 11000 \;I$, m\'as de diez veces
m\'as intenso (el c\'alculo preciso, sin la aproximaci\'on $a \gg
d$, da exactamente un factor diez de diferencia).

Volvamos ahora a los resultados de la figura
\ref{fig:datos:JcJxvarias}, y fij\'emonos en una curva $\Jx(H)$
t\'{\i}pica, por ejemplo la que a campos aplicados bajos comienza
teniendo valores $\Jx=4\Exund{6}{\Acms}$. A esta densidad de
corriente \emph{aparente} aplicada (que corresponde a una
corriente real de $I=60\und{mA}$), el \Index{autocampo} de
corriente uniforme es $\mH_u \sim 6\Exund{-3}{T}$, mientras que el
de corriente distribuida en los laterales con $\lambda=d$ es
$\mH_l \sim 7\Exund{-2}{T}$. Entre ambos valores se sit\'ua el
campo $\mH \sim 1\Exund{-2}{T}$ en que la corriente de salto
$\Jx(H)$ comienza a sentir la influencia del campo aplicado
externamente. Parece as\'{\i} confirmarse nuestra intuici\'on
primera: el campo externo no hace notar sus efectos sobre \Jx\
hasta que supera en intensidad al \Index{autocampo} generado por
la corriente inyectada; mientras sea de menor valor, ser\'a el
autocampo el que prevalezca. Esto mismo parece apreciarse en el
comportamiento de $\Jcpin(H)$, aunque la resoluci\'on experimental
del campo aplicado no permita medirlo con tanto detalle: como
$\Jcpin$ es, a campos bajos, unas tres veces menor que $\Jx$, el
autocampo que se genera es tambi\'en tres veces menos (suponiendo
que la distribuci\'on de corriente fuese la misma en ambos
reg\'{\i}menes), de modo que el punto en que el campo externo
supera al autocampo se sit\'ua en torno a $\mH \sim
3\Exund{-3}{T}$, lo que para nuestra ventana experimental equivale
a campo aplicado nulo.

\figDatosExJxHTA \figDatosExHTAfit

Una vez hechas estas consideraciones, volvamos al an\'alisis de
los par\'ametros cr\'{\i}ticos. En la figura
\ref{fig:datos:ExJxHTA} vemos los valores de \Ex\ y, de nuevo,
\Jx\ correspondientes a la muestra \fA, a varias temperaturas, en
funci\'on del campo aplicado. Como hemos visto, los datos que
est\'an m\'as afectados de incertidumbre, debido al calentamiento,
son los de \Ex: se muestran para una de sus curvas las barras de
\Index{error}, crecientes para campos mayores. En la figura
\ref{fig:datos:ExHTAfit} tomamos dos de estas curvas y las
ajustamos al modelo \BS\ (BS) \cite{Bezuglyj92,Xiao98}, que
establece los valores de \Ex\ y de \Jx\ como \be \Ex=\Ez
\;\frac{(1-t)\; (3t+1)}{2\; \sqrt{2} \;t^{3/4} \;(3t-1)^{1/2}},
\label{ec:BSEx}\ee \be \Jx=\Jz \;\frac{2\; \sqrt{2} \; t^{3/4}\;
(3t-1)^{1/2}}{3t+1}, \label{ec:BSJx}\ee
% Ex=E0*(1-t)*(3*t+1)/(2*sqrt(2)*t^0.75)/(3*t-1)^0.5;
% Jx=J0*1e6*2*sqrt(2)*t^0.75*(3*t-1)^0.5/(3*t+1);
donde \[t=\frac{1+b+\sqrt{b^2+8b+4}}{3\;(1+2b)}\] y
\[ b=B/B_T. \]
$B=\mH$ es el campo magn\'etico aplicado, y $B_T$ un par\'ametro
del modelo BS vinculado al \Index{tiempo de relajaci\'on}
energ\'etica de los portadores $\tau_{\epsilon}$
---una combinaci\'on de los \index{tiempo de dispersi\'on} tiempos
de dispersi\'on electr\'on-electr\'on y electr\'on-fon\'on,
$\tau_{\epsilon} = (\tau_{ee}^{-1}+\tau_{ef}^{-1})^{-1}$
 \cite{Bezuglyj92}---, a la resistividad en el estado normal \rhon\ y
a la resistencia de transferencia t\'ermica entre filme y entorno
$h$:
\be B_T=0.374\;\frac{e_0\;h\;\rhon\;\tau_{\epsilon}}{\kB\;d} \label{ec:BT}\ee
En esta \'ultima ecuaci\'on, \kB\ es la constante de Boltzmann, $e_0$
la carga del electr\'on, y $d$ el espesor del filme.

Los dos ajustes que se muestran en la figura
\ref{fig:datos:ExHTAfit} emplean distintos valores de $B$. En el
ajuste que se separa de los datos de $\Ex(H)$ a campos bajos
(l\'{\i}neas de trazo discontinuo), tomamos $B=\mH$, siendo $H$ el
campo magn\'etico aplicado externamente. Como resultado, obtenemos
que los mejores par\'ametros son $\Jz=6.08\Exund{6}{\Acms}$,
$\Ez=13.3\und{V/cm}$, y $B_T=0.31\und{T}$.

En el ajuste que recoge por completo la tendencia de los datos
$\Ex(H)$ en todos los rangos de campo aplicado, sin perder por
esto la fidelidad a la tendencia de $\Jx(H)$ (l\'{\i}neas
continuas), incluimos los efectos del \Index{autocampo} generado
por la corriente, haciendo $B=\mu_0\,(H+c\;I^*)$, donde $c$ es un
par\'ametro libre del ajuste. Obtenemos $\Ez=12.0\und{V/cm}$,
$\Jz=6.48\Exund{6}{\Acms}$, $B_T=0.27\und{T}$, y
$c=3133\und{Oe/A}$. Con este cambio, el par\'ametro que m\'as
var\'{\i}a en t\'erminos relativos es $B_T$, que lo hace un 14\%
aproximadamente. En general, sin variar mucho los par\'ametros
f\'{\i}sicos implicados, el ajuste mejora considerablemente si
tenemos en cuenta los efectos del \Index{autocampo} generado por
la corriente.

Empleando los par\'ametros caracter\'{\i}sticos de nuestra
muestra\footnote{$\rhon \sim 75\und{\mOcm}$, $h \sim
1000\und{W/cm^2\,K}$, $d \sim 150\und{nm}$}, obtenemos de la
ecuaci\'on \ref{ec:BT} un valor para el \Index{tiempo de
relajaci\'on} energ\'etica de las part\'{\i}culas portadoras
$\taue \sim 1.2\Exund{-12}{s}$ para esta temperatura
$T=76.2\und{K}$, lo que es del todo compatible con los resultados
de otro tipo de experimentos, como de medidas de impedancia
superficial en el rango de las microondas
\cite{Flik92,Bonn93,Gao93}. Este es un valor que se escapa del
rango de validez del modelo \LO, por lo que el empleo de dicho
modelo para la extracci\'on de \taue\ resulta ineficaz, como se ha
visto en trabajos previos, en los que \taue\ es algunos \'ordenes
de magnitud menor \cite{Doettinger94,Xiao96,Doettinger97}.
Aparentemente el modelo \BS, en contraposici\'on, arroja valores
m\'as realistas de este par\'ametro, pero en este punto se
requerir\'{\i}a un estudio m\'as sistem\'atico que escapa a las
pretensiones actuales del presente trabajo.

\figDatosExJxHvarias

La aparente validez del modelo \BS\ se pone en evidencia tambi\'en
en la figura \ref{fig:datos:ExJxHTA}, donde los puntos
experimentales de $\Jx(H)$ est\'an ajustados, para todas las
temperaturas mostradas, con la ecuaci\'on \ref{ec:BSJx} incluyendo
el \Index{autocampo}\footnote{El par\'ametro $c$ es el mismo en
todas las curvas, y los par\'ametros que var\'{\i}an para las
otras temperaturas son $\Jz=4.28\Exund{6}{\Acms}$,
$B_T=0.27\und{T}$ (para $T=80.4\und{K}$) y
$\Jz=2.39\Exund{6}{\Acms}$, $B_T=0.33\und{T}$ (para
$T=84.1\und{K}$). Estos ajustes requieren de bastante trabajo para
lograr acuerdos simult\'aneos de las expresiones de \Jx\ y \Ex, y
s\'olo el mostrado previamente para $T=76.2\und{K}$ se hizo con
rigor. Estos otros, para temperaturas mayores, se han hecho s\'olo
aproximadamente, para ver si se conservaba un acuerdo general, y
la tendencia observada es que la bondad del ajuste se pierde
parcialmente para los valores de \Ex\ conforme aumenta la
temperatura.}. La figura \ref{fig:datos:ExJxHvarias} recoge estas
mismas tendencias de $\Jx(H)$ y $\Ex(H)$ para varias muestras y
diferentes temperaturas. Estos efectos del autocampo podr\'{\i}an
ser, por tanto, la causa de que los modelos que explican la
corriente de salto se desv\'{\i}en de los datos experimentales a
campos magn\'eticos aplicados bajos; de ser as\'{\i}, se
podr\'{\i}an descartar otros mecanismos m\'as complejos que se
pueden encontrar en la bibliograf\'{\i}a \cite{Chiaverini00}.

Por otra parte, el valor obtenido para el par\'ametro
$c=3133\und{Oe/A}$, que da $\Hsf=c \;I$, es intermedio a los que
calcul\'abamos para el \Index{autocampo} con corriente
uniformemente distribuida en la secci\'on ($c \simeq
1000\und{Oe/A}$, ec. \ref{ec:campoUnif}) y con corriente
concentrada en los laterales de la muestra $\lambda=d$ ($c \simeq
11000\und{Oe/A}$, ec. \ref{ec:campoLateral}). Seg\'un esta
\'ultima ecuaci\'on, al valor $c$ obtenido de nuestros ajustes le
corresponde una longitud de penetraci\'on de la corriente en el
interior de la muestra $\lambda \simeq 8\;d = 0.12\;a =
1.2\Exund{-6}{m}$, de manera que la densidad de corriente
\emph{real} ser\'{\i}a m\'as de cuatro veces superior a la
aparente (esta \'ultima obtenida de considerar que la corriente
inyectada se distribuye uniformemente por toda la secci\'on $
a\times d$). Esto no es m\'as que una estimaci\'on muy cruda, pues
la ecuaci\'on \ref{ec:campoLateral} da el valor del campo justo en
la superficie, y lo que en realidad necesitar\'{\i}amos calcular
es una especie de valor efectivo del campo en toda la regi\'on por
la que pasa la corriente. De hacerlo as\'{\i}, encontrar\'{\i}amos
que la densidad de corriente para un campo efectivo $\Hsf = c\;I$
es incluso mayor, con campos m\'as intensos en la superficie, y
con $\lambda$ m\'as peque\~na. Es importante se\~nalar
aqu\'{\i}que esta posibilidad acercar\'{\i}a notablemente los
valores de \Jx\ a los predichos te\'oricamente para la
\Index{corriente de desapareamiento} \Jd\
\cite{Kunchur93,Curras00b,Curras01}, por lo que habr\'{\i}a que
considerar con m\'as detalle la posibilidad de que la transici\'on
abrupta al estado normal en \Jx\ se produjese por ruptura de pares
de Cooper, demasiado acelerados como para que su existencia siga
compensando desde un punto de vista energ\'etico. Esta posibilidad
de que la corriente circule en realidad concentrada en los
laterales de la muestra se ha confirmado, al menos en parte, de
manera experimental, y hay algunos modelos que intentan explicar
este proceso \cite{Fuchs98,Fuchs98B,Zhang01,Liu02}. Estas
consideraciones se han hecho aqu\'{\i} de manera breve, para
apuntar su posible relevancia en estos mecanismos y en
pel\'{\i}culas delgadas. Un estudio m\'as detallado se lleva a
cabo en la tesis doctoral de M.~T. Gonz\'alez \cite{Gonzalez03}.

\figDatosJxJcNA

Para terminar este apartado del estudio de las dependencias de los
par\'ametros cr\'{\i}ticos con el campo aplicado, sometemos
nuestros datos a un escalamiento propuesto para dicha dependencia
en algunas publicaciones. En 1990, Xu, Shi y Fox \cite{Xu90}
propusieron un modelo generalizado para la dependencia de la
corriente cr\'{\i}tica $\Jc(T,H)$, que unificaba todos los modelos
previos: el de Bean \cite{Bean62}, el de Anderson
\cite{Anderson62,Anderson64}, y otros posteriores. Este modelo
generalizado separa las dependencias respecto de $T$ y de $H$, y
tiene la forma \be
\Jc(H,T)=\Jc(T)\;\Jc(H)=\frac{\Jc(T)}{(1+H/H_0(T))^n}.
\label{ec:JcXu} \ee En algunos trabajos se emple\'o esta misma
dependencia, con el par\'ametro $H_0(T)$ constante, para ajustar
el comportamiento tambi\'en de \Jx\ \cite{Xiao96,Xiao98}. Como ya
hemos visto en la secci\'on \ref{ssec:depT}, la variaci\'on de
$\Jc(T)$ con la temperatura es la misma que la de $\Jx(T)$,
dependiendo de la temperatura reducida con un exponente 3/2 (ec.
\ref{ec:JT32}, p. \pageref{ec:JT32}). Veamos si las dependencias
respecto de $H$ son tambi\'en similares.

En la figura \ref{fig:datos:JxJcNA} vemos el resultado de dividir
\Jx\ y \Jcpin\ entre los valores a campo aplicado nulo que
corresponden a cada temperatura, de modo que el resultado es la
dependencia respecto de $H$ seg\'un la ecuaci\'on \ref{ec:JcXu}.
Las curvas de \Jx\ coinciden notablemente en una \'unica
dependencia, que se describe con la funcionalidad propuesta y con
$H_0$ constante. Por el contrario, $\Jcpin(H)$ no se ajusta a esa
misma funci\'on, ni siquiera haciendo $H_0$ y $n$ dependientes de
la temperatura, y las curvas no coinciden en una misma
dependencia. Normalizar \Jc\ en vez de \Jcpin\ no var\'{\i}a estos
resultados, \Jc\ y \Jcpin\ normalizadas tienen valores muy
similares.

Como hemos visto, las variaciones del valor de \Jx\ con $H$ y $T$
presentan notables similitudes con las de \Jc\ (o \Jcpin) y \Jd,
aunque tambi\'en se pueden describir aparentemente bien con
modelos espec\'{\i}ficos que contemplan mecanismos completamente
diferentes de los que originan \Jc\ y \Jd. No queda claro, por
tanto, a qu\'e mecanismos atribuir la transici\'on abrupta al
estado normal en \Jx, y se hace necesario m\'as trabajo en esta
direcci\'on.

\subsection{Dependencia respecto de la duraci\'on de la medida}

\figDatosEtJ

En la figura \ref{fig:datos:EtJ} vemos los resultados de un
experimento realizado aplicando corrientes constantes, de valores
muy parecidos entre s\'{\i}, y cercanas a la corriente de salto.
Medimos con la DAQ la evoluci\'on temporal del voltaje a corriente
constante. En todas las curvas se observa una variaci\'on de la
se\~nal, relativamente suave, en los primeros instantes de la
medida, atribuible como hemos visto a una elevaci\'on de la
temperatura de la muestra generada por la disipaci\'on. Pero
adem\'as, pasado un cierto tiempo, estas corrientes cercanas a
\Jx\ hacen transitar abruptamente la muestra al estado normal. La
corriente de $6.09\und{\Acms}$ lleva a un voltaje aparentemente
estable en tiempos mayores de 1~ms, pero la de 6.13 (un 0.6\%
mayor) causa una \Index{avalancha} tan pronto se aplica (dentro de
la resoluci\'on temporal de la medida). Los valores intermedios de
corriente conllevan una transici\'on en momentos diferentes, m\'as
pronto cuanto m\'as alta es $J$. Esto se traduce en una aparente
dependencia del valor de \Jx\ respecto del tiempo que dure la
medida: un experimento que no mida la evoluci\'on temporal del
voltaje con esta resoluci\'on, sino simplemente un valor promedio
$V$ durante un \Index{pulso} de corriente de duraci\'on $t$,
ver\'a que una misma corriente $J$ ser\'a cr\'{\i}tica o no
dependiendo de dicha duraci\'on $t$. Algunos trabajos previamente
publicados informaban, sin embargo, de que no se apreciaban
diferencias en el valor de \Jx\ en funci\'on de $t$
\cite{Doettinger94,Doettinger95}. Veremos en cap\'{\i}tulos
posteriores (sec. \ref{ssec:depJxt}) que esto puede ser debido a
que en todo momento sus experimentos se encuadraban en un rango
temporal excesivamente largo. Por el contrario otros trabajos, que
empleaban tiempos mucho m\'as cortos, s\'{\i} aprecian diferencias
en la medida dependiendo de la duraci\'on del \Index{pulso}
\cite{Jakob00}, corroborando la observaci\'on de la figura
\ref{fig:datos:EtJ}.

\section{Variaci\'on de la potencia disipada antes del salto}

\figDatosWxT

\figDatosWxHA

En las figuras \ref{fig:datos:WxT} y \ref{fig:datos:WxHA} vemos la
variaci\'on que se produce en el valor de la potencia disipada
justo antes del salto $\Wx=\Ex\times\Jx$, al variar la temperatura
inicial del \Index{ba\~no} y el campo magn\'etico aplicado,
respectivamente. Aunque el estudio sistem\'atico de la dependencia
respecto de temperatura no es uno de los prop\'ositos de este
trabajo, en todas las muestras en que hemos medido curvas \EJ\ a
diferentes temperaturas (muestras \fA, \fD\ y \fE) hemos observado
esta tendencia, de que la potencia \Wx\ disminuya conforme nos
acercamos a \Tc. Esta misma tendencia se encuentra en otras
muestras de la bibliograf\'{\i}a \cite{Xiao98,Curras00b}, aunque
excepcionalmente se puede encontrar alguna cuyo comportamiento sea
justo el opuesto, siendo $\Wx(T)$ mayor a temperaturas m\'as
elevadas \cite{Kamm00,Curras01}.

La tendencia de la potencia $\Wx(H)$ a aumentar con el campo
externo es, por el contrario, universal hasta el momento, tanto en
nuestras muestras como en otras cuyos datos se han publicado
\cite{Xiao98,Xiao99,Pauly00}. Es conocido que aplicar campo
ºmagn\'etico tiene el efecto de disminuir la temperatura
cr\'{\i}tica \Tc, as\'{\i} que si aumentamos el campo lo que
hacemos es aproximar la muestra a la transici\'on, de manera
parecida a si aument\'asemos directamente la temperatura. El que
la potencia disipada antes del salto sea mayor cuanto m\'as nos
acercamos a \Tc\ parecer\'{\i}a indicar que el mecanismo implicado
en la transici\'on al estado normal nada tiene que ver con efectos
t\'ermicos. En principio, si la causa del salto fuese que el calor
eleva la temperatura de la muestra hasta por encima de \Tc, se
podr\'{\i}a esperar que la potencia necesaria para generar la
transici\'on disminuyese con el \index{campo magn\'etico} campo
aplicado o, en todo caso, fuese independiente de \'el
\cite{Xiao99}, y desde luego que no aumente con la temperatura.

M\'as adelante (sec. \ref{ssec:WxTH}) veremos, sin embargo, que
este razonamiento es incorrecto cuando hacemos un detallado
an\'alisis de la disipaci\'on del calor desde la muestra al
entorno. Concluiremos que no s\'olo el valor de \Wx\ es suficiente
para establecer si un estado dado es cr\'{\i}tico o no, y que hay
que implicar otros factores.
  % cap 5
\clearemptydoublepage
%%%%%%%%%%%%%%%%%%%%%%%%%%%%%%%%%%%%%%%%%%%%%%%%%%%%%%%%%%%%%%%%%%
\chapter{Modelo de calentamiento en avalancha}
%\chaptermark{Modelo de calentamiento en avalancha}
%\addtocontents{toc}{\protect\vspace{0.2cm}}
\label{cap:cal}
%%%%%%%%%%%%%%%%%%%%%%%%%%%%%%%%%%%%%%%%%%%%%%%%%%%%%%%%%%%%%%%%%%

%-%-%-%-%-%-%-%-%-%-%-%-%-%-%-%-%-%-%-%-%-%-%-%-  figura -%-%-%-%-%-%-%-%

\newcommand{\figCalEJP}{  % alias

\figura {fig:cal:EJP}      % label
{fig/cal/EJP}              % file
{Ejemplos de curvas \EJ\ para distintas temperaturas en uno de los
filmes de \YBCO\ estudiados en este trabajo. El \'area de los
rect\'angulos es la potencia disipada por unidad de volumen. Estos
ejemplos ilustran c\'omo un relativamente peque\~no aumento de la
temperatura, a corriente constante, aumenta considerablemente la
disipaci\'on en el filme.}                  % caption
{Aumento considerable de la potencia disipada con el incremento de la temperatura, a densidad de corriente $J$ constante.}                  % toc
{\stfigw}                  % width \textwidth
}
% - - - - - - - - - - - - - - - - - - - - - - - - - - - - - - - - - - - %

%-%-%-%-%-%-%-%-%-%-%-%-%-%-%-%-%-%-%-%-%-%-%-%-  figura -%-%-%-%-%-%-%-%

\newcommand{\figCalET}{  % alias

\figura {fig:cal:ET}      % label
{fig/cal/ET}              % file
{Valor del campo $E$ en funci\'on de la temperatura, a distintas
densidades de
    corriente $J$ constantes, para la muestra \fA.}                  % caption
{Valor del campo $E$ en funci\'on de la temperatura, a distintas
densidades de
    corriente $J$ constantes.}                  % toc
{\stfigw}                  % width \textwidth
}
% - - - - - - - - - - - - - - - - - - - - - - - - - - - - - - - - - - - %

%-%-%-%-%-%-%-%-%-%-%-%-%-%-%-%-%-%-%-%-%-%-%-%-  figura -%-%-%-%-%-%-%-%

\newcommand{\figCalFilmsubs}{  % alias

\figura {fig:cal:filmsubs}      % label
{fig/cal/filmsubs}              % file
{Corte transversal del filme y el substrato (esquema). Se indica
el volumen de substrato que se ha calentado pasado un tiempo $t$
desde que comenz\'o la generaci\'on de calor en la muestra. El resto
del substrato no ha tenido tiempo de sufrir ning\'un calentamiento.}                  % caption
{Corte transversal del filme y el substrato (esquema).}                  % toc
{.8}            % width \textwidth
}
% - - - - - - - - - - - - - - - - - - - - - - - - - - - - - - - - - - - %

%-%-%-%-%-%-%-%-%-%-%-%-%-%-%-%-%-%-%-%-%-%-%-%-  figura -%-%-%-%-%-%-%-%

\newcommand{\figCalDTs}{  % alias

\figura {fig:cal:DTs}      % label
{fig/cal/DTs}              % file
{Aumento de la temperatura del substrato caliente \DTs\ en funci\'on
del tiempo, normalizado, suponiendo potencia disipada constante
(ecuaci\'on \ref{ec:cal:DtsDtmax}).}
% caption
{Aumento de la temperatura del substrato caliente \DTs\ en funci\'on
del tiempo, a potencia disipada constante.}                  % toc
{\stfigw}            % width \textwidth
}
% - - - - - - - - - - - - - - - - - - - - - - - - - - - - - - - - - - - %

%-%-%-%-%-%-%-%-%-%-%-%-%-%-%-%-%-%-%-%-%-%-%-%-  figura -%-%-%-%-%-%-%-%

\newcommand{\figCalDTsZoom}{  % alias

\figura {fig:cal:DTsZoom}      % label
{fig/cal/DTsZoom}              % file
{Aumento de la temperatura del substrato caliente \DTs\ en funci\'on
del tiempo, normalizado, suponiendo potencia disipada constante.
(Detalle de la figura \ref{fig:cal:DTs}).}                  % caption
{Detalle de la figura \ref{fig:cal:DTs}.}                  % toc
{\stfigw}            % width \textwidth
}
% - - - - - - - - - - - - - - - - - - - - - - - - - - - - - - - - - - - %

%-%-%-%-%-%-%-%-%-%-%-%-%-%-%-%-%-%-%-%-%-%-%-%-  figura -%-%-%-%-%-%-%-%

\newcommand{\figCaldDTsdt}{  % alias

\figura {fig:cal:dDTsdt}      % label
{fig/cal/dDTsdt}              % file
{Derivada del aumento de la temperatura del substrato caliente
$\indiff{\DTs}{t}$ en funci\'on del tiempo, dividida por $P^{-1}$,
suponiendo potencia disipada $P$ constante.}                  %caption
{Derivada del aumento de la temperatura del substrato caliente $\indiff{\DTs}{t}$ en funci\'on del tiempo, a potencia disipada $P$ constante.}                  % toc
{\stfigw}            % width \textwidth
}
% - - - - - - - - - - - - - - - - - - - - - - - - - - - - - - - - - - - %

%-%-%-%-%-%-%-%-%-%-%-%-%-%-%-%-%-%-%-%-%-%-%-%-  figura -%-%-%-%-%-%-%-%

\newcommand{\figCalAlpha}{  % alias

\figura {fig:cal:alpha}      % label
{fig/cal/alpha}              % file
{Funci\'on $\alp(t)$ que se emplea para corregir la distribuci\'on de temperatura en el substrato, dada por la ecuaci\'on \ref{ec:alpha}.}                  % caption
{Funci\'on $\alp(t)$ que se emplea para corregir la distribuci\'on de temperatura en el substrato.}                  % toc
{\stfigw}            % width \textwidth
}
% - - - - - - - - - - - - - - - - - - - - - - - - - - - - - - - - - - - %

%-%-%-%-%-%-%-%-%-%-%-%-%-%-%-%-%-%-%-%-%-%-%-%-  figura -%-%-%-%-%-%-%-%

\newcommand{\figCaldvsvs}{  % alias

\figura {fig:cal:dvsvs}      % label
{fig/cal/dvsvs}              % file
{T\'ermino entre par\'entesis de la ecuaci\'on \ref{ec:dTfdtPcte}. Como no se anula ni siquiera a potencia constante, la derivada $\indiff{\Tf}{t}$ nunca es cero, por lo que estrictamente no existe estabilidad.}                  % caption
{T\'ermino entre par\'entesis de la ecuaci\'on \ref{ec:dTfdtPcte}, que caracteriza el comportamiento de $\indiff{\Tf}{t}$ a $P$ constante.}                  % toc
{\stfigw}            % width \textwidth
}
% - - - - - - - - - - - - - - - - - - - - - - - - - - - - - - - - - - - %

%-%-%-%-%-%-%-%-%-%-%-%-%-%-%-%-%-%-%-%-%-%-%-%-  figura -%-%-%-%-%-%-%-%

\newcommand{\figCalfunodos}{  % alias

\figura {fig:cal:f1f2}      % label
{fig/cal/f1f2}              % file
{T\'ermino que determina la condici\'on de avalancha a potencia $P$ constante, seg\'un la ecuaci\'on \ref{ec:condAvaPcte}. Como siempre es negativo, la condici\'on de avalancha nunca se satisface.}                  % caption
{T\'ermino que determina la condici\'on de avalancha a potencia $P$ constante, seg\'un la ecuaci\'on \ref{ec:condAvaPcte}.}                  % toc
{\stfigw}            % width \textwidth
}
% - - - - - - - - - - - - - - - - - - - - - - - - - - - - - - - - - - - %

%-%-%-%-%-%-%-%-%-%-%-%-%-%-%-%-%-%-%-%-%-%-%-%-  figura -%-%-%-%-%-%-%-%

\newcommand{\figCalsDTtbasic}{  % alias

\figura {fig:cal:sDTtbasic}      % label
{fig/cal/sDTtbasic}              % file
{Resultados de una simulaci\'on con realimentaci\'on, a
$\Tb=76.2\und{K}$ y $B=0$, aplicando tres corrientes de valores
diferentes pero cercanos entre s\'{\i} (7.70, 7.83 y
$7.90\Exund{6}{\Acms}$). Se representa el incremento de
temperatura del micropuente $\DTf=\Tf-\Tb$, calculado seg\'un el
modelo simple de las ecuaciones \ref{ec:Tfsimple} y previas. La
corriente m\'as baja genera un estado casi estacionario tras el
incremento inicial de temperatura, pero las otras dos causan
avalanchas t\'ermicas (m\'as pronto cuanto mayor es la corriente)
que elevan la temperatura de la muestra por encima de \Tc, al
estado
normal.}                  % caption
{Resultados de una simulaci\'on con realimentaci\'on, calculada seg\'un el modelo simple de las ecuaciones \ref{ec:Tfsimple} y previas.}                  % toc
{\stfigw}            % width \textwidth
}
% - - - - - - - - - - - - - - - - - - - - - - - - - - - - - - - - - - - %

%-%-%-%-%-%-%-%-%-%-%-%-%-%-%-%-%-%-%-%-%-%-%-%-  figura -%-%-%-%-%-%-%-%

\newcommand{\figCalComPcte}{  % alias

\figura{fig:cal:comPcte}      % label
{fig/cal/comPcte}              % file
{Comparaci\'on de una simulaci\'on con realimentaci\'on
(l\'{\i}nea continua) con otra en la que la potencia permanece
constante (l\'{\i}nea a trazos). Se representa el aumento de
temperatura $\DT=\DTf=\DTs$ (ecuaciones \ref{ec:DTs} y
\ref{ec:Tfsimple}). La temperatura inicial y la corriente aplicada
son las mismas en ambos casos, pero en el de realimentaci\'on se
tiene en cuenta que la potencia $P$ aumenta cuando aumenta la
temperatura. El que estabiliza antes y a temperatura menor es el
de potencia constante, con resultados an\'alogos a los mostrados
en
la figura \ref{fig:cal:DTs}.}                  % caption
{Comparaci\'on de una simulaci\'on con realimentaci\'on con otra en la que la potencia permanece constante.}                  % toc
{\stfigw}            % width \textwidth
}
% - - - - - - - - - - - - - - - - - - - - - - - - - - - - - - - - - - - %

%-%-%-%-%-%-%-%-%-%-%-%-%-%-%-%-%-%-%-%-%-%-%-%-  figura -%-%-%-%-%-%-%-%

\newcommand{\figCalJxTsimple}{  % alias

\figura {fig:cal:JxTsimple}      % label
{fig/cal/JxTsimple}              % file
{Comparaci\'on con el experimento de los valores de \Jx\
calculados con la ecuaci\'on \ref{ec:Tfsimple} y previas,
introduciendo realimentaci\'on en la potencia en la ecuaci\'on
\ref{ec:Qt}. La l\'{\i}nea que une los puntos calculados es s\'olo
una gu\'{\i}a visual. El modelo, debido a su simplicidad, predice
valores de \Jx\ que exceden en un 30\% los reales. Introduciremos
correcciones al modelo sencillo para mejorar las predicciones,
pero lo relevante ya se encuentra aqu\'{\i}: existen corrientes
que hacen transitar a la muestra hasta el estado normal s\'olo con
mecanismos t\'ermicos, y la dependencia de \Jx\ con la temperatura
es id\'entica a la
experimental.}                  % caption
{Comparaci\'on con el experimento de los valores de \Jx\ calculados con el modelo m\'as simple.}                  % toc
{\stfigw}            % width \textwidth
}
% - - - - - - - - - - - - - - - - - - - - - - - - - - - - - - - - - - - %

%-%-%-%-%-%-%-%-%-%-%-%-%-%-%-%-%-%-%-%-%-%-%-%-  figura -%-%-%-%-%-%-%-%

\newcommand{\figCalsDTt}{  % alias

\figura {fig:cal:sDTt}      % label
{fig/cal/sDTt}              % file
{Simulaci\'on de la evoluci\'on temporal de la temperatura de un
micropuente, aplicando distintas corrientes de valores pr\'oximos
entre s\'{\i} (6.70, 6.73, 6.74 y $6.84\Exund{6}{\Acms}$), a
$B=0$. Entre la primera y la \'ultima hay un 2\% de diferencia,
pero conllevan resultados muy diferentes. La temperatura inicial
es
$\Tb=75.6\und{K}$.}                  % caption
{Simulaci\'on de la evoluci\'on temporal de la temperatura de un micropuente, a distintas corrientes pr\'oximas entre s\'{\i}. $\Tb=75.6\und{K}$, $B=0$.}                  % toc
{\stfigw}            % width \textwidth
}
% - - - - - - - - - - - - - - - - - - - - - - - - - - - - - - - - - - - %

%-%-%-%-%-%-%-%-%-%-%-%-%-%-%-%-%-%-%-%-%-%-%-%-  figura -%-%-%-%-%-%-%-%

\newcommand{\figCalsEtJ}{  % alias

\bfig
  \centering
  \includegraphics[width=\stfigw\textwidth,clip=true]{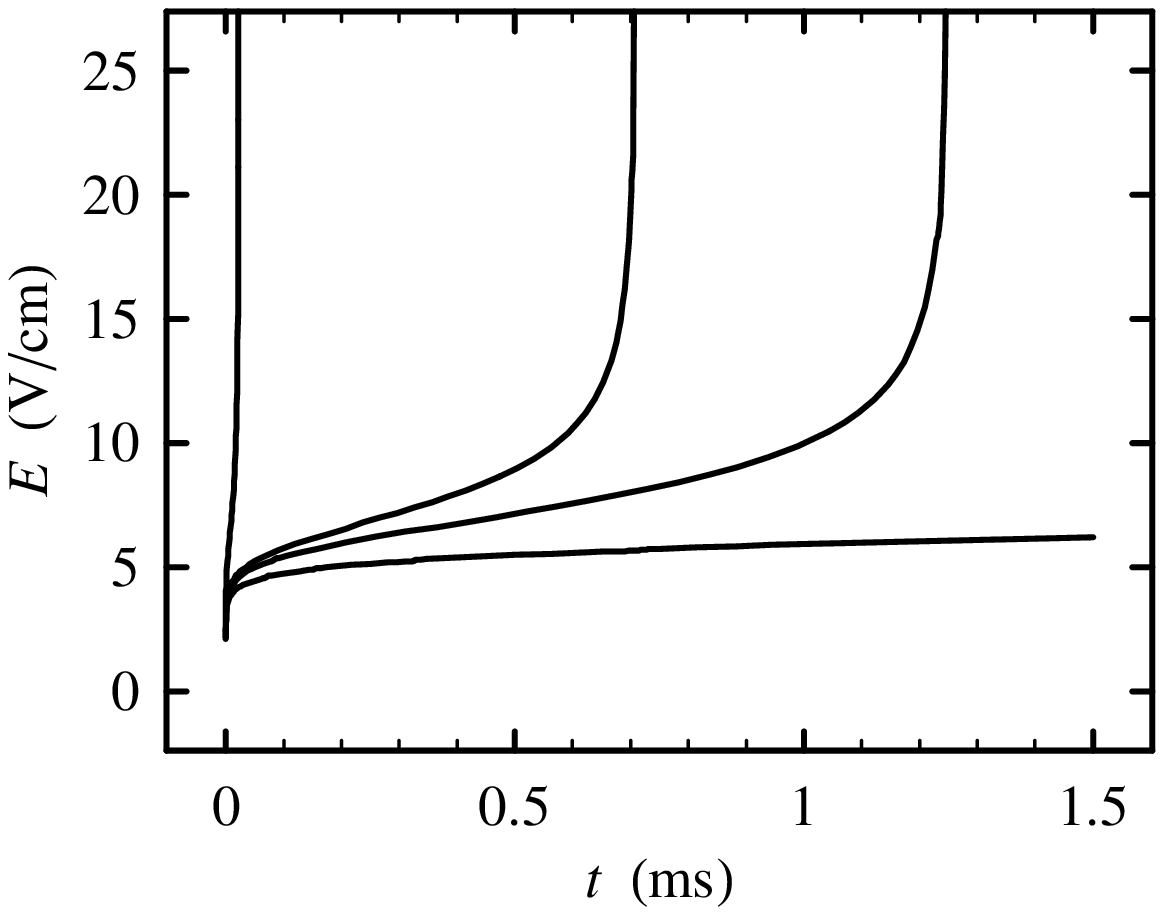}
  \includegraphics[width=\stfigw\textwidth,clip=true]{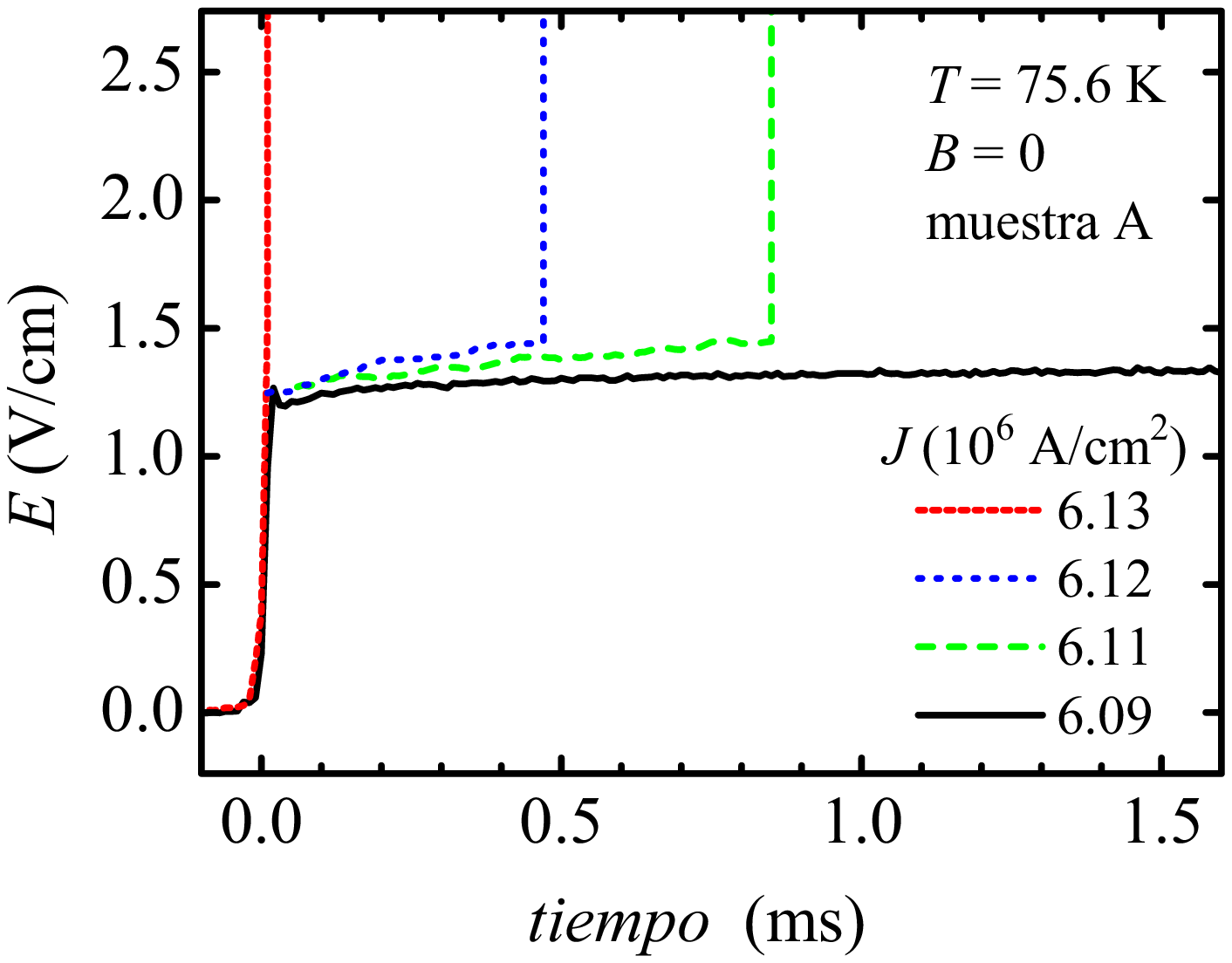}
  \caption[Evoluci\'on temporal del campo $E$ en un micropuente a
  distintas corrientes constantes, correspondiente a la simulaci\'on
  de la figura \ref{fig:cal:sDTt}, y comparaci\'on con el experimento. $\Tb=75.6\und{K}$,
  $B=0$.]{Arriba, evoluci\'on temporal del campo $E$ en un micropuente a
  distintas corrientes constantes (6.70, 6.73, 6.74 y $6.84\Exund{6}{\Acms}$), correspondiente a la simulaci\'on
  de la figura \ref{fig:cal:sDTt}. Abajo, por comparaci\'on, los resultados experimentales
  ya mostrados previamente en la figura \ref{fig:datos:EtJ} (p.\pageref{fig:datos:EtJ}), con 6.09, 6.11, 6.12 y $6.13\Exund{6}{\Acms}$. $\Tb=75.6\und{K}$,
  $B=0$. Las transiciones del modelo no son tan abruptas, ni a las mismas corrientes que las experimentales, pero el comportamiento es semejante.}
  \label{fig:cal:sEtJ}

\efig %%%%%%%%%%%%%%%%%%%%%%

}
% - - - - - - - - - - - - - - - - - - - - - - - - - - - - - - - - - - - %

%-%-%-%-%-%-%-%-%-%-%-%-%-%-%-%-%-%-%-%-%-%-%-%-  figura -%-%-%-%-%-%-%-%

%-%-%-%-%-%-%-%-%-%-%-%-%-%-%-%-%-%-%-%-%-%-%-%-  figura -%-%-%-%-%-%-%-%

\newcommand{\figCalsDTtEtJSetSeis}{  % alias

\bfig
  \centering
  \includegraphics[width=\stfigw\textwidth,clip=true]{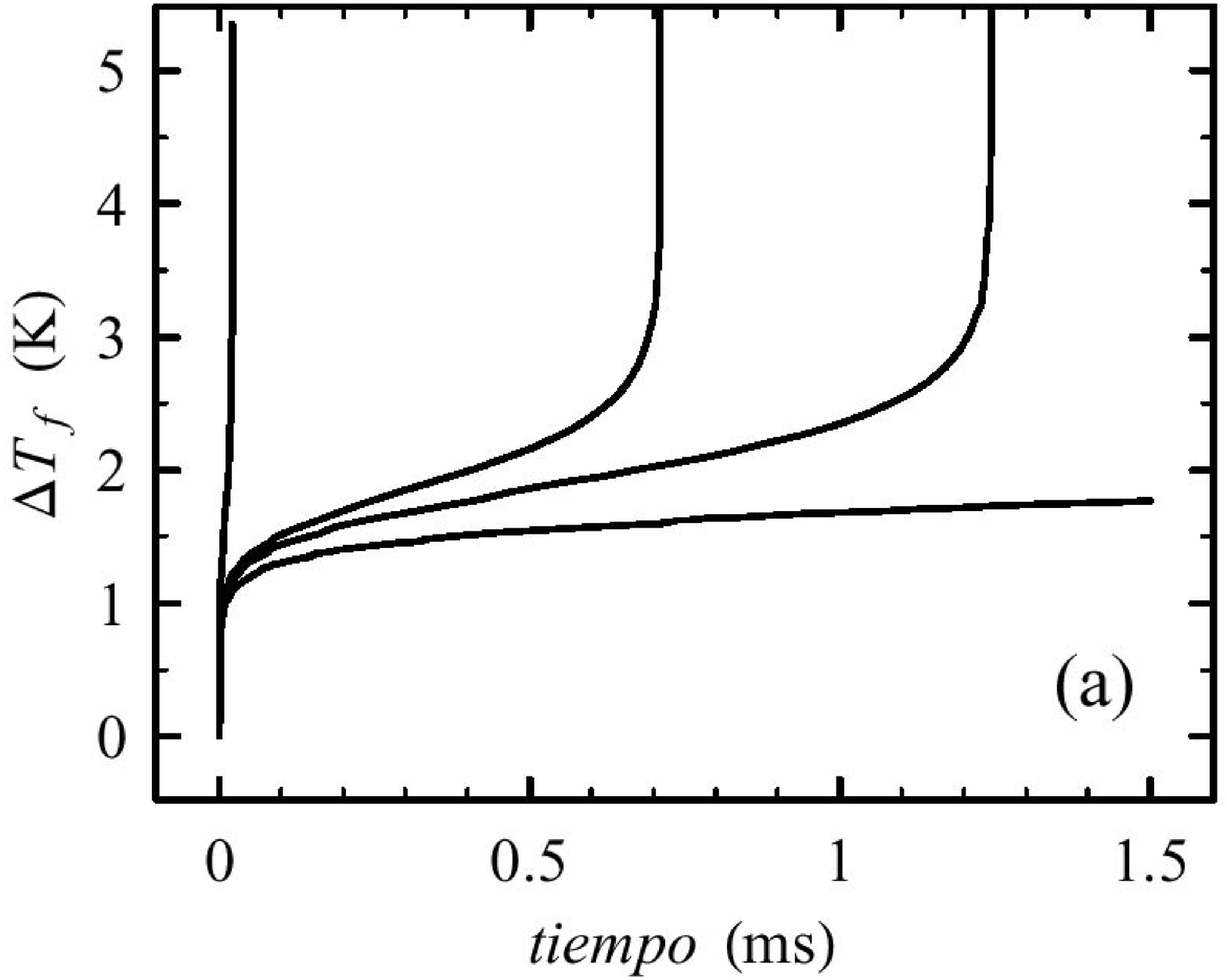}
  \includegraphics[width=\stfigw\textwidth,clip=true]{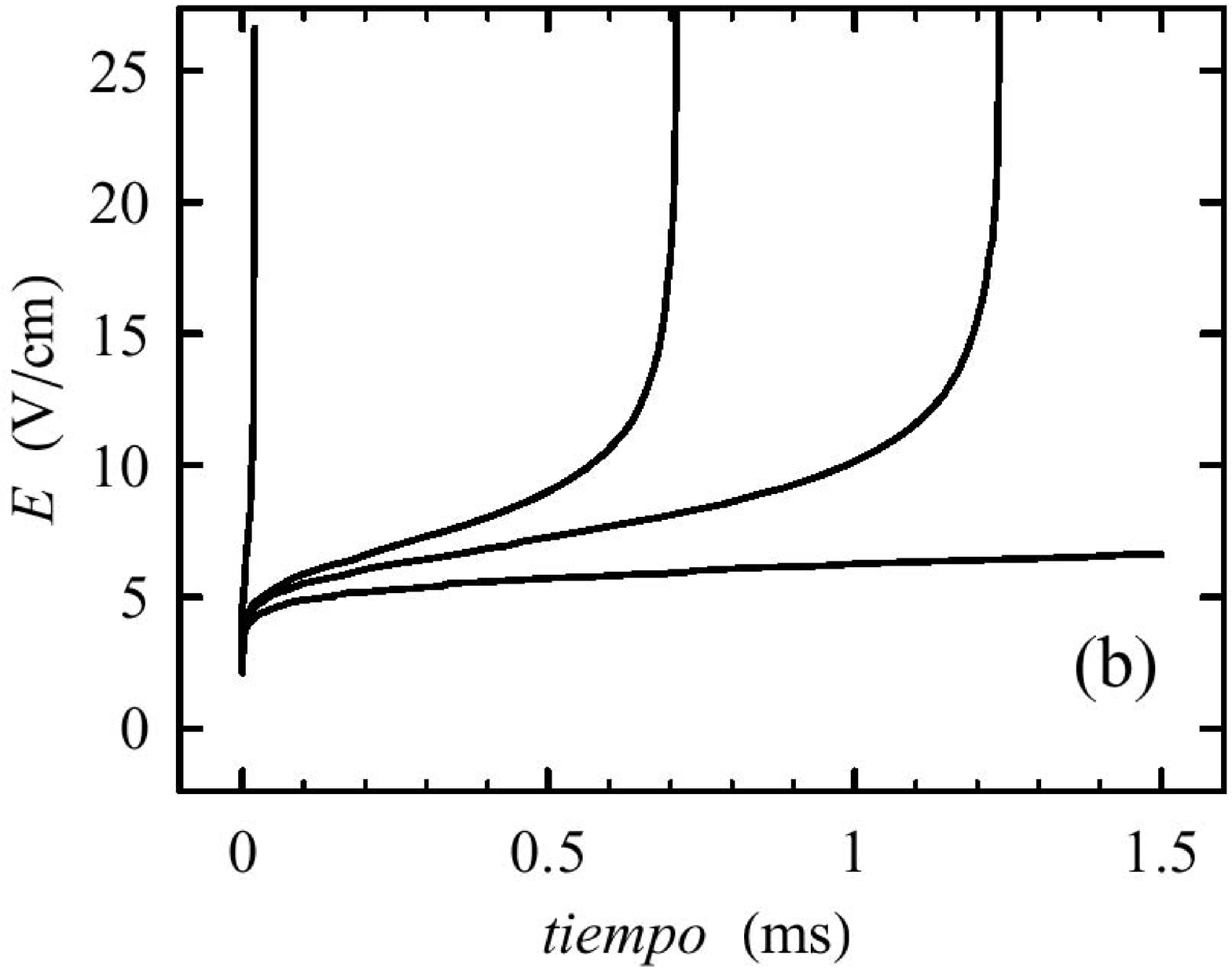}
  \caption[Simulaci\'on de la evoluci\'on temporal de la temperatura y del campo el\'ectrico,
  a distintas corrientes pr\'oximas entre s\'{\i}. $\Tb=76.2\und{K}$,
  $B=0$.]{Simulaci\'on de la evoluci\'on temporal de la temperatura (a) y del campo el\'ectrico (b),
  aplicando distintas corrientes de valores pr\'oximos entre s\'{\i} (6.38, 6.41, 6.42 y $6.52\Exund{6}{\Acms}$), a $B=0$. Entre la
  primera y la \'ultima hay un 2\% de diferencia, pero conllevan
  resultados muy diferentes. La diferencia con la simulaci\'on de
  las
  figuras \ref{fig:cal:sDTt} y \ref{fig:cal:sEtJ} es que la temperatura inicial es ahora
  de $\Tb=76.2\und{K}$. Los resultados son parecidos, pero
  necesitaremos estos para posteriores comparaciones.}
  \label{fig:cal:sDTtEtJ76}

\efig %%%%%%%%%%%%%%%%%%%%%%

}
% - - - - - - - - - - - - - - - - - - - - - - - - - - - - - - - - - - - %

%-%-%-%-%-%-%-%-%-%-%-%-%-%-%-%-%-%-%-%-%-%-%-%-  figura -%-%-%-%-%-%-%-%

\newcommand{\figCalsrhotWtSetSeis}{  % alias

\bfig
  \centering
  \includegraphics[width=\stfigw\textwidth,clip=true]{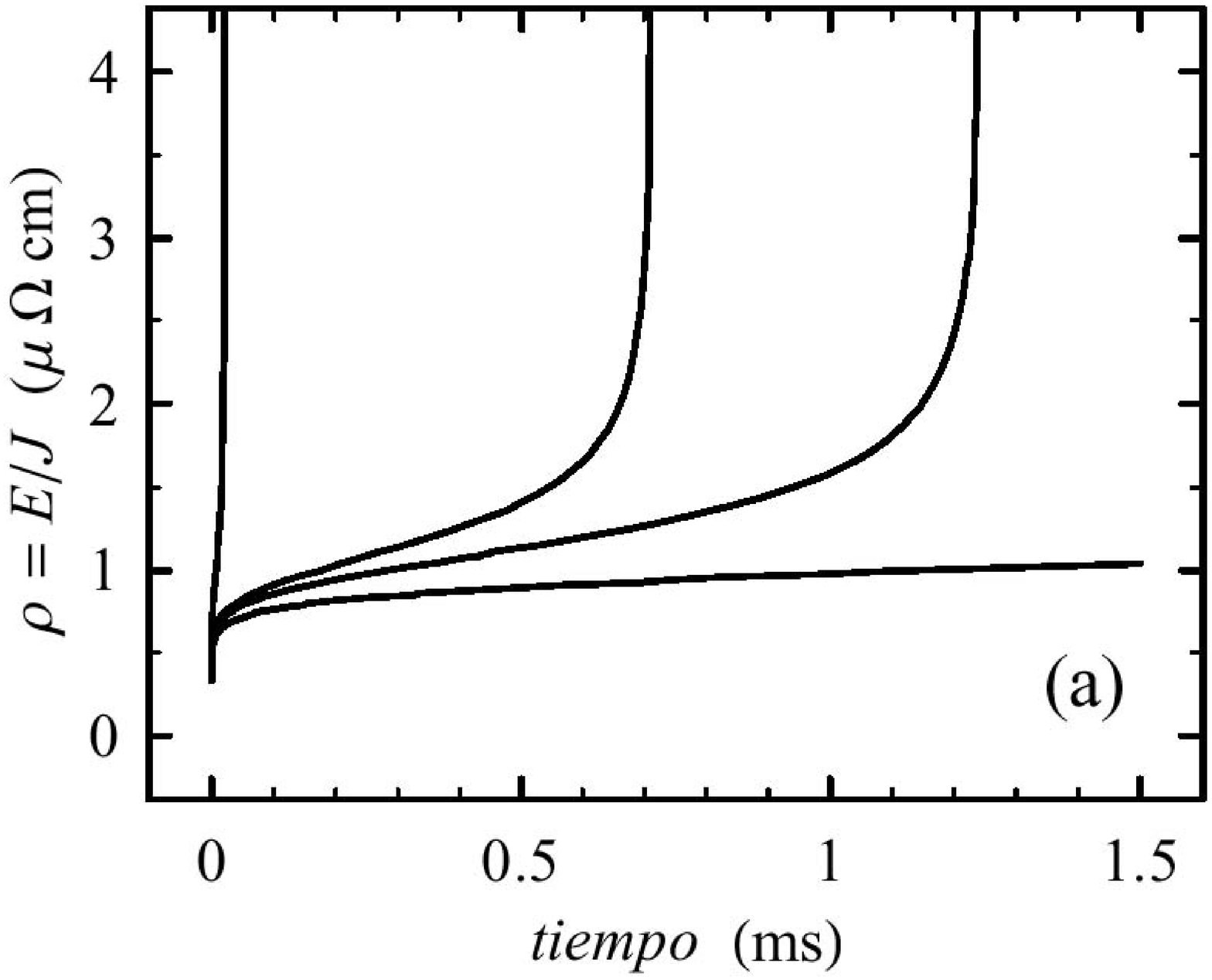}
  \includegraphics[width=\stfigw\textwidth,clip=true]{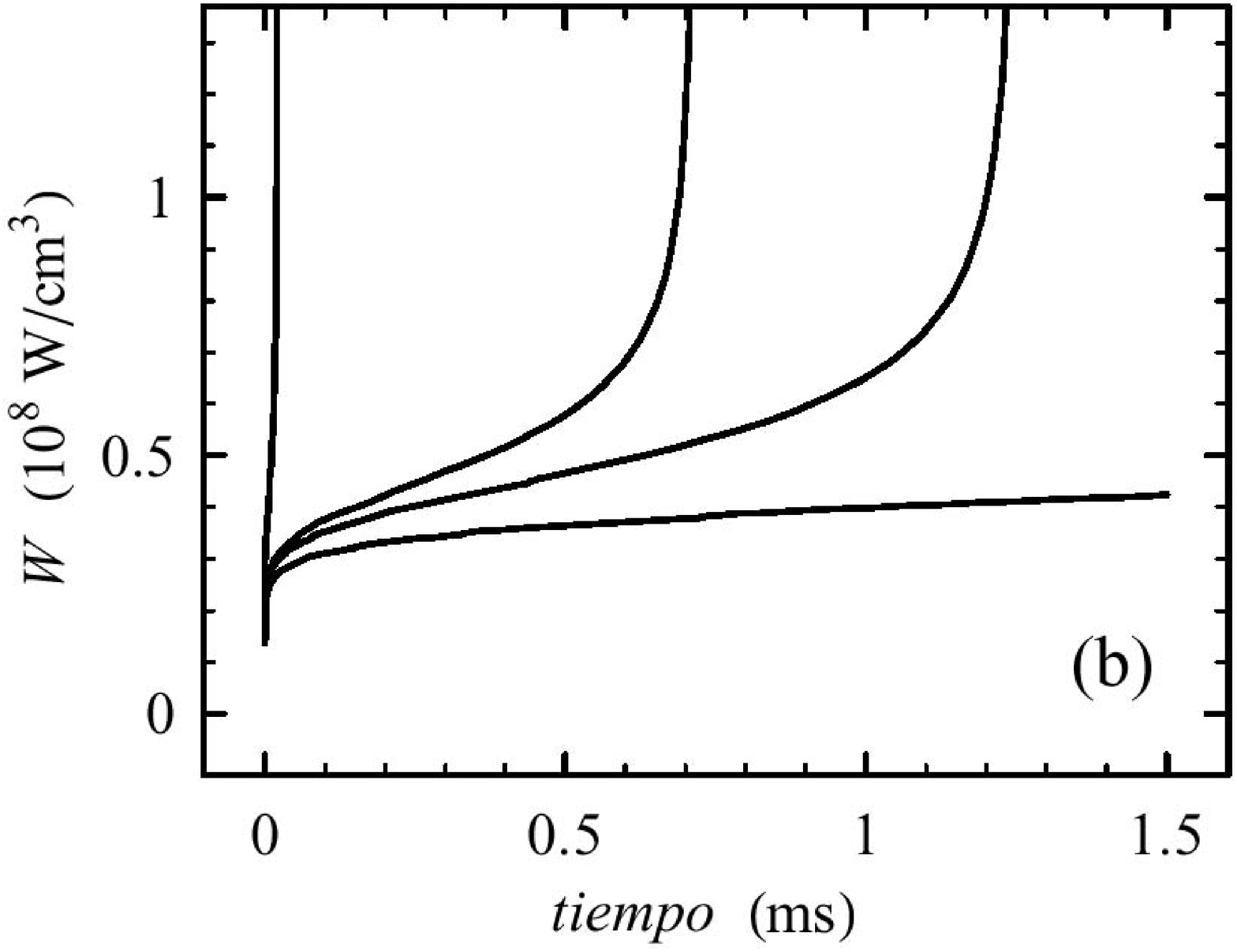}
  \caption{Evoluci\'on temporal de la resistividad $\rho$ del
  micropuente (a) y de la potencia disipada $W$ (b) a distintas
  corrientes constantes, correspondientes a la simulaci\'on de la
  figura \ref{fig:cal:sDTtEtJ76}. $\Tb=76.2\und{K}$, $B=0$.}
  \label{fig:cal:srhotWt76}

\efig %%%%%%%%%%%%%%%%%%%%%%

}
% - - - - - - - - - - - - - - - - - - - - - - - - - - - - - - - - - - - %

%-%-%-%-%-%-%-%-%-%-%-%-%-%-%-%-%-%-%-%-%-%-%-%-  figura -%-%-%-%-%-%-%-%

\newcommand{\figCalsDTtWtOchDos}{  % alias

\bfig
  \centering
  \includegraphics[width=\stfigw\textwidth,clip=true]{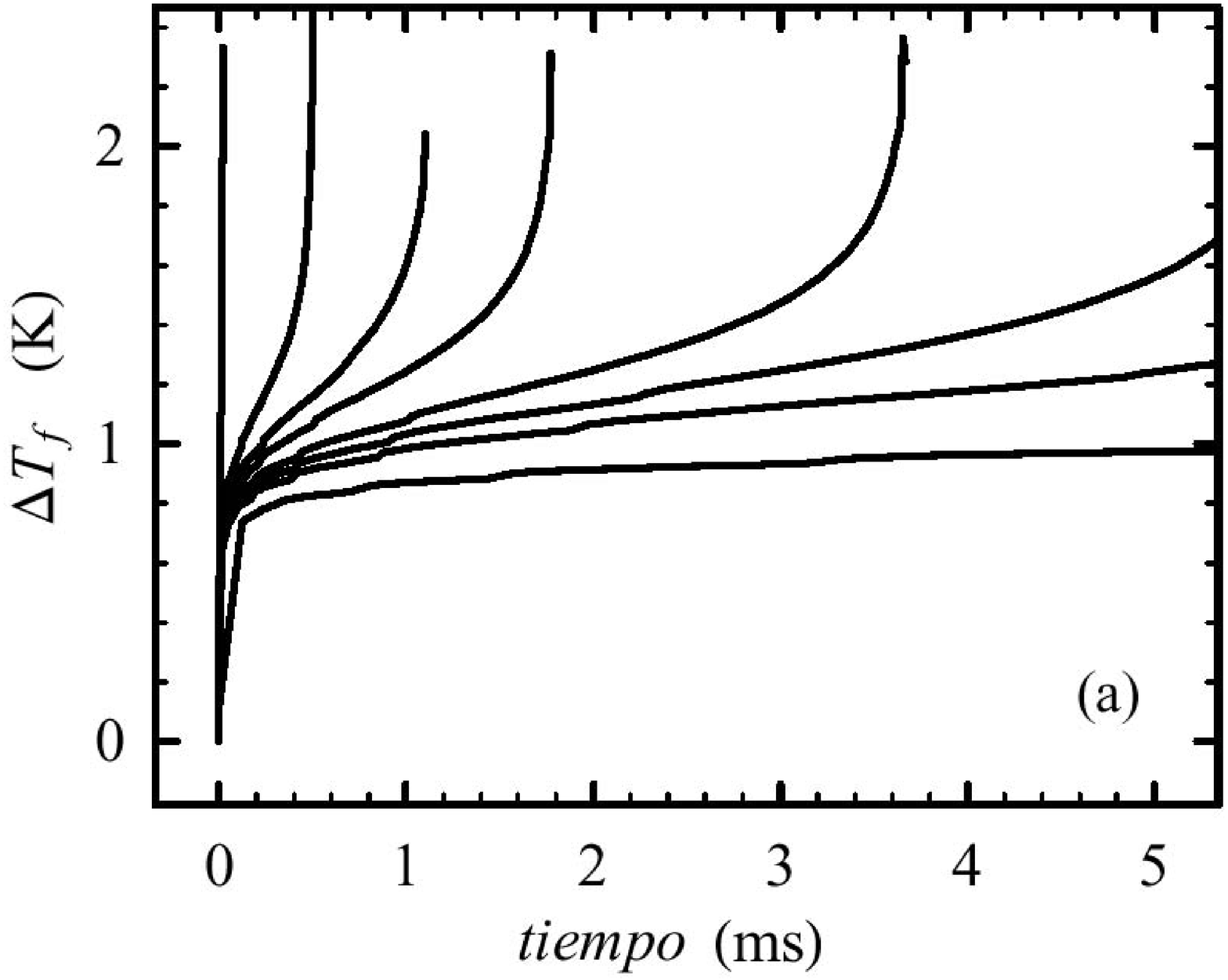}
  \includegraphics[width=\stfigw\textwidth,clip=true]{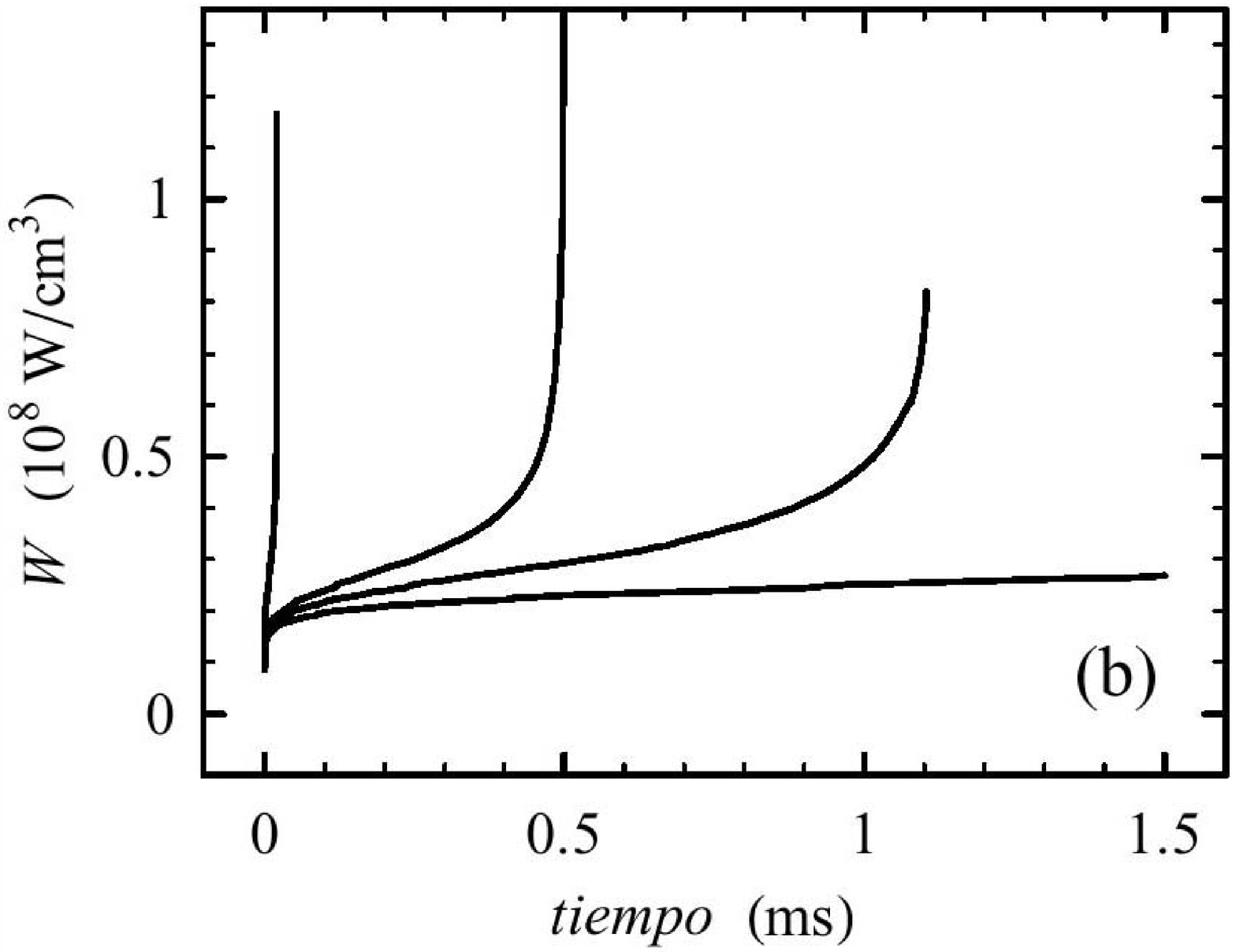}
  \caption[Simulaci\'on de la evoluci\'on temporal
 de la temperatura y de la potencia disipada, a
 distintas corrientes pr\'oximas entre s\'{\i}.
 $\Tb=82.0\und{K}$, $B=0$.]{Arriba, simulaci\'on
 de la evoluci\'on temporal de la temperatura,
 aplicando distintas corrientes en el rango de
 3.47--$3.55\Exund{6}{\Acms}$, a $B=0$. La
 temperatura inicial es de $\Tb=82.0\und{K}$.
 Abajo, evoluci\'on temporal de la potencia
 disipada $W=E \times J$ a algunas de esas
 mismas corrientes. La figura (a) abarca un
 rango temporal m\'as amplio que el resto de las
 figuras mostradas, para apreciar las tendencias
 a tiempos mayores.}
  \label{fig:cal:sDTtWt82}

\efig %%%%%%%%%%%%%%%%%%%%%%

}
% - - - - - - - - - - - - - - - - - - - - - - - - - - - - - - - - - - - %

%-%-%-%-%-%-%-%-%-%-%-%-%-%-%-%-%-%-%-%-%-%-%-%-  figura -%-%-%-%-%-%-%-%

\newcommand{\figCalsDTtOchDos}{  % alias

\figura {fig:cal:sDTt82}      % label
{fig/cal/sDTt82}              % file
{Simulaci\'on de la evoluci\'on temporal de la temperatura, aplicando
distintas corrientes en el rango de 3.47--$3.55\Exund{6}{\Acms}$,
a $B=0$. La temperatura inicial es de $\Tb=82.0\und{K}$.} % caption
{Simulaci\'on de la evoluci\'on temporal de la temperatura, a distintas corrientes pr\'oximas entre s\'{\i}. $\Tb=82.0\und{K}$, $B=0$.}                  % toc
{\stfigw}            % width \textwidth
}
% - - - - - - - - - - - - - - - - - - - - - - - - - - - - - - - - - - - %

%-%-%-%-%-%-%-%-%-%-%-%-%-%-%-%-%-%-%-%-%-%-%-%-  figura -%-%-%-%-%-%-%-%

\newcommand{\figCalsWtOchDos}{  % alias

\figura {fig:cal:sWt82}      % label
{fig/cal/sWt82}              % file
{Evoluci\'on temporal de la potencia disipada $W=E \times J$ a distintas corrientes, correspondiente a la simulaci\'on de la figura \ref{fig:cal:sDTt82}. $\Tb=82.0\und{K}$, $B=0$.}                  % caption
{Evoluci\'on temporal de la potencia disipada $W=E \times J$ a distintas corrientes, correspondiente a la simulaci\'on de la figura \ref{fig:cal:sDTt82}. $\Tb=82.0\und{K}$, $B=0$.}                  % toc
{\stfigw}            % width \textwidth
}
% - - - - - - - - - - - - - - - - - - - - - - - - - - - - - - - - - - - %

%-%-%-%-%-%-%-%-%-%-%-%-%-%-%-%-%-%-%-%-%-%-%-%-  figura -%-%-%-%-%-%-%-%

\newcommand{\figCalsDTtB}{  % alias

\figura {fig:cal:sDTtB1}      % label
{fig/cal/sDTtB1}              % file
{Simulaci\'on de la evoluci\'on temporal de la temperatura,
aplicando distintas corrientes de valores pr\'oximos entre s\'{\i}
(2.68, 2.70, 2.71 y $2.80\Exund{6}{\Acms}$). La diferencia con la
simulaci\'on de la figura \ref{fig:cal:sDTtEtJ76}-a es que el
campo aplicado es ahora
$\mH=1\und{T}$.}                  % caption
{Simulaci\'on de la evoluci\'on temporal de la temperatura, a
distintas corrientes pr\'oximas entre s\'{\i}. $\Tb=76.2\und{K}$,
$\mH=1\und{T}$.}                  % toc
{\stfigw}            % width \textwidth
}
% - - - - - - - - - - - - - - - - - - - - - - - - - - - - - - - - - - - %

%-%-%-%-%-%-%-%-%-%-%-%-%-%-%-%-%-%-%-%-%-%-%-%-  figura -%-%-%-%-%-%-%-%

\newcommand{\figCalsEtJWtB}{  % alias
\bfig
  \centering
  \includegraphics[width=\stfigw\textwidth,clip=true]{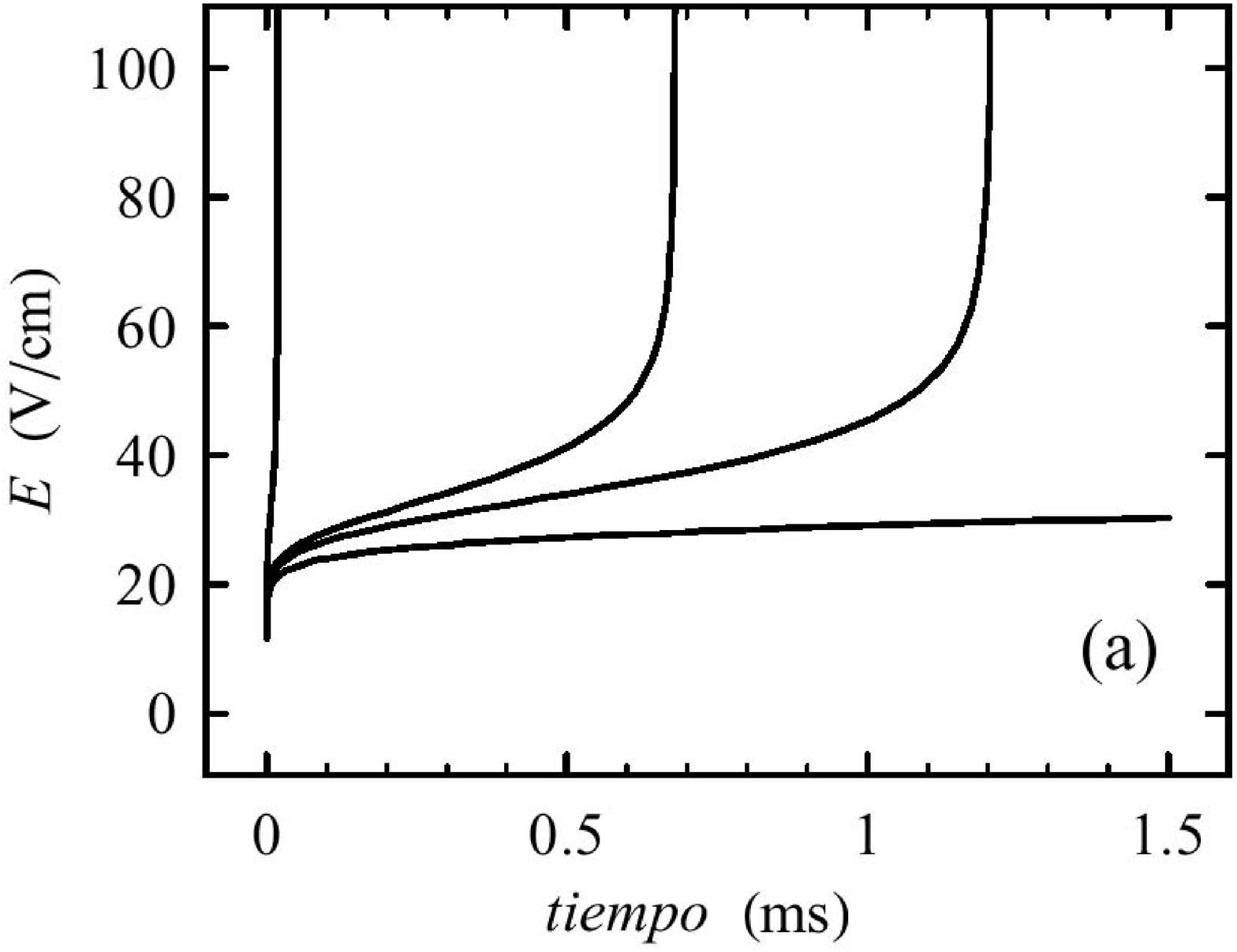}
  \includegraphics[width=\stfigw\textwidth,clip=true]{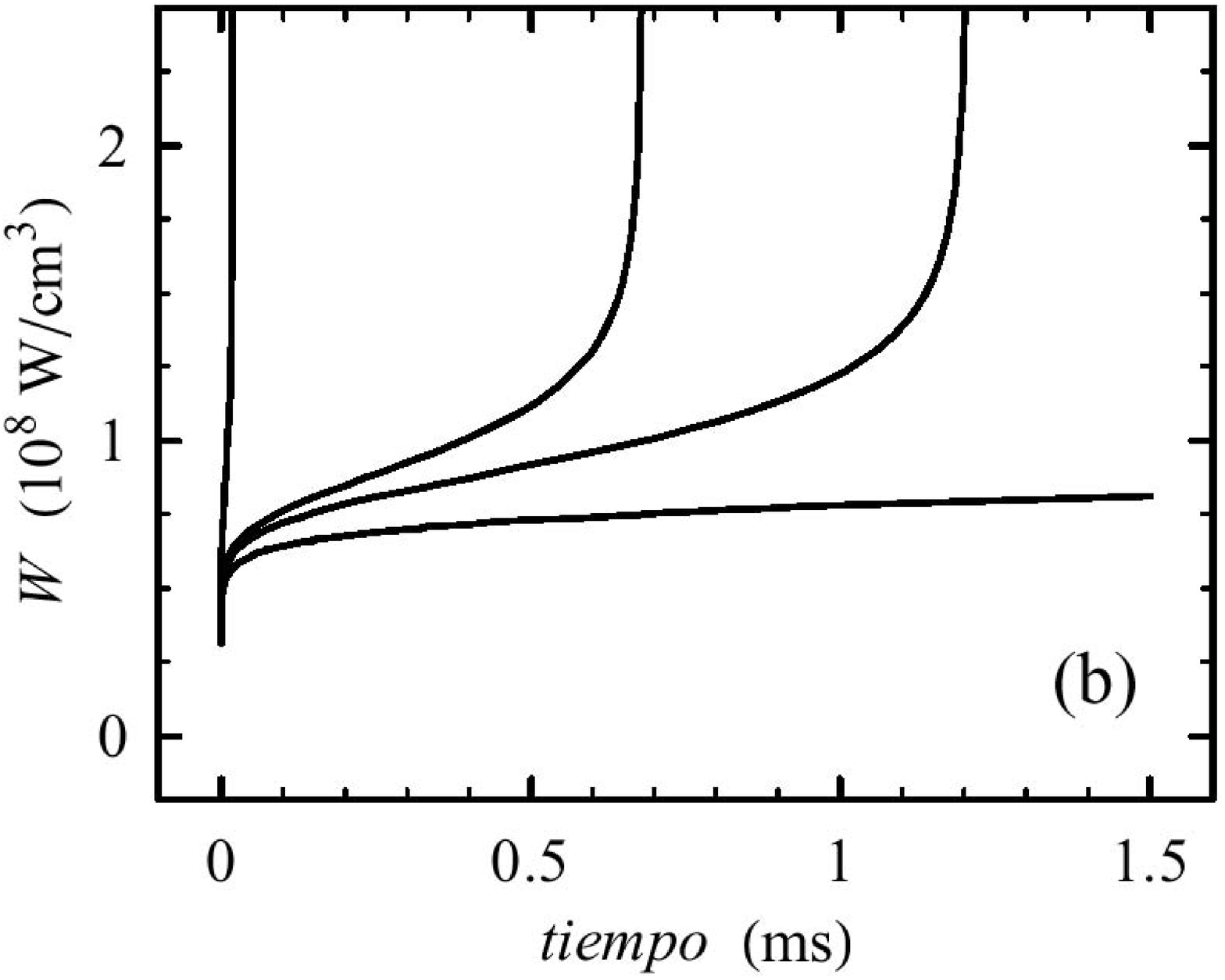}
  \caption{Evoluci\'on temporal del campo $E$ (a) y de la potencia
  disipada (b) a distintas corrientes, correspondiente a la
  simulaci\'on de la figura \ref{fig:cal:sDTtB1}. $\Tb=76.2\und{K}$,
  $\mH=1\und{T}$.}
  \label{fig:cal:sEtJWtB1}

\efig %%%%%%%%%%%%%%%%%%%%%%
}
% - - - - - - - - - - - - - - - - - - - - - - - - - - - - - - - - - - - %

%-%-%-%-%-%-%-%-%-%-%-%-%-%-%-%-%-%-%-%-%-%-%-%-  figura -%-%-%-%-%-%-%-%

\newcommand{\figCalsEtJB}{  % alias

\figura {fig:cal:sEtJB1}      % label
{fig/cal/sEtJB1}              % file
{Evoluci\'on temporal del campo $E$ a distintas corrientes, correspondiente a la simulaci\'on de la figura \ref{fig:cal:sDTtB1}. $\Tb=76.2\und{K}$, $\mH=1\und{T}$.}                  % caption
{Evoluci\'on temporal del campo $E$ a distintas corrientes, correspondiente a la simulaci\'on de la figura \ref{fig:cal:sDTtB1}. $\Tb=76.2\und{K}$, $\mH=1\und{T}$.}                  % toc
{\stfigw}            % width \textwidth
}
% - - - - - - - - - - - - - - - - - - - - - - - - - - - - - - - - - - - %

%-%-%-%-%-%-%-%-%-%-%-%-%-%-%-%-%-%-%-%-%-%-%-%-  figura -%-%-%-%-%-%-%-%

\newcommand{\figCalsWtB}{  % alias

\figura {fig:cal:sWtB1}      % label
{fig/cal/sWtB1}              % file
{Evoluci\'on temporal de la potencia disipada $W=E \times J$ a distintas corrientes, correspondiente a la simulaci\'on de la figura \ref{fig:cal:sDTtB1}. $\Tb=76.2\und{K}$, $\mH=1\und{T}$.}                  % caption
{Evoluci\'on temporal de la potencia disipada $W=E \times J$ a distintas corrientes, correspondiente a la simulaci\'on de la figura \ref{fig:cal:sDTtB1}. $\Tb=76.2\und{K}$, $\mH=1\und{T}$r.}                  % toc
{\stfigw}            % width \textwidth
}
% - - - - - - - - - - - - - - - - - - - - - - - - - - - - - - - - - - - %

%-%-%-%-%-%-%-%-%-%-%-%-%-%-%-%-%-%-%-%-%-%-%-%-  figura -%-%-%-%-%-%-%-%

\newcommand{\figCaltJxSTO}{  % alias

\figura {fig:cal:tJxSTO}      % label
{fig/cal/tJxSTO}              % file
{Correlaci\'on entre el valor de \Jx\ y el tiempo $t$ de duraci\'on de
la medida, a una temperatura inicial $\Tb=82.0\und{K}$, obtenida
de la simulaci\'on de la figura \ref{fig:cal:sDTtWt82}-a: se
representa el tiempo $t$ que una corriente $J$ tarda en inducir
una avalancha, contado desde que se aplica el pulso de corriente.}
% caption
{Correlaci\'on entre el valor de \Jx\ y el tiempo $t$ de duraci\'on de la medida.}                  % toc
{\stfigw}            % width \textwidth
}
% - - - - - - - - - - - - - - - - - - - - - - - - - - - - - - - - - - - %

%-%-%-%-%-%-%-%-%-%-%-%-%-%-%-%-%-%-%-%-%-%-%-%-  figura -%-%-%-%-%-%-%-%

\newcommand{\figCaltJxcompMgO}{  % alias

\figura {fig:cal:tJxcompMgO}      % label
{fig/cal/tJxcompMgO}              % file
{Comparaci\'on de los resultados de la figura \ref{fig:cal:tJxSTO} (en trazo continuo) con los obtenidos de simulaciones equivalentes considerando que el substrato tiene una conductividad el\'ectrica 15 veces mayor (en trazo discontinuo; se pretende simular as\'{\i} el empleo de substratos de \MgO\ en vez de \STOf). Se representa el tiempo $t$ que una corriente $J$ tarda en generar una avalancha, contado desde que se aplica el pulso de corriente.}                  % caption
{Comparaci\'on de los resultados de la figura \ref{fig:cal:tJxSTO} con los obtenidos de simulaciones equivalentes considerando que el substrato tiene una conductividad el\'ectrica 15 veces mayor (como si en vez de \STOf\ emple\'asemos \MgO).}                  % toc
{\stfigw}            % width \textwidth
}
% - - - - - - - - - - - - - - - - - - - - - - - - - - - - - - - - - - - %

\newcommand{\tabCalWTJ}{

\begin{table}
  \centering
  \caption[Valores de la potencia estacionaria \Wx\ y la derivada
$\indiff{W}{T}$ para los valores estables previos al salto de
algunas simulaciones.]{Valores de la potencia estacionaria \Wx\ y
la derivada $\indiff{W}{T}$ para los valores estables previos al
salto de las simulaciones de las figuras \ref{fig:cal:sDTtEtJ76}
-\ref{fig:cal:sEtJWtB1}, as\'{\i} como de otros campos
intermedios. La potencia estable previa al salto decrece con la
temperatura inicial, pero aumenta con el campo. }\label{tab:WTJ}

\begin{tabular}{ccccc}
    \\[-8pt]
 \hline
  $B$  &  $T=\Tb+\DTf$ & $J$ & \Wx  &  $\indiff{W}{T}$
    \\ {\footnotesize T} & {\footnotesize K} & {\footnotesize $10^6\;\Acms$}
     & {\footnotesize$10^7\und{\Wcmc}$ }& {\footnotesize
     $10^7\und{W/cm^{3}\,K}$}      \\ \hline
  0 & $76.2+1.8$ & $6.38$ & 4.32 & 3.34
  \\
  0 &   $82.0+1.0$ & 3.49  & 2.43 & 2.66
  \\
  0.1 & $76.2+2.8$ & 5.80 & 6.52 & 2.21
  \\
  0.4 & $76.2+3.2$ & 3.80 & 7.35 & 2.16
  \\
  0.7 & $76.2+3.4$ & 3.18 & 7.66 & 1.99
  \\
  1.0 & $76.2+3.5$ & 2.68 & 8.35 &  1.40
  \\ \hline
\end{tabular}

\end{table}
}

\comenta{ Los valores de la derivada $\indiff{W}{T}$ tambi\'en
var\'{\i}an, de modo que no se satisface la condici\'on de
avalancha de la ecuaci\'on \ref{ec:condAvaSimple} y el sistema
tiende a valores aparentemente estacionarios.}

%-%-%-%-%-%-%-%-%-%-%-%-%-%-%-%-%-%-%-%-%-%-%-%-  figura -%-%-%-%-%-%-%-%

\newcommand{\figCalPydP}{  % alias

\figura {fig:cal:PydP}      % label
{fig/cal/PydP}              % file
{Resultados de simulaciones para la muestra \fA\ a
$T=76.2\und{K}$, en funci\'on del campo magn\'etico aplicado. En (a),
potencia disipada por unidad de volumen justo antes del salto \Wx\
---la potencia $W$ a la temperatura estacionaria m\'as alta
posible---. El modelo predice la misma tendencia de \Wx\ creciente
con $H$ que los experimentos (comp\'arese con la fig.
\ref{fig:datos:WxHA}). En (b), la derivada de la potencia respecto
de la temperatura $\indiff{W}{T}$, en ese mismo punto previo al
salto. Mientras que \Wx\ aumenta con el campo aplicado, su
derivada, sin embargo, decrece.
(Las l\'{\i}neas son gu\'{\i}as para los ojos).}                  % caption
{Simulaciones para la dependencia de \Wx\ y $\indiff{W}{T}$ respecto del campo aplicado $H$}                  % toc
{\stfigw}            % width \textwidth
}
% - - - - - - - - - - - - - - - - - - - - - - - - - - - - - - - - - - - %

%-%-%-%-%-%-%-%-%-%-%-%-%-%-%-%-%-%-%-%-%-%-%-%-  figura -%-%-%-%-%-%-%-%

\newcommand{\figCalsEJSetSeis}{  % alias

\figura {fig:cal:sEJ76}      % label
{fig/cal/sEJ76data}              % file
{Comparaci\'on de la curva \EJ\ que predicen las simulaciones con pulsos de 1~ms (trazo continuo) con los datos experimentales correspondientes (c\'{\i}rculos negros, unidos por l\'{\i}nea de puntos). $\Tb=76.2\und{K}, B=0$. Se muestra tambi\'en la tendencia de los datos con menos disipaci\'on (l\'{\i}nea a trazos discontinuos), dada por el modelo de la ecuaci\'on \ref{ec:EJT} y que es la base de nuestros c\'alculos. La predicci\'on del modelo se aleja de esta tendencia y establece una \Jx, una \emph{singularidad} en la que la muestra transita al estado normal, en $J=6.4\Exund{6}{\Acms}$, valor que difiere de la corriente de salto experimental en apenas un 7\%. La tendencia de trazo discontinuo no tiene una singularidad, y crece continuamente.}                  % caption
{Comparaci\'on de la curva \EJ\ que predicen las simulaciones con los datos experimentales. $\Tb=76.2\und{K}, B=0$.}                  % toc
{\stfigw}            % width \textwidth
}
% - - - - - - - - - - - - - - - - - - - - - - - - - - - - - - - - - - - %

%-%-%-%-%-%-%-%-%-%-%-%-%-%-%-%-%-%-%-%-%-%-%-%-  figura -%-%-%-%-%-%-%-%

\newcommand{\figCalsEJSetSeisB}{  % alias

\figura {fig:cal:sEJ76B1}      % label
{fig/cal/sEJ76B1data}              % file
{Comparaci\'on de la curva \EJ\ que predicen las simulaciones con pulsos de 1~ms (trazo continuo) con los datos experimentales correspondientes (c\'{\i}rculos negros, unidos por l\'{\i}nea de puntos: la flecha indica el salto al estado normal), con campo magn\'etico aplicado. $\Tb=76.2\und{K}, \mH=1\und{T}$. La l\'{\i}nea a trazos discontinuos es la tendencia de los datos con menos disipaci\'on, dada por el modelo de la ecuaci\'on \ref{ec:EJTB1}. La predicci\'on del modelo se aleja de esta tendencia y establece una \Jx, una \emph{singularidad} en la que la muestra transita al estado normal, en $J=2.8\Exund{6}{\Acms}$, valor que difiere de la corriente de salto experimental en menos de un 8\%.}                  % caption
{Comparaci\'on de la curva \EJ\ que predicen las simulaciones con los datos experimentales. $\Tb=76.2\und{K}, \mH=1\und{T}$.}                  % toc
{\stfigw}            % width \textwidth
}
% - - - - - - - - - - - - - - - - - - - - - - - - - - - - - - - - - - - %

%-%-%-%-%-%-%-%-%-%-%-%-%-%-%-%-%-%-%-%-%-%-%-%-  figura -%-%-%-%-%-%-%-%

\newcommand{\figCalsDTW}{  % alias

\figura {fig:cal:sDTW}      % label
{fig/cal/sDTW}              % file
{Correlaci\'on entre potencia disipada y aumento de temperatura de
la muestra, seg\'un las predicciones de nuestro modelo t\'ermico
(c\'{\i}rculos negros) para pulsos de $1\und{ms}$. Comp\'arese con
los resultados experimentales de la figura \ref{fig:resul:DTW} (p.
\pageref{fig:resul:DTW}). La tendencia que mostraban esos datos se
muestra aqu\'{\i} tambi\'en con una l\'{\i}nea recta.}                  % caption
{Predicci\'on de la correlaci\'on entre potencia disipada y aumento de temperatura de la muestra, seg\'un nuestro modelo t\'ermico, para pulsos de $1\und{ms}$.}                  % toc
{\stfigw}            % width \textwidth
}
% - - - - - - - - - - - - - - - - - - - - - - - - - - - - - - - - - - - %

%-%-%-%-%-%-%-%-%-%-%-%-%-%-%-%-%-%-%-%-%-%-%-%-  figura -%-%-%-%-%-%-%-%

\newcommand{\figCalsDTWdos}{  % alias

\figura {fig:cal:sDTW2}      % label
{fig/cal/sDTW2}              % file
{Correlaci\'on entre potencia disipada y aumento de temperatura de
la muestra, seg\'un las predicciones de nuestro modelo t\'ermico
(c\'{\i}rculos negros). La diferencia con la figura
\ref{fig:cal:sDTW} es que ahora empleamos pulsos de duraci\'on
diez veces menor, de
$100\und{\mu s}$.}                  % caption
{Predicci\'on de la correlaci\'on entre potencia disipada y aumento de temperatura de la muestra, seg\'un nuestro modelo t\'ermico, para pulsos de $100\und{\mu s}$.}                  % toc
{\stfigw}            % width \textwidth
}
% - - - - - - - - - - - - - - - - - - - - - - - - - - - - - - - - - - - %

%-%-%-%-%-%-%-%-%-%-%-%-%-%-%-%-%-%-%-%-%-%-%-%-  figura -%-%-%-%-%-%-%-%

\newcommand{\figCalJxTexpcalc}{  % alias

\figura {fig:cal:JxTexpcalc}      % label
{fig/cal/JxTexpcalc}              % file
{Variaci\'on con la temperatura de la corriente de salto \Jx. En
c\'{\i}rculos blancos, los resultados experimentales ya mostrados
en el cap\'{\i}tulo \ref{cap:resul}. En cuadrados negros, los
calculados seg\'un nuestro modelo de calentamiento en avalancha.
La coincidencia en valores absolutos es notable. Pero lo que es
m\'as importante: se reproduce la misma tendencia con la
temperatura, cosa que hasta el momento ning\'un modelo t\'ermico
hab\'{\i}a logrado. Las l\'{\i}neas de trazo continuo son ajustes
a los datos empleando una funci\'on con exponente 3/2 en
temperaturas reducidas (ec. \ref{ec:JT32}, p. \pageref{ec:JT32}),
mientras que la l\'{\i}nea de puntos es el mejor ajuste con
exponente 1/2 (ec. \ref{ec:JT12}).}                  % caption
{Variaci\'on con la temperatura de la corriente de salto \Jx: comparaci\'on de las predicciones del modelo con los experimentos.}                  % toc
{\stfigw}            % width \textwidth
}
% - - - - - - - - - - - - - - - - - - - - - - - - - - - - - - - - - - - %

%-%-%-%-%-%-%-%-%-%-%-%-%-%-%-%-%-%-%-%-%-%-%-%-  figura -%-%-%-%-%-%-%-%

\newcommand{\figCalJxTcompara}{  % alias

\figura {fig:cal:JxTcompara}      % label
{fig/cal/JxTcompara}              % file
{Variaci\'on con la temperatura de la corriente de salto \Jx. En
c\'{\i}rculos blancos, y en cuadrados negros, los mismos puntos de
la figura \ref{fig:cal:JxTexpcalc}. Por compararlos con estos, se
incluyen los resultados de emplear el modelo t\'ermico con
sucesivamente menos correcciones: en tri\'angulos, los obtenidos
con la ecuaci\'on \ref{ec:Tf} (p. \pageref{ec:Tf}); en rombos, los
de la ecuaci\'on \ref{ec:Tfsimple} (p. \pageref{ec:Tfsimple}) ya
mostrados previamente en la figura \ref{fig:cal:JxTsimple}. Las
l\'{\i}neas que unen los puntos son s\'olo gu\'{\i}as visuales.}
% caption
{Variaci\'on con la temperatura de la corriente de salto \Jx: comparaci\'on de las predicciones del modelo con distintas correcciones.}                  % toc
{\stfigw}            % width \textwidth
}
% - - - - - - - - - - - - - - - - - - - - - - - - - - - - - - - - - - - %

\newcommand{\figCalsDTtDos}{  % alias

\figura {fig:cal:sDTtDos}      % label
{fig/cal/sDTt}              % file
{Ilustraci\'on de los dos posibles tipos de evoluciones temporales
de la temperatura: estabilidad o avalancha. Es la misma
simulaci\'on ya mostrada en la figura \ref{fig:cal:sDTt} (p.
\pageref{fig:cal:sDTt}), que incluimos de nuevo aqu\'{\i} por
comodidad. Se aplican distintas corrientes de valores pr\'oximos
entre s\'{\i} (6.70, 6.73, 6.74 y $6.84\Exund{6}{\Acms}$), a $B=0$
y $\Tb=75.6\und{K}$. Entre la primera y la \'ultima hay un 2\% de
diferencia, pero conllevan resultados muy diferentes. En la
evoluci\'on que conduce a la avalancha, llega un momento en que la
derivada de \DT\ en el tiempo (la recta tangente a la curva)
crece, mientras que en el caso estable dicha derivada es
\emph{siempre} cada vez menor.}                  % caption
{Ilustraci\'on de las dos posibles evoluciones temporales de la temperatura: estabilidad o avalancha.}                  % toc
{\stfigw}            % width \textwidth
}

%-%-%-%-%-%-%-%-%-%-%-%-%-%-%-%-%-%-%-%-%-%-%-%-  figura -%-%-%-%-%-%-%-%

\newcommand{\figCaldfLtJ}{  % alias

\figura {fig:cal:dfLtJ}      % label
{fig/cal/dfLtJ}              % file
{Diagrama de estados $t$--$J$ para $\Tf=78\und{K}$ y $B=0$, establecido por la condici\'on de avalancha de la ecuaci\'on \ref{ec:condAvaSencillo}. La regi\'on gris representa estados estables (signos negativos de la ecuaci\'on), y la blanca estados cr\'{\i}ticos que conllevan una avalancha t\'ermica (signos positivos).}                  % caption
{Diagrama de estados estables y cr\'{\i}ticos $t$--$J$ para $\Tf=78\und{K}$ y $B=0$.}                  % toc
{\stfigw}            % width \textwidth
}
% - - - - - - - - - - - - - - - - - - - - - - - - - - - - - - - - - - - %

%-%-%-%-%-%-%-%-%-%-%-%-%-%-%-%-%-%-%-%-%-%-%-%-  figura -%-%-%-%-%-%-%-%

\newcommand{\figCaldfJTTc}{  % alias

\figura {fig:cal:dfJTTc}      % label
{fig/cal/dfJTTc}              % file
{Diagrama de estados $J$--$\Tf/\Tc$ para $t=0.1\und{ms}$ y $B=0$, establecido por la condici\'on de avalancha de la ecuaci\'on \ref{ec:condAvaSencillo}. La regi\'on gris representa estados estables (signos negativos de la ecuaci\'on), y la blanca estados cr\'{\i}ticos que conllevan una avalancha t\'ermica (signos positivos). Dividir $t$ por un factor diez o multiplicarlo por cualquier factor no var\'{\i}a los resultados, as\'{\i} que este diagrama de estados es v\'alido para cualquier $t>10\und{\mu s}$.}                  % caption
{Diagrama de estados $J$--$\Tf/\Tc$ para $t>10\und{\mu s}$ y $B=0$.}                  % toc
{\stfigw}            % width \textwidth
}
% - - - - - - - - - - - - - - - - - - - - - - - - - - - - - - - - - - - %

%-%-%-%-%-%-%-%-%-%-%-%-%-%-%-%-%-%-%-%-%-%-%-%-  figura -%-%-%-%-%-%-%-%

\newcommand{\figCaldfJTTcns}{  % alias

\figura {fig:cal:dfJTTcns}      % label
{fig/cal/dfJTTcns}              % file
{Diagrama de estados $J$--$\Tf/\Tc$ para $t=1\und{ns}$ y $B=0$, establecido por la condici\'on de avalancha de la ecuaci\'on \ref{ec:condAvaSencillo}. La regi\'on gris representa estados estables, y la blanca estados cr\'{\i}ticos que conllevan una avalancha t\'ermica.}                  % caption
{Diagrama de estados $J$--$\Tf/\Tc$ para $t=1\und{ns}$ y $B=0$.}                  % toc
{\stfigw}            % width \textwidth
}
% - - - - - - - - - - - - - - - - - - - - - - - - - - - - - - - - - - - %

%-%-%-%-%-%-%-%-%-%-%-%-%-%-%-%-%-%-%-%-%-%-%-%-  figura -%-%-%-%-%-%-%-%

\newcommand{\figCalsdfJxTTc}{  % alias

\figura {fig:cal:sdfJxTTc}      % label
{fig/cal/sdfJxTTc}              % file
{Dependencia de la corriente de salto respecto de la temperatura,
para $B=0$. Comparaci\'on de los resultados extra\'{\i}dos de la
condici\'on de avalancha con los c\'alculos m\'as detallados de
apartados previos. La l\'{\i}nea continua es la frontera que
separa la regi\'on de estabilidad de la cr\'{\i}tica en la figura
\ref{fig:cal:dfJTTc}. La l\'{\i}nea de trazos discontinuos es la
que ajusta los mismos datos que se representaron como
rect\'angulos negros en la figura \ref{fig:cal:JxTexpcalc} (p.
\pageref{fig:cal:JxTexpcalc}). Destacamos la similitud de ambas
curvas en sus dependencias funcionales respecto de la temperatura.
No se piense que las diferencias provienen de simplificaciones de
los c\'alculos: en realidad ambas curvas representan cosas
ligeramente distintas. Ver texto para
detalles.}                  % caption
{Dependencia de la corriente de salto respecto de la temperatura, para $B=0$: comparaci\'on de la condici\'on de avalancha con los c\'alculos m\'as detallados de apartados previos.}                  % toc
{\stfigw}            % width \textwidth
}
% - - - - - - - - - - - - - - - - - - - - - - - - - - - - - - - - - - - %

%-%-%-%-%-%-%-%-%-%-%-%-%-%-%-%-%-%-%-%-%-%-%-%-  figura -%-%-%-%-%-%-%-%

\newcommand{\figCaldfLtJB}{  % alias

\figura {fig:cal:dfLtJB1}      % label
{fig/cal/dfLtJB1}              % file
{Diagrama de estados $t$--$J$ para $\Tf=78\und{K}$ y $\mH=1\und{T}$, establecido por la condici\'on de avalancha de la ecuaci\'on \ref{ec:condAvaSencillo}.}                  % caption
{Diagrama de estados estables y cr\'{\i}ticos $t$--$J$ para $\Tf=78\und{K}$ y $\mH=1\und{T}$.}                  % toc
{\stfigw}            % width \textwidth
}
% - - - - - - - - - - - - - - - - - - - - - - - - - - - - - - - - - - - %

%-%-%-%-%-%-%-%-%-%-%-%-%-%-%-%-%-%-%-%-%-%-%-%-  figura -%-%-%-%-%-%-%-%

\newcommand{\figCaldfJTTcB}{  % alias

\figura {fig:cal:dfJTTcB1}      % label
{fig/cal/dfJTTcB1}              % file
{Diagrama de estados $J$--$\Tf/\Tc$ para $t=0.1\und{ms}$ y $\mH=1\und{T}$, establecido por la condici\'on de avalancha de la ecuaci\'on \ref{ec:condAvaSencillo}. La regi\'on gris representa estados estables, y la blanca estados cr\'{\i}ticos que conllevan una avalancha t\'ermica.}                  % caption
{Diagrama de estados $J$--$\Tf/\Tc$ para $t=0.1\und{ms}$ y $\mH=1\und{T}$.}                  % toc
{\stfigw}            % width \textwidth
}
% - - - - - - - - - - - - - - - - - - - - - - - - - - - - - - - - - - - %

%-%-%-%-%-%-%-%-%-%-%-%-%-%-%-%-%-%-%-%-%-%-%-%-  figura -%-%-%-%-%-%-%-%

\newcommand{\figCalsdfJxTTcB}{  % alias

\figura {fig:cal:sdfJxTTcB1}      % label
{fig/cal/sdfJxTTcB1}              % file
{Variaci\'on de la corriente de salto con la temperatura, para
$\mH=1\und{T}$. Comparaci\'on de los resultados extra\'{\i}dos de
la condici\'on de avalancha con los datos experimentales. La
l\'{\i}nea continua es la frontera que separa la regi\'on de
estabilidad de la cr\'{\i}tica en la figura
\ref{fig:cal:dfJTTcB1}. La l\'{\i}nea de trazos discontinuos es la
que ajusta a los datos experimentales. Ver
texto para detalles.}                  % caption
{Dependencia de la corriente de salto respecto de la temperatura, para $\mH=1\und{T}$: comparaci\'on de la condici\'on de avalancha con los datos experimentales.}                  % toc
{\stfigw}            % width \textwidth
}
% - - - - - - - - - - - - - - - - - - - - - - - - - - - - - - - - - - - %

%-%-%-%-%-%-%-%-%-%-%-%-%-%-%-%-%-%-%-%-%-%-%-%-  figura -%-%-%-%-%-%-%-%

\newcommand{\figCalManta}{  % alias

\figura {fig:cal:manta}      % label
{fig/cal/manta2}              % file
{Frontera entre las regiones de estabilidad e inestabilidad
determinadas por la ecuaci\'on \ref{ec:condAvaSencillo}, para campo
aplicado nulo. Esta frontera corresponde a los valores de \Jx. Por
debajo, el sistema es estable y alcanza temperaturas casi
estacionarias. Por encima, se cumple el criterio de la ecuaci\'on
\ref{ec:condAvaSencillo} y el sistema es inestable, obteni\'endose
avalanchas t\'ermicas que llevan la muestra al estado normal.} % caption
{Frontera en el espacio $J$--\Tf--$t$ ($B=0$) entre las regiones de estabilidad e inestabilidad determinadas por la condici\'on de avalancha.}                  % toc
{\stfigw}            % width \textwidth
}
% - - - - - - - - - - - - - - - - - - - - - - - - - - - - - - - - - - - %

%%%%%%%%%%%%%%%%%%%%%%%%%%%%%%%%%%%%%%%%
\section{Introducci\'on}
%%%%%%%%%%%%%%%%%%%%%%%%%%%%%%%%%%%%%%%%

En este cap\'{\i}tulo proponemos un modelo sencillo de
\Index{calentamiento} en avalancha que puede explicar la
transici\'on abrupta de un filme superconductor al estado normal
inducida por una corriente el\'ectrica. Recordemos que a la
densidad de corriente que induce la transici\'on la denominamos
\Jx\ y, como hemos visto, en los filmes que hemos estudiado
depende apreciablemente de la temperatura $T$
---en general, de la temperatura reducida $T/\Tc$--- y de la
amplitud del campo magn\'etico aplicado $H$. La pretensi\'on de
nuestro modelo es no s\'olo justificar la existencia de \Jx, sino
tambi\'en su dependencia respecto de $T$ y $H$, y dar cuenta, en
general, de los resultados observados cerca del salto.

Nuestro modelo se basa en que por encima de la corriente
cr\'{\i}tica \Jc, definida como es habitual como aquella a partir
de la cual hay disipaci\'on, dicha disipaci\'on es fuertemente
dependiente de la temperatura. Esta dependencia bien conocida
puede causar, en funci\'on de las condiciones de refrigeraci\'on
del filme, un notable aumento de la temperatura que puede llegar a
inducir el salto al estado normal. Subrayemos ya aqu\'{\i} que,
como veremos en este cap\'{\i}tulo, la efectividad de este proceso
depende crucialmente de las condiciones de refrigeraci\'on, es
decir, del balance energ\'etico entre la generaci\'on de calor en
el filme y su evacuaci\'on al entorno. Este tipo de
\emph{\Index{calentamiento} en cascada} se suele dar por supuesto
para explicar $\Jx(T,H)$ en superconductores masivos, sin entrar
en mayores consideraciones \cite{Yang99,Tournier00}, quiz\'a
incluso precipitadamente. Se pueden encontrar en la
bibliograf\'{\i}a an\'alisis t\'ermicos de este tipo de muestras
en los que se considera detalladamente el calentamiento como causa
de la transici\'on \cite{Elschner99}. Sin embargo, la posibilidad
de que el calentamiento sea quien genere la transici\'on tambi\'en
en pel\'{\i}culas delgadas ha sido sucesivamente descartada, por
una incorrecta estimaci\'on del balance energ\'etico: se estimaba
que la evacuaci\'on del calor era demasiado buena como para que
\'este tuviera algo que ver. Mostraremos aqu\'{\i} que una
estimaci\'on m\'as realista del reparto de energ\'{\i}a en las
pel\'{\i}culas delgadas que hemos estudiado permite explicar la
existencia de \Jx, y su dependencia respecto de $T$ y $H$,
invocando s\'olo un mecanismo puramente t\'ermico
\index{extr\'{\i}nseco} (extr\'{\i}nseco). Naturalmente nuestra
propuesta no excluye que en la transici\'on al estado normal
inducida por una corriente \emph{supercr\'{\i}tica} est\'en
tambi\'en involucrados mecanismos \index{intr\'{\i}nseco}
intr\'{\i}nsecos, como los mencionados en el cap\'{\i}tulo
\ref{cap:datos} \index{corriente de desapareamiento}
(desapareamiento de pares de Cooper, cambio de r\'egimen del
movimiento de \Index{v\'ortices}\ldots). Sin embargo, incluso en
el caso de que dichos mecanismos intr\'{\i}nsecos est\'en
presentes, nuestros resultados sugieren que el \emph{calentamiento
en cascada} debe de desempe\~nar un papel importante tanto en los
valores de \Jx\ medidos como en su dependencia respecto de $T$ y
$H$.

%%%%%%%%%%%%%%%%%%%%%%%%%%%%%%%%%%%%%%%%
\section{Aspectos b\'asicos del modelo}
%%%%%%%%%%%%%%%%%%%%%%%%%%%%%%%%%%%%%%%%

Los dos ingredientes esenciales del modelo son: i) Por un lado,
tener en cuenta que la potencia disipada es altamente dependiente
de la propia temperatura del micropuente. ii) Considerar de manera
adecuada el intercambio t\'ermico entre el filme y su entorno, a
trav\'es del substrato.

La combinaci\'on de ambos elementos nos permitir\'a apreciar que
la disipaci\'on puede, en determinadas circunstancias, aumentar
abruptamente y generarse as\'{\i} una avalancha t\'ermica.
Explicamos cada uno de estos aspectos en un subapartado
independiente.

% ----------------------------------- %
\subsection{Variaci\'on de la disipaci\'on con la temperatura}
\label{sec:PT}
% ----------------------------------- %

Para ilustrar la bien conocida dependencia de la disipaci\'on
respecto de la temperatura podemos utilizar algunas de las curvas
\EJ\ obtenidas en uno de nuestros filmes (muestra \fA),
representadas en la figura \ref{fig:cal:EJP}. Estas curvas
corresponden a campo magn\'etico aplicado nulo. Como ejemplo, para
una densidad de corriente $J=4.75~\eAcms$, que es superior a $J_c$
en todas las isotermas mostradas, la disipaci\'on por unidad de
volumen $W=E \times J$ a cada temperatura es equivalente a las
\'areas de los rect\'angulos mostrados. En cada rect\'angulo se
inscribe la potencia que se disipa en su correspondiente isoterma.
Esta figura pone pues en evidencia el aspecto fundamental que se
pretende enfatizar aqu\'{\i}: la potencia disipada aumenta
considerablemente cuando, a corriente constante, aumenta la
temperatura del filme. El incremento de la potencia disipada no es
lineal con el de la temperatura, sino que crece de manera
exponencial. Para destacar esta dependencia, en la figura
\ref{fig:cal:ET} representamos el campo el\'ectrico $E$ en
funci\'on de la temperatura para distintos valores de $J$, con
datos extra\'{\i}dos de las curvas \EJ. Como a corriente constante
la potencia disipada $W = E \times J$ es proporcional a $E$, la
variaci\'on de la potencia con la temperatura se muestra
impl\'{\i}citamente en esa figura. La representaci\'on es
logar\'{\i}tmica para apreciar lo fuertemente que var\'{\i}a $E$
---y por lo tanto $W$--- con peque\~nos cambios de temperatura.

\figCalEJP

\figCalET

Se pone as\'{\i} de manifiesto lo relevante que es considerar con
atenci\'on cualquier posible \Index{calentamiento} del filme, por
peque\~no que sea. El calor generado aumenta de manera tan notable
con peque\~nos incrementos de temperatura que si no se evacua
eficientemente puede, como veremos, producirse una
\Index{avalancha} t\'ermica.

El calor total que se genera en la muestra por unidad de tiempo es
$P = W\;\vf = E\;J\;\vf$, donde \vf\ es el volumen del filme. Un
punto relevante en los c\'alculos que aqu\'{\i} se mostrar\'an es
considerar que esta potencia var\'{\i}a, a corriente $J$
constante, debido a la variaci\'on del campo $E$ con la
temperatura, como se acaba de mostrar con el ejemplo de la figura
\ref{fig:cal:ET}. Esto es lo que en este contexto denominaremos
\emph{\Index{realimentaci\'on}}: cuando la temperatura de la
muestra aumenta, aunque sea levemente, la potencia $P$ tambi\'en
se ve incrementada, y en algunos casos no tan levemente; esto
motiva un nuevo incremento de temperatura, que vuelve a causar un
aumento de disipaci\'on, y as\'{\i} sucesivamente. No hacer estas
consideraciones implica subestimar el calor generado.

Algunas publicaciones han tenido en cuenta este mecanismo de
\Index{realimentaci\'on} (en muestras masivas \cite{Elschner99} y en
capas delgadas \cite{Skokov93,Gurevich87}), pero no basta s\'olo con
esto para hacer una correcta descripci\'on de la evoluci\'on t\'ermica
de estas muestras. Hay que considerar de manera adecuada tambi\'en
la capacidad de evacuaci\'on de calor desde el micropuente al
entorno. En el siguiente subapartado mostramos c\'omo lo hace
nuestro modelo, e indicamos qu\'e aspectos no se tuvieron en cuenta
en los trabajos citados.

% ----------------------------------- %
\subsection{Intercambio t\'ermico de la muestra}
% ----------------------------------- %

De estar el filme completamente aislado del entorno (caso
adiab\'atico), la potencia t\'{\i}pica generada cerca del salto
(en torno a $10^7 \Wcmc$) tardar\'{\i}a $1 \mus$ en aumentar 10~K
la temperatura, y pasados $0.1\und{ms}$ esta llegar\'{\i}a hasta
unos $1000\und{K}$. Evidentemente esto no ocurre (supondr\'{\i}a
la destrucci\'on instant\'anea de la muestra), porque la
pel\'{\i}cula evacua calor al entorno con mucha eficiencia.

La aproximaci\'on habitual al problema del \Index{calentamiento}
consiste en resolver el balance energ\'etico del sistema
\cite{Gurevich87}. Mientras se pueda evacuar todo el calor que se
genera, el sistema tender\'a a estabilizarse. Cuando la generaci\'on
de calor $P$ supere la evacuaci\'on al entorno \Qp, se producir\'a un
progresivo calentamiento de la muestra. La cuesti\'on est\'a en
calcular correctamente estos t\'erminos.

Es habitual encontrar en la bibliograf\'{\i}a ciertas
simplificaciones que llevan a resultados err\'oneos, por hacer un
estimaci\'on a la baja del primer t\'ermino $P$, y excesiva del
segundo t\'ermino \Qp. Ya hemos visto en el apartado \ref{sec:PT}
que si no tenemos en cuenta la \emph{\Index{realimentaci\'on}} en
$P$ podemos estar subestimando el calor generado.

En lo relativo a la evacuaci\'on \Qp, lo normal es suponer un
t\'ermino de transferencia entre dos medios a trav\'es de una
frontera, del tipo
\be
\Qp = \gamma \; (\DT)^n = h\; A\; (\DT)^n \label{ec:evacuaDT}
\ee donde \DT\ es la diferencia de temperatura entre la muestra y
su entorno, $h$ es el coeficiente de transferencia de calor de la
frontera, $A$ el \'area de contacto entre muestra y entorno, y $n$
es un exponente del orden de 1.

Esta generalidad dar\'a cuenta realmente del calor evacuado si
definimos correctamente \DT\ y $h$, es decir, si establecemos de
manera oportuna cu\'al es el entorno al que se va el calor desde
la muestra. Esta forma de expresar la evacuaci\'on ha dado buenos
resultados en el an\'alisis t\'ermico de muestras masivas dentro
de un medio l\'{\i}quido \cite{Elschner99}, cuando se consideraba
que el entorno al que se evacua el calor de la muestra es el
\Index{ba\~no} circundante con temperatura constante, con $h$
tambi\'en constante. El caso de pel\'{\i}culas delgadas es
diferente: como ya vimos en la parte experimental (ver p.
\pageref{par:substratogas}), el calor se transmite mejor al
substrato que al gas o al l\'{\i}quido circundante, as\'{\i} que
es habitual considerar que el entorno del filme es exclusivamente
su substrato. El problema ahora surge cuando se hacen
simplificaciones del tipo $\DT = \Tf - \Tb$, bastante frecuentes
\cite{Gurevich87,Bezuglyj92,Skocpol74,Skokov93}, que suponen que
el entorno de la muestra, es decir, el propio substrato, est\'a a
temperatura constante ---en este caso termalizado con el
\Index{ba\~no} externo a temperatura \Tb---, mientras que la
muestra se encuentra a una temperatura superior \Tf. En este
esquema, $h$ es el coeficiente de transferencia de calor de la
frontera entre filme y substrato, y se suele considerar constante.
Estas simplificaciones sobreestiman la evacuaci\'on real de calor
desde la muestra al entorno. En realidad el calor se transmite a
trav\'es del substrato a velocidad finita, generando en su seno
una distribuci\'on no uniforme de temperatura que en general
requiere de complejos c\'alculos num\'ericos para su exacta
determinaci\'on. De esta manera, la parte del substrato en
contacto directo con el filme ya no est\'a a temperatura \Tb, sino
a otra mayor que esta, por lo que la eficiencia de la evacuaci\'on
de calor se ve mermada. No tener en cuenta esta distribuci\'on no
uniforme de temperatura en el substrato supone sobreestimar la
capacidad de evacuaci\'on de calor de la muestra, y conduce a
conclusiones err\'oneas. En nuestro modelo vamos a tenerlo en
cuenta, de manera sencilla, calculando la temperatura a la que se
encuentra la capa de substrato adyacente al filme\footnote{Una
aproximaci\'on alternativa a una correcta descripci\'on del
sistema ser\'{\i}a, si se quiere definir $\DT = \Tf - \Tb$ en la
ecuaci\'on \ref{ec:evacuaDT}, hacer que el coeficiente $h$ sea
dependiente de la temperatura \Tf: de este modo el entorno de la
muestra ser\'{\i}a el ba\~no a temperatura constante \Tb, y todo
el substrato har\'{\i}a de \emph{frontera} entre ambos medios, con
un coeficiente efectivo $h$ que depender\'{\i}a de la
distribuci\'on de temperatura en su seno. En todo caso, siguiendo
esta l\'{\i}nea no se puede mantener $h$ constante e igual al
coeficiente de transferencia entre filme y substrato si se quiere
describir correctamente el sistema. Este planteamiento alternativo
con un $h$ efectivo variable es equivalente al modelo que
propondremos aqu\'{\i}, y se podr\'{\i}a extraer de \'el con
algunos c\'alculos.}.

\subsubsection{Modelizaci\'on de la geometr\'{\i}a}
\label{sec:geometria}

Vamos a hacer una modelizaci\'on muy elemental de esta distribuci\'on
no uniforme de temperatura. En el modelo que proponemos ya no
tendremos un sistema compuesto de dos fases (una muestra inmersa
en un entorno). Consideraremos por el contrario un \'unico sistema
continuo, la muestra m\'as el substrato, en cuyo seno hay una
distribuci\'on no uniforme de temperatura que evoluciona en el
tiempo, con generaci\'on variable de calor en uno de sus extremos,
el que ocupa la muestra. Desde este punto de vista el t\'ermino de
evacuaci\'on de calor \Qp\ ya no se expresa f\'acilmente con una
funci\'on de transferencia a trav\'es de una frontera entre dos medios
a temperaturas constantes, del tipo de la \refec{ec:evacuaDT}. Al
no tener una expresi\'on de \Qp\ no podremos compararla con $P$ y
establecer un criterio de estabilidad, pero veremos que podemos
hacerlo de otra forma.

Para modelizar apropiadamente la muestra con el substrato
recordemos c\'omo es su geometr\'{\i}a. En la figura
\ref{fig:cal:filmsubs} representamos un corte transversal del
micropuente superconductor depositado sobre el substrato
cristalino. Las dimensiones no est\'an a escala (el micropuente
tiene en realidad un espesor $d$ mucho m\'as delgado en
proporci\'on a su ancho $a$): en la figura se muestra solamente un
esquema conceptual.

Recordemos que el coeficiente de transferencia de calor $h$ entre
la muestra y el substrato es muy superior al que presentan la
muestra o el substrato con la atm\'osfera circundante, ya sea
\'esta l\'{\i}quida o gaseosa. Esto se entiende si recordamos que
el m\'etodo de crecimiento de la pel\'{\i}cula hace que \'esta sea
una continuaci\'on de la estructura cristalina del substrato. Los
valores que se extraen de la bibliograf\'{\i}a son de $1000~{\rm
W/cm^2\,K}$ para filme-substrato \cite{Nahum91}, y de $1~{\rm
W/cm^2\,K}$ para filme-entorno cuando este \'ultimo es nitr\'ogeno
l\'{\i}quido \cite{Mosqueira93,Mosqueira93b,Mosqueira94}, o
incluso menos si es gas. (Consideraremos que el del
substrato-entorno es tambi\'en igual a \'este \'ultimo valor).
Podemos por lo tanto despreciar la evacuaci\'on del calor desde la
muestra a la atm\'osfera, considerando que a tiempos cortos todo
el calor se transmite al substrato.

\figCalFilmsubs

Tambi\'en podemos despreciar la evacuaci\'on de calor a la
atm\'osfera desde el substrato, debido a que el coeficiente de
transferencia de calor del substrato-entorno es igualmente muy
peque\~no. Recu\'erdese que el substrato, un monocristal, tiene
una gran capacidad de conducci\'on ($\kappa$) de calor. Este
encontrar\'a por lo tanto menos resistencia para circular por el
substrato que para salir a la atm\'osfera. Por supuesto, si el
tiempo de la medida fuese muy largo, el substrato se
calentar\'{\i}a notablemente y entrar\'{\i}a en juego la
evacuaci\'on a la atm\'osfera, pero en una primera estimaci\'on
podemos obviar esta posibilidad.

\comenta{El substrato es $10^9$ veces m\'as voluminosos que la
pel\'{\i}cula, y el calor que genera \'esta, a\'un siendo mucho
para ella misma, es en comparaci\'on insignificante para todo el
substrato. La potencia disipada t\'{\i}pica de $10^7 {\rm
W/cm^3}$, que en caso adiab\'atico aumentaba la temperatura del
filme en 1000~K pasados 0.1~ms, har\'{\i}a que el substrato se
calentase menos de $10^{-5} \rm K$. Pero ojo: el calor no se
reparte por todo el substrato instant\'aneamente... }

Considerando que la longitud del micropuente es mayor que el ancho
o el espesor, el problema se puede en principio reducir a dos
dimensiones, las mostradas en la figura \ref{fig:cal:filmsubs}. El
sistema es sim\'etrico por traslaci\'on a lo largo del eje que
determina el puente\footnote{Esta simplificaci\'on pierde validez
para describir nuestro sistema si el micropuente \emph{no es}
mucho m\'as largo que ancho (si $a \sim l$); esta posibilidad
requerir\'{\i}a un c\'alculo m\'as complejo que el que
aqu\'{\i}mostramos, en tres dimensiones. En nuestro caso, la
relaci\'on t\'{\i}pica de nuestros puentes es s\'olo $l/a=5$, con
lo que quiz\'a estamos muy al l\'{\i}mite de la validez de la
resoluci\'on en 2D. Pero como primera aproximaci\'on es razonable
y sencilla, y veremos que los resultados del modelo son muy
acordes con el experimento. De todos modos apuntaremos m\'as
adelante la necesidad de hacer c\'alculos m\'as refinados, por
\'este y otros motivos.}. La \Index{difusividad t\'ermica} en el
seno del substrato, $D=\kappa / c_p$, lleva a que pasado un tiempo
$t$ el calor haya penetrado en \'este una distancia
\cite{Chapman84,Gupta93}
\be
 r(t)=2 \sqrt{D t},
\label{ec:rt} \ee representada tambi\'en en la figura
\ref{fig:cal:filmsubs}. El valor de la difusividad para el \STOf
\index{substrato} es $D=0.18\und{cm^2/s}$
\cite{Gupta93,Marshall93}, por lo que el calor alcanzar\'a la pared
inferior de nuestro substrato, separada $1\und{mm}$ del
micropuente, en unos $14\und{ms}$. Para medidas de duraci\'on menor,
el modelo que proponemos tiene plena validez, e incluso a tiempos
m\'as largos si el substrato no se calienta mucho.

Consideraremos que el volumen del sistema que est\'a dentro de ese
radio (variable con el tiempo) se ha calentado en mayor o menor
medida hasta una temperatura promedio \Ts, y que el resto
permanece a la temperatura inicial del ba\~no \Tb. De esta manera
simplificamos la distribuci\'on no uniforme de temperatura en el
substrato dividi\'endolo en s\'olo dos partes: una termalizada con el
ba\~no externo a \Tb, y otra a temperatura $\Ts > \Tb$, m\'as cerca de
la muestra.

% (j-n) nota de Antonio: esto nos permite calcular \DT\ en \ref{ec:evacuaDT} con $\DT=\Tf-\Ts$.

El calor que se genera en la muestra (asumamos que uniformemente,
por simplificar) la calienta s\'olo a ella cuando $t \tiende 0$,
pero luego se va repartiendo por el resto del substrato conforme
pasa el tiempo, seg\'un la ecuaci\'on \ref{ec:rt}. Una buena
estimaci\'on del volumen \vs\ que se ha calentado con el paso del
tiempo es, por tanto (largo $\times$ ancho $\times$ alto)
\be
\vs(t) = l \times (2\,r(t) + a) \times (r(t) + d)
 \label{ec:vst}
\ee donde $l$, $a$ y $d$ son la longitud, el ancho y espesor de
micropuente, respectivamente. Esta ecuaci\'on del volumen total
calentado tiende al volumen de la muestra $v_f=l \times a \times
d$ a tiempos cortos.

Tal y como lo hemos definido, el volumen $\vs(t)$ tiene
geometr\'{\i}a prism\'atica, que se podr\'{\i}a dividir entre
$4/\pi \sim 1.27$ para hacer una aproximaci\'on a una forma
cil\'{\i}ndrica, que es el caso l\'{\i}mite ideal.
\label{par:cilindro} Emplearemos este factor posteriormente para
obtener resultados algo m\'as realistas con los c\'alculos de
nuestro modelo. No es correcto en el l\'{\i}mite de $t$ tendiendo
a cero, en que el volumen \vs\ coincide con el del micropuente
\vf, que es prism\'atico, pero la correcci\'on es apropiada a
tiempos largos, cuando $t \tiende \infty$. Cuando hagamos
c\'alculos tendremos en cuenta este factor de manera adecuada,
viendo qu\'e tiempos consideraremos como muy largos, dentro de los
rangos de validez de este modelo.

\subsubsection{Aumento de temperatura}

En cualquier momento dado, el aumento \emph{promedio} de
temperatura de este volumen caliente \vs\ (de film m\'as substrato,
pero que llamo de substrato por simplificar) est\'a determinado por
el calor total generado hasta ese momento, $Q(t)$, que es la
integral en el tiempo de la potencia disipada $P(t)$:
\be
 Q(t)=\int_0^t P(\tau) \, \ud \tau,
 \label{ec:Qt}
 \ee
\be
\DTs(t)=\Ts-\Tb=\frac{Q(t)}{c_p \, \vs(t)}.
 \label{ec:DTs}
 \ee
donde $c_p$ es la capacidad calor\'{\i}fica por unidad de volumen
del substrato, y que consideramos constante con la temperatura.
Como la $c_p$ del \STOf \index{substrato}\ es pr\'acticamente
coincidente con la del \YBCO, resulta factible unificar
micropuente y substrato como un volumen \'unico, a efectos de esta
ecuaci\'on. Tomamos $c_p = 1\;{\rm J/K\,cm^3}$
\cite{Gupta93,Touloukian70}.
% cita en Gupta: Y.S. Touloukian, ed. Thermophysical Properties of Matter, vols. 2 and 5 (IFI/Plenum, NY 1970)

\figCalDTs

\figCalDTsZoom

Veamos qu\'e ocurre en el caso concreto de que la potencia
disipada \emph{fuese constante}. La dependencia de $\DTs(t)$ en
este caso se representa en la figura \ref{fig:cal:DTs}. Observamos
que, a tiempos largos, cuando el volumen del filme se hace
despreciable comparado con el del substrato caliente, el
incremento de temperatura alcanza un r\'egimen casi estacionario.
El calor disipado $Q(t) = P\;t$ ser\'{\i}a lineal con el tiempo,
pero el volumen caliente cil\'{\i}ndrico tambi\'en tiene un
l\'{\i}mite lineal en $t$ a tiempos largos, de modo que el
cociente tiende a ser constante: el aumento de temperatura
m\'aximo, a $t \tiende \infty$, es \be \DT_{s-m\acute{a}x} =
\frac{P}{8\;l\;D}. \label{ec:DTmax} \ee De este modo, la
ecuaci\'on que se representa en la figura \ref{fig:cal:DTs} es \be
\frac{\DTs(t)}{\DT_{s-m\acute{a}x}}=\frac{8\;l\;D\;t}{c_p \,
\vs(t)} \label{ec:cal:DtsDtmax}. \ee Por dar algunos valores
t\'{\i}picos, tomemos la isoterma m\'as baja de la figura
\ref{fig:datos:EJmodelo} (p. \pageref{fig:datos:EJmodelo}). En esa
muestra y a esa temperatura, el salto ocurre con \Jx\ rondando los
$6\Exund{6}{\Acms}$. La correspondiente potencia disipada es
$W=E\;J= 7.5 \Exp{6} \Wcmc$, $P=W\;\vf= 5.6 \Exund{-4}{W}$. Si
aplicamos un \Index{pulso} de esa corriente durante $t=1\und{ms}$,
el volumen de substrato implicado es $\vs(1 \und{ms})=7.3
\Exund{-6}{cm^3}$, de manera que $\DTs=P\,t/(c_p\,\vs) \simeq 8
\Exund{-2}{K}$, pr\'acticamente coincidente con el valor de
$\DT_{s-m\acute{a}x}$ que se extrae de la ecuaci\'on
\ref{ec:DTmax}. Es un peque\~no aumento de temperatura, en
apariencia despreciable. Uno podr\'{\i}a pensar que con tan poco
incremento de temperatura tan cerca del salto el
\Index{calentamiento} no puede ser la causa de la transici\'on.
Este \DTs\ no es suficiente para llevar al sistema por encima de
\Tc.  Pero si recordamos que la potencia disipada $P$ no se
mantiene en realidad constante, sino que su variaci\'on depende
fuertemente del incremento de temperatura, deber\'{\i}amos
considerarla en nuestros c\'alculos.

El 95\% del aumento m\'aximo $\DT_{s-m\acute{a}x}$ se alcanza, con
independencia del valor concreto, hacia los 100~$\mu s$. Este es
el \Index{tiempo caracter\'{\i}stico} de calentamiento \taui
\index{tiempo de inercia t\'ermica} de nuestro sistema.
\label{sssec:DT} Es mucho mayor que el \Index{tiempo
bolom\'etrico} de enfriamiento
\be
\tau_B= \frac{\cp\;l\;a\;d}{h\;l\;a}=\frac{\cp\;d}{h} \simeq
15\und{ns} \label{tbolom}
\ee
porque se refiere a los efectos de inercia del substrato, y no a
la resistencia t\'ermica de frontera que determina \tauB. En
cualquier caso es un tiempo muy breve, apenas diez veces mayor que
la resoluci\'on temporal de nuestro sistema de adquisici\'on de datos.

En la figura \ref{fig:cal:DTsZoom} representamos una visi\'on
ampliada de la gr\'afica \ref{fig:cal:DTs}, para tiempos m\'as
breves. Vemos que para restringir el aumento de temperatura a
s\'olo un 10\% del m\'aximo habr\'{\i}a que aplicar pulsos de unos
$10\und{ns}$. La derivada de este aumento de temperatura,
$\indiff{\DTs}{t}$ (figura \ref{fig:cal:dDTsdt}) refleja lo
abruptamente que \DTs\ crece en los primeros instantes, y c\'omo
lo hace m\'as suavemente pasado el tiempo.

\comenta{ Si no hubiese transferencia de calor al substrato y el
filme estuviese aislado, esa tendencia inicial de calentamiento
tan dr\'astico se mantendr\'{\i}a todo el tiempo.}

\figCaldDTsdt

En una primera aproximaci\'on, muy simplista pero que m\'as tarde
refinaremos, podemos hacer que el aumento de temperatura del
micropuente \DTf\ sea igual al incremento promedio \DTs\ de la
ecuaci\'on \ref{ec:DTs}. De este modo, la temperatura del
micropuente ser\'{\i}a
\be
\Tf(t)=\Tb+\DTf(t)=\Tb+\DTs(t) \label{ec:Tfsimple}. \ee A este
modelo geom\'etrico tan elemental, que pretende dar cuenta de la
distribuci\'on de temperatura dentro del conjunto de micropuente
m\'as substrato, se le pueden a\~nadir algunas correcciones muy
razonables que lo aproximen algo m\'as al sistema real. Lo veremos
en secciones posteriores. Pero por el momento ya podemos hacer
algunos c\'alculos con lo que hasta aqu\'{\i} hemos explicado, y
ver en qu\'e medida afecta el tener en cuenta la
\Index{realimentaci\'on}.

% ----------------------------------- %
\subsection{Generaci\'on de calor en avalancha en el filme}
\label{sec:Cal:avalancha}
% ----------------------------------- %

El modelo geom\'etrico que acabamos de exponer en la secci\'on
\ref{sec:geometria} no es original nuestro. Se puede encontrar por
ejemplo aplicado en la referencia \cite{Gupta93} y en posteriores
publicaciones. Lo que en este trabajo hacemos de original es
combinar esta geometr\'{\i}a con la \Index{realimentaci\'on} de la
potencia al variar la temperatura. Juntando los dos elementos
---la variaci\'on de la potencia disipada con la temperatura, y la
distribuci\'on de temperatura en el entorno de la muestra---,
calculamos la evoluci\'on de la temperatura del micropuente a lo
largo de una medida.

\figCalsDTtbasic

En la figura \ref{fig:cal:sDTtbasic} mostramos los resultados de
una simulaci\'on de un experimento t\'{\i}pico, empleando la
ecuaci\'on \ref{ec:Tfsimple} y previas, y teniendo en cuenta en la
ecuaci\'on \ref{ec:Qt} que la potencia disipada aumenta al
aumentar la temperatura ---$P$ var\'{\i}a con $\Tf$ como
explicamos en la secci\'on \ref{sec:PT}---. El experimento
consiste en que, a una temperatura inicial, determinada por el
\Index{ba\~no} en que est\'a inmersa la muestra, aplicamos a
nuestro micropuente un \Index{pulso} breve de corriente $J$
constante \footnote{Se podr\'{\i}a simular igualmente la
aplicaci\'on de rampas en que var\'{\i}e la corriente, pero el
caso de pulsos con corrientes constantes resulta m\'as
clarificador, por ser m\'as simple, de los procesos t\'ermicos que
tienen lugar durante la medida.} en el instante $t=0$. Vemos que
la temperatura del micropuente aumenta con el paso del tiempo,
causando as\'{\i}un aumento del campo el\'ectrico y de la se\~nal
de voltaje que se mide. Esta variaci\'on del campo conlleva una
variaci\'on de la potencia disipada, lo que a su vez acelera el
aumento de temperatura. Experimentalmente no podemos medir la
temperatura del micropuente, pero a corriente constante vemos su
variaci\'on a trav\'es de la se\~nal de voltaje, como en la figura
\ref{fig:datos:EtJ} de la p\'agina \pageref{fig:datos:EtJ}. Vemos
que los resultados experimentales de aquella figura y de las
simulaciones que ahora se presentan muestran tendencias
semejantes: \emph{toda} corriente aplicada causa un aumento de
temperatura de la muestra, pero se distinguen dos reg\'{\i}menes
esencialmente distintos. Hay un rango de corrientes con las que,
pasado un tiempo inicial de crecimiento r\'apido, se obtiene una
aparente estabilizaci\'on de la temperatura a tiempos largos; por
el contrario, hay otras corrientes que producen calentamientos en
cascada, elevando la temperatura de la muestra, abruptamente, por
encima de la temperatura cr\'{\i}tica y haci\'endola transitar
as\'{\i} al estado normal. El cambio de r\'egimen estacionario a
cr\'{\i}tico ocurre en un rango estrecho de corrientes: corrientes
estables y cr\'{\i}ticas difieren experimentalmente de menos del
1\%, mientras que nuestro modelo sencillo da un valor que ronda el
2\%.

\figCalComPcte

En la figura \ref{fig:cal:comPcte} comparamos una simulaci\'on con
realimentaci\'on en la potencia ---variando $P$ con $\Tf$ en la
ecuaci\'on \ref{ec:Qt} convenientemente\mbox{---,} con su
correspondiente resultado a potencia constante, sin introducir
\Index{realimentaci\'on} ---haciendo $P$ constante en la
ecuaci\'on \ref{ec:Qt}, e igual al valor inicial en $t=0$---.
Representamos la curva con corriente m\'as baja de la figura
\ref{fig:cal:sDTtbasic}, y la que le corresponde haciendo $P$
constante, tal y como hicimos en la figura \ref{fig:cal:DTs}. Los
par\'ametros son iguales en ambos casos. Vemos con esta
comparaci\'on que al introducir \Index{realimentaci\'on}, incluso
en los casos en que se tiende pr\'acticamente a una temperatura
estable, se alcanzan temperaturas m\'as elevadas que cuando la
potencia se mantiene (artificialmente) constante. Pero el
resultado m\'as importante no es este, sino el de que al
introducir \Index{realimentaci\'on} se consiguen, en ciertas
condiciones, avalanchas t\'ermicas que no aparec\'{\i}an en el
caso de $P$ constante (fig. \ref{fig:cal:sDTtbasic}). Para tener
esta posibilidad es necesario que la potencia aumente con el paso
del tiempo: la \Index{realimentaci\'on} es fundamental, y como
vemos puede tener efectos notables.

Dependiendo de la temperatura inicial y de la corriente aplicada,
los resultados de las simulaciones se dividen, como acabamos de
ver en la figura \ref{fig:cal:sDTtbasic}, en dos categor\'{\i}as:
por un lado est\'an los que, tras un aumento inicial, r\'apido, de
temperatura (en un \Index{tiempo caracter\'{\i}stico} de unos
$100\und{\mu s}$, como el del caso a potencia constante de la
figura \ref{fig:cal:DTs}) predicen una temperatura (y por lo tanto
una se\~nal de voltaje) \emph{pr\'acticamente} estable a tiempos
largos. Por otro lado est\'an los que conllevan una
\Index{avalancha} abrupta que lleva a la muestra por encima de su
temperatura cr\'{\i}tica \Tc, causando la transici\'on al estado
normal. Esta transici\'on ocurre en tiempos muy breves, tambi\'en
de unos $100\und{\mu s}$, tras un cambio brusco en la tendencia
del aumento de temperatura.

\figCalJxTsimple

Los resultados de estas simulaciones que hemos mostrado
---calculando la temperatura del filme con la ecuaci\'on
\ref{ec:Tfsimple} y empleando los par\'ametros del muestra \fA---,
dicen que, a $\Tb=76.2\und{K}$, la corriente de
$7.70\Exund{6}{\Acms}$ causa un estado casi estacionario, y que
s\'olo si elevamos la corriente hasta unos $7.80\Exund{6}{\Acms}$
tendremos una transici\'on al estado normal dentro del tiempo que
dura la medida. Seg\'un estos resultados, el valor de \Jx\ ronda
por tanto los $7.80\Exund{6}{\Acms}$. En comparaci\'on con el
resultado experimental de $6\Exund{6}{\Acms}$, tenemos una
predicci\'on te\'orica que se equivoca por exceso en un 30\%, lo
que no est\'a nada mal dada la sencillez del modelo que empleamos
para describir la complejidad del sistema real. Los valores de
\Jx\ calculados an\'alogamente para otras temperaturas iniciales
se muestran en la figura \ref{fig:cal:JxTsimple}. El exceso de la
predicci\'on en un 30\% se mantiene para todas las temperaturas.
En las siguientes secciones emplearemos alguna correcciones muy
razonables a la ecuaci\'on \ref{ec:Tfsimple} que nos permitir\'an
acercar m\'as las predicciones a los resultados experimentales.
Pero lo esencial ya se recoge en esto que hasta aqu\'{\i} hemos
expuesto: cuando se tienen en cuenta tanto la
\Index{realimentaci\'on} en la potencia disipada (lo que hace
aumentar notablemente el calor generado) como la distribuci\'on de
temperaturas no uniforme en el entorno de la muestra (lo que
disminuye la capacidad de evacuaci\'on del calor) se pueden
generar avalanchas t\'ermicas que elevan la temperatura de la
muestra por encima de \Tc\ muy r\'apidamente, causando
as\'{\i}transiciones abruptas al estado normal.

Los c\'alculos los hemos realizado con un programa escrito en
\emph{Mathematica} que mostramos con detalle en el ap\'endice
\ref{cap:appMath}. En resumen, lo que el programa hace es dividir
la duraci\'on de la medida en peque\~nos intervalos de tiempo, de
manera que en cada momento se calcula la temperatura del
micropuente \Tf\ con las ecuaciones \ref{ec:Tfsimple} y previas. A
una corriente $J$ aplicada, la temperatura calculada \Tf\
determina un campo el\'ectrico $E$ (que introducimos en el programa
empleando una funci\'on que ajusta a nuestros datos experimentales:
explicaremos esto con detalle en la secci\'on \ref{sec:detalles}),
lo que nos proporciona una se\~nal de voltaje que podemos comparar
con la experimental. Pero adem\'as nos sirve para calcular la
potencia disipada $P=E\;J\;\vf$, lo que nos permite, en el
siguiente intervalo de tiempo, calcular de nuevo la temperatura
del micropuente convenientemente: en cada intervalo, la potencia
$P$ que conlleva una cierta temperatura de la muestra est\'a a su
vez determinada por la temperatura \Tf\ del intervalo previo. De
este modo introducimos la \Index{realimentaci\'on}.

Destaquemos aqu\'{\i} otro resultado importante, que ya
apreci\'abamos en los experimentos de la figura
\ref{fig:datos:EtJ} y que ahora hemos reproducido con nuestro
modelo te\'orico: el \Index{tiempo caracter\'{\i}stico} de
estabilizaci\'on de la temperatura \taui, en los casos a $J<\Jx$
donde no se produce avalancha, es de unos $100\und{\mu s}$.
Recordemos, como ya dijimos en la secci\'on \ref{sssec:DT} (p.
\pageref{sssec:DT}) que este tiempo caracter\'{\i}stico es mayor
que el \Index{tiempo bolom\'etrico} \tauB\ de enfriamiento porque
tiene en cuenta sobre todo la inercia t\'ermica del substrato,
cosa que \tauB\ no hace. Si este \taui\ es el tiempo
caracter\'{\i}stico de inercia t\'ermica de nuestro sistema, es
comprensible que cualquier medida que se realice en escalas
temporales mucho mayores no muestre, por debajo de \Jx, una
dependencia respecto de duraci\'on de la medida. Si por ejemplo
medimos experimentalmente curvas \EJ\ cerca de la transici\'on al
estado normal (pero manteni\'endonos siempre por debajo de \Jx),
en sentidos ascendente y descendente de la corriente, con pulsos
para cada valor de la corriente de, digamos, decenas de
milisegundos, no es de esperar que apreciemos efectos de
\Index{hist\'eresis}: la duraci\'on de la medida de cada
\Index{pulso} es \'ordenes de magnitud mayor que el tiempo de
calentamiento o de enfriamiento. Pero que no veamos efectos de
\Index{hist\'eresis} no significa, como hemos visto aqu\'{\i}, que
no haya calentamiento implicado: lo que ocurre es que no medimos,
en este ejemplo de experimento, con la resoluci\'on temporal
suficiente como para apreciarla.

Pues bien: diversas publicaciones, donde se mostraban experimentos
en que se empleaban pulsos de decenas de milisegundos
\cite{Curras01} o incluso de segundos \cite{Xiao96}, descartaban
la implicaci\'on de fen\'omenos de \Index{calentamiento}
relevantes debido precisamente a la ausencia de
\Index{hist\'eresis} en sus medidas. Con lo que hasta
aqu\'{\i}hemos mostrado desmontamos este argumento: si en esos
experimentos no se apreciaban efectos de \Index{hist\'eresis} era
debido a la poca resoluci\'on temporal de la medida, y no
necesariamente a que los efectos t\'ermicos fuesen despreciables.
Como corroboraci\'on experimental de lo que acabamos de explicar,
resultados que empleaban mayor resoluci\'on temporal empezaban a
apreciar dependencia del experimento respecto de la duraci\'on de
la medida s\'olo cuando sus pulsos se reduc\'{\i}an por debajo del
milisegundo, hasta 100 o $50\und{\mu s}$ \cite{Jakob00}.

%%%%%%%%%%%%%%%%%%%%%%%%%%%%%%%%%%%%%%%%
\section{Detalles del c\'alculo y mejoras al modelo}
\label{sec:detalles}
%%%%%%%%%%%%%%%%%%%%%%%%%%%%%%%%%%%%%%%%

En esta secci\'on explicamos c\'omo introducimos la disipaci\'on
en funci\'on de la temperatura dentro de nuestras simulaciones, y
proponemos un par de mejoras muy razonables al modelo sencillo de
la ecuaci\'on \ref{ec:Tfsimple}, que acercar\'an m\'as
todav\'{\i}a los resultados te\'oricos a los experimentales.

% ------------------------------------ %
\subsection{Disipaci\'on en funci\'on de la temperatura}
% ------------------------------------ %

En los c\'alculos para simular la evoluci\'on temporal del sistema
necesitamos conocer cu\'al es concretamente el campo $E$ que se
tiene cuando var\'{\i}a la temperatura, para introducir con ello
la \Index{realimentaci\'on}. Esta variaci\'on de $E$ con \Tf\ a
una corriente $J$ dada la podr\'{\i}amos obtener como
interpolaci\'on de nuestros datos experimentales, pero es m\'as
pr\'actico buscar una \'unica funci\'on $E(J,\Tf)$ que englobe
toda esa informaci\'on. Esta funci\'on la obtenemos de los ajustes
a los datos experimentales que vimos en el cap\'{\i}tulo
\ref{cap:datos}, empleando los modelos habituales que describen
las curvas \EJ. Para campo aplicado nulo, usamos la dependencia de
la ecuaci\'on \ref{ec:EJT} (p. \pageref{ec:EJT}), mientras que
para campo aplicado usamos la ecuaci\'on \ref{ec:EJTB1}. De esta
manera obtenemos una \'unica funci\'on para cada caso que nos da
el valor de $E(J,\Tf)$, y que empleamos en nuestros c\'alculos
para introducir realimentaci\'on, variando el valor de la potencia
al variar la temperatura de la muestra. Para ver los detalles de
estos c\'alculos remitimos al ap\'endice \ref{cap:appMath}.

% ------------------------------------ %
\subsection{Mejoras al modelo sencillo}
\label{ssec:mejoras}
% ------------------------------------ %

En el apartado \ref{sec:Cal:avalancha} hemos visto c\'omo el
sencillo modelo t\'ermico de las ecuaciones \ref{ec:Tfsimple} y
previas reproduce la existencia de una corriente cr\'{\i}tica que
genera una avalancha, elevando as\'{\i} la temperatura de la
muestra de manera abrupta y haci\'endola transitar al estado
normal. Dada la sencillez del modelo propuesto, la predicci\'on
te\'orica de \Jx\ difer\'{\i}a del resultado experimental en
aproximadamente un 30\% (fig. \ref{fig:cal:JxTsimple}). Con la
intenci\'on de reducir esta diferencia introducimos aqu\'{\i}, y
para todos los c\'alculos consecuentes, un par de mejoras al
modelo sencillo que modifican los resultados exclusivamente de
manera cuantitativa, sin alterar las ideas b\'asicas. Como
veremos, estas mejoras son muy razonables, y se aplican con la
intenci\'on de hacer una descripci\'on m\'as realista del sistema.
Est\'an basadas en las correcciones que ya se encontraban en la
referencia \cite{Gupta93}, pero aqu\'{\i} se han adaptado para
tener en cuenta las dependencias temporales.

\paragraph{Distribuci\'on no uniforme de temperatura.}
% ------------------------------------ %

Lo que nos interesa en \'ultima instancia es la temperatura del
micropuente, no la del substrato. Si queremos introducir
\Index{realimentaci\'on} en el modelo y calcular una nueva potencia
disipada teniendo en cuenta la variaci\'on de temperatura a lo largo
del tiempo de la medida, es la temperatura del filme \Tf\ la que
es relevante. En una primera aproximaci\'on, la m\'as simple, hemos
considerado que $\Tf=\Ts$ (ec. \ref{ec:Tfsimple}), lo que implica
que la temperatura en el volumen caliente es uniforme e igual al
promedio que calcul\'abamos con la ecuaci\'on \ref{ec:DTs}. Para
conseguir resultados m\'as realistas, sin embargo, podemos tener en
cuenta que en el seno del volumen \vs\ al que ha llegado el calor
en un tiempo $t$ hay una distribuci\'on no uniforme de temperatura,
con $T \tiende \Tf$ cerca de la muestra y $T \tiende \Tb$ en el
extremo opuesto. Para lo que nos interesa es suficiente conocer
cu\'al es la temperatura en el extremo cercano al filme, y no tanto
c\'omo es en detalle la distribuci\'on de temperatura dentro de \vs.
El aumento de temperatura en ese punto ser\'a mayor que el promedio
\DTs\ y, en general, lo podemos tomar como un factor \alp\ mayor
que \DTs:
\be
\DTf(t)=\alp\;\DTs(t) \label{ec:DTf}, \ee de manera que la
temperatura de filme queda
\be
\Tf(t)=\Tb+\DTf(t)=\Tb+\alp\;\DTs(t). \label{ec:Tf} \ee En el caso
m\'as simple que consideramos anteriormente, haciendo la temperatura
del filme igual al promedio \Ts, \alp\ tomaba el valor 1.

Este factor \alp\ se puede hacer m\'as complejo, con la
intenci\'on de aproximarnos m\'as al sistema real. Hacer \alp\
mayor que 1 implica corregir, en lo que a la temperatura del film
se refiere, la simplificaci\'on que supone dividir el substrato
caliente s\'olo en dos regiones de temperaturas uniformes: una
termalizada con el ba\~no a \Tb\ y la otra a una temperatura \Ts\
algo mayor. La regi\'on del substrato pegada al film tiene en
realidad una temperatura m\'as elevada que \emph{la media} \Ts\ de
todo el volumen caliente, y este factor \alp\ dar\'{\i}a cuenta de
esto.

Para tener en cuenta el valor de \alp\ sin profundizar en la
soluci\'on completa de la ecuaci\'on de conducci\'on del calor,
que ser\'{\i}a bastante compleja, podemos hacer unas estimaciones
de su comportamiento en casos l\'{\i}mite que nos permitan hacer
c\'alculos. A tiempos cortos, cuando $t \tiende 0$, es decir,
cuando el $\vs(t)$ coincide con el volumen del micropuente \vf, o
es de su orden de magnitud, tiene mucho sentido hacer la
temperatura del film igual a la del volumen total caliente: son
casi coincidentes. En este caso l\'{\i}mite se puede hacer $\alp =
1$.

Por el contrario, cuanto m\'as pasa el tiempo menos es v\'alida la
aproximaci\'on $\alp=1$, ya que la geometr\'{\i}a tiende a la de
un semicilindro calentado por su eje, con distribuci\'on no
uniforme de temperatura. El micropuente, en el eje, calienta a
todo el cilindro, y es en el propio filme donde se alcanza la
m\'axima temperatura de la distribuci\'on. En el caso ideal,
cuando $t \tiende \infty$, \alp\ toma el valor $4/\sqrt{\pi}
\simeq 2.26$ \cite{Gupta93}.

Podemos adem\'as incluir en este factor aquel que corrige la forma
del volumen caliente, tendiendo a hacerla cil\'{\i}ndrica a
tiempos largos (ver sec. \ref{sec:geometria}, p.
\pageref{par:cilindro}). De este modo, el valor l\'{\i}mite de
\alp\ con $t \tiende \infty$ ser\'a en realidad $4/\sqrt{\pi}
\times 4/\pi \simeq 2.26 \times 1.27 = 2.87$, pero seguir\'a
valiendo 1 en tiempos cercanos al cero.

\figCalAlpha

Tomaremos en nuestros c\'alculos un factor \alp\ dependiente del
tiempo que tenga los l\'{\i}mites que hemos se\~nalado. De forma
sencilla usamos como aproximaci\'on al valor real de \alp\ una
funci\'on que tienda suavemente a 2.87, representada en la figura
\ref{fig:cal:alpha}, de la forma
\be
\alp(t) = 1 + 1.87\; (1 - e^{-t/t_c}) \label{ec:alpha} \ee donde
$t_c$ es un \Index{tiempo caracter\'{\i}stico} que estimamos
comparando el volumen caliente de substrato $\vs(t)$ con el del
filme \vf. Como primera aproximaci\'on, tomamos el tiempo $t$ en
que $\vs(t)$ se hace grande comparado con \vf, digamos diez veces
mayor. En nuestro caso, usando los par\'ametros propios de nuestro
sistema y empleando la ecuaci\'on \ref{ec:vst}, ese tiempo es $t_c
\simeq 15 \und{ns}$. M\'as adelante haremos estimaciones de qu\'e
pasar\'{\i}a si el substrato tuviese una \Index{conductividad
t\'ermica} quince veces mayor (como si emple\'asemos \MgO
\index{substrato}\ en vez de \STOf
\cite{Marshall93,Doettinger94}); en este otro caso $t_c \simeq
0.25 \und{ns}$.

\paragraph{Frontera entre filme y substrato.}
% ------------------------------------ %

N\'otese que, a pesar del anterior factor corrector de la
geometr\'{\i}a, seguimos estimando por defecto la temperatura del
film: al poner al filme en un continuo con el substrato y sin
m\'as consideraciones, hacemos que el film y la capa de substrato
inmediatamente pegada a \'el est\'en a la misma temperatura; es
decir, que el coeficiente de transferencia de calor entre ambos
sea infinito. En esta imagen, el film y el substrato son un
continuo de caracter\'{\i}sticas t\'ermicas iguales. En realidad
el film est\'a a una temperatura algo mayor que la capa de
substrato a que est\'a adherido; en un caso estacionario esta
diferencia ser\'{\i}a de $\DT = P /(l\;a\;h)$, donde $P$ es la
potencia estacionaria, el producto $l\; a$ es el \'area de
contacto, y $h$ es el coeficiente de transferencia de calor entre
substrato y film. Como no estamos en el estacionario este
l\'{\i}mite es inexacto: $P$ var\'{\i}a con el tiempo y la
temperatura, y la resoluci\'on del problema se complica.

Veremos que obviar esta cuesti\'on, como ya hemos mencionado, no
afecta cualitativamente a los principales resultados, pero, para
acercar m\'as las predicciones del modelo al experimento, en los
resultados finales que mostraremos se incluye un t\'ermino que da
cuenta de este incremento de temperatura extra, mejorando
cuantitativamente las predicciones. Juntando las dos correcciones
que hemos mencionado, la temperatura del micropuente resulta
\be
\Tf(t)=\Tb+\DTf(t)=\Tb+\alp(t)\;\DTs(t)+ \beta(t) \; \frac{P}
{l\;a\;h}. \label{ec:Tfcorr}
\ee

Para la funci\'on $\beta(t)$ emplearemos una dependencia temporal
an\'aloga a la que hemos expuesto para $\alp(t)$ en la ecuaci\'on
\ref{ec:alpha}, de modo que tambi\'en $\beta(t)$ transite
suavemente de un l\'{\i}mite al otro, pero teniendo en cuenta que
los l\'{\i}mites son ahora $\beta=0$ para $t \tiende 0$ y
$\beta=1$ para $t \tiende \infty$. Tambi\'en haremos, como primera
aproximaci\'on, el par\'ametro $t_c$ diez veces menor que en el
caso de \alp: la resistencia a la transferencia de calor entre
medios se pone de manifiesto desde los primeros momentos de la
evoluci\'on de la medida, no hay que esperar a que el calor llegue
a grandes vol\'umenes de substrato. Teniendo en cuenta estos
l\'{\i}mites, haremos que
\be
\beta(t) =  1 - e^{-t \;10/t_c}. \label{ec:beta} \ee En el
siguiente cap\'{\i}tulo compararemos las simulaciones, teniendo en
cuenta estos dos t\'erminos correctores, con los resultados
experimentales, y veremos que las similitudes son notables,
apoyando la validez de dichas correcciones.
    % cap 6
\clearemptydoublepage
%%%%%%%%%%%%%%%%%%%%%%%%%%%%%%%%%%%%%%%%%%%%%%%%%%%%%%%%%%%%%%%%%%
\chapter[Predicciones del modelo de calentamiento]{Predicciones del modelo de calentamiento en avalancha}
%\chaptermark{Modelo de calentamiento en avalancha}
%\addtocontents{toc}{\protect\vspace{0.2cm}}
\label{cap:calpr}
%%%%%%%%%%%%%%%%%%%%%%%%%%%%%%%%%%%%%%%%%%%%%%%%%%%%%%%%%%%%%%%%%%

En este cap\'{\i}tulo veremos las principales predicciones que se
extraen del modelo de calentamiento en avalancha que hemos
propuesto, y las compararemos con los datos experimentales.
Tambi\'en sugeriremos una \Index{condici\'on de avalancha} que
explica por qu\'e el sistema, bajo ciertas condiciones, alcanza
estados aparentemente estacionarios, mientras que en otras
condiciones se produce una avalancha t\'ermica que lleva a la
muestra al estado normal. Esta condici\'on de avalancha nos
permitir\'a interpretar algunos resultados experimentales dentro
del contexto del calentamiento.

%%%%%%%%%%%%%%%%%%%%%%%%%%%%%%%%%%%%%%%
\section{Predicciones del modelo y comparaci\'on con los experimentos}
%%%%%%%%%%%%%%%%%%%%%%%%%%%%%%%%%%%%%%%

% ------------------------------------ %
\subsection{Variaci\'on del valor de \Jx\ en funci\'on de la duraci\'on de la medida}
\label{ssec:depJxt}
% ------------------------------------ %

En la figura \ref{fig:datos:EtJ} de la p\'agina
\pageref{fig:datos:EtJ} ve\'{\i}amos como la \Index{avalancha} se
alcanzaba en distintos momentos a partir del inicio de la medida
dependiendo del valor concreto de la corriente que aplic\'asemos.
De este modo, el valor experimental de \Jx\ depend\'{\i}a del
tiempo que durase la medida. Algunos valores de corriente aplicada
daban una aparente estabilizaci\'on, sin causar avalancha. Por el
contrario otros valores, a partir de una cierta corriente,
conduc\'{\i}an siempre a una avalancha, aunque en tiempos m\'as
breves cuanto mayor fuese el valor de $J$.

\figCalsDTt

Esto mismo lo predicen nuestras simulaciones, como ya hemos
mencionado. Con las correcciones explicadas en el apartado previo,
en la figura \ref{fig:cal:sDTt} vemos la evoluci\'on en el tiempo
del incremento de temperatura \DTf\ del micropuente al aplicar
distintas corrientes de valores cercanos entre s\'{\i} (6.70,
6.73, 6.74 y $6.84\Exund{6}{\Acms}$), partiendo de una temperatura
inicial de $75.6\und{K}$, igual a la del experimento de la figura
\ref{fig:datos:EtJ} con el que hacemos la comparaci\'on. Entre la
primera, que lleva a una temperatura aparentemente estable dentro
de los tiempos de la medida, y la \'ultima, que produce una
\Index{avalancha} inmediata, hay un 2\% de diferencia. Pero entre
la primera y la segunda de las corrientes aplicadas, que ya dan
resultados distintos dentro del tiempo que dura la medida, hay
s\'olo un 0.5\% de diferencia. Esto da cuenta, al menos
cualitativamente, de los resultados de la tesis de S.~R. Curr\'as
\cite{Curras00b} publicados en la referencia \cite{Curras01}, en
los que una cierta corriente aplicada generaban un estado
estacionario prolongable varios minutos, pero aumentar la
corriente en un 0.2\% hac\'{\i}a transitar la muestra al estado
normal. Entonces se utiliz\'o este resultado experimental para
descartar que el \Index{calentamiento} fuese exclusivamente el
causante de la transici\'on al estado normal: muy peque\~nas
diferencias de la potencia disipada conllevaban estados
notablemente diferentes, as\'{\i}que parec\'{\i}a m\'as probable
que el motivo fuese alg\'un \index{intr\'{\i}nseco} mecanismo
intr\'{\i}nseco. Ahora vemos, sin embargo, que la
\Index{realimentaci\'on} en la potencia disipada justifica estas
diferencias a pesar de hacer el experimento con potencias
iniciales notablemente parecidas. No podemos, por tanto, descartar
que el calentamiento est\'e directamente involucrado en la ruptura
de la superconductividad de la muestra.

\figCalsEtJ

De estas simulaciones obtenemos tambi\'en la evoluci\'on del campo
el\'ectrico $E(J,\Tf)$, que se representa en la figura
\ref{fig:cal:sEtJ} juntamente con los resultados experimentales ya
mostrados en la figura \ref{fig:datos:EtJ}, y que se repiten
aqu\'{\i}para facilitar la comparaci\'on. Los saltos de la
simulaci\'on no son tan abruptos como los experimentales, pero el
comportamiento con la duraci\'on de la medida es an\'alogo. (Que
los saltos no ocurran exactamente en el mismo instante no es
significativo: se han elegido estos valores de corriente para la
simulaci\'on porque muestran tendencias semejantes, pero no se ha
buscado que la transici\'on sea \emph{exactamente} en el mismo
instante temporal). Probablemente el modelo es demasiado pobre
para dar cuenta de cambios tan r\'apidos en tan poco tiempo, pero
lo interesante es destacar que incluso en la sencillez del modelo
se predice una avalancha que lleva a la muestra al estado normal.

% (j-n) nota de Antonio: introducir algunos s\'{\i}mbolos de los datos
% experimentales para "ver" las diferencias.
% Digo: no, porque son "demasiado" diferentes, y no me conviene:
% a pesar de las correcciones ocurren a \Jx\ distintas, y por lo
% tanto los valores del voltaje son mayores.

\figCalsDTtEtJSetSeis

% \figCalsEtJSetSeis

\figCalsrhotWtSetSeis

% \figCalsWtSetSeis

Los resultados para la temperatura inicial de $75.6\und{K}$ no
difieren significativamente de los de $76.2\und{K}$, pero
mostramos tambi\'en estos \'ultimos (figura
\ref{fig:cal:sDTtEtJ76}) porque m\'as adelante los necesitaremos
para compararlos con las simulaciones obtenidas al aplicar campo
magn\'etico, y con otros datos experimentales. Las corrientes que
ahora, al incrementar un poco la temperatura inicial, producen la
avalancha a tiempos iguales son algo menores (6.38, 6.41, 6.42 y
$6.52\Exund{6}{\Acms}$), como cab\'{\i}a esperar. Para esta misma
simulaci\'on, las evoluciones de la \Index{resistividad}
$\rho=E/J$ y de la potencia disipada $W=E \times J$ se muestran en
la figura \ref{fig:cal:srhotWt76}.

Resaltemos aqu\'{\i} el efecto que ha tenido el introducir las
correcciones al modelo sugeridas en el apartado
\ref{ssec:mejoras}. De las simulaciones de la figura
\ref{fig:cal:sDTtbasic} se extra\'{\i}a un valor te\'orico de \Jx\
de $7.8\Exund{6}{\Acms}$, que exced\'{\i}a el experimental en un
30\%. Con las correcciones, vemos ahora en la figura
\ref{fig:cal:sDTtEtJ76} que el valor predicho desciende hasta
rondar los $6.4\Exund{6}{\Acms}$, lo que supone una discrepancia
de s\'olo el 7\% respecto al valor experimental de
$6.0\Exund{6}{\Acms}$. En todo caso la mejora es solamente
cuantitativa: los principales resultados ya aparec\'{\i}an
cualitativamente en el modelo sin correcciones.

En la figura \ref{fig:cal:sDTtEtJ76} vemos c\'omo el
sobrecalentamiento \DTf\ evoluciona de manera relativamente lenta
hasta aproximadamente $\DTf = 2.5\und{K}$, y c\'omo a partir de
ese momento se acelera el incremento de temperatura de modo que
casi al instante se produce una transici\'on. Estos $2.5\und{K}$
de sobrecalentamiento antes de la transici\'on seguramente no son
tantos en el experimento real: ya hemos dicho que el salto que se
predice aqu\'{\i} no es tan abrupto como el que se mide, y es casi
seguro que en la realidad la temperatura de la muestra no aumenta
tanto antes de transitar. Pero estos valores aqu\'{\i} predichos
para campo magn\'etico aplicado nulo nos ser\'an de utilidad
cuando los comparemos con los obtenidos al aplicar campo, para
interpretar la tendencia de los experimentos. El incremento
m\'aximo de temperatura estacionario es de unos $1.8\und{K}$

Tambi\'en estaremos interesados en comparar el valor de la
potencia disipada en torno al salto. N\'otese, en la figura
\ref{fig:cal:srhotWt76}.b, que el mayor valor estable que se
obtiene de la potencia ronda los $4\Exund{7}{\Wcmc}$. Valores
mayores siempre conllevan una avalancha, y no podr\'{\i}an ser
medidos experimentalmente como puntos estacionarios.

\figCalsDTtWtOchDos

% \figCalsWtOchDos

Mostramos tambi\'en los resultados de las simulaciones para otra
temperatura inicial notablemente mayor, de $\Tb=82.0\und{K}$,
ahora con m\'as corrientes diferentes aplicadas, en el rango de
3.47--$3.55\Exund{6}{\Acms}$. Centramos nuestro inter\'es en las
evoluciones de la temperatura y de la potencia disipada, que se
muestran en la figura \ref{fig:cal:sDTtWt82}. Ahora la avalancha
se alcanza con menor sobrecalentamiento que en la simulaci\'on a
76.2~K (figura \ref{fig:cal:sDTtEtJ76}), de s\'olo unos
$1.6\und{K}$: al elevar la temperatura inicial (la del
\Index{ba\~no}) necesitamos menos sobrecalentamiento de la muestra
para entrar en r\'egimen cr\'{\i}tico. Las correspondientes curvas
de potencia disipada disminuyen su valor estacionario previo al
salto, comparadas con las obtenidas a menor temperatura inicial.
El l\'{\i}mite estable ronda ahora los $2.5\Exund{7}{\Wcmc}$, un
40\% menos.

\figCalsDTtB \figCalsEtJWtB
%\figCalsWtB

En las figuras \ref{fig:cal:sDTtB1} y \ref{fig:cal:sEtJWtB1} se
muestran resultados an\'alogos a estos, a temperatura inicial
$\Tb=76.2\und{K}$, pero para \index{campo magn\'etico} campo
aplicado $\mH=1\und{T}$. Se obtienen empleando para las
simulaciones la funci\'on \ref{ec:EJTB1} para describir las curvas
experimentales. Las corrientes a las que se produce la avalancha
son ahora menores: los valores empleados en los resultados que se
muestran son 2.68, 2.70, 2.71 y $2.80\Exund{6}{\Acms}$. El
incremento de temperatura en el momento en que se produce la
avalancha es de unos $5\und{K}$, el doble que en el caso sin campo
magn\'etico a la misma temperatura. El incremento m\'aximo de
temperatura estacionario es de unos $3.5\und{K}$. Tambi\'en aumenta
la potencia estacionaria m\'as alta que se puede alcanzar,
situ\'andose en torno a los $8\Exund{7}{\Wcmc}$. Analizamos estos
resultados en el siguiente apartado.

\figCaltJxSTO \figCaltJxcompMgO

En la figura \ref{fig:cal:tJxSTO} se recoge la correlaci\'on entre
corriente aplicada y tiempo $t$ que dicha corriente tarda en
causar una \Index{avalancha}. Esta informaci\'on se ha
extra\'{\i}do de la simulaci\'on de la figura
\ref{fig:cal:sDTtWt82}.a. Como ya hemos visto, una cierta
corriente aplicada $J$ puede inducir, o no, una avalancha
t\'ermica dependiendo de cu\'anto tiempo la mantengamos.
Equivalentemente, nuestra predicci\'on nos dice que el valor
experimental de la corriente de salto \Jx\ depender\'a de la
duraci\'on del \Index{pulso} de medida, pudiendo alcanzarse
valores m\'as altos si se mide en pulsos m\'as breves.

La primera conclusi\'on que podemos sacar de esta representaci\'on
es que conforme aumenta la duraci\'on de la medida $t$, la
corriente \Jx\ que medimos tiende a un valor m\'{\i}nimo, en este
caso unos $3.45\Exund{6}{\Acms}$. Corrientes menores dar\'an
resultados casi estacionarios, sin llegar a causar nunca una
transici\'on al estado normal, por mucho tiempo que se mantenga
aplicada la corriente. En realidad estos c\'alculos, como sabemos,
suponen que el substrato es infinito, y que por mucho que pase el
tiempo el calor no sale de \'el. La realidad es m\'as compleja, y
habr\'{\i}a quiz\'a que tener en cuenta, para estos $t$ tan
largos, correcciones relativas al entorno del substrato
(\Index{portamuestras} y atm\'osfera circundante). Pero como
primera aproximaci\'on esto resulta suficiente, y en cualquier
caso coincide con los resultados experimentales en el sentido de
que se puede aplicar una corriente $J \lesssim  \Jx$ y permanecer
mucho tiempo en ella sin causar un salto, como ya hemos mencionado
\cite{Curras00b,Curras01}.

Destaquemos de esta figura que el tiempo $t$ en el que el valor de
\Jx\ comienza a ser independiente de la duraci\'on del \Index{pulso}
coincide con el tiempo en que el volumen caliente de substrato se
hace grande comparado con el volumen de filme. Recordemos que \vs\
es unas diez veces mayor que \vf\ hacia los 15 ns (ver la ecuaci\'on
\ref{ec:alpha} y su explicaci\'on en la p\'agina \pageref{ec:alpha}).

En la figura \ref{fig:cal:tJxcompMgO} se comparan los resultados
anteriores con los que an\'alogamente se obtienen suponiendo que
la \Index{conductividad t\'ermica} del substrato es quince veces
mayor, y dejando igual los dem\'as par\'ametros. De esta manera
pretendemos simular qu\'e sucede cuando se emplean substratos
cristalinos de \MgO \index{substrato}\ en vez de \STOf, como se ha
hecho en algunos experimentos de otros grupos
\cite{Doettinger94,Doettinger95}. Las pel\'{\i}culas delgadas que
se crecen sobre estos substratos de \MgO\ suelen tener una
temperatura cr\'{\i}tica m\'as baja que las crecidas sobre \STOf,
debido a una menor compatibilidad entre las constantes de red
\index{par\'ametros de red} substrato-filme, as\'{\i} que no
variar ning\'un par\'ametro m\'as que la conductividad t\'ermica
es probablemente una descripci\'on optimista, pero nos sirve en
primera aproximaci\'on para contrastar los resultados.

A pesar de las diferencias, en ambos casos la tendencia que
muestra \Jx\ con tiempos de medida largos lleva a concluir que una
vez que los pulsos aplicados superan una cierta duraci\'on (del
orden de decenas de microsegundo de duraci\'on en el \STOf
\index{substrato}, y de nanosegundos en este supuesto \MgO\
ideal), todos los experimentos son pr\'acticamente equivalentes a
efectos de establecer la corriente cr\'{\i}tica, dentro de un
mismo tipo de substrato. Esto es coincidente con resultados
experimentales en los que habi\'endose empleados pulsos muy largos
(de decenas de milisegundos) no se apreciaban diferencias entre
multiplicar o dividir su duraci\'on por factores del orden de 100
\cite{Doettinger94,Doettinger95}. En otros experimentos en que se
mide, sin embargo, con mejor resoluci\'on temporal, se aprecian
notables diferencias en los resultados cuando el tiempo de medida
se reduce al orden de los microsegundos \cite{Jakob00}.

El valor m\'{\i}nimo de \Jx, como era de esperar, aumenta al
mejorar la conductividad t\'ermica del substrato, pero poco m\'as
de un 10\%. En estos c\'alculos, por simplicidad, no hemos
cambiado el coeficiente de transferencia de calor entre
pel\'{\i}cula y substrato al calcular con \STOf\ o con \MgO: en
todo caso ambos valores son del mismo orden de magnitud
\cite{Doettinger94}.

N\'otese tambi\'en que, cuando el tiempo del \Index{pulso} es tan
breve que el substrato apenas interviene en el fen\'omeno, las
\Jx\ obtenidas en cada caso tienden a valores muy semejantes.
Recordemos que al simular el comportamiento del \MgO
\index{substrato}\ el volumen de substrato caliente se hace grande
comparado con el del filme bastante antes que con el \STOf: \vs\
es diez veces mayor que \vf a los $0.25\und{ns}$. Esto tambi\'en
se refleja en esta figura \ref{fig:cal:tJxcompMgO}, en la que el
valor de \Jx\ comienza a variar mucho con $t$ a tiempos menores
para el \MgO \index{substrato} que para el \STOf. Es la relaci\'on
entre volumen de filme y volumen de substrato que interviene en el
fen\'omeno lo que establece estos tiempos caracter\'{\i}sticos.

Otra conclusi\'on interesante es que, efectivamente, el valor
experimental de la corriente de salto \Jx\ puede aumentar si
medimos suficientemente r\'apido, aunque para ver apreciables
variaciones de \Jx\ hay que reducir mucho la duraci\'on del
\Index{pulso} $t$. En el caso del \STOf \index{substrato}, para
aumentar el valor m\'{\i}nimo de la corriente de salto en un 10\%
(llevarla hasta $3.8\Exund{6}{\Acms}$) hay que bajar $t$ hasta
menos de 100~ns. Seg\'un este modelo, s\'olo a partir de ese
momento comienzan a verse diferencias m\'as notables con la
variaci\'on del pulso.

La tendencia indica que cuando $t \tiende 0$ no se produce
avalancha t\'ermica para ninguna corriente aplicada, de manera que
se podr\'{\i}a ver una transici\'on al estado normal causada, de
existir, por otros fen\'omenos distintos de los t\'ermicos. Si
hubiese un fen\'omeno \index{intr\'{\i}nseco} intr\'{\i}nseco a
una corriente \Jxi\ mayor que las \Jx\ m\'{\i}nimas a se que
tiende en la figura \ref{fig:cal:tJxcompMgO}, dicho fen\'omeno
estar\'{\i}a entremezclado con el \Index{calentamiento} y no se
discriminar\'{\i}a midiendo con pulsos de larga duraci\'on. El
fen\'omeno intr\'{\i}nseco se ver\'{\i}a alterado, adelantado a
corriente m\'as bajas debido al calentamiento, hasta que
reduj\'esemos tanto el tiempo de medida $t$ que la \Jx\
determinada por el calentamiento superase ese valor \Jxi. A partir
de ese momento ver\'{\i}amos siempre un valor constante de
corriente de salto \Jxi\ con independencia de sucesivas
reducciones de la duraci\'on del pulso, pues el fen\'omeno
intr\'{\i}nseco ser\'{\i}a dominante. En este tipo de medidas que
aqu\'{\i} simulamos, s\'olo podr\'{\i}amos empezar a discriminar
\emph{n\'{\i}tidamente} ambos efectos (el del calentamiento y el
posible intr\'{\i}nseco que hubiese) cuando reduj\'esemos la
duraci\'on de la medida a menos de 1~ns, momento en el que el
valor de \Jx\ comienza a variar apreciablemente.

Pero esta dificultad para discriminar los efectos
\index{extr\'{\i}nseco} extr\'{\i}nsecos de calentamiento de los
posibles \index{intr\'{\i}nseco} intr\'{\i}nsecos que est\'en
presentes (como los mencionados en el cap\'{\i}tulo
\ref{cap:datos}) no quiere decir que sea imposible separarlos en
alguna medida. El calentamiento puede afectar a la determinaci\'on
precisa de una posible \Jxi, pero medidas detalladas de la
evoluci\'on temporal del salto podr\'{\i}an permitir detectar el
efecto intr\'{\i}nseco, aunque estuviese precipitado por el
calentamiento de la muestra. Esta podr\'{\i}a ser una
explicaci\'on para la discrepancia entre las simulaciones y los
resultados experimentales, puesta de manifiesto de la figura
\ref{fig:cal:sEtJ}: el experimento muestra una transici\'on mucho
m\'as abrupta de la que somos capaces de reproducir con nuestras
simulaciones. Quiz\'a se deba a la simplicidad de nuestro modelo,
pero sin hacer c\'alculos m\'as precisos no podemos descartar que
haya un fen\'omeno \index{intr\'{\i}nseco} intr\'{\i}nseco m\'as
r\'apido que el calentamiento (como un desapareamiento de los
pares de Cooper, por ejemplo), y que se anticipa a su valor
esperado debido al incremento de temperatura en avalancha de la
muestra. Una interesante v\'{\i}a de continuaci\'on del presente
trabajo ser\'{\i}a hacer estos c\'alculos de calentamiento con
mayor detalle, mejorando el realismo del modelo.

% ------------------------------------ %
\subsection{Variaci\'on de la potencia disipada previa al salto \Wx\ con la temperatura y el campo}
\label{ssec:WxTH}
% ------------------------------------ %

\tabCalWTJ

\figCalPydP

Haciendo simulaciones como las del apartado anterior podemos
conocer en todo momento cu\'al es la temperatura de la muestra, a
corriente y campo magn\'etico aplicados contantes. Con esta
temperatura as\'{\i} calculada
---que debido al calentamiento es, en general, mayor que la inicial del ba\~no---,
y haciendo uso de la ecuaci\'on \ref{ec:EJT} que da el valor del
campo $E$ (y consecuentemente la potencia $W=E \times J$ a
corriente constante) obtenemos, para cualquier instante de tiempo,
los valores precisos de $W$ y su derivada $\indiff{W}{T}$. En la
tabla \ref{tab:WTJ} se muestran los valores de \Wx, y de la
derivada $\indiff{W}{T}$, que se alcanzan en los resultados
estacionarios de las figuras \ref{fig:cal:srhotWt76}.b,
\ref{fig:cal:sDTtWt82}.b y \ref{fig:cal:sEtJWtB1}.b (los m\'as altos
que no producen avalancha en el tiempo de la medida). En esta
tabla se incluyen tambi\'en otros valores, calculados de la misma
manera, para m\'as campos magn\'eticos aplicados de valores
intermedios. Para esta muestra, aumentar la temperatura inicial
(en el caso calculado, desde 76.2 hasta 82.0~K) implica disminuir
el valor de la potencia disipada antes del salto. Es la misma
tendencia que mostraban los experimentos (ver fig.
\ref{fig:datos:WxT}, p. \pageref{fig:datos:WxT}). Pero la
tendencia es la contraria si aumentamos el campo magn\'etico
aplicado.

En la figura \ref{fig:cal:PydP} representamos gr\'aficamente los
valores de $W$ y $\indiff{W}{T}$ en funci\'on del campo
magn\'etico aplicado que se incluyen en dicha tabla, calculados
seg\'un nuestro modelo de calentamiento. Comp\'arese con los
resultados experimentales correspondientes, mostrados previamente
en la figura \ref{fig:datos:WxHA} (p. \pageref{fig:datos:WxHA}).
Aunque los c\'alculos te\'oricos dan valores algo mayores para la
potencia estacionaria disipada antes del salto, el orden de
magnitud es el mismo que en el experimento. Y lo que es m\'as
importante: predecimos la misma tendencia de aumento de la
potencia con el campo aplicado que se observa experimentalmente,
tanto en los datos que hemos mostrado en el cap\'{\i}tulo
\ref{cap:resul} como en los que se encuentran en algunas
publicaciones \cite{Xiao98,Xiao99,Pauly00}. Pero mientras que la
potencia disipada justo antes del salto \Wx\ \emph{crece}, el
valor de la derivada de la potencia respecto de la temperatura
$\indiff{W}{T}$ en ese instante \emph{decrece} con el campo
aplicado.

Ya hemos apuntado que el que \Wx\ aumente con $H$ se emple\'o como
argumento para descartar que los efectos t\'ermicos fuesen los
causantes directos de la transici\'on al estado normal. Si \'estos
eran la causa, resultaba parad\'ojico que al aumentar el campo se
necesitase m\'as potencia disipada $W$ para hacer transitar a la
muestra. (Como \Tc\ disminuye al aumentar $H$, un aumento de campo
aplicado equivale, a una temperatura constante dada, a aproximar
m\'as la muestra a su temperatura de transici\'on, por lo que en
principio har\'{\i}a falta menos calor para hacerla transitar).
Veremos en la secci\'on \ref{sec:Cal:condAva} que este
razonamiento no es correcto en el contexto del modelo t\'ermico
que proponemos: no es s\'olo el valor de $W$ lo que causa una
avalancha, tambi\'en su derivada con la temperatura est\'a
implicada en que el sistema sea cr\'{\i}tico o estable. Los
resultados de las simulaciones aqu\'{\i}mostrados confirman que un
modelo t\'ermico puede explicar los resultados experimentales.

%El valor de la derivada $\indiff{W}{T}$ tambi\'en var\'{\i}a, de manera
%m\'as notable cuando aumenta el campo magn\'etico: en este caso la
%potencia se multiplica por dos respecto al valor sin campo a la
%misma temperatura, mientras que la derivada se divide por m\'as de
%dos. En los casos a $B=0$, en que es s\'olo la temperatura inicial
%lo que causa las variaciones, la potencia disminuye un factor 1.77
%mientras que la derivada lo hace s\'olo un 1.25. En estos ejemplos,
%las tres potencias $W$ que se muestran en la tabla llevan a
%estados estables, a pesar de lo distintas que son entre s\'{\i}.

\comenta{Esto va en la l\'{\i}nea de lo que establec\'{\i}a
nuestra condici\'on de avalancha en la ecuaci\'on
\ref{ec:condAvaSimple}: para que la condici\'on no se cumpla y
valores diferentes de la potencia lleven a estados estables, sus
derivadas con la temperatura deben variar tambi\'en. En estos
ejemplos, las tres potencias $W$ que se muestran en la tabla
llevan a estados estables, a pesar de lo distintas que son entre
s\'{\i}. Es la combinaci\'on de la potencia y su derivada, seg\'un
establecen las ecuaciones \ref{ec:condAvaSimple} y
\ref{ec:condAvaSencillo}, lo que determina que el sistema sea
cr\'{\i}tico o no. Veremos esto ampliamente m\'as adelante.}

% ------------------------------------ %
\subsection{Curvas caracter\'{\i}sticas \EJ\ afectadas de calentamiento}
\label{ssec:EJcal}
% ------------------------------------ %

De simulaciones como las mostradas, por ejemplo, en la figura
\ref{fig:cal:sDTtEtJ76}, podemos reproducir una curva \EJ,
haciendo c\'alculos para una misma temperatura inicial y aplicando
distintas corrientes. Tras un tiempo $t$ tendremos un campo
$E(J,t)$ derivado de nuestros c\'alculos, que podemos comparar con
el resultado experimental de hacer una medida con un \Index{pulso}
de duraci\'on $t$. En general, el valor del campo $E$ ser\'a tambi\'en
dependiente de la temperatura inicial y del campo magn\'etico
aplicado. Simular de este modo una curva \EJ\ equivale a dar una
informaci\'on recopilada de muchas simulaciones independientes como
las mostradas en la figura \ref{fig:cal:sDTtEtJ76}, una para cada
punto.

\figCalsEJSetSeis

\figCalsEJSetSeisB

En la figura \ref{fig:cal:sEJ76} se muestran los resultados para
una temperatura inicial $\Tb=76.2\und{K}$, y a campo magn\'etico
nulo, empleando pulsos de 1~ms. Se puede ver c\'omo la
simulaci\'on predice un distanciamiento de la tendencia que
muestran los datos en la regi\'on de baja disipaci\'on y, lo que
es m\'as importante, la existencia de una \emph{singularidad}: una
corriente m\'axima que se puede aplicar sin que la muestra
transite al estado normal. La singularidad no est\'a
impl\'{\i}cita en la tendencia de trazo discontinuo (ec.
\ref{ec:EJT}, p. \pageref{ec:EJT}) que es la base de los
c\'alculos: esta crece continuamente. Como ya hemos venido
observando en resultados previos, esta separaci\'on de la
simulaci\'on respecto de la tendencia previa no es tan abrupta
como en el experimento (el salto de los datos experimentales se
muestra en la l\'{\i}nea de puntos), pero nuestro modelo, a pesar
de su sencillez, predice el valor de \Jx\ con un \Index{error} de
menos del 7\%.

El buen acuerdo entre modelo y experimento es todav\'{\i}a m\'as
llamativo en la figura \ref{fig:cal:sEJ76B1}, donde se muestran
resultados an\'alogos para el experimento con campo magn\'etico
aplicado $\mH=1\und{T}$. Se aprecia no s\'olo la predicci\'on de
una corriente de salto (con un \Index{error} de menos del 8\%),
sino tambi\'en mucho mejor la desviaci\'on que los datos
experimentales muestran respecto de la tendencia que tienen en el
rango de menos disipaci\'on. Con nuestro modelo t\'ermico
justificamos esta desviaci\'on exclusivamente por el incremento de
temperatura de la muestra en esta regi\'on de alta
disipaci\'on\footnote{Este incremento de temperatura lo
estim\'abamos en la figura \ref{fig:cal:sDTtB1}, quiz\'a un poco
al alza, en unos $3.5\und{K}$ cerca de la transici\'on (m\'axima
temperatura estacionaria). Pero no nos interesa tanto la exactitud
de este valor, como su comparaci\'on con el del caso sin campo
magn\'etico de la figura \ref{fig:cal:sDTtEtJ76}: la relaci\'on es
de dos a uno. Con campo aplicado se produce un incremento de
temperatura, antes de la transici\'on por avalancha, mayor que sin
campo aplicado.}. Este es un resultado importante, pues reafirma
la validez del modelo propuesto, sugiriendo que la singularidad en
\Jx\ no se introduce forzadamente, de manera artificial: surge
como consecuencia inevitable del \Index{calentamiento}, el mismo
calentamiento que tan bien explica los resultados experimentales
antes del salto. En el siguiente apartado mostraremos otro
resultado que va en esta misma l\'{\i}nea, dando fuerza a la
validez de nuestro modelo t\'ermico.

% ------------------------------------ %
\subsection{Variaci\'on de la corriente de salto \Jx\ con la temperatura}
\label{ssec:cal:JxT}
% ------------------------------------ %

\figCalJxTexpcalc

\figCalJxTcompara

En la figura \ref{fig:cal:JxTexpcalc} se muestra la dependencia de
la corriente de salto \Jx\ respecto de la temperatura,
compar\'andose los resultados experimentales (c\'{\i}rculos
blancos) con las predicciones te\'oricas de nuestro modelo de
calentamiento en avalancha (cuadrados negros). Lo m\'as relevante
no es tanto la proximidad entre los valores reales y los
predichos, como que estos \'ultimos tengan la misma dependencia
funcional con un exponente 3/2 en temperaturas reducidas (ver sec.
\ref{ssec:depT}, y ec. \ref{ec:JT32}). En la figura
\ref{fig:cal:JxTcompara} se muestran las predicciones con
sucesivamente menos correcciones, y se aprecia que todas ellas
muestran igual tendencia: las correcciones simplemente afinan el
valor absoluto del resultado. Esta dependencia la muestran incluso
los resultados que se obtienen del modelo m\'as sencillo, carente
de correcciones, empleando la ecuaci\'on \ref{ec:Tfsimple}, como
queda reflejado en esta figura (rombos) y como ya hab\'{\i}amos
visto previamente en la figura \ref{fig:cal:JxTsimple}.

Si este resultado es importante es porque, hasta ahora, los
modelos que se hab\'{\i}an propuesto para describir el
comportamiento t\'ermico de este tipo de muestras
\cite{Skocpol74,Gurevich87} no eran capaces de describir la
tendencia experimental y universal de \Jx\ con la temperatura
($\Jx(\eps) \propto \eps^{3/2}$, siendo $\eps \simeq 1-T/\Tc$).
Por este motivo, frecuentemente se descart\'o el
\Index{calentamiento} como causa directa de la transici\'on en
este r\'egimen de altas densidades de corriente, incluso en
trabajos previos de nuestro laboratorio
\cite{Curras00b,Curras01,Xiao96}. Aqu\'{\i} hemos puesto de
manifiesto que nuestro modelo t\'ermico s\'{\i} puede explicar
esta dependencia respecto de la temperatura.

% ------------------------------------ %
\subsection{Correlaci\'on entre potencia disipada y aumento de temperatura}
\label{ssec:cal:DTW}
% ------------------------------------ %

Para finalizar este apartado vamos a hacer una \'ultima comparaci\'on
con los datos experimentales que es m\'as bien una verificaci\'on de
la coherencia del modelo: queremos comprobar si nuestra
descripci\'on t\'ermica del sistema es correcta, y que no estamos
introduciendo un sobrecalentamiento artificioso y generando con
ello una avalancha t\'ermica irreal.

Ya hemos visto que conocer la temperatura real del micropuente es
dif\'{\i}cil experimentalmente, al menos en el estado
superconductor. Pero en la figura \ref{fig:resul:DTW} de la
p\'agina \pageref{fig:resul:DTW} mostr\'abamos los resultados de
un experimento en que correlacion\'abamos la potencia disipada
$W=E \times J$ con el incremento de temperatura \DT\ de la
muestra, en la regi\'on normal por encima de \Tc. Recordemos que
este experimento se hac\'{\i}a necesariamente a $T>\Tc$ debido a
la imposibilidad de medir directamente la temperatura de la
muestra. En esta regi\'on en que la muestra est\'a en estado
normal pod\'{\i}amos atribuir cualquier diferencia del voltaje
experimental, seg\'un aplic\'asemos una corriente que provoque
disipaci\'on elevada u otra en que la disipaci\'on sea
despreciable, exclusivamente a efectos t\'ermicos, pues en
principio no hay fen\'omenos \index{intr\'{\i}nseco}
intr\'{\i}nsecos que dependan notablemente de la corriente
aplicada. De este modo estim\'abamos el aumento de temperatura
\DT. Esta correlaci\'on de \DT\ con la potencia disipada $W$, a
pesar de haberse medido por encima de \Tc, es en principio
extrapolable a la regi\'on superconductora, de la que s\'olo
distamos unos pocos grados: aunque la superconductividad modifique
algo los par\'ametros t\'ermicos del filme, los par\'ametros del
substrato, que son los m\'as relevantes, no se alteran por ello.
As\'{\i}podremos comprobar la fiabilidad de nuestro modelo, viendo
qu\'e correlaci\'on te\'orica entre \DT\ y $W$ se extrae de \'el,
y compar\'andola con la experimental.

\figCalsDTW

En la figura \ref{fig:cal:sDTW} vemos, en c\'{\i}rculos negros,
las predicciones que el modelo propuesto aqu\'{\i} ---calculando
la temperatura del filme con la ecuaci\'on \ref{ec:Tfcorr}, e
introduciendo realimentaci\'on en la ecuaci\'on \ref{ec:DTs}
mediante la funci\'on \ref{ec:EJT}--- hace para dicha
correlaci\'on. Son resultados para pulsos de 1~ms, a
$\Tb=76.2\und{K}$, aplicando corrientes entre 2.5 y
$6.4\Exund{6}{\Acms}$. Para facilitar la comparaci\'on con los
resultados experimentales mostramos la misma l\'{\i}nea recta que
se inclu\'{\i}a en aquella figura, de ecuaci\'on $\DT=0.047 W$
(con la potencia $W$ en $\und{MW/cm^3}$), y que recoge la
tendencia de los datos.

Los resultados son notablemente similares, aunque la tendencia de
que se extrae de los c\'alculos parece indicar que podemos estar
subestimando te\'oricamente el \Index{calentamiento} a corrientes
altas. Esto coincidir\'{\i}a con que nuestras predicciones,
basadas en un modelo muy sencillo, no dan exactamente los valores
esperados de \Jx, ni producen transiciones tan abruptas como las
experimentales. Pero tampoco hay que tener demasiado en cuenta
estos puntos tan cercanos al salto. Como ya explicamos al hacer
este experimento, este tipo de representaci\'on s\'olo tiene algo
de sentido cuando el voltaje (y por tanto la potencia, y la
temperatura) est\'a determinado sin ambig\"uedad. No es el caso de
los puntos m\'as cercanos al salto, ya que ah\'{\i} la temperatura
presenta tendencias de crecimiento que no permiten establecer un
valor concreto de \DT. Esta ambig\"uedad se traslada
exponencialmente a $E(J,T)$, y por \'el a $W$, de manera que no es
extra\~no que los puntos de la gr\'afica con potencias m\'as altas
sean menos de fiar.

N\'otese una vez m\'as que no es s\'olo la potencia lo que causa
una \Index{avalancha} t\'ermica: en el experimento realizado en el
estado normal pod\'{\i}amos llegar a aplicar potencias de m\'as de
$75\Exund{6}{W/cm^3}$ sin que esto supusiese un aumento de
temperatura de m\'as de 4~K, y en todo caso obteniendo estados
casi estacionarios. Sin embargo, cuando esta muestra se encuentra
en estado superconductor, se produce una avalancha al llegar a esa
misma potencia (el \'ultimo c\'{\i}rculo negro que se muestra en
la figura \ref{fig:cal:sDTW} es muy pr\'oximo a \Jx). No es que la
temperatura haya aumentando en los instantes previos hasta cerca
de \Tc\ (lo que supondr\'{\i}a un aumento \DT\ de m\'as de 10~K),
sino que en ese punto la variaci\'on de la potencia disipada con
la temperatura se hace tan notable que se tiene una fuerte
\Index{realimentaci\'on} positiva en lapsos de tiempo breves. Esto
lo entenderemos mejor cuando estudiemos, en el apartado
\ref{sec:Cal:condAva}, la condici\'on de avalancha que se extrae
de nuestro modelo t\'ermico.

\figCalsDTWdos

Por comparaci\'on, en la figura \ref{fig:cal:sDTW2} mostramos
nuevamente unos resultados simulados para las mismas corrientes
que antes, pero acortando la duraci\'on del \Index{pulso} en un
factor diez. La tendencia es la misma aunque, por supuesto, los
puntos se sit\'uen en coordenadas diferentes (hay menos disipaci\'on,
y menor aumento de temperatura).

Se concluye de estos resultados que el modelo t\'ermico que
proponemos no debe de estar generando una transici\'on en \Jx\ de
manera artificiosa, por una sobreestimaci\'on del calentamiento
real. Al contrario, el modelo da buena cuenta de los resultados
experimentales que acotan este calentamiento.

%%%%%%%%%%%%%%%%%%%%%%%%%%%%%%%%%%%%%%%
\section{Condici\'on de avalancha t\'ermica}
\label{sec:Cal:condAva}
%%%%%%%%%%%%%%%%%%%%%%%%%%%%%%%%%%%%%%%

Analizaremos ahora qu\'e condiciones hacen que la evoluci\'on de la
temperatura tome un camino u otro, el de la estabilidad o el de la
avalancha.

Ya hemos mencionado que dentro de este modelo t\'ermico que
proponemos no hay estrictamente una evacuaci\'on desde la muestra a
un medio externo: lo que estamos calculando es una distribuci\'on no
uniforme de temperatura en un medio continuo, siendo este la suma
del micropuente m\'as el substrato. Desde este punto de vista no es
f\'acil hacer consideraciones fij\'andonos en las inestabilidades que
se producen cuando el calor generado $P$ es mayor que el evacuado
\Qp, porque \emph{en este modelo} no hay dos medios que
intercambien calor. Hacer estas consideraciones entra\~na notables
dificultades. Pero se pueden hacer otras igualmente interesantes.

Tomemos, para simplificar, la temperatura del micropuente que
mostr\'abamos en la ecuaci\'on \ref{ec:Tf}:
\[\Tf(t)=\Tb+\DTf(t)=\Tb+\alp\;\DTs(t). \]
Supongamos adem\'as que \alp\ es independiente del tiempo, para no
complicar demasiado los resultados. La temperatura del ba\~no \Tb\
tambi\'en es una constante. La derivada de la temperatura del
micropuente resulta por tanto
\be
\diff{\Tf(t)}{t} = \frac{\alp}{{\cp \;\vs(t)}}\;\left( {P(t) -
\frac{{\vs'(t)}}{{\vs(t)}}\;Q(t)} \right). \label{ec:dTfdt} \ee
Recu\'erdese que \[Q(t)=\int_0^t P(\tau) \, \ud \tau.\] La
condici\'on estricta de estabilidad es $\indiff{\Tf}{t} = 0$. En
realidad esta condici\'on no se cumple nunca en nuestro modelo (ni
en el experimento), salvo como caso l\'{\i}mite tendiendo a
tiempos muy largos. Podemos verlo incluso en el l\'{\i}mite de
potencia constante, en cuyo caso esta derivada tendr\'{\i}a la
forma siguiente:
\be
\left[\diff{\Tf(t)}{t}\right]_{P = cte} = \frac{\alp}{{\cp
\;\vs(t)}}\;P\;\left( {1 - \frac{{\vs'(t)}}{{\vs(t)}}\;t} \right).
\label{ec:dTfdtPcte} \ee El t\'ermino dentro del par\'entesis en
el lado derecho no se anula nunca, salvo en el l\'{\i}mite $t
\tiende \infty$. Puede verse su representaci\'on en la figura
\ref{fig:cal:dvsvs}. Esta misma dependencia temporal se mostr\'o
ya en la figura \ref{fig:cal:dDTsdt} (p.
\pageref{fig:cal:dDTsdt}). Si la derivada de la temperatura no se
anula a potencia constante, menos a\'un si la potencia crece con
el tiempo.

\figCaldvsvs \figCalsDTtDos

Expresado esto en t\'erminos habituales de disipaci\'on y
evacuaci\'on desde la perspectiva del filme, tenemos que siempre
$P$ es mayor que \Qp, por lo que la temperatura siempre aumenta.
La diferencia entre una medida sin avalancha y otra con avalancha
no est\'a por tanto determinada porque la temperatura se
estabilice o no. Lo relevante es en realidad la tendencia con que
crece la temperatura. En la figura \ref{fig:cal:sDTtDos} mostramos
un resultado t\'{\i}pico, ya conocido: la temperatura siempre
crece un poco, pero puede tender a una \emph{aparente}
estabilizaci\'on cuando crece lentamente, o desbocarse y crecer
abruptamente. Se puede apreciar que la \Index{avalancha} se
distingue de la estabilidad por la forma en que var\'{\i}a la
temperatura en el tiempo. En la evoluci\'on que conduce a la
avalancha, llega un momento en que la derivada de \DT\ en el
tiempo ---la recta tangente a la curva en cuesti\'on--- crece,
mientras que en el caso estable la derivada $\indiff{\DT}{t}$ es
\emph{siempre} cada vez menor.

Formalmente hablando, en el caso que desemboca en avalancha, la
derivada segunda en el tiempo de la temperatura del micropuente se
hace positiva. Esto nos lleva a establecer una \Index{condici\'on de
avalancha} con la siguiente expresi\'on:
\be
\diffn{\Tf}{t}{2}>0. \label{ec:condDT2} \ee Desarrollando esta
derivada segunda a partir de la ecuaci\'on \ref{ec:dTfdt}, tendremos
avalancha si
\be
\diffn{\Tf}{t}{2} = \frac{\alp}{{\cp
\;\vs(t)}}\;\left(\diff{P(t)}{t} - 2\;P(t)\;f_1(t) + Q(t)\;f_2(t)
\right) > 0. \label{ec:condDT2Compl}
\ee Las funciones $f_1(t)$ y $f_2(t)$ son
dependientes del tiempo y de los par\'ametros de la muestra y el
substrato, pero independientes de sus temperaturas y de la
potencia disipada. Son siempre positivas, y consisten en
combinaciones de la funci\'on $\vs(t)$ y sus derivadas primera y
segunda. Sus expresiones concretas, aunque no son demasiado
relevantes para los razonamientos que siguen, se muestran a
continuaci\'on:
\be f_1(t)=\frac{\vs'(t)}{\vs(t)}, \label{ec:f1}\ee %= \frac{D ( 2\;d +8\;\sqrt{D\;t} +a)}{\sqrt{D\;t}\;(d+2\;\sqrt{D\;t})\;(4\;\sqrt{D\;t}+a)},
\be f_2(t)=2\;\frac{\vs'(t)^2}{\vs(t)^2}-\frac{\vs''(t)}{\vs(t)},
\label{ec:f2} \ee donde \vs(t) es el que se mostr\'o en la ecuaci\'on
\ref{ec:vst},
\[ \vs(t) = l\; (d + 2\;\sqrt{D\; t})\; (a + 4\;\sqrt{D\; t}),\]
y
\be \vs'(t)=\frac{D\; l\; (a + 2d + 8\;\sqrt{D\; t})}{\sqrt{D\;t}}, \ee
\be \vs''(t)=-\frac{D^2\; l\; (a + 2 d)}{2 \; (D\; t)^{3/2}}. \ee

Los l\'{\i}mites de $f_1$ y de $f_2$ son $\infty$ para $t \tiende
0$ y 0 para $t \tiende \infty$. A tiempos cortos $f_2$ es mayor
que $f_1$, pero decrece m\'as r\'apidamente, de modo que a tiempos
largos esta relaci\'on se invierte.

Siguiendo con nuestro ejemplo elemental de potencia constante,
tenemos que si $\indiff{P}{t}=0$ la \Index{condici\'on de
avalancha} queda reducida a \be t \; f_2(t)- 2 \; f_1(t) > 0,
\label{ec:condAvaPcte}\ee que nunca se satisface pues es siempre
negativo. En la figura \ref{fig:cal:f1f2} se representa el
t\'ermino de la izquierda de la desigualdad, en escala
logar\'{\i}tmica para apreciar mejor su dependencia. En este caso,
como ya venimos viendo por diversos resultados, la temperatura,
aunque siempre aumenta, lo hace cada vez m\'as lentamente
tendiendo a una estabilizaci\'on a tiempos largos, y no se produce
avalancha t\'ermica.

\figCalfunodos

En realidad, el sumando que incluye a la funci\'on $f_2(t)$ en la
ecuaci\'on \ref{ec:condDT2Compl} es pr\'acticamente irrelevante en
los rangos que manejamos: a tiempos muy cortos, $Q$ (la integral
de $P$ en el tiempo) es peque\~no, y a tiempos largos es la propia
funci\'on $f_2(t)$ la que se vuelve despreciable. Eliminando pues
ese sumando, y prescindiendo de t\'erminos que no influyen en el
signo, la condici\'on de avalancha queda \be\diff{P(t)}{t} - 2 \;
P(t) \; f_1(t) > 0. \label{ec:condAva} \ee Se puede ver
as\'{\i}que no s\'olo el valor concreto de la potencia $P$ en un
momento dado determina una avalancha: para causar este fen\'omeno
es tambi\'en relevante su derivada $\indiff{P}{t}$. Para
simplificar imaginemos que aplicamos un \Index{pulso} cuadrado de
corriente $J$ constante, de manera que la variaci\'on de $P$ se
deba exclusivamente a c\'omo var\'{\i}e la temperatura \Tf\ de la
muestra. De este modo, $P(t) = P(\,\Tf(t)\,)$, y la expresi\'on se
convierte en \be \diff{P(\Tf)}{\Tf}\;\diff{\Tf}{t} -
2\;P(\Tf)\;f_1(t) > 0. \label{ec:condAvaSimple} \ee A este
\'ultimo resultado es al que prestaremos m\'as atenci\'on, viendo
qu\'e conclusiones se pueden extraer de la condici\'on que impone.

\comenta{Subrayemos una vez m\'as que la resoluci\'on de este problema
en general, y el estudio de esta \'ultima expresi\'on en particular,
debe hacerse con m\'etodos num\'ericos iterativos: en todo momento la
potencia disipada y la temperatura de la muestra est\'an ligadas, de
modo que incrementos de la temperatura causan aumentos de la
disipaci\'on, que a su vez vuelve a aumentar la temperatura, y as\'{\i}sucesivamente.}  % ojo, que no es cierto para la condici\'on de avalancha

\comenta{Estas consideraciones que siguen hay que revisarlas con
calma para darles un correcto sentido f\'{\i}sico. De momento no
lo entiendo muy bien.}

En la ecuaci\'on \ref{ec:condAvaSimple} se pone de manifiesto lo
crucial que resulta no s\'olo el valor de $P$ para tener una
avalancha, sino tambi\'en el valor de su derivada con la
temperatura. Entre otras cosas, como ya hemos venido observando,
si $P$ fuese constante nunca se cumplir\'{\i}a la desigualdad
\ref{ec:condAvaSimple} y no habr\'{\i}a avalancha
($\indiff{P}{\Tf}$ valdr\'{\i}a cero y la expresi\'on ser\'{\i}a
siempre negativa). Es necesario que la potencia var\'{\i}e. Esta
condici\'on explica ciertas paradojas aparentes que se observan en
los experimentos: si, por ejemplo por la aplicaci\'on de un campo
magn\'etico, modificamos la forma de la curva \EJ, podemos llegar
a una situaci\'on en que para iguales valores de $P$ tuvi\'esemos
otros menores de $\indiff{P}{\Tf}$. En ese caso podr\'{\i}an no
observarse avalanchas con potencias disipadas $P$ que antes, con
campo nulo, eran suficientes para inducir la transici\'on. Si no
se hacen las consideraciones que aqu\'{\i} hemos mostrado dentro
de nuestro modelo t\'ermico, esta observaci\'on experimental suele
llevar a que se descarte el \Index{calentamiento} como causa del
salto \cite{Curras00b,Curras01,Xiao99} sin que sea necesariamente
correcto.

Incluyendo en la \'ultima expresi\'on \ref{ec:condAvaSimple} la
derivada de la temperatura calculada en la ecuaci\'on
\ref{ec:dTfdt}, y haciendo para simplificar $Q \tiende 0$
(condici\'on de pulsos breves), tenemos la \Index{condici\'on de
avalancha} expresada como \be
\diff{P(\Tf)}{\Tf}\;\frac{\alp}{\cp\;\vs(t)}\;P(\Tf) -
2\;P(\Tf)\;f_1(t)  > 0. \label{ec:condAvaSencillo} \ee Emplearemos
esta expresi\'on para estimar qu\'e valores de los par\'ametros
que caracterizan nuestro sistema conducen irremediablemente a una
avalancha. De momento podemos interpretar la informaci\'on que
aporta esta ecuaci\'on: dada una funcionalidad de $P(\Tf)$ ---que
en general depender\'a tambi\'en del valor concreto de $J$ y $B$
aplicados, no s\'olo de la temperatura---, si pasado un tiempo $t$
desde que empezamos a disipar calor se consigue llevar la muestra
hasta una temperatura \Tf\ de manera que el signo de la
expresi\'on sea positivo, tendremos inevitablemente una avalancha
pocos instantes despu\'es, salvo que paremos la medida y
eliminemos la corriente. Si el signo es negativo, sin embargo,
nada nos garantiza que momentos m\'as tarde no pueda cambiar a
positivo y desembocar tambi\'en en una avalancha. La ecuaci\'on
nos dice por tanto si un estado concreto es \emph{suficiente} para
generar una avalancha, pero no dice nada de que un estado estable
vaya \emph{necesariamente} a mantenerse siempre as\'{\i}. En
cualquier caso podemos hacer el estudio de esta expresi\'on a
tiempos muy largos, con $t \tiende \infty$, de manera que cubramos
los rangos de tiempo t\'{\i}picos experimentales, y obtener de
este modo un mapa de las regiones cr\'{\i}ticas: si para unos
par\'ametros dados la condici\'on \ref{ec:condAvaSencillo} no se
cumple dentro de los rangos de tiempo del experimento, se tiene
que el sistema no conduce a una avalancha.

En cualquier caso insistimos en el resultado m\'as relevante que
nos dan estas ecuaciones de \Index{condici\'on de avalancha}:
\emph{no s\'olo el valor de la potencia disipada} determina que un
cierto estado conduzca a una avalancha, \emph{tambi\'en la
derivada de la potencia} $\indiff{P}{T}$ est\'a implicada. Puede
haber estados que, a igualdad de potencia disipada, tengan la
derivada diferente, de modo que unos sean estables y otros
cr\'{\i}ticos.

La ecuaci\'on \ref{ec:condAvaSencillo}, como las anteriores de
este apartado, est\'an expresadas para una $J$ y un $B$ constantes
dados, y por eso incluyen s\'olo dependencias respecto de \Tf.
Pero en funci\'on del valor de $J$, \Tf\ y $B$ concretos tendremos
curvas \EJ\ diferentes, y de ah\'{\i} distintos valores de $P$ y
de su derivada, y por lo tanto tambi\'en distintos signos para la
expresi\'on \ref{ec:condAvaSencillo} que nos da el criterio de
avalancha. De esta manera, dicha ecuaci\'on divide el espacio de
estados tetradimensional $J$--\Tf--$B$--$t$ en regiones de dos
tipos, una en la que la expresi\'on tiene signo negativo (estados
casi estacionarios) y otra en la que tiene signo positivo (estados
cr\'{\i}ticos que causan avalancha).

\figCalManta

Veamos en primer lugar qu\'e ocurre a $\mH = 0$. Limitando
as\'{\i}el espacio de estados a tres dimensiones ($J$--\Tf--$t$),
vemos en la figura \ref{fig:cal:manta} una representaci\'on de la
frontera entre la regi\'on cr\'{\i}tica ---que cumple la
condici\'on de la ecuaci\'on \ref{ec:condAvaSencillo}--- y la
estable, que no la cumple
---pues el signo es negativo---. Por encima de esta frontera
tenemos la regi\'on inestable: ah\'{\i} siempre se producir\'a una
avalancha t\'ermica que hace transitar la muestra al estado
normal. Por debajo de la frontera siempre se alcanzan estados casi
estacionarios, que tienden a una estabilizaci\'on de la
temperatura.

\figCaldfLtJ

Veamos algunos cortes significativos de este espacio de estados.
Para $\Tf= 78\und{K}$, por ejemplo, tenemos el diagrama de estados
$t$--$J$ (tiempo del pulso--densidad de corriente aplicada) que se
muestra en la figura \ref{fig:cal:dfLtJ}. La regi\'on de sombreado
oscuro indica un signo negativo de la ecuaci\'on
\ref{ec:condAvaSencillo} (lo que implica que no se genera una
avalancha dentro del tiempo $t$), y la de color blanco un signo
positivo (lo que implica un estado cr\'{\i}tico que conduce
inevitablemente a la avalancha). En el eje vertical se representa
el logaritmo decimal del tiempo, para apreciar qu\'e ocurre en
distintos \'ordenes de magnitud. La frontera entre ambas regiones
se puede interpretar como las coordenadas de \Jx\ en funci\'on de
$t$, para $\Tf= 78\und{K}$ y $\mH = 0$. La tendencia de aumento de
\Jx\ conforme decrece la duraci\'on del \Index{pulso} es
logar\'{\i}tmica en la regi\'on de pulsos cortos: hay que dividir
la duraci\'on del pulso por mil para conseguir un aumento de \Jx\
del 50\%.

N\'otese la similitud de esta dependencia con los resultados de la
figura \ref{fig:cal:tJxSTO} (p. \pageref{fig:cal:tJxSTO}), aunque
aquella fuese para una temperatura algo mayor. Recu\'erdese que se
hab\'{\i}a obtenido de manera completamente diferente, simulando
la evoluci\'on temporal de los experimentos con c\'alculos m\'as
complejos. Ahora, a pesar de las simplificaciones que hemos
incorporado para llegar a la ecuaci\'on \ref{ec:condAvaSencillo}
(de manera destacada, la aproximaci\'on de pulsos cortos para
hacer $Q \tiende 0$, que en principio fallar\'{\i}a para $t$
grande), los resultados coincide notablemente, en lo que a la
forma de la dependencia respecto de $t$ se refiere, aunque estos
\'ultimos tengan menos precisi\'on. Las reflexiones que se
hicieron all\'{\i} acerca de la dependencia de $\Jx$ respecto de
la duraci\'on de la medida son aplicables igualmente para estos
nuevos resultados.

\figCaldfJTTc \figCaldfJTTcns

En la figura \ref{fig:cal:dfJTTc} mostramos un diagrama de estados
a $t$ constante, estudiando el espacio de estados en funci\'on de la
densidad de corriente y la temperatura. \comenta{$J$--$\Tf/\Tc$}
Los resultados mostrados son para $t=0.1\und{ms}$, pero en
realidad pr\'acticamente son id\'enticos para cualquier $t>10\und{\mu
s}$. S\'olo si disminuimos $t$ por debajo de este umbral comenzamos
a apreciar diferencias, siendo esto acorde con los resultados de
la figura \ref{fig:cal:dfLtJ}. Por ejemplo, en la figura
\ref{fig:cal:dfJTTcns} se muestra el diagrama an\'alogo para
$t=1\und{ns}$, que ya es notablemente diferente al anterior: la
regi\'on de estabilidad (en oscuro) aumenta, dando para las mismas
temperaturas una regi\'on mayor de estabilidad.

\figCalsdfJxTTc

La frontera entre la regi\'on de estabilidad (oscura) y la
regi\'on cr\'{\i}tica (en blanco) de la figura
\ref{fig:cal:dfJTTc} nos da de nuevo los valores de \Jx, en este
caso para tiempos de medida que en general sean m\'as largos de
$10\und{\mu s}$. Esta frontera se muestra como una l\'{\i}nea
continua en la figura \ref{fig:cal:sdfJxTTc}, en la
representaci\'on logar\'{\i}tmica habitual. La l\'{\i}nea inferior
de esa figura, en trazo discontinuo, es la que ajusta a los
c\'alculos con nuestro modelo en apartados previos, con duraci\'on
de pulso de 1~ms (los rect\'angulos negros en la figura
\ref{fig:cal:JxTexpcalc} de la p\'agina
\pageref{fig:cal:JxTexpcalc}). La similitud entre las dos curvas
es notable. Vemos que los resultados calculados empleando el
criterio de avalancha tambi\'en reproducen la misma tendencia en
temperatura reducida que se ve en los experimentos, y que ya
reprodujimos con nuestros c\'alculos detallados. Podr\'{\i}a
pensarse que las diferencias en valores absolutos de ambas curvas
se deben a las simplificaciones que hemos ido adoptando para
llegar a nuestra \Index{condici\'on de avalancha}, pero en
realidad lo que ocurre es que representan cosas distintas:

La curva de trazo discontinuo est\'a calculada empleando
\Index{realimentaci\'on} y temperatura variable partiendo de una
temperatura inicial \Tb, la del ba\~no. En esta representaci\'on, \Tf\
es en realidad, para esta curva, la temperatura inicial de la
muestra \Tb, no la que tiene en el momento de la avalancha. Esto
se debe a que es una simulaci\'on de los experimentos: como en estos
no se puede medir la temperatura real de la muestra, las
representaciones frente a la temperatura emplean inevitablemente
la temperatura inicial, la del ba\~no.

La otra curva, en trazo continuo, representa sin embargo los
valores de \Jx\ en funci\'on de la temperatura \emph{real} de la
muestra en el instante $t$ que hemos considerado en el c\'alculo
del diagrama de estados. Esta curva da los valores m\'{\i}nimos de
corriente aplicada que hacen que la muestra vaya inevitablemente
hacia una avalancha t\'ermica, siempre que la muestra alcance la
temperatura \Tf\ en alg\'un instante $t>10\und{\mu s}$. Con
corrientes que est\'en por encima de esa curva siempre se produce
una avalancha. Si est\'an por debajo, el sistema se mantiene
estable, en un estado casi estacionario.

Por lo tanto, podemos interpretar las diferencias entre estas dos
curvas, calculadas con dos m\'etodos distintos, como desplazamientos
horizontales, en temperaturas, de una respecto de la otra. La
curva de trazo discontinuo tiene que desplazarse hacia
temperaturas m\'as altas para coincidir con la de trazo continuo,
pero este desplazamiento es distinto en cada punto. Por ejemplo,
en $\Tf=76.2\und{K}$ ($\Tf/\Tc=0.85$) hay que correrla +2.7 K para
hacerla coincidir con la otra ($\DTf/\Tc=0.03$). En
$\Tf=82.0\und{K}$ ($\Tf/\Tc=0.91$), basta con correrla +1.8 K, es
decir, $\DTf/\Tc=0.02$.

Estos desplazamientos en temperaturas, seg\'un lo antedicho, pueden
interpretarse de este modo: la medida que parte inicialmente de
una temperatura de 76.2~K, y que se realiza aplicando justamente
el valor \Jx\ que da la curva de trazo discontinuo para esta
temperatura, lleva a la muestra a una temperatura 2.7~K m\'as
elevada dentro del tiempo que dura la medida, e induce instantes
despu\'es una avalancha. Si partimos de una temperatura de 82.0~K y
aplicamos el correspondiente valor de \Jx, el calentamiento ser\'a
de 1.8~K en el momento en que se satisface la \Index{condici\'on de
avalancha}, y se tendr\'a igualmente una cascada t\'ermica.

N\'otese que estos valores de sobrecalentamiento antes del salto
son aproximadamente los mismos que encontr\'abamos en las curvas
de la evoluci\'on de la temperatura en el tiempo calculadas en las
figuras \ref{fig:cal:sDTtEtJ76} y \ref{fig:cal:sDTtWt82}
(p\'aginas \pageref{fig:cal:sDTtEtJ76} y
\pageref{fig:cal:sDTtWt82}), para temperaturas iniciales de 76.2 y
82.0~K, aplicando distintas corrientes. En aquel caso ve\'{\i}amos
que hab\'{\i}a un aumento de temperatura de unos 2.5~K en un caso,
y de 1.6~K en el otro, antes de que la tendencia de aumento de
temperatura se desbocase en cascada. Esta coincidencia constituye
un \'exito notable de nuestra \Index{condici\'on de avalancha} de
la ecuaci\'on \ref{ec:condAvaSencillo}, derivada de criterios de
estabilizaci\'on de la temperatura.

\figCaldfLtJB \figCaldfJTTcB

Para terminar, y como \'ultima corroboraci\'on de la validez de
nuestro modelo y de la \Index{condici\'on de avalancha} que de
\'el se deriva, veamos qu\'e ocurre al aplicar campo magn\'etico.
Subir el campo magn\'etico a $\mH = 1\und{T}$ implica, como hemos
visto, variar la forma de las curvas \EJ, y por lo tanto el valor
de $P$ y de su derivada a dem\'as par\'ametros iguales.
Introduciendo en nuestros c\'alculos las nuevas curvas \EJ\
(ecuaci\'on \ref{ec:EJTB1}) tenemos un nuevo diagrama de estados
en el espacio $J$--$t$, a temperatura constante de 78 K, que
mostramos en la figura \ref{fig:cal:dfLtJB1}. Comp\'arese con la
figura \ref{fig:cal:dfLtJ}: la regi\'on de estabilidad disminuye,
y el valor de \Jx\ (determinado por la frontera entre regiones) se
hace menor para experimentos en igualdad de condiciones. El
correspondiente diagrama $J$--$\Tf/\Tc$, mostrado en la figura
\ref{fig:cal:dfJTTcB1}, tambi\'en ve aumentada su regi\'on
cr\'{\i}tica.

\figCalsdfJxTTcB

Igual que para el caso $B=0$, comparamos en la figura
\ref{fig:cal:sdfJxTTcB1} la curva de \Jx\ en funci\'on de la
temperatura que se extrae de la frontera entre ambas regiones, con
la que proviene del experimento. En este caso la curva inferior,
de trazo discontinuo, no es una simulaci\'on, sino la
experimental, para ahorrarnos los c\'alculos. Mostramos
as\'{\i}m\'as destacadamente no s\'olo las distintas conclusiones
del modelo, sino tambi\'en su similitud con el experimento. En
todo caso recordemos que la simulaci\'on detallada de $\Jx(\Tf)$
de apartados previos coincide notablemente bien con el
experimento. Lo que se quiere poner de manifiesto aqu\'{\i} es que
ahora ambas curvas est\'an desplazadas horizontalmente una de la
otra mucho m\'as que antes, con campo aplicado nulo. Concretamente
en 76.2~K, por comparar con los resultados previos, el
desplazamiento necesario para hacerlas coincidir es de +6.7~K  (un
\DTf/\Tc\ de unos 0.07), de nuevo un resultado comparable al que
obten\'{\i}amos como sobrecalentamiento antes de la avalancha en
la evoluci\'on de la temperatura para $\mH = 1\und{T}$, en la
figura \ref{fig:cal:sDTtB1} (p\'agina \pageref{fig:cal:sDTtB1}).
Ya mencion\'abamos entonces que estos sobrecalentamientos que se
deducen de nuestro modelo sencillo no son, probablemente, tan
elevados en la realidad: lo que nos interesa resaltar no son tanto
sus valores absolutos, como la tendencia de variaci\'on que
muestran con el \index{campo magn\'etico} campo aplicado y la
temperatura inicial. Concretamente, en estos c\'alculos, como en
aquellos, vemos que las avalanchas con campo magn\'etico aplicado
se producen tras un mayor sobrecalentamiento, y con una potencia
disipada mayor, que en el caso sin \index{campo magn\'etico} campo
aplicado a la misma temperatura. Recordemos una vez m\'as, aun a
riesgo de ser demasiado insistentes, que este comportamiento
llev\'o a pensar que la transici\'on no pod\'{\i}a estar
directamente relacionada con fen\'omenos t\'ermicos, pues de ser
as\'{\i}se esperaba que la potencia disipada necesaria para causar
una transici\'on disminuyese al aumentar $H$ o, en el mejor de los
casos, que fuese independiente de \'el \cite{Xiao99,Curras01}. Sin
embargo todo lo que hemos venido calculado hasta ahora, basado
exclusivamente en calentamientos de las muestras, predice
exactamente esa misma tendencia, entre otros contundentes
resultados que apoyan la validez el modelo. No podemos, por lo
tanto, descartar que la causa real de la transici\'on est\'e
vinculada a mecanismos t\'ermicos. Al contrario: parece m\'as bien
que dichos mecanismos est\'an \'{\i}ntimamente relacionados con la
transici\'on y, si no son ellos exclusivamente los que la inducen,
al menos s\'{\i} que est\'an entremezclados con otros posibles
\index{intr\'{\i}nseco} intr\'{\i}nsecos.
    % cap 7
\clearemptydoublepage

%%%%%%%%%%%%%%%%%%%%%%%%%%%%%%%%%%%%%%%%%%%%%%%%%%%%%%%%%%%%%%%%%%
\chapter{Resumen y conclusiones de la parte I}
%\chaptermark{Conclusiones}
%\addtocontents{toc}{\protect\vspace{0.2cm}}
\label{cap:concl}
%%%%%%%%%%%%%%%%%%%%%%%%%%%%%%%%%%%%%%%%%%%%%%%%%%%%%%%%%%%%%%%%%%

A continuaci\'on hacemos un resumen de los principales resultados
de esta primera parte de la memoria, dedicada al estudio de la
transici\'on al estado normal en el r\'egimen de altas densidades
de corriente. Recogemos tambi\'en las principales conclusiones que
se extraen de dichos resultados, as\'{\i} como los nuevos
interrogantes planteados y las posibles l\'{\i}neas de
continuaci\'on de estos trabajos.

%%%%%%%%%%%%%%%%%%%%%%%%%%%%%%%%%%%%%%%%
\section{Resumen}
%%%%%%%%%%%%%%%%%%%%%%%%%%%%%%%%%%%%%%%%

%%%%%%%%%%%%%%%%%%%%%%%%%%%%%%%%%%%%%%%%
\subsection{Del experimento}
%%%%%%%%%%%%%%%%%%%%%%%%%%%%%%%%%%%%%%%%

Empleando un banco de pulverizaci\'on cat\'odica, se crecieron
numerosas pel\'{\i}culas de \YBCOf\ de espesor alrededor de
 $150\und{nm}$ sobre substratos cristalinos de \STOf\ de $1\und{cm^2}$ de
superficie. Sobre estas pel\'{\i}culas se grabaron, mediante
fotolitografiado, micropuentes de dimensiones t\'{\i}picas
$l=50\und{\mu m}$ y $a=10\und{\mu m}$. Este tipo de muestras
presentan, frente a otros posibles, dos ventajas fundamentales: su
peque\~na secci\'on permite aplicar elevadas densidades de
corriente con corrientes absolutas bajas, y presentan una muy
buena evacuaci\'on del calor, debido a la elevada relaci\'on
superficie/volumen que tienen, y a que est\'an en muy buen
contacto t\'ermico con el substrato cristalino, mucho m\'as
masivo.

Las pel\'{\i}culas y los micropuentes resultantes se
caracterizaron mediante difractometr\'{\i}a de rayos-x (XRD),
perfilometr\'{\i}a, microscop\'{\i}a de fuerza at\'omica (AFM), y
medidas de resistividad en funci\'on de la temperatura y el campo
magn\'etico. Las muestras de mejor calidad ---con temperatura
cr\'{\i}tica \Tc\ m\'as alta, menor resistividad en el estado
normal a $T>\Tc$ y con transici\'on resistiva m\'as estrecha--- se
reservaron para las medidas m\'as precisas de curvas
caracter\'{\i}sticas y de par\'ametros cr\'{\i}ticos, mientras que
las de peor calidad se emplearon para experimentos m\'as agresivos
y potencialmente destructivos, de caracterizaci\'on t\'ermica a
tiempos largos o a muy altas disipaciones. Se aplicaron
\index{contactos el\'ectricos} contactos el\'ectricos mediante
microsoldadura por \Index{ultrasonidos}, logrando resistencias de
contacto despreciables frente a las resistencias
caracter\'{\i}sticas de las muestras en estudio.

Con estos micropuentes se realizaron medidas de curvas
caracter\'{\i}sticas \EJ\ (curvas \VI\ escaladas a las dimensiones
geom\'etricas de las muestras), en funci\'on de la temperatura
---entre aproximadamente la temperatura del nitr\'ogeno l\'{\i}quido y \Tc ---
y del campo magn\'etico aplicado perpendicular a las capas
---desde campo aplicado nulo hasta 1~T---. Se emple\'o una
configuraci\'on est\'andar de cuatro hilos en l\'{\i}nea, una
fuente de corriente programable de precisi\'on, y un
nanovolt\'{\i}metro (en el r\'egimen de baja disipaci\'on, donde
los efectos t\'ermicos son despreciables). Para las medidas \VI\
hasta la corriente de salto \Ix\ se utiliz\'o una tarjeta de
adquisici\'on (DAQ) que permit\'{\i}a hacer medidas r\'apidas (en
el rango del milisegundo, con resoluci\'on de $10\und{\mu s}$) y
minimizar los efectos t\'ermicos. Las curvas caracter\'{\i}sticas
medidas son comparables a las de los mejores filmes de \YBCO\
\'optimamente dopado que se encuentran en la bibliograf\'{\i}a.
Las muestras se alojaban en un criostato refrigerado con flujo de
vapor de nitr\'ogeno l\'{\i}quido, cuya temperatura se regulaba
electr\'onicamente.

Tambi\'en se hicieron medidas de caracterizaci\'on t\'ermica de
las muestras, de las que podemos concluir que, al menos en tiempo
breves, el intercambio t\'ermico del micropuente ocurre
principalmente con el substrato, estableci\'endose una inercia
t\'ermica del sistema con tiempos caracter\'{\i}sticos que rondan
las d\'ecimas de milisegundo.

%%%%%%%%%%%%%%%%%%%%%%%%%%%%%%%%%%%%%%%%
\subsection{Del an\'alisis}
%%%%%%%%%%%%%%%%%%%%%%%%%%%%%%%%%%%%%%%%

De las diferentes curvas \EJ\ se extrajeron los valores de los
par\'ametros que las caracterizan: la corriente \Jc\ en que se
alcanza un cierto \Index{umbral} de disipaci\'on dentro de estado
mixto
---t\'{\i}picamente $10\und{\mu V/ cm}$---, la corriente \Jcpin\ en que
los \Index{v\'ortices} superan un cierto umbral de velocidad $v=E/B$
---alrededor de $10^{-3}\und{m/s}$---, y la corriente \Jx\ de
transici\'on al estado normal. Otros par\'ametros que se estudiaron
son el voltaje \Ex\ a que ocurre la transici\'on en \Jx, y la
potencia disipada $\Wx=\Ex \times \Jx$ en ese mismo punto.

Estos par\'ametros se analizaron en funci\'on de la temperatura
inicial de la muestra (la temperatura antes de inyectar corriente;
entre aproximadamente la temperatura del nitr\'ogeno l\'{\i}quido
y \Tc), y en funci\'on del campo magn\'etico aplicado
perpendicular a los filmes (con valores \mH\ entre 0 y 1~T). La
variaci\'on con la temperatura, tanto de \Jc\ como de \Jx, es la
habitual en estas muestras: ambas densidades de corriente
presentan una relaci\'on funcional con la temperatura reducida
$\eps \simeq (1-T/\Tc)$ con un exponente 3/2. Estos resultados
sugieren que quiz\'a \Jx\ se origine por mecanismos similares a
los de \Jc\
---disipaci\'on por movimiento de v\'ortices--- o a los de la
\index{corriente de desapareamiento} corriente de desapareamiento
\Jd\ ---en la que los pares de Cooper se destruyen por ser
demasiado energ\'eticos---.

La relaci\'on de \Jx\ con el campo magn\'etico se describe
notablemente bien con un modelo propuesto inicialmente para \Jc,
que tiene la forma $(1-H/H_0)^{-n}$. Por el contrario, ni \Jc\ ni
\Jcpin\ siguen de manera tan notable esta dependencia con $H$,
aunque se aproximan bastante. Tanto la corriente de salto \Jx\
como el voltaje \Ex\ parecen ajustarse bastante bien al modelo
\BS\ (BS) \cite{Bezuglyj92}, que es una extensi\'on del modelo
\LO\ (LO) \cite{Larkin76} con correcciones t\'ermicas (el modelo
LO no proporciona buenos resultados para este tipo de muestras
\cite{Doettinger94,Xiao96,Doettinger97}). Ambos modelo atribuyen
la transici\'on en \Jx\ a un cambio dr\'astico de la din\'amica de
los v\'ortices magn\'eticos en el estado mixto, debido al valor
finito del \Index{tiempo de relajaci\'on} energ\'etica de los
portadores \taue. El valor que, de unos an\'alisis preliminares
con el modelo BS, hemos extra\'{\i}do para una de nuestras
muestras, coincide de manera notable con los valores de \taue\
obtenidos de otros experimentos ($\taue \sim 1.2\Exund{-12}{s}$ en
$T=76.2\und{K}$), aunque se hacen necesarios m\'as estudios en
esta l\'{\i}nea para ser concluyentes.

Para poder aplicar el modelo BS al r\'egimen de campos bajos que
hemos estudiado hemos tenido que tener en cuenta el
\Index{autocampo} generado por las elevadas corrientes que se
inyectan en la muestra. El autocampo de la corriente se superpone
al campo aplicado externamente, de modo que este \'ultimo hace
notar sus efectos s\'olo cuando supera el valor de aquel. Estos
efectos del autocampo podr\'{\i}an ser la causa de que los modelos
que explican la corriente de salto se desv\'{\i}en de los datos
experimentales a campos magn\'eticos aplicados bajos; de ser
as\'{\i}, se podr\'{\i}an descartar otros mecanismos m\'as
complejos que se pueden encontrar en la bibliograf\'{\i}a
\cite{Chiaverini00}. De estos an\'alisis parece concluirse que la
corriente inyectada en el micropuente puede estar
distribuy\'endose de manera no uniforme, concentrada en los
laterales de la muestra, de modo que la densidad de corriente real
ser\'{\i}a superior a la aparente (calculando \'esta con
distribuci\'on de corriente uniforme). Esta posibilidad
acercar\'{\i}a de manera considerable los valores de \Jx\ a los
predichos te\'oricamente para \Jd, \index{corriente de
desapareamiento} la corriente de desapareamiento de los pares de
Cooper \cite{Kunchur93,Curras00b,Curras01}, por lo que la causa de
la transici\'on al estado normal podr\'{\i}a tener que ver con
este mecanismo.

El valor de la corriente \Jx\ depende, por debajo del rango de
varios milisegundos, del tiempo que dure la medida, de modo que
ciertos valores de la corriente aplicada no causan transici\'on si
se aplican durante poco tiempo, pero acaban induci\'endola si se
mantienen m\'as tiempo. Los experimentos en donde se realizan
medidas en rangos temporales suficientemente cortos, por debajo de
los tiempos t\'ermicos caracter\'{\i}sticos de d\'ecimas de
milisegundo, aprecian notables diferencias dependiendo de la
duraci\'on de los pulsos de corriente \cite{Jakob00}. Pulsos mucho
m\'as largos que estos son, por el contrario, en todo momento
equivalentes a corrientes continuas
\cite{Curras01,Xiao96,Doettinger94,Doettinger95}.

En nuestras muestras, la potencia disipada antes del salto \Wx\
disminuye al aumentar la temperatura inicial del ba\~no, cuanto
m\'as cerca de \Tc\ est\'e. Esta tendencia se observa tambi\'en en
las muestras de la bibliograf\'{\i}a \cite{Xiao98,Curras00b},
aunque con algunas excepciones \cite{Kamm00,Curras01}. Como
contraposici\'on, \Wx\ aumenta, siempre, de manera notable al
aplicar campos magn\'eticos externos cada vez mayores. Este
comportamiento lo presentan todas las muestras de las que tenemos
referencias \cite{Xiao98,Xiao99,Pauly00}, sin excepci\'on, y en
ocasiones se ha empleado como argumento para descartar el origen
t\'ermico de la transici\'on en \Jx\ \cite{Xiao99}.

%%%%%%%%%%%%%%%%%%%%%%%%%%%%%%%%%%%%%%%%
\subsection{Del modelo de calentamiento en avalancha}
%%%%%%%%%%%%%%%%%%%%%%%%%%%%%%%%%%%%%%%%

Los trabajos recientes en los que se descartaba el
\Index{calentamiento} como causa principal de la transici\'on
abrupta al estado normal lo hac\'{\i}an bas\'andose en argumentos
que aqu\'{\i} se han refutado. La variaci\'on experimental de \Jx\
con la temperatura \cite{Curras01,Xiao96}, por ejemplo, se predice
con el modelo sencillo de calentamiento que hemos propuesto en el
cap\'{\i}tulo \ref{cap:cal}. Con la \Index{condici\'on de
avalancha} que de \'el se deriva, y que hemos deducido en el
cap\'{\i}tulo \ref{cap:calpr}, tambi\'en se explica la aparente
paradoja de que la potencia disipada en el instante previo al
salto \Px\ sea m\'as grande cuanto mayor es el campo magn\'etico
aplicado ---lo que podr\'{\i}a quiz\'as extenderse a las muestras
en las que la potencia tambi\'en aumenta con la temperatura del
ba\~no---. Lo relevante para producir una transici\'on debida a
una \Index{avalancha} t\'ermica, como se ha puesto de manifiesto,
no es s\'olo el valor concreto de la potencia disipada, sino
tambi\'en c\'omo var\'{\i}a \'este con la temperatura. La
formalizaci\'on de esta idea, a corriente y campo aplicado
constantes, involucra no s\'olo al valor de $P(T)$, sino tambi\'en
al de su derivada con la temperatura $\indiff{P}{T}$.

La ausencia de \Index{hist\'eresis} en las curvas
caracter\'{\i}sticas \EJ\ \cite{Curras01,Xiao96}, o la invariancia
de los experimentos con la longitud del pulso de corriente
\cite{Doettinger94,Doettinger95} que ven algunos autores se
justifica porque los tiempos caracter\'{\i}sticos del sistema son
muy breves, rondando las d\'ecimas de milisegundo. Cualquier
experimento que trabaje con pulsos de corriente de longitud mayor
ser\'a comparable a uno que use corriente continua. Confirmando
esta idea, ciertos autores han visto experimentalmente que la
transici\'on se suaviza, y deja de ser abrupta, (con la
consiguiente aparici\'on de efectos de \Index{hist\'eresis})
cuando se emplean pulsos de decenas de microsegundos, aun en
muestras depositadas sobre substratos mejores conductores del
calor \cite{Jakob00}.

El modelo propuesto aqu\'{\i}, basado en supuestos muy razonables,
predice con notable bondad los principales los resultados
experimentales observados en nuestras muestras. Sucesivos
refinamientos mejorar\'{\i}an seguramente la calidad de las
predicciones, sobre todo en lo que se refiere a lo abrupto de las
transiciones, pero podr\'{\i}a llegar a perderse intuici\'on sobre
lo que sucede si el modelo se complica. En todo caso se hace
necesaria una resoluci\'on num\'erica del problema basada en
elementos finitos, para ratificar con m\'as contundencia lo que
aqu\'{\i} se ha mostrado.

De cualquier manera, nuestros resultados experimentales y los
correspondientes an\'alisis sugieren que el calentamiento es un
efecto de gran importancia en el tipo de filmes que hemos
estudiado en nuestros experimentos. De haber cualquier otro
posible fen\'omeno \index{intr\'{\i}nseco} intr\'{\i}nseco como
causante o acelerador de la transici\'on, ser\'{\i}a muy
dif\'{\i}cil en estas condiciones acceder experimentalmente a
\'el, discrimin\'andolo \emph{n\'{\i}tidamente} de los efectos
t\'ermicos. Se observar\'{\i}a m\'as bien una anticipaci\'on del
fen\'omeno intr\'{\i}nseco respecto de sus valores esperados,
causada por el r\'apido incremento de temperatura de la muestra en
la avalancha: ambos efectos, el intr\'{\i}nseco y el
extr\'{\i}nseco del calentamiento estar\'{\i}an entremezclados.
Esta podr\'{\i}a ser la explicaci\'on de que las predicciones de
nuestro modelo t\'ermico, que no tiene en cuenta ning\'un
mecanismo intr\'{\i}nseco, den transiciones m\'as suaves que las
experimentales. Se hace necesaria una resoluci\'on m\'as detallada
del balance energ\'etico del sistema para ver si el
\Index{calentamiento}, exclusivamente, podr\'{\i}a dar cuenta de
todo el fen\'omeno de la transici\'on en \Jx, o si por el
contrario transiciones tan abruptas requieren necesariamente un
mecanismo intr\'{\i}nseco, m\'as r\'apido que el calentamiento,
para producirse.

%%%%%%%%%%%%%%%%%%%%%%%%%%%%%%%%%%%%%%%%
\section{Conclusiones}
%%%%%%%%%%%%%%%%%%%%%%%%%%%%%%%%%%%%%%%%

Las principales conclusiones de esta primera parte, extra\'{\i}das
de lo que se acaba de resumir, son:

\begin{enumerate}

\item El intercambio t\'ermico de estos micropuentes de \YBCOf\
ocurre principalmente con el substrato, siendo despreciable, en
tiempos breves, la interacci\'on con la atm\'osfera circundante.
Esta interacci\'on del filme con el substrato establece una
inercia t\'ermica del sistema con tiempos caracter\'{\i}sticos de
d\'ecimas de milisegundo, mayores que los \index{tiempo
bolom\'etrico} tiempos bolom\'etricos de enfriamiento.

\item El modelo \BS\ (BS) explica aparentemente bien la dependencia
de los par\'ametros cr\'{\i}ticos \Ex\ y \Jx\ en la regi\'on de
campos bajos estudiada, proporcionando un \Index{tiempo de
relajaci\'on} energ\'etica de los portadores que coincide con el
de otros experimentos. Esto parece indicar que la transici\'on al
estado normal se produce, siguiendo el modelo BS, por una
avalancha en la din\'amica de los v\'ortices magn\'eticos.

\item Por otra parte, la variaci\'on de la corriente de salto \Jx\
presenta las mismas dependencias funcionales de la temperatura y
el campo magn\'etico que la corriente cr\'{\i}tica \Jc, lo que
sugiere que ambos fen\'omenos ---la aparici\'on de disipaci\'on en
el estado mixto, y la transici\'on abrupta al estado normal---
podr\'{\i}an tener su causa en mecanismos f\'{\i}sicos an\'alogos:
alteraciones en la din\'amica de los v\'ortices.

\item Sin embargo, para poder aplicar el modelo BS en la regi\'on de campos bajos
estudiada hemos tenido que tener en cuenta el autocampo generado
por las altas corrientes inyectadas. Este an\'alisis sugiere que
la densidad de corriente real podr\'{\i}a ser mayor que la
aparente, si la corriente se distribuyese de manera no uniforme,
concentr\'andose en los laterales de la muestra. De ser as\'{\i},
los valores reales de \Jx\ podr\'{\i}an coincidir con los de
desapareamiento \index{corriente de desapareamiento} de los pares
de Cooper, siendo este desapareamiento la causa de la transici\'on
al estado normal.

\item A mayores, el modelo t\'ermico que hemos propuesto para
describir al sistema de micropuente m\'as substrato, considerando
la \Index{realimentaci\'on} en la disipaci\'on, predice
notablemente bien los principales resultados experimentales,
incluso de manera cuantitativa. De esto podemos concluir que,
aunque pueda existir otro fen\'omeno m\'as intr\'{\i}nseco
relacionado con la transici\'on abrupta
---no excluido por los resultados del modelo sencillo---, los efectos
t\'ermicos tienen en estos experimentos m\'as relevancia de la que
hasta ahora se hab\'{\i}a considerado.

\end{enumerate}

%%%%%%%%%%%%%%%%%%%%%%%%%%%%%%%%%%%%%%%%
\section{L\'{\i}neas de continuaci\'on}
%%%%%%%%%%%%%%%%%%%%%%%%%%%%%%%%%%%%%%%%

Las conclusiones que acabamos de relacionar no dan respuesta a
todos los interrogantes relacionados con la causa de la
transici\'on abrupta al estado normal en \Jx; antes al contrario,
plantean a su vez otros nuevos. Mencionamos los m\'as destacados,
as\'{\i} como las posibles l\'{\i}neas de investigaci\'on para
intentar resolverlos:

\begin{enumerate}

\item La posibilidad de que la transici\'on abrupta en \Jx\ se deba a
ruptura de pares de Cooper ---por estar la corriente muy
concentrada en los laterales de la muestra, y alcanzarse
as\'{\i}la corriente de desapareamiento \Jd ---, podr\'{\i}a
estudiarse repitiendo estos experimentos con micropuentes de
diferentes anchuras o espesores, del orden de los
aqu\'{\i}empleados y a\'un menores, y viendo de qu\'e manera
dependen los valores de \Ix\ y \Jx\ de estos par\'ametros
geom\'etricos.

\item El estudio experimental detallado de la transici\'on abrupta en \Jx\
requiere de dispositivos todav\'{\i}a m\'as r\'apidos de los
aqu\'{\i}empleados. Por un lado, para seguir el voltaje a lo largo
de la transici\'on se hace necesario un sistema de medida m\'as
r\'apido a\'un que la tarjeta adquisidora, que aumente la
resoluci\'on temporal por encima de los $10\und{\mu s}$. Por otro,
es imprescindible mayor flexibilidad y rapidez en la aplicaci\'on
de la corriente. Por ejemplo, se necesitan valores estables de
corriente en tiempos muy breves, muy por debajo de los tiempos
caracter\'{\i}sticos t\'ermicos del sistema (d\'ecimas de
milisegundo), para poder estudiar con precisi\'on los
reg\'{\i}menes temporales en que los efectos t\'ermicos
todav\'{\i}a afectan poco.

\item Desde un punto de vista te\'orico, ser\'{\i}a conveniente mejorar
el modelo t\'ermico tan sencillo que aqu\'{\i} se ha propuesto,
para tener en cuenta de manera realista los grandes cambios de
disipaci\'on que se producen en tiempos tan breves en torno a la
transici\'on en \Jx. Un an\'alisis detallado mediante simulaci\'on
por elementos finitos, por ejemplo, podr\'{\i}a ayudar a
discriminar mejor si la transici\'on abrupta que se mide a altas
densidades de corriente puede explicarse exclusivamente por
mecanismos t\'ermicos, y por lo tanto extr\'{\i}nsecos, o si por
el contrario la velocidad del fen\'omeno requiere para su
explicaci\'on mecanismos electr\'onicos m\'as intr\'{\i}nsecos.

\end{enumerate}
    % cap 8
\clearemptydoublepage

\part{Paraconductividad: tiempos de relajaci\'on y altas temperaturas reducidas}

%%%%%%%%%%%%%%%%%%%%%%%%%%%%%%%%%%%%%%%%%%%%%%%%%%%%%%%%%%%%%%%%%%
\chapter{Introducci\'on a la parte II}
\label{cap:introds}
%%%%%%%%%%%%%%%%%%%%%%%%%%%%%%%%%%%%%%%%%%%%%%%%%%%%%%%%%%%%%%%%%%

%-%-%-%-%-%-%-%-%-%-%-%-%-%-%-%-%-%-%-%-%-%-%-%-  figura -%-%-%-%-%-%-%-%

\newcommand{\figDserror}{  % alias

\figura {fig:ds:dserror}      % label
{fig/ds/dserror}              % file
{Incertidumbre en la determinaci\'on de la paraconductividad. Los
datos son de la paraconductividad en funci\'on de la temperatura
reducida, para varias muestras (cristales y capas delgadas). La
zona sombreada cubre la regi\'on en que se encuadran todos los
resultados experimentales, de estas y otras muestras, y con el
empleo de diferentes fondos para la resistividad normal. Gr\'afica
extra\'{\i}da de un art\'{\i}culo incluido en el cap\'{\i}tulo
siguiente
(Phys. Rev. B {\bf 63}, 144515), en su figura 2.}                  % caption
{Incertidumbre en la determinaci\'on de la paraconductividad.}                  % toc
{\stfigw}            % width \textwidth
}
% - - - - - - - - - - - - - - - - - - - - - - - - - - - - - - - - - - - %

%%%%%%%%%%%%%%%%%%%%%%%%%%%%%%%%%%%%%%%%
\section{Definici\'on de paraconductividad}
%%%%%%%%%%%%%%%%%%%%%%%%%%%%%%%%%%%%%%%%

Adem\'as de su relativamente elevada temperatura cr\'{\i}tica \Tc,
una caracter\'{\i}stica de gran importancia que poseen los
superconductores de altas temperaturas (HTSC) es la peque\~nez, en
todas las direcciones, de la denominada longitud de correlaci\'on
o de coherencia superconductora $\xi(T)$. Esta longitud de
coherencia es aproximadamente la dimensi\'on espacial de los pares
de Cooper, y por tanto la distancia entre los dos electrones
apareados que los conforman. En t\'erminos de la funci\'on de onda
cu\'antica superconductora $\Psi(r)$, $\xi(T)$ es la longitud
caracter\'{\i}stica a la cual $\Psi(r)$ puede variar
apreciablemente sin necesidad de pagar un elevado precio en
energ\'{\i}a ($\xi(T)$ es la dimensi\'on m\'{\i}nima del paquete
de onda).

Aplicando el principio de incertidumbre de Heisenberg, podemos ver
cualitativamente que en torno a la transici\'on superconductora
$\xi$ y \Tc\ est\'an interconectadas entre s\'{\i} \cite[cap.
1.3]{Tinkham96}: \be \xi \; \Delta p \geq \hbar,
\label{ec:ds:indet} \ee siendo $\Delta p \simeq \kB \; \Tc / v_F$
el momento de los portadores, $v_F$ su velocidad de Fermi, y \kB\
la constante de Boltzmann. Vemos pues que, crudamente, \be \xi
\simeq \hbar \; v_f / (\kB \; \Tc). \label{ec:ds:xhi} \ee Esta
relaci\'on, naturalmente muy aproximada \cite{Tinkham96}, nos
indica que, dado que no es de esperar que $v_F$ difiera mucho en
los HTSC de los valores que toma en los superconductores
convencionales con baja \Tc, es previsible que $\xi$ sea mucho
m\'as peque\~na en los HTSC que en los superconductores
met\'alicos convencionales.

Una consecuencia importante de que simult\'aneamente \Tc\ sea alta
y $\xi(T)$ sea peque\~na en los HTSC es que las fluctuaciones
termodin\'amicas son, cerca de \Tc, muy apreciables en estos
materiales: por un lado la energ\'{\i}a de agitaci\'on t\'ermica
$\kB \; \Tc$ ---siempre disponible \emph{gratuitamente}--- es
elevada, y por otro lado la energ\'{\i}a necesaria para crear y
destruir pares de Cooper es peque\~na. Esta \'ultima energ\'{\i}a
es, de modo aproximado, la necesaria para crear un par de Cooper
multiplicada por el n\'umero de pares que se encuentran en el
volumen coherente\footnote{Esta multiplicaci\'on se debe a que la
coherencia superconductora no permite crear o destruir un solo par
de Cooper, lo que ser\'{\i}a equivalente a variar $\Psi(r)$ a
escalas espaciales menores que $\xi$.}. Pero ese n\'umero de pares
ser\'a proporcional al volumen coherente, es decir, a $\xi(T)^D$,
siendo $D = 3,2,1$ la dimensionalidad de la muestra. Dado que en
los HTSC $\xi(T)$ es peque\~na, sea cual sea la dimensionalidad el
volumen coherente ser\'a peque\~no; pero adem\'as estos materiales
son laminares, y en general $3 \leq D \leq 2$, lo que contribuye a
hacer que el volumen coherente sea todav\'{\i}a menor. Todo esto
conduce a que la creaci\'on y destrucci\'on de pares de Cooper por
fluctuaciones t\'ermicas sea un fen\'omeno relevante y de gran
inter\'es en los HTSC.

Una de las magnitudes m\'as adecuadas, tanto desde el punto de
vista de la facilidad y precisi\'on de su medida como de su
sensibilidad a estos efectos, para observar experimentalmente las
fluctuaciones t\'ermicas, es la \Index{resistividad} el\'ectrica
$\rho(T)$: la creaci\'on de pares de Cooper por fluctuaciones
t\'ermicas hace disminuir apreciablemente $\rho(T)$, incluso para
temperaturas varios grados por encima de \Tc. De hecho la posible
importancia de estos efectos en los HTSC fue mencionada ya por
Bednorz y M\"uller en su publicaci\'on pionera \cite{Bednorz86}.
La magnitud convencional que recoge estos efectos es la denominada
\emph{paraconductividad}, el exceso de \Index{conductividad
el\'ectrica} debido a fluctuaciones, que se define como \be \Delta
\sigma(T) = \frac{1}{\rho(T)} - \frac{1}{\rho_{bg}(T)}.
\label{ec:ds:ds} \ee En esta expresi\'on $\rho(T)$ es la
resistencia experimental a cada temperatura, mientras que
$\rho_{bg}(T)$ es la denominada \Index{resistividad} de fondo
(\emph{background}), la resistividad que habr\'{\i}a alrededor de
\Tc\ si no hubiese transici\'on superconductora.

%%%%%%%%%%%%%%%%%%%%%%%%%%%%%%%%%%%%%%%%
\section{Medida experimental}
%%%%%%%%%%%%%%%%%%%%%%%%%%%%%%%%%%%%%%%%

% ------------------------------------ %
\subsection{Determinaci\'on y errores}
% ------------------------------------ %

Las medidas de la paraconductividad que se muestran en esta
segunda parte se realizaron con el mismo montaje experimental
descrito en el cap\'{\i}tulo \ref{cap:disps}, salvo por los
aparatos empleado para la determinaci\'on del voltaje en los
puentes superconductores. En este caso, para medir con buena
resoluci\'on la \Index{resistividad} en el estado normal, se
emple\'o un amplificador lock-in EG\&G (Princeton Applied
Research) 5302, con filtro pasa-banda en 37.4~Hz. La corriente se
inyect\'o con la propia fuente de voltaje interna del lock-in
---funcionando a esta misma frecuencia--- y una resistencia
tamp\'on en serie con la muestra.

En la determinaci\'on experimental de la paraconductividad \ds\
existe una incertidumbre, que es constante en t\'erminos relativos
para toda temperatura, y que puede llegar a ser mayor que del
20\%. Esta incertidumbre constante se debe a la dificultad de
establecer con precisi\'on las dimensiones de las muestras, en
pel\'{\i}culas (ver sec. \ref{par:errorgeometrico}, p.
\pageref{par:errorgeometrico}), pero sobre todo en cristales,
debido al tama\~no proporcionalmente grande de los
\index{contactos el\'ectricos} contactos. Adem\'as, a este
\Index{error} constante en valores relativos, se le suma otro que
es creciente en temperaturas, debido a la ambigua determinaci\'on
de la \Index{resistividad} de fondo \rbgt. Para \'esta, el \'unico
acceso experimental conocido y com\'unmente empleado consiste en
la extrapolaci\'on del comportamiento de \rt\ desde regiones de
altas temperaturas, supuestamente libre de fluctuaciones, hasta
las regiones de m\'as baja temperatura. Las ambig\"uedades
inherentes a esta extrapolaci\'on en los mejores casos se
mantienen por debajo del 10\% s\'olo para temperaturas
reducidas\footnote{En la regi\'on por encima de la transici\'on
$T>\Tc$, la temperatura reducida se define con signo opuesto a la
que hemos empleado en la primera parte de este trabajo: $\eps
\equiv \ln(T/\Tc) \simeq T/\Tc - 1$.} $\eps\leq10^{-1}$, y crecen
dram\'aticamente para mayores temperaturas, conforme \ds\ tiende a
anularse, pudiendo llegar hasta el 700\% ya en $\eps
\sim5\cdot10^{-1}$.

\figDserror

Estas ambig\"uedades experimentales y del an\'alisis se recogen en
la figura \ref{fig:ds:dserror}: la zona sombreada cubre la
regi\'on en que se encuadran todos los resultados experimentales,
para diversas muestras y con el empleo de distintos fondos. Pero
incluso siendo esta dispersi\'on experimental, como mencionamos,
tan proporcionalmente amplia, ninguno de los modelos te\'oricos
propuestos antes de la publicaci\'on de nuestros trabajos
consegu\'{\i}a explicar la tendencia descendente de los resultados
experimentales de la region $\eps \geq10^{-1}$. A pesar de estas
enormes incertidumbre, todav\'{\i}a es posible discriminar entre
los modelos previos, que no segu\'{\i}an la tendencia
experimental, y el nuevo modelo que aqu\'{\i} se propone para
explicar esta regi\'on de alta temperatura (modelo de \emph{corte
en energ\'{\i}a total}, que incluye la energ\'{\i}a de
localizaci\'on de los pares de Cooper en el volumen coherente.),
que se ajusta notablemente a la tendencia de los datos.

% ------------------------------------ %
\subsection{Resoluci\'on experimental}
% ------------------------------------ %

\begin{table}
  \centering
    \caption[Dimensiones t\'{\i}picas de las muestras empleadas en las
  medidas de paraconductividad.]{Dimensiones t\'{\i}picas de las muestras empleadas en las
  medidas de paraconductividad. Distancia entre contactos de
  voltaje $l$ ---no es el largo de la muestra---, ancho $a$ y
  espesor $d$.}\label{tab:ds:muest}
  \begin{tabular}{cccc}
    \\[-8pt]
  \hline
  Tipo & $l$ & $a$ & $d$ \\
       & (mm) & (mm) & (${\rm \mu m}$) \\ \hline
  cristal & 0.5 & 0.5 & 70 \\
  filme & 1 & 0.1 & 0.2 \\ \hline
\end{tabular}

\end{table}

Hagamos una estimaci\'on de cu\'al es la resoluci\'on de la medida
(la separaci\'on m\'{\i}nima que pueden tener dos valores
pr\'oximos de la resistividad para que sean distinguibles) en
estos experimentos de determinaci\'on de paraconductividad. Esta
estimaci\'on es relevante debido a la manera en que determinamos
esta magnitud, como resta de valores de resistividad: es necesario
conocer hasta qu\'e l\'{\i}mite de proximidad entre valores de
fondo y valores de resistividad con fluctuaciones tenemos acceso
experimental.

Las dimensiones t\'{\i}picas de las muestras empleadas en estas
medidas se muestran en la tabla \ref{tab:ds:muest}. El ruido de
fondo t\'{\i}pico en la medida es $\delta V_1 \approx 25\und{nV}$,
mientras que la resoluci\'on del lock-in $\delta V_2$ es de una
parte en $10^4$ del m\'aximo de escala que se emplee en cada
momento. Veamos cual de estas dos fuentes de incertidumbre es la
que resulta m\'as limitante.

La resistencia de contacto \index{contactos el\'ectricos} de los
hilos de corriente es $R_{cont} \approx 40\und{m\Omega}$. Como
queremos limitar la disipaci\'on por contacto a menos de un cierto
l\'{\i}mite, digamos de $10^{-8}~W$, la corriente m\'axima que
podemos aplicar es
\[  I^2 \cdot R_{cont} \lsim  10^{-8} \und{W} , \]
\[ I^2 \lsim \frac{10^{-8} W}{40 \Exund{-3}{\Omega}} = 2.5 \Exund{-8}{A^2} ,\]
\[ I_{m\acute ax} \approx 5 \Exund{-4}{A}.\]
Esto implica una densidad de corriente m\'axima, en cristales, de
\[ J_{m\acute ax}^{c} = \frac {5 \Exund{-4}{A}} {0.5 \cdot 0.07
\cdot 10^{-2}\und{cm^2}} \approx 1.5\und{A/cm^2}, \] y en filmes,
de
\[ J_{m\acute ax}^{f} = \frac{5 \Exund{-4}{A}}{0.1 \cdot 0.0002
\cdot 10^{-2}\und{cm^2}} = 2500\und{A/cm^2}.\]

Suponiendo que el material tiene, a una cierta temperatura, una
resistividad $\rho=100\und{\mu\Omega\;cm}$, y que aplicamos a las
muestras sus correspondientes densidades de corriente m\'aximas, el
voltaje que mide el lock-in es, en cristales,
\[ V = E \cdot l = J_{m\acute ax}^{c} \cdot \rho \cdot l =
1.5\und{A/cm^2} \times 10^{-4}\und{\Omega\;cm} \cdot
5\Exund{-2}{cm} = 7.5\und{\mu V},
\]
mientras que en los filmes es
\[ V = 2500\und{A/cm^2} \cdot 10^{-4} \und{\Omega~cm} \cdot 1\Exund{-1}{cm} = 25\und{mV}. \]

El ruido de fondo $\delta V_1$ afecta a la medida de estos
voltajes de la siguiente manera:

En el cristal,
\[  \frac{\delta \rho_{1}}{\rho} = \frac{\delta V_1}{V} =
\frac{2.5 \Exund{-8}{V}}{7.5 \Exund{-6}{V}} = 3.3 \times
10^{-3},\]
\[ \delta \rho_{1} = \rho \cdot \frac{\delta V_1}{V} =
\frac{\delta V_1}{J_{m\acute ax} \cdot l} = \rho \cdot 3.3 \times
10^{-3} = 0.33\und{\mu\Omega\;cm}.\]

En el filme,
\[  \frac{\delta \rho_{1}}{\rho} = \frac{\delta V_1}{V} =
\frac{2.5 \Exund{-8}{V}}{25 \Exund{-3}{V}} = 1.0 \times 10^{-6},\]
\[ \delta \rho_{1} = \rho \cdot \frac{\delta V_1}{V} =
\frac{\delta V_1}{J_{m\acute ax} \cdot l} = \rho \cdot 1.0 \times
10^{-6} = 1.0 \times 10^{-4}\und{\mu\Omega\;cm}.\]

Como lo que resuelve el lock-in es una parte en diez mil
$(10^{-4})$ del m\'aximo de escala, tenemos que:

En el cristal, los $7.5\und{\mu V}$ que se obtienen se
medir\'{\i}an en la escala de $V_{m\acute ax} = 10\und{\mu V}$, lo
que implica una resoluci\'on de
\[\delta V_{2} = V_{m\acute ax} \cdot 10^{-4} = 10^{-9}\und{V}, \]
que se traduce en
\[  \frac{\delta \rho_{2}}{\rho} = \frac{\delta V_{2}}{V}
= \frac{10^{-9}\und{V}}{7.5 \Exund{-6}{V}} = 1.3 \times 10^{-4}
,\]
\[ \delta \rho_{2} = 1.3 \times 10^{-4} \cdot \rho =
0.013\und{\mu\Omega\;cm}.\]

Este resultado es menor que el de su correspondiente ruido de
fondo $\delta \rho_{1}$, as\'{\i} que tanta resoluci\'on del
lock-in no sirve para gran cosa, y el factor limitante es dicho
ruido. La resoluci\'on final de la medida es, en este caso,
$\delta \rho_{1}^{crist} = 0.33~\mu\Omega ~cm$.

En el filme, de manera an\'aloga, los $25\und{mV}$ de se\~nal que
se obtienen se medir\'{\i}an en la escala de $V_{m\acute ax} =
50\und{mV}$, lo que implica una resoluci\'on del lock-in de
\[\delta V_{2} = V_{m\acute ax} \cdot 10^{-4} = 5 \Exund{-6}{V}. \]
Esto lleva a
\[  \frac{\delta \rho_{2}}{\rho} = \frac{\delta V_{2}}{V}
= \frac{5 \Exund{-6}{V}}{25 \Exund{-3}{V}} = 2.0 \times 10^{-4},\]
\[ \delta \rho_{2}^{film} = 2.0 \times 10^{-4} \cdot \rho =
0.02\und{\mu\Omega\;cm}. \] El lock-in no es capaz de apreciar, en
esta escala, los voltajes de ruidos de fondo, as\'{\i} que el
factor limitante para la medici\'on de la resistividad viene
determinado por la propia instrumentaci\'on. De este modo, y salvo
mejoras experimentales, la resoluci\'on de la medida es $\delta
\rho_{2}^{film} = 0.02~\mu\Omega ~cm$, mejor en todo caso que la
que se obtiene para los cristales.

Como hemos visto, la resoluci\'on de la medida viene determinada
en principio por el ruido de fondo en la medida del voltaje, que
se presupone constante y peque\~no. Variando las dimensiones de la
muestra y la corriente aplicada se puede hacer que este ruido se
haga despreciable frente a la se\~nal medida, y as\'{\i} la
resoluci\'on estar\'a determinada por la que establezcan los
dispositivos de medida. En los cristales t\'{\i}picos que
empleamos el factor limitante es el ruido de fondo, y la buena
resoluci\'on que proporciona el lock-in se desperdicia
parcialmente. En las pel\'{\i}culas delgadas, por el contrario, se
obtienen se\~nales muy altas con relativamente poca corriente, lo
que lleva a aprovechar por completo la resoluci\'on que brinda el
lock-in, siendo \'esta el factor limitante. Los valores de la
paraconductividad que se determinen a temperaturas muy altas, de
manera que la diferencia entre la resistividad de fondo y la
experimental con fluctuaciones sea menor o del orden de la
resoluci\'on de la medida, carecen de fiabilidad.

En el caso de los filmes, si pudi\'esemos aplicar un voltaje de
compensaci\'on constante para contrarrestar la se\~nal experimental
---cosa que pueden hacer dispositivos como los lock-ins, sin
variar el voltaje en la muestra, sino s\'olo internamente en su
lectura--- podr\'{\i}amos medir en escalas de $V_{m\acute ax}$
menores, reduciendo con ello $\delta V_{2} = V_{m\acute ax} \cdot
10^{-4}$. Har\'{\i}amos el voltaje de compensaci\'on muy cercano
en valor absoluto a la se\~nal de voltaje experimental para una
temperatura dada (en 100~K, $V_{offset} = 25 \und{mV}$) pero de
signo contrario, para que se contrarresten, de modo que
medir\'{\i}amos las diferencias, ya muy peque\~nas, con respecto a
este voltaje al variar la temperatura. Lograr\'{\i}amos de esta
manera acercarnos a la resoluci\'on m\'{\i}nima determinada por el
ruido de fondo, $\delta \rho_{1, f} = 1 \times 10^{-4} ~\mu\Omega
~cm$. La dificultad experimental est\'a, por supuesto, en que este
voltaje de compensaci\'on tendr\'{\i}a que ser estable por lo
menos hasta la cifra $\delta V_{2} = 5 \Exund{-6}{V}$.

Se debe se\~nalar aqu\'{\i}, como apunte complementario, que el
ruido $\delta V_1 = 25\ nV$ es el que se observa s\'olo cuando la
resistividad se anula tras la transici\'on superconductora, y
podemos medir en la escala m\'{\i}nima del lock-in. En el estado
normal, sin embargo, se manifiestan dispersiones de los puntos
experimentales del orden de la propia resoluci\'on del lock-in: es
``la \'ultima cifra que oscila'', lo que implica un ruido de fondo
real comparable a la resoluci\'on de escala $\delta V_2$.
Adem\'as, esto sucede s\'olo cuando la temperatura es
aproximadamente estable. Cuando la temperatura var\'{\i}a y se
realizan medidas en deriva t\'ermica (ya sea natural o
controlada), se aprecian dispersiones de los puntos experimentales
respecto de la tendencia que son algo mayores que las que
generar\'{\i}a simplemente esta \emph{oscilaci\'on} de la \'ultima
cifra del aparato. As\'{\i}, por ejemplo, en la regi\'on donde uno
de los filmes\footnote{Muestra LYS50(1).} tiene
\Index{resistividad} $\rho = 100\ \mu\Omega\ cm$ (en $T=142\ K$)
las desviaciones son cercanas a cuatro partes en $10^4$, cuatro
veces la resoluci\'on del lock-in. Esto apunta en la direcci\'on
de que, para conseguir llegar a resoluciones tan bajas como la que
generar\'{\i}a exclusivamente el ruido de fondo $\delta V_1$, se
necesitar\'{\i}a, adem\'as del antedicho voltaje de compensaci\'on
altamente estable, una excelente estabilizaci\'on en temperatura
(o una deriva \emph{infinitamente} lenta) que evitase esta fuente
de error a\~nadida.

\clearemptydoublepage

%%%%%%%%%%%%%%%%%%%%%%%%%%%%%%%%%%%%%%%%%%%%%%%%%%%%%%%%%%%%%%%%%%
\chapter{Resultados y conclusiones de la parte II (separatas de los art\'{\i}culos publicados)}
\chaptermark{Resultados y conclusiones de la parte II}
%\addtocontents{toc}{\protect\vspace{0.2cm}}
\label{cap:pubds}
%%%%%%%%%%%%%%%%%%%%%%%%%%%%%%%%%%%%%%%%%%%%%%%%%%%%%%%%%%%%%%%%%%

A modo de resumen, se incluyen en las p\'aginas siguientes las
publicaciones relacionadas con la paraconductividad en las que el
autor de esta memoria particip\'o a lo largo del periodo de
doctorado.

En las dos primeras publicaciones se estudia el \Index{tiempo de
relajaci\'on} de los pares de Cooper fluctuantes. En las tres
siguientes se estudia el comportamiento de la paraconductividad a
\emph{altas} temperaturas reducidas, con $\eps \equiv \ln(T/\Tc)
\gtrsim 0.1$. El principal resultado de este \'ultimo estudio es la
sugerencia de que a altas temperaturas reducidas las fluctuaciones
superconductoras est\'an dominadas por el principio de incertidumbre
de Heisenberg, que limita el \emph{encogimiento} de la funci\'on de
onda superconductora cuando $T$ aumenta por encima de \Tc.

Las aportaciones del autor de esta memoria a dichos estudios sobre
la paraconductividad consistieron principalmente en las medidas de
la \Index{resistividad} en funci\'on de la temperatura, la
extracci\'on de la paraconductividad a partir de esas medidas, y
el an\'alisis de los diversos errores experimentales y de las
incertidumbres asociadas con las estimaciones del fondo
(\emph{background}) y de la temperatura cr\'{\i}tica \Tc.

\subsection*{Nota para esta versi\'on en arXiv}

Por motivos de copyright no se han incluido en esta versi\'on de
la tesis, especial para {arXiv.org}, las mencionadas
publicaciones. \'Estas se encuentran relacionadas en la p\'agina
\pageref{listads}. Se conserva a continuaci\'on la numeraci\'on de
p\'aginas original de esta tesis, por lo que se saltan los
n\'umeros que correspond\'ian a las p\'aginas omitidas.

\clearemptydoublepage

\addtocounter{page}{36}  % sumar tantas p\'aginas como se vayan a insertar luego

\phantomsection \addcontentsline{toc}{part}{Ap\'endice}
\part*{Ap\'endice} \clearemptydoublepage

\appendix

\setlength{\parskip} {1ex plus 0.5ex minus 0.2ex}

%%%%%%%%%%%%%%%%%%%%%%%%%%%%%%%%%%%%%%%%%%%%%%%%%%%%%%%%%%%%%%%%%%
% \Appendix{A}{C\'alculos con \emph{Mathematica}}{C\'alculos con
\chapter{Herramientas de edici\'on y c\'alculo}
%\chaptermark{C\'alculos con \emph{Mathematica}}
\addtocontents{toc}{\protect\vspace{0.2cm}}
\label{cap:appMath}
%%%%%%%%%%%%%%%%%%%%%%%%%%%%%%%%%%%%%%%%%%%%%%%%%%%%%%%%%%%%%%%%%%

\section{Herramientas de edici\'on}

Para la elaboraci\'on del documento original se emplearon las
siguientes herramientas inform\'aticas.

Maquetaci\'on: PDF\TeX\ 3.14159-1.10a (de la distribuci\'on
MiK\TeX\ 2.2) junto con WinEdt 1.414 (Aleksander Simonic), y
Acrobat 4.0 (Adobe). C\'alculos, an\'alisis de datos, y
representaci\'on gr\'afica: Origin 7.0 (OriginLab) y Mathematica
4.1 (Wolfram Research). Tratamiento de imagen digital: Photoshop
5.5 (Adobe) e IrfanView 3.61 (Irfan Skiljan). Dise\~no de
gr\'aficos vectoriales: CorelDraw 8 (Corel). Gesti\'on de
bibliograf\'{\i}a en Bib\TeX: BibDB 2.2 (Eyal Doron). Extractos en
versi\'on HTML: T\raisebox{-0.1cm}{T}H 2.2 (Ian Hutchinson).

Este documento se escribi\'o en \LaTeX~$2\varepsilon$, empleando los
paquetes \textsf{graphicx}, \textsf{hyperref}, \textsf{amsmath},
\textsf{amssymb}, \textsf{afterpage}, \textsf{float},
\textsf{fancyheadings}, \textsf{calc}, \textsf{pifont},
\textsf{enumerate}, \textsf{array}, \textsf{makeidx},
\textsf{babel}, \textsf{natbib}, \textsf{hypernat}, e
\textsf{inputenc}.

Se emple\'o Acrobat 4.0 (Adobe) para insertar, en el documento
final, las publicaciones de la segunda parte y los ficheros
Mathematica de este ap\'endice, a partir de sus versiones en PDF
(\emph{Portable Document Format}). Para escribir estos \'ultimos en
PDF, se imprimieron desde Mathematica a un archivo, empleando el
controlador de impresora Adobe Distiller 4.2.4.

Para la versi\'on de arXiv se emplearon im\'agenes EPS de tama\~no
peque\~no, lo que oblig\'o en algunos casos a prescindir del color
y de altas resoluciones.

\section{C\'alculos con \emph{Mathematica}}

Se incluyen a continuaci\'on dos ficheros del entorno de c\'alculo
\emph{Mathematica} (Wolfram Research), en su versi\'on 4.1. En el
primero de ellos se explica c\'omo se realizaron los c\'alculos de
los cap\'{\i}tulos \ref{cap:cal} y \ref{cap:calpr}. Se muestra el
programa de c\'alculo iterativo para la determinaci\'on con
\Index{realimentaci\'on} de la temperatura del micropuente, que
resuelve la ecuaci\'on \ref{ec:Tfcorr}, o las ecuaciones
\ref{ec:Tf} y \ref{ec:Tfsimple} si se desean menos correcciones al
modelo. Se muestran tambi\'en las funciones complementarias
necesarias para este programa, y algunos ejemplos de uso
elementales.

En el otro fichero que se incluye, se detalla c\'omo se le dio
formato a las figuras que, hechas con \emph{Mathematica}, se
incorporaron al presente documento.

\newpage
\includegraphics[width=\textwidth]{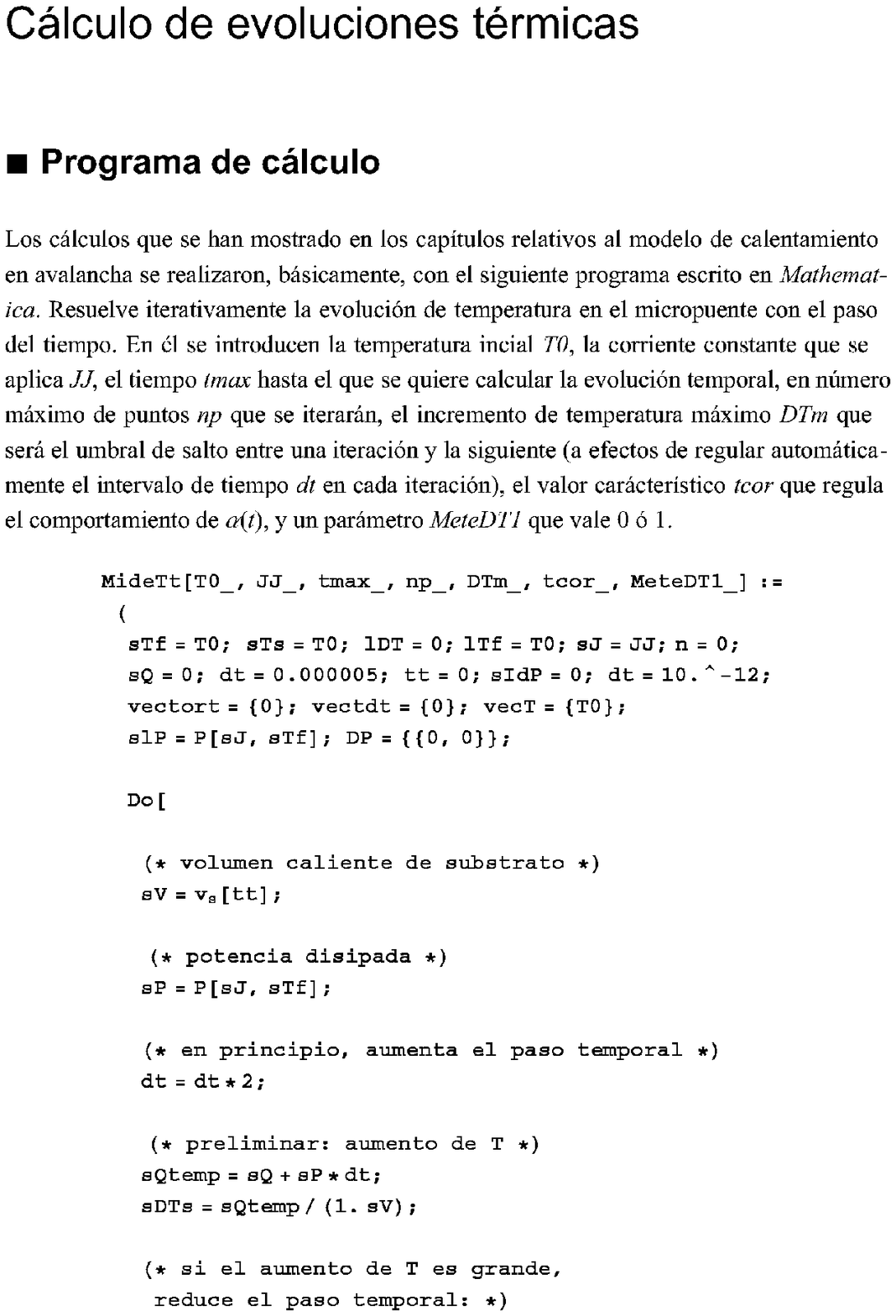}
\newpage
\includegraphics[width=\textwidth]{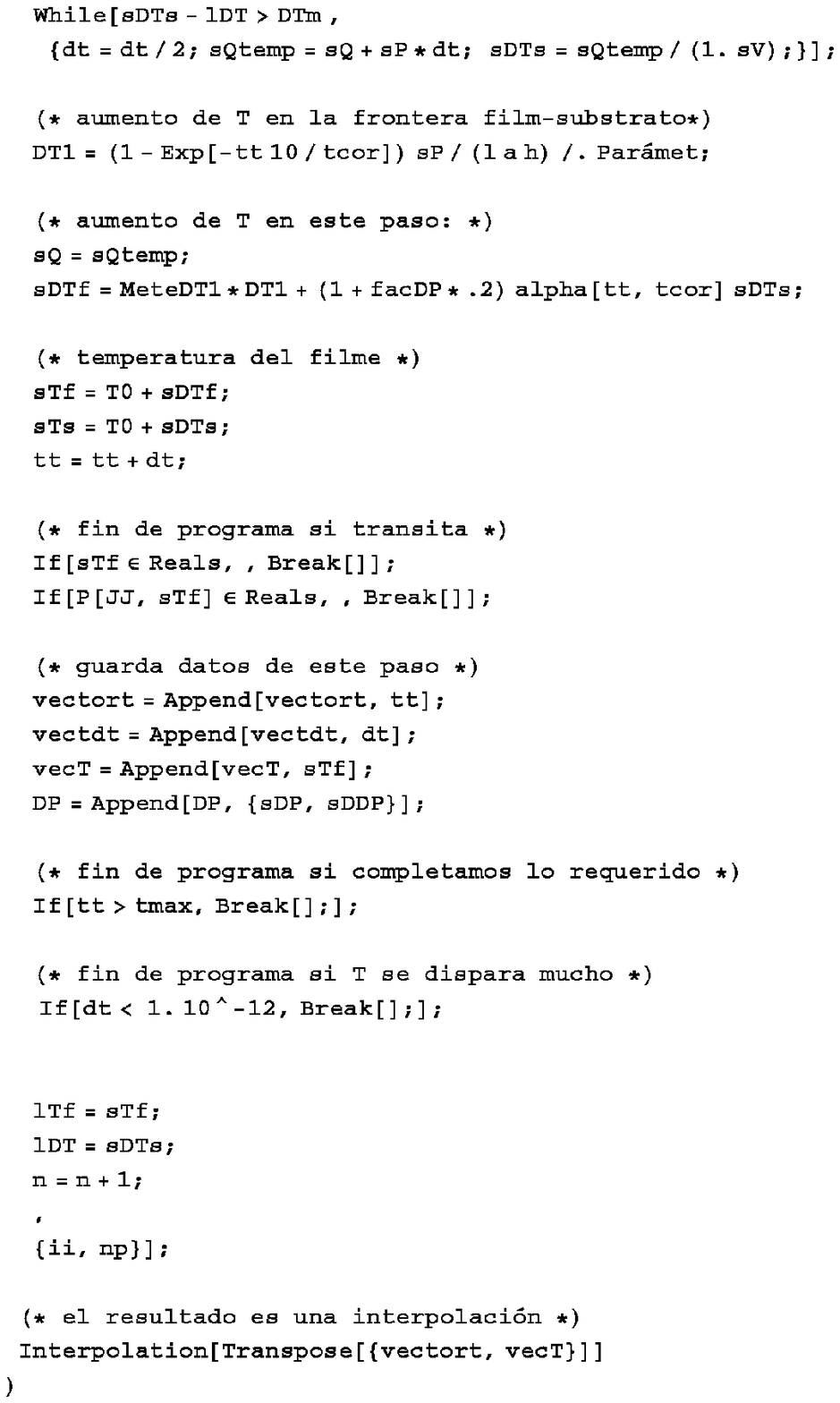}
\newpage
\includegraphics[width=\textwidth]{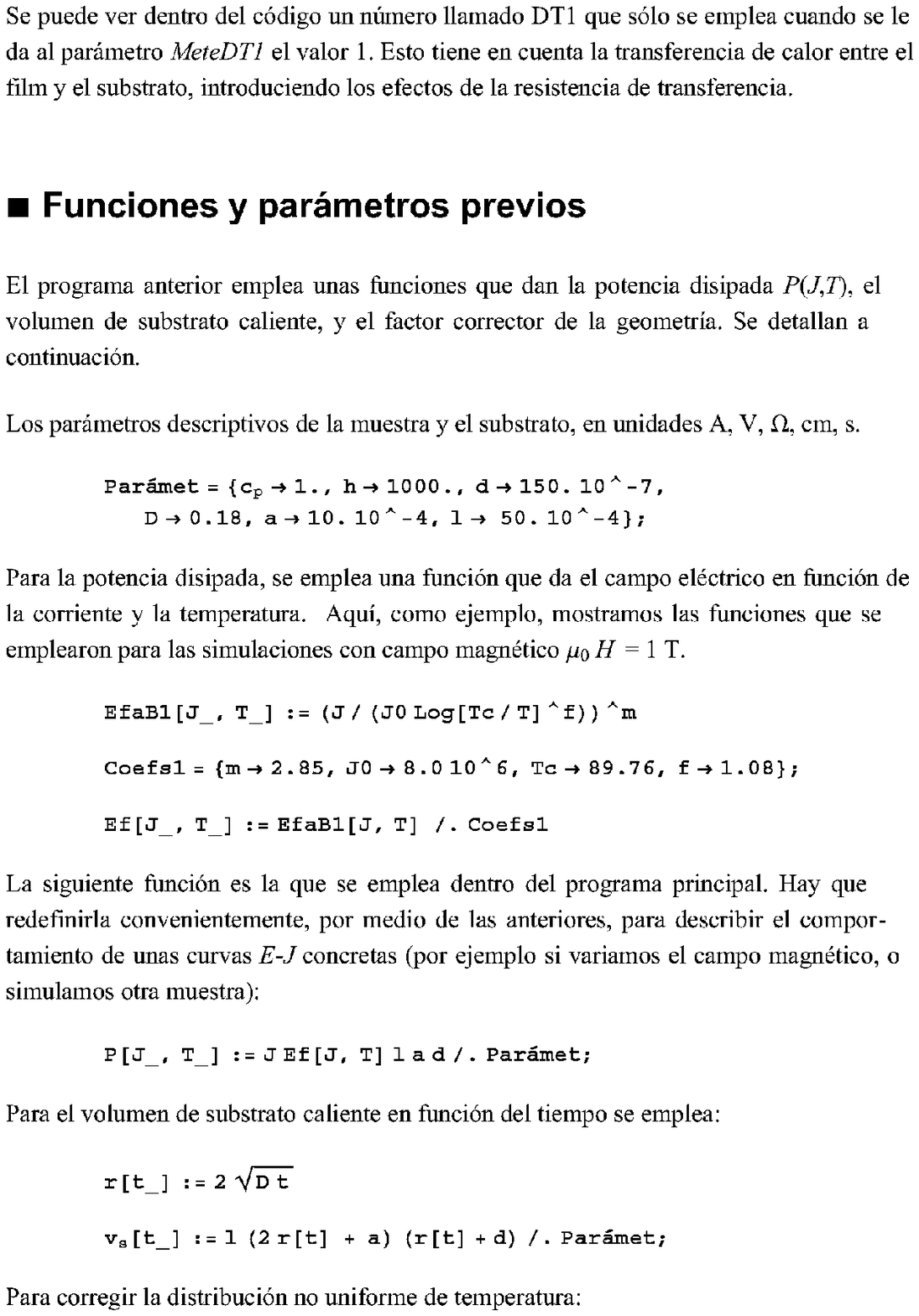}
\newpage
\includegraphics[width=\textwidth]{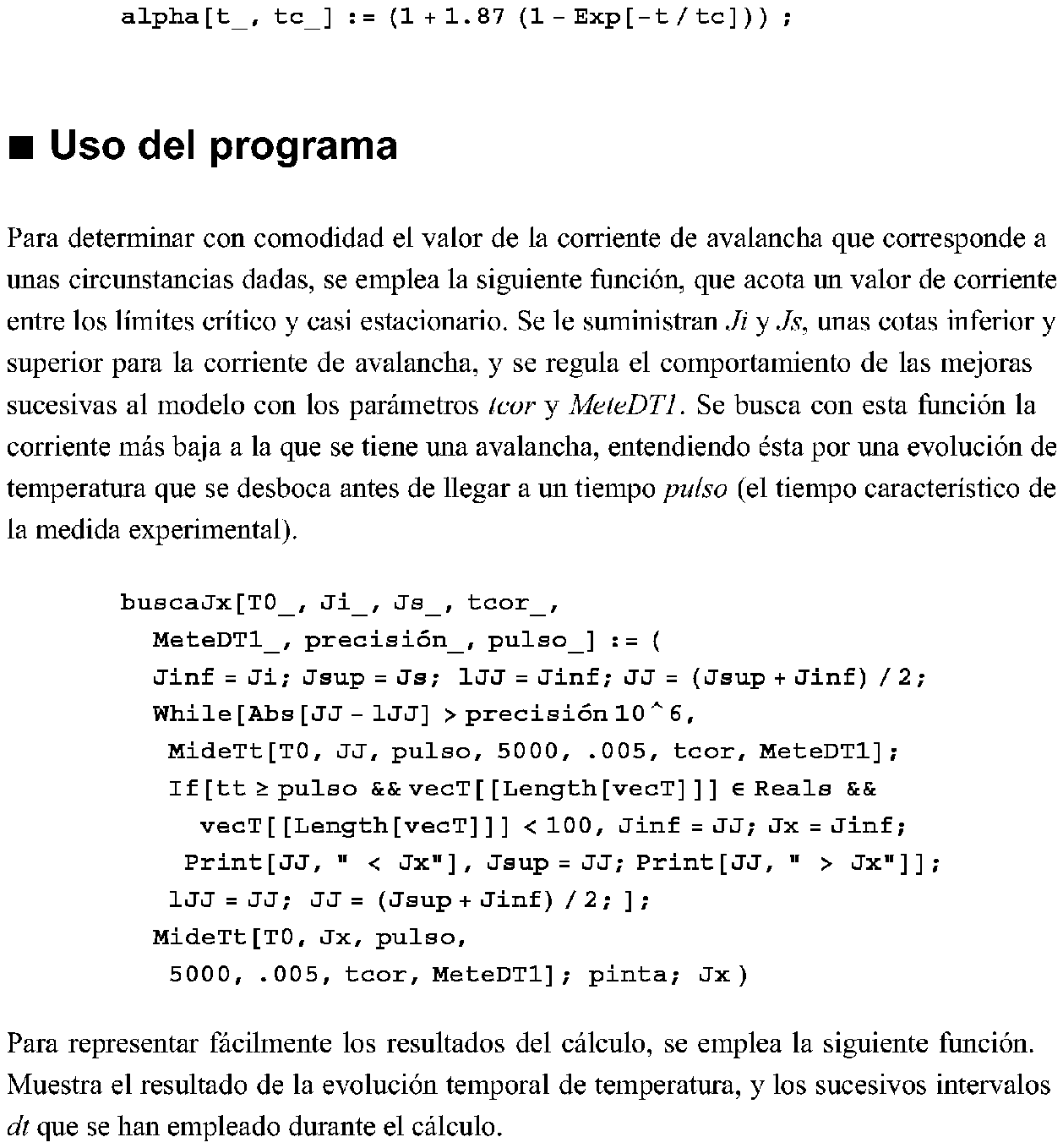}
\newpage
\includegraphics[width=\textwidth]{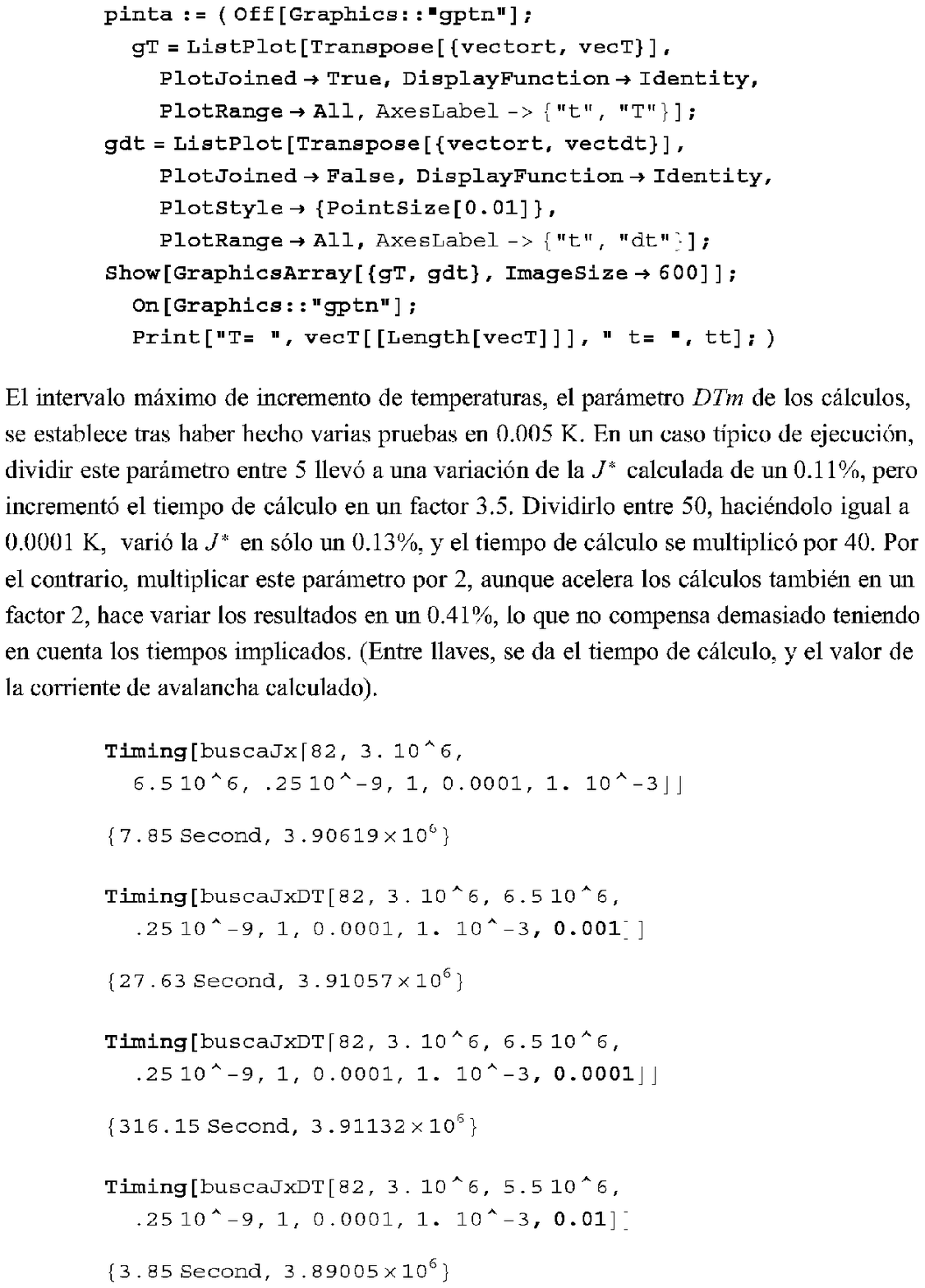}
\newpage
\includegraphics[width=\textwidth]{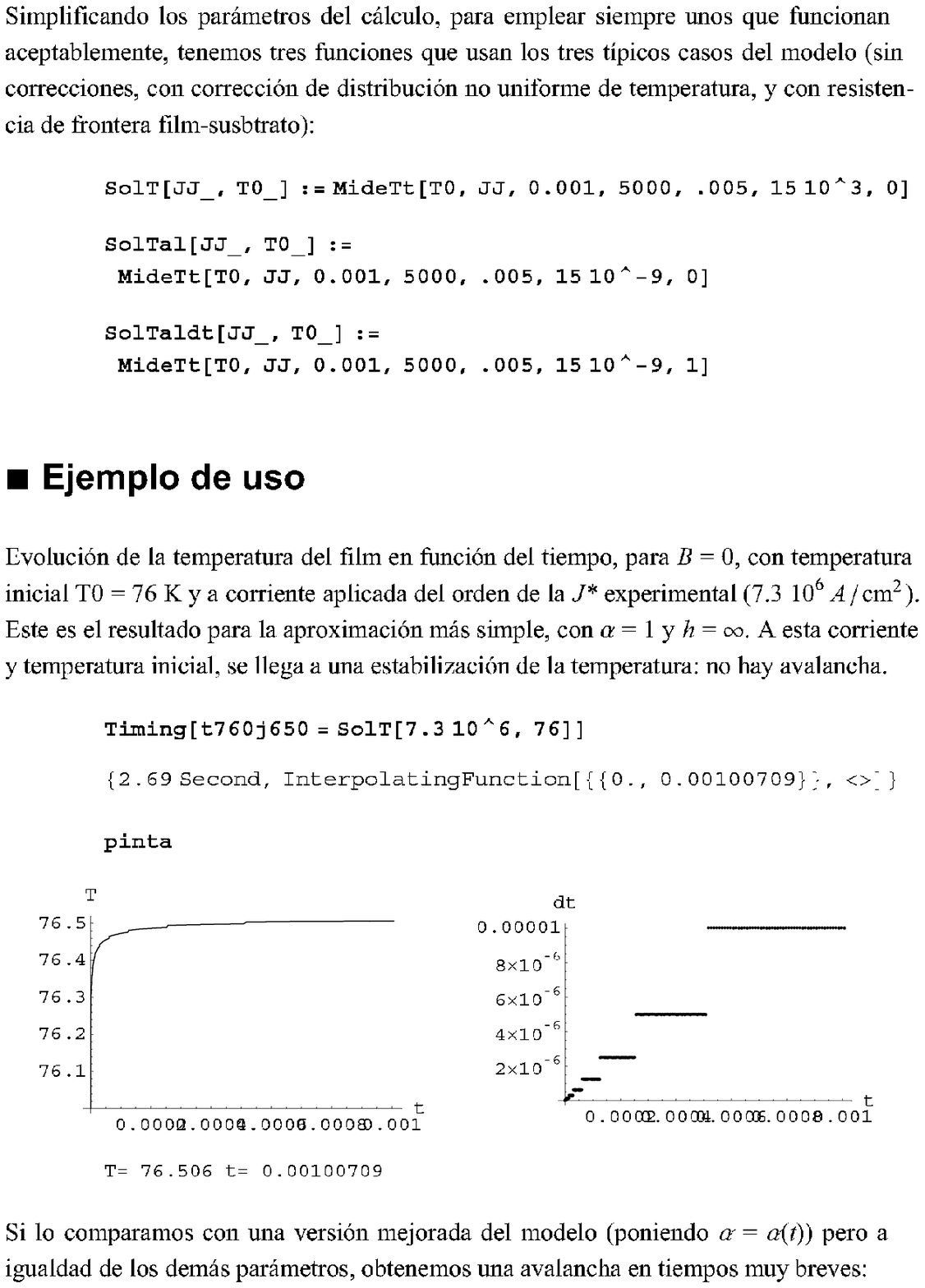}
\newpage
\includegraphics[width=\textwidth]{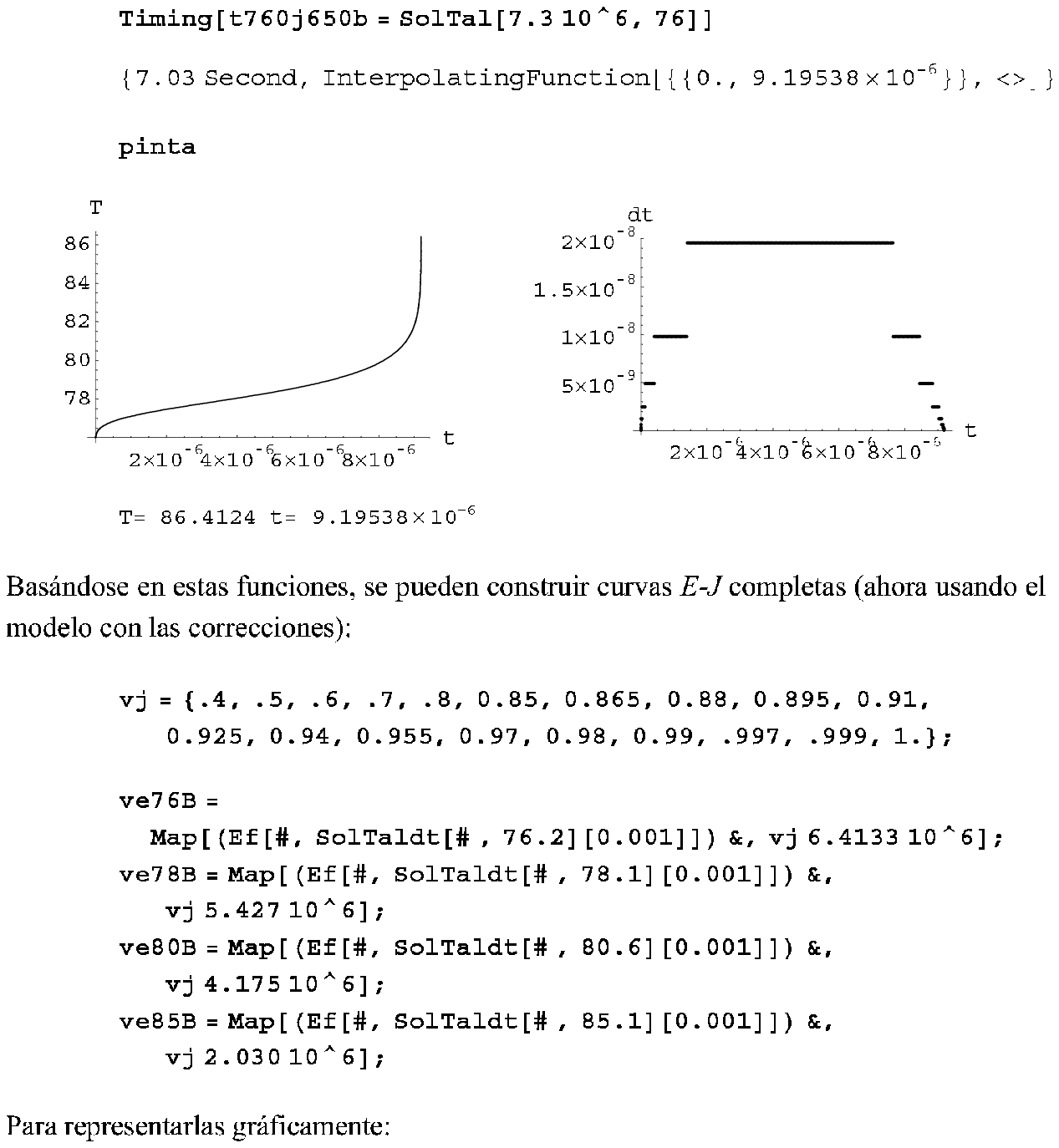}
\newpage
\includegraphics[width=\textwidth]{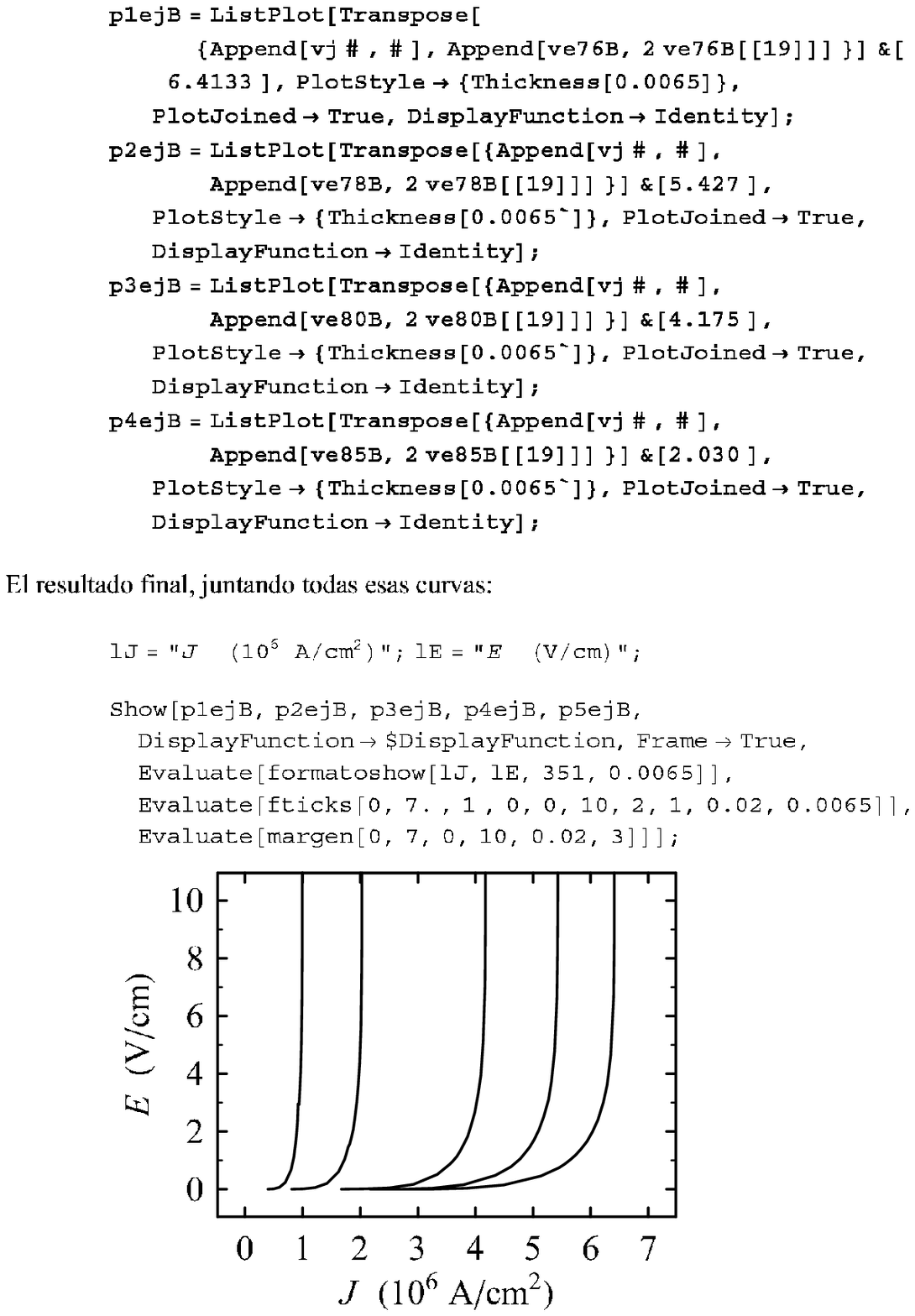}
\newpage
\includegraphics[width=\textwidth]{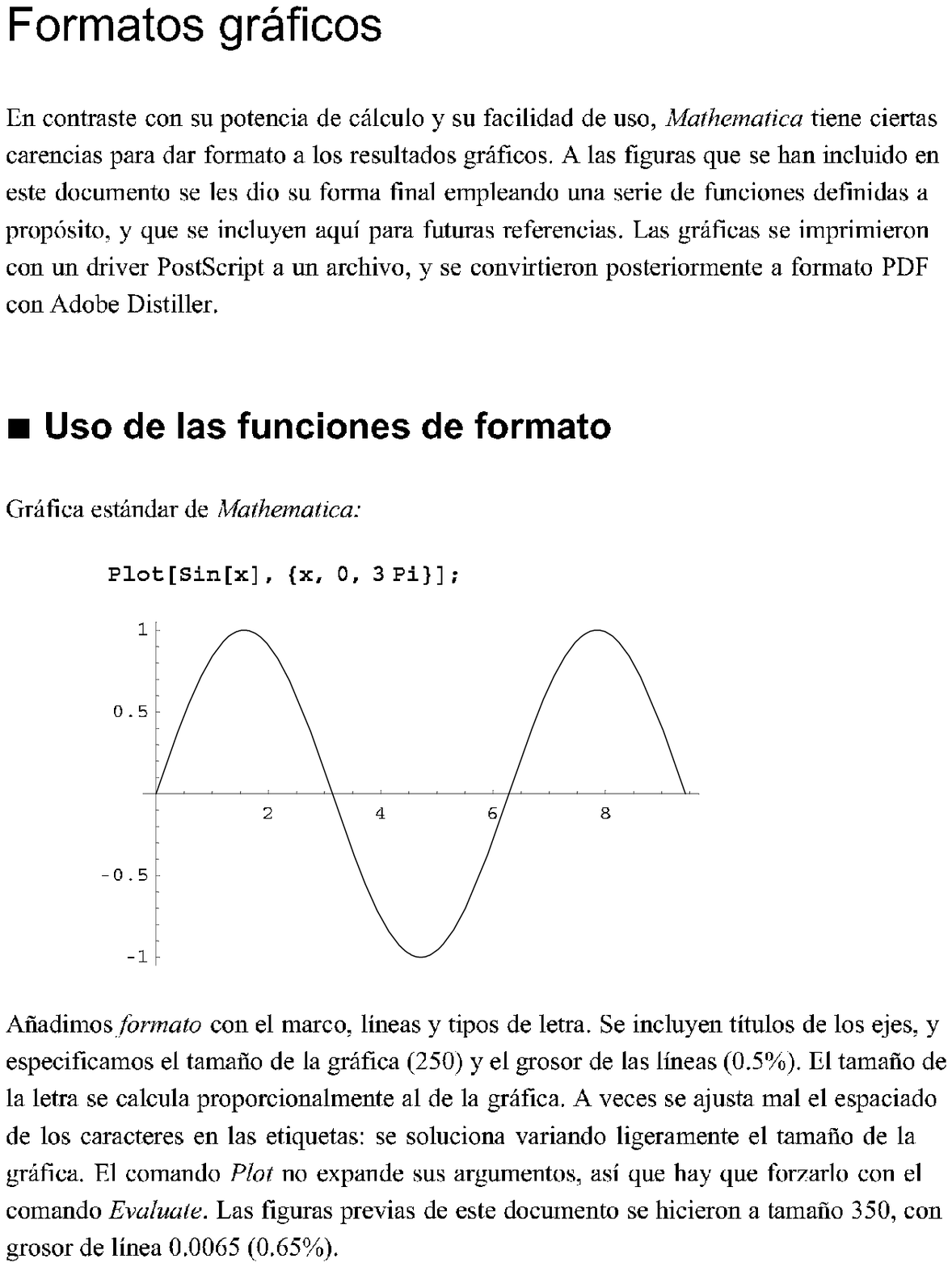}
\newpage
\includegraphics[width=\textwidth]{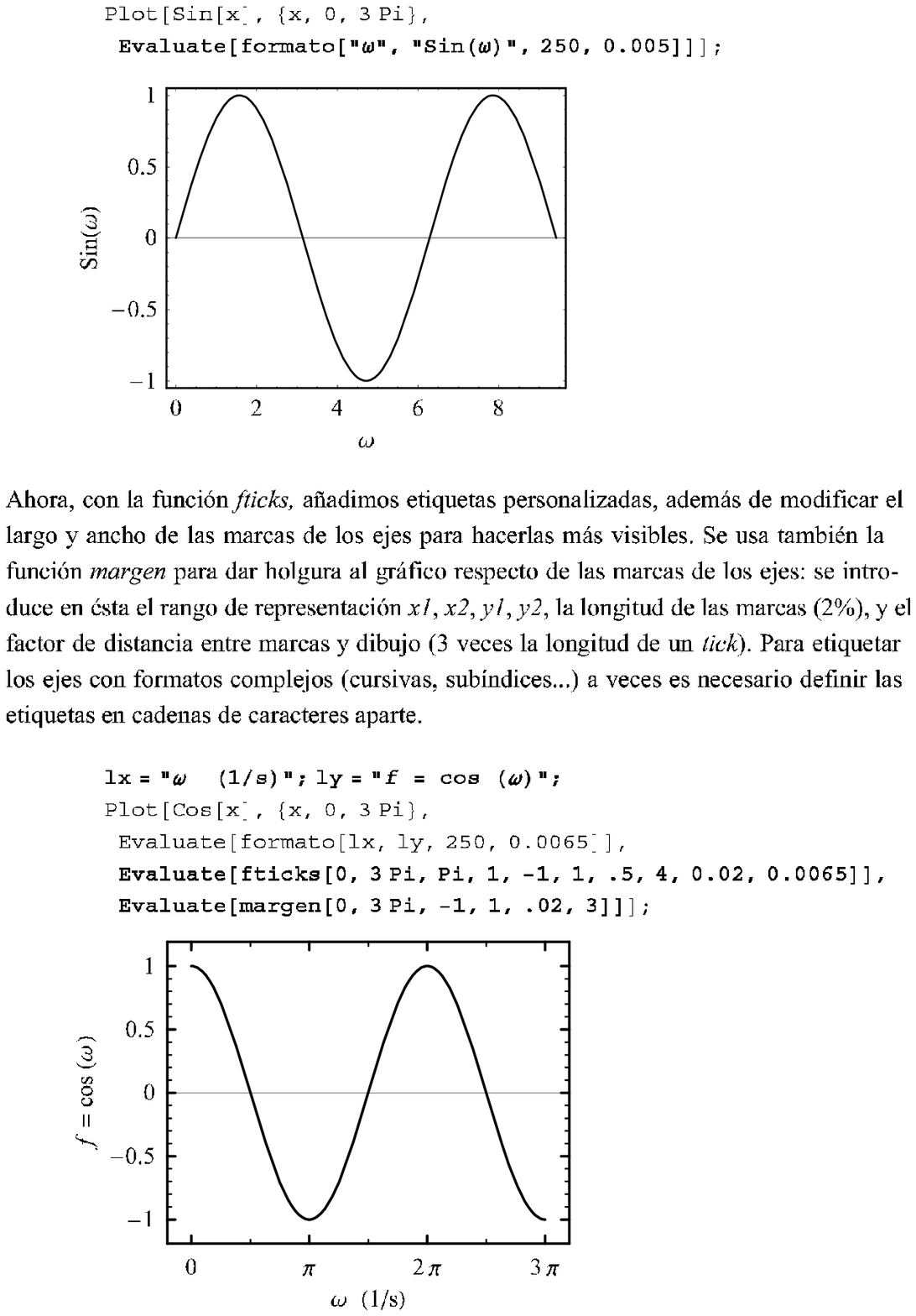}
\newpage
\includegraphics[width=\textwidth]{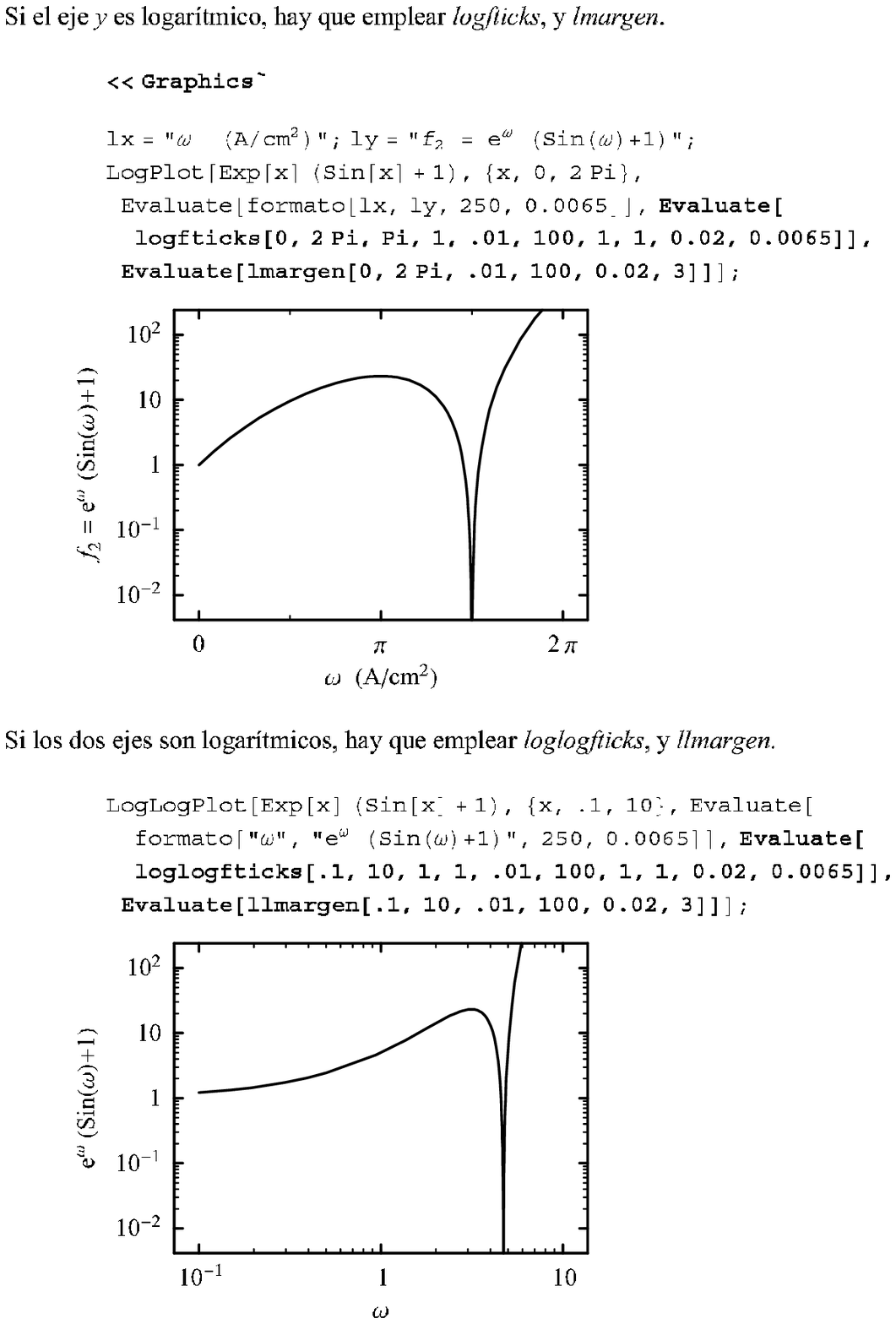}
\newpage
\includegraphics[width=\textwidth]{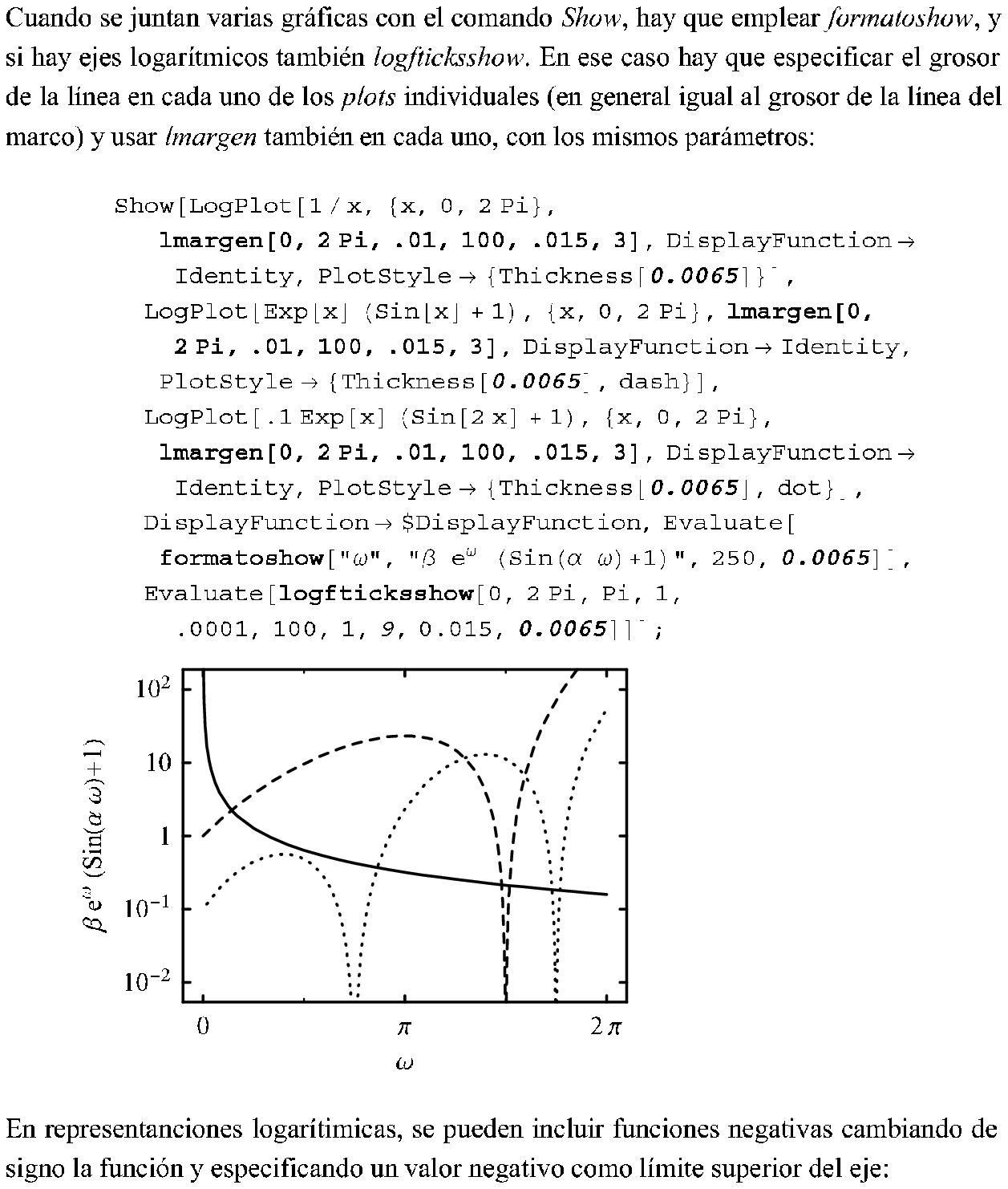}
\newpage
\includegraphics[width=\textwidth]{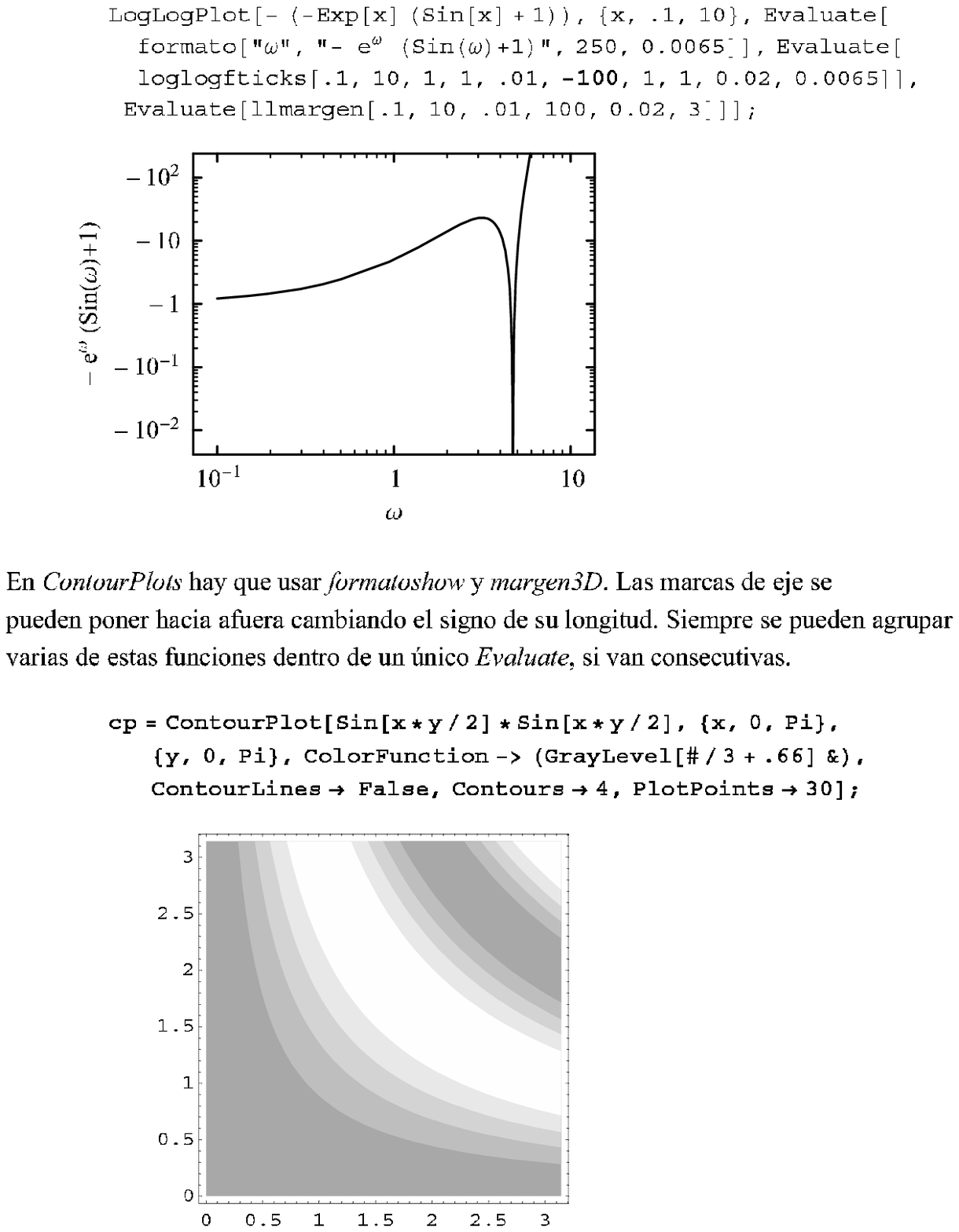}
\newpage
\includegraphics[width=\textwidth]{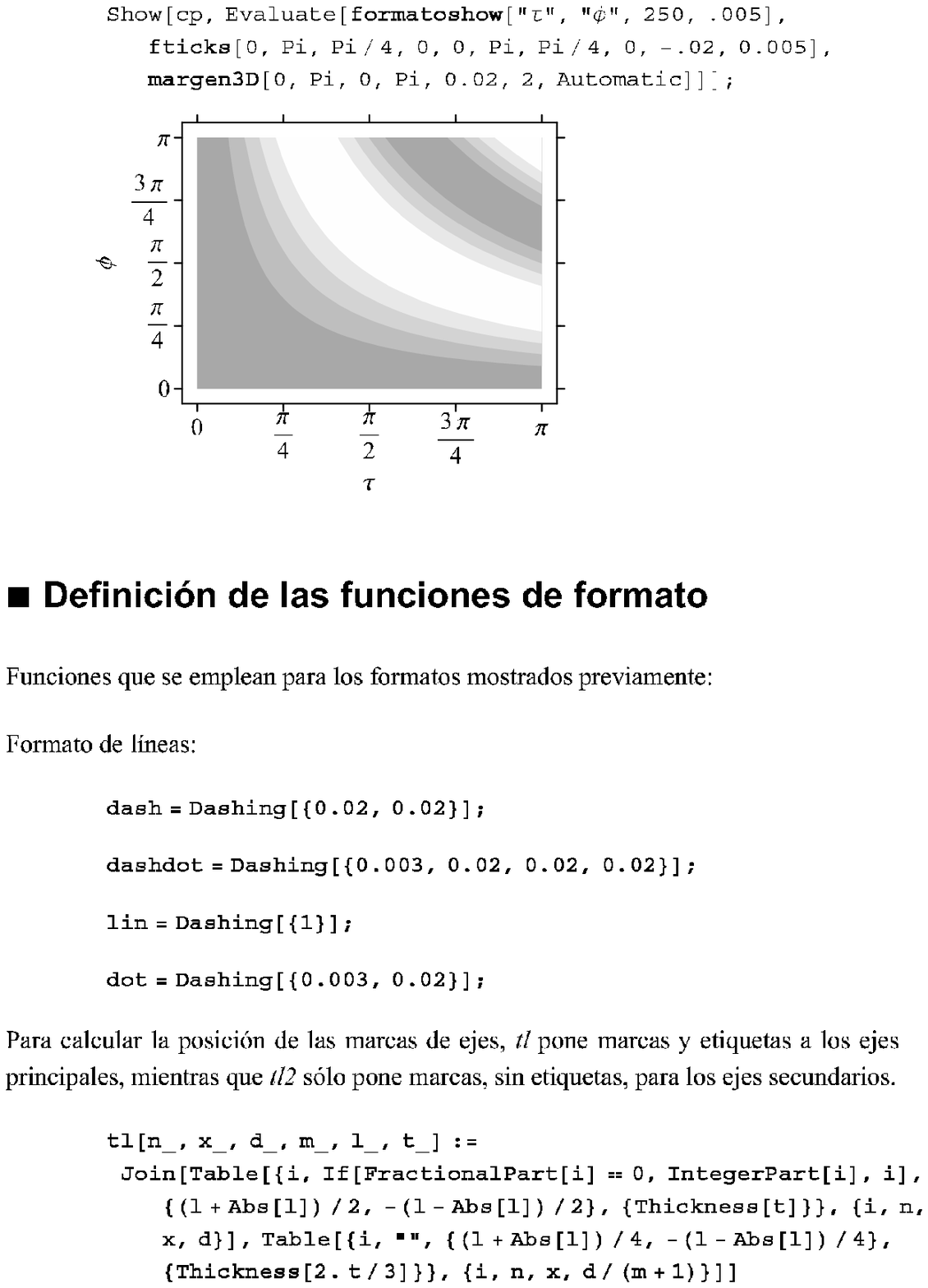}
\newpage
\includegraphics[width=\textwidth]{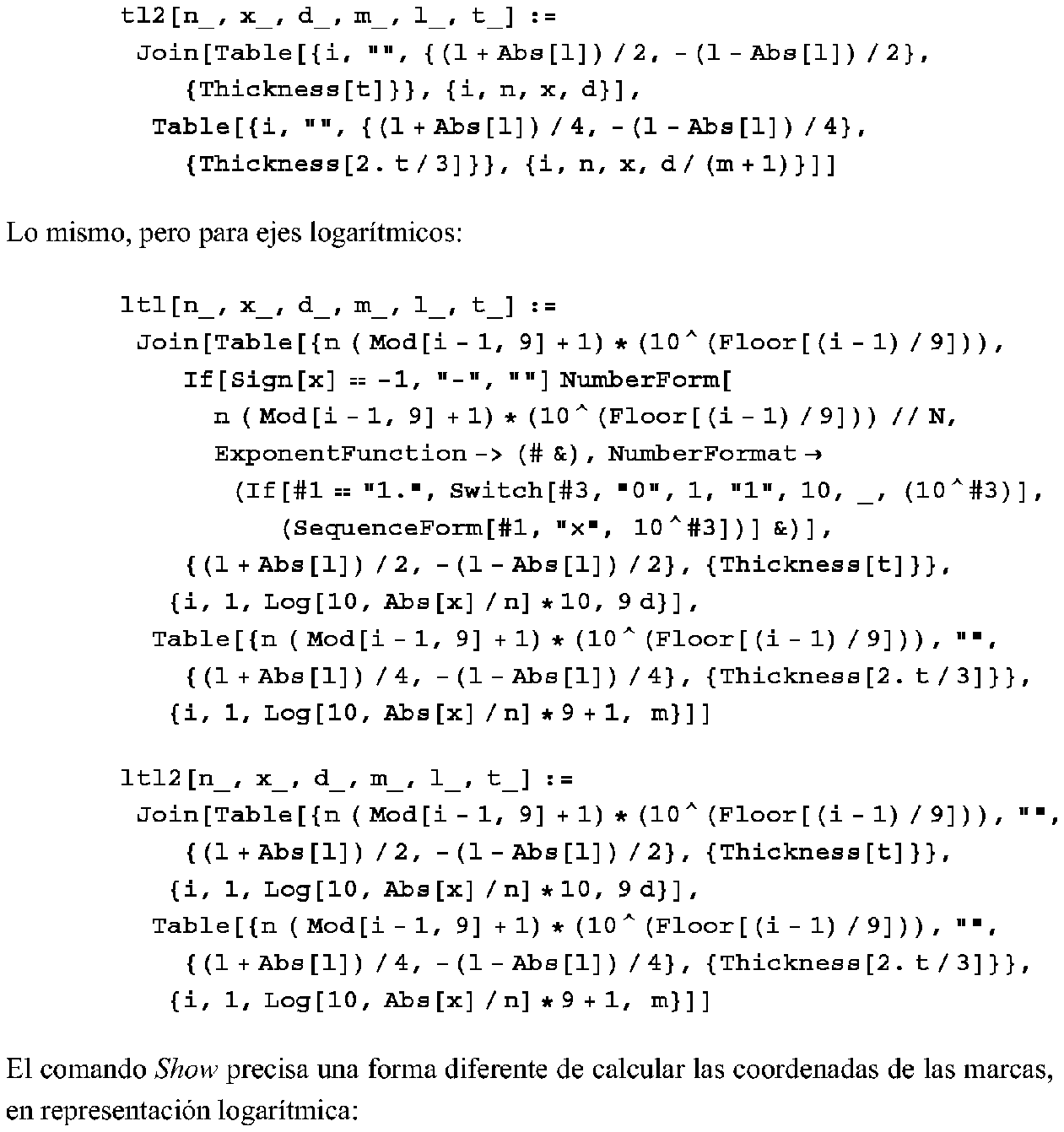}
\newpage
\includegraphics[width=\textwidth]{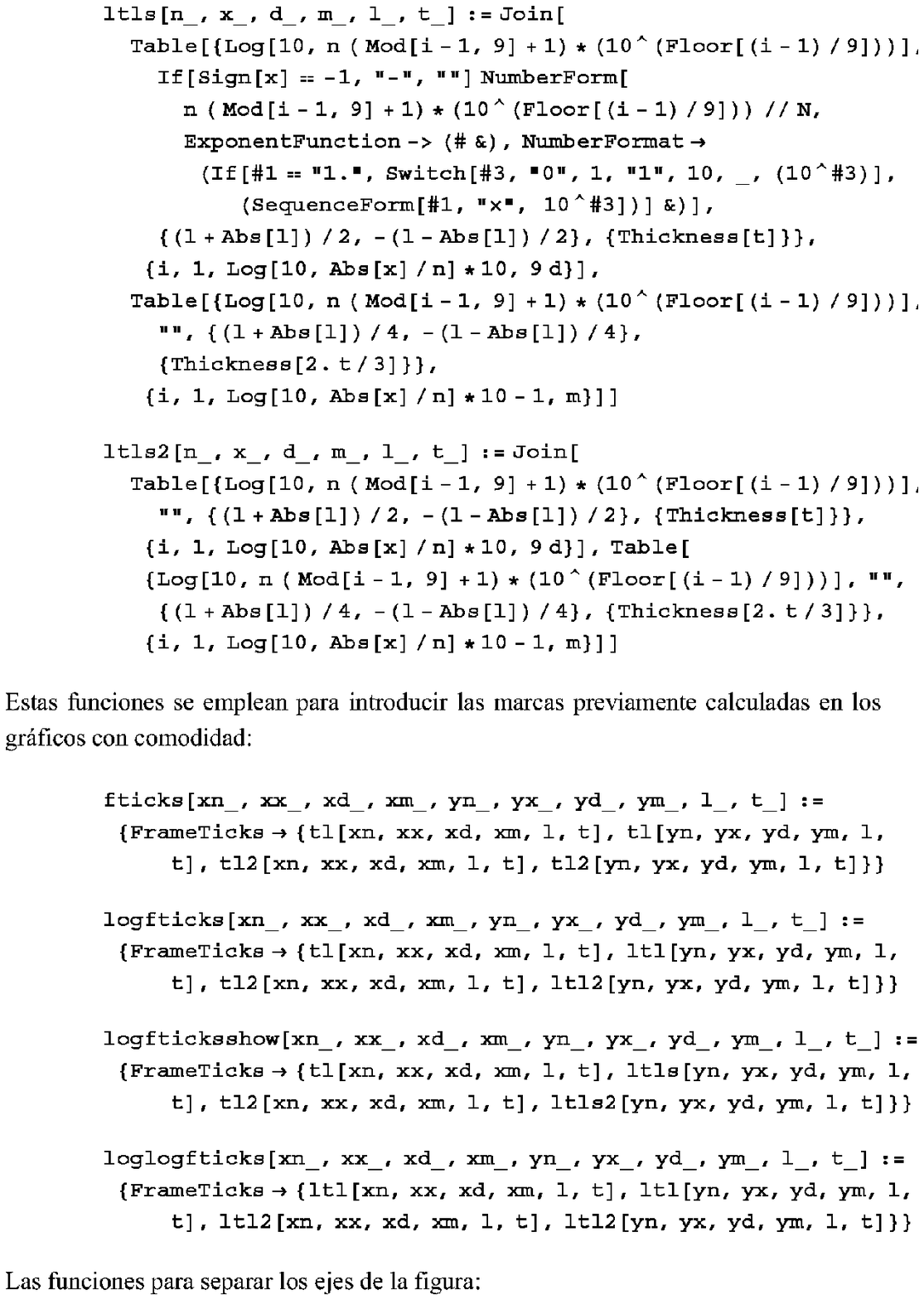}
\newpage
\includegraphics[width=\textwidth]{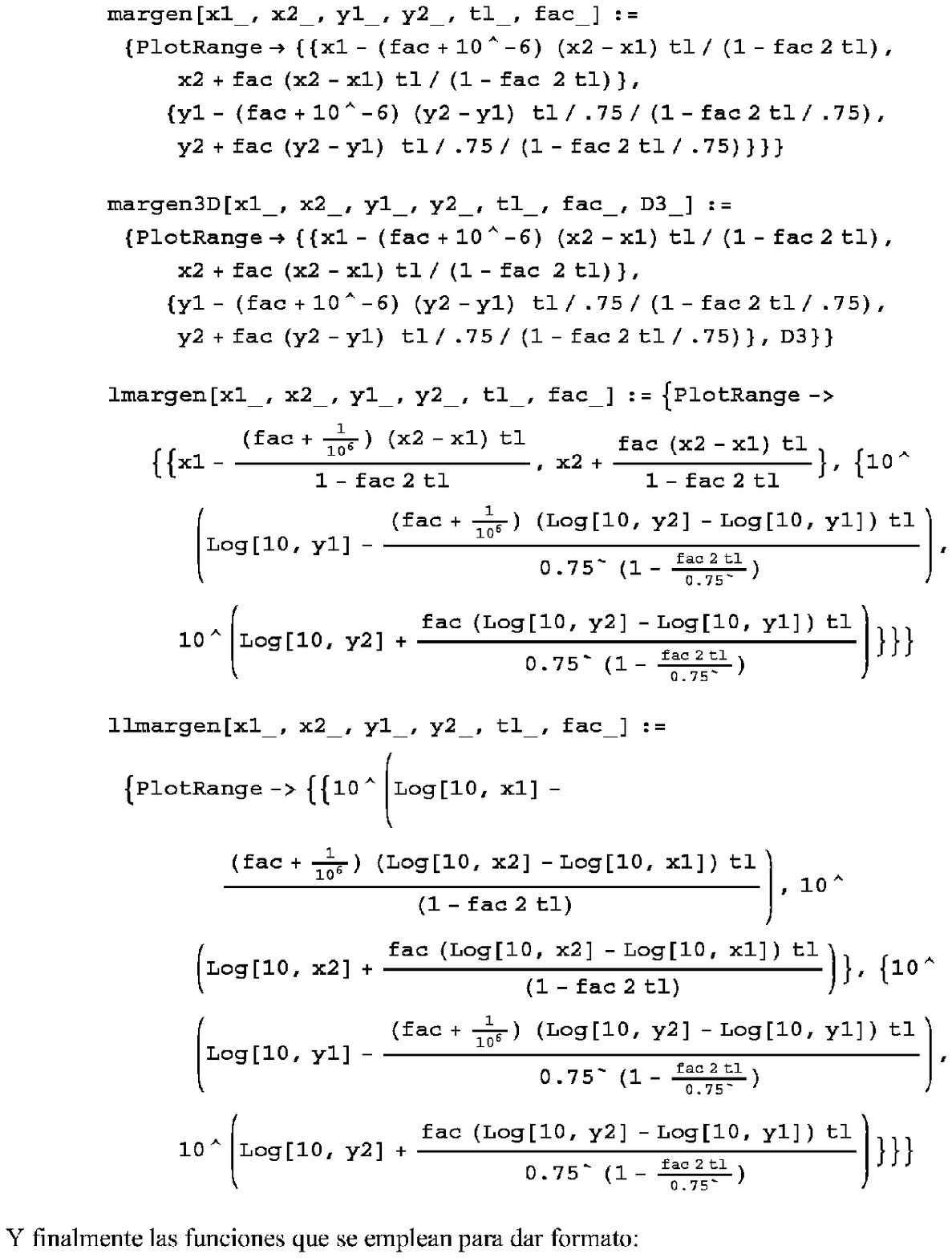}
\newpage
\includegraphics[width=\textwidth]{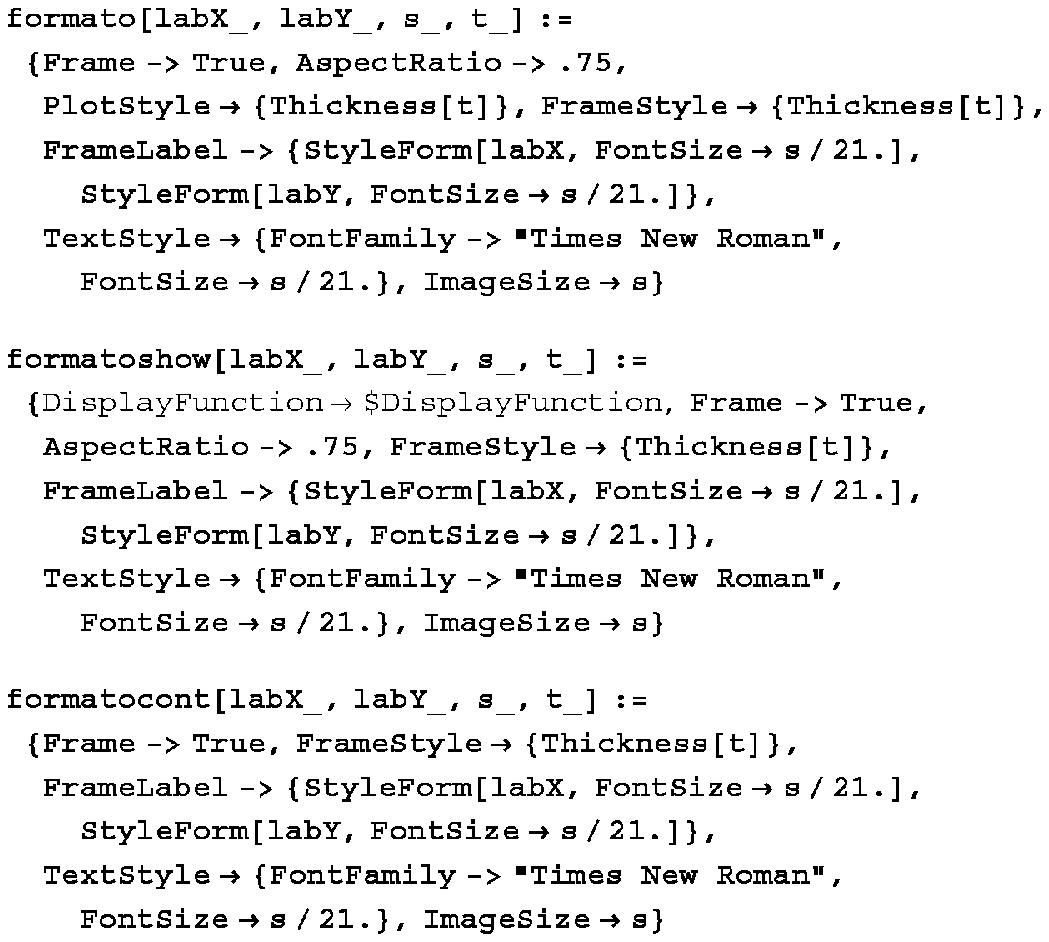}

% \addtocounter{page}{18} % sumar tantas p\'aginas como se vayan a insertar luego
\clearemptydoublepage

\setlength{\parskip}{0ex plus 0.5ex} \phantomsection
\addcontentsline{toc}{chapter}{Bibliograf\'{\i}a}
\bibliographystyle{hispacit}   % Estilo est\'andar, pero en espa\~nol
\bibliography{} %otro/jose,otro/DataTex

\begin{thebibliography}{100}

\bibitem{Bednorz86}
J.~G. Bednorz y K.~A. M{\"u}ller.
\newblock Z. Phys. B {\bf 64}, 189 (1986).

\bibitem{Ginsberg89}
D.~M. Ginsberg (ed.).
\newblock {\em Physical Properties of High Temperature Superconductors\/}
  (World Scientific, Singapur, 1989).

\bibitem{Narlikar}
A.~V. Narlikar (ed.).
\newblock {\em Studies of High Temperature Superconductors (Advances in
  Research and Applications)\/}, tomos 13, 22, 31, 32 (Nova Science, New York,
  1994, 1997, 2000).

\bibitem{Poole95}
C.~P. Poole, Jr., H.~A. Farach y R.~J. Creswick.
\newblock {\em Superconductivity\/} (Academic Press, San Diego, 1995).

\bibitem{Klein85}
W.~Klein, R.~P. Huebener, S.~Gauss y J.~Parisi.
\newblock J. Low Temp. Phys. {\bf 61}, 413  (1985).

\bibitem{Kunchur02}
M.~N. Kunchur.
\newblock Phys. Rev. Lett. {\bf 89}, 137005 (2002).
\newblock {cond-mat/0208256}.

\bibitem{Doettinger94}
S.~G. Doettinger, R.~P. Huebener, R.~Gerdemann, A.~K{\"u}hle, S.~Anders, T.~G.
  Tr{\"a}uble y J.~C. Vill{\'e}gier.
\newblock Phys. Rev. Lett. {\bf 73}, 1691  (1994).

\bibitem{Xiao99}
Z.~L. Xiao, P.~Voss-de Haan, G.~Jakob, T.~Kluge, P.~Haibach, H.~Adrian y E.~Y.
  Andrei.
\newblock Phys. Rev. B {\bf 59}, 1481  (1999).

\bibitem{Chiaverini00}
J.~Chiaverini, J.~N. Eckstein, I.~Bozovic, S.~Doniach y A.~Kapitulnik.
\newblock {\em Nonlinear I-V, high flux-flow velocity instability and evidence
  for quantum fluctuations in {BSCCO} superconducting films\/} (2000).
\newblock {cond-mat/0007479}.

\bibitem{Curras01}
S.~R. Curr{\'a}s, P.~Wagner, M.~Ruibal, J.~Vi{\~n}a, M.~R. Osorio, M.~T.
  Gonz{\'a}lez, J.~A. Veira, J.~Maza y F.~Vidal.
\newblock Supercond. Sci. Technol. {\bf 14}, 748  (2001).

\bibitem{Curras00b}
S.~Rodr{\'{\i}}guez~Curr{\'a}s.
\newblock {\em Contribuci{\'o}n al estudio de la interrelaci{\'o}n entre la
  densidad de corriente cr\'{\i}tica y la resistividad normal en muestras
  granulares y pel\'{\i}culas delgadas de {YBa$_2$Cu$_3$O$_{7-\delta}$}\/}.
\newblock Tesis Doctoral, Universidade de Santiago de Compostela (2000).

\bibitem{Gonzalez02b}
M.~Gonz{\'a}lez, S.~Vidal, J.~Vi{\~n}a, M.~R. Osorio, J.~Maza y F.~Vidal.
\newblock Physica C {\bf 372-376}, 1852 (2002).

\bibitem{Gonzalez02}
M.~T. Gonz{\'a}lez, J.~Vi{\~n}a, S.~R. Curr{\'a}s, J.~Maza y F.~Vidal.
\newblock {\em Normal-superconducting transition induced by high current
  densities in YBa$_2$Cu$_3$O$_{7-\delta}$ melt-textured samples and thin
  films: Similarities and differences\/} (2002).
\newblock En tr\'amites de publicaci\'on.
\newblock {cond-mat/0305020}.

\bibitem{Gonzalez03}
M.~T. Gonz{\'a}lez.
\newblock {\em On the electrical transport properties of YBa$_2$Cu$_3$O$_{7-\delta}$
  superconductors textured by top seeded melt growth\/}.
\newblock Tesis Doctoral, Universidade de Santiago de Compostela (2003).

\bibitem{Xiao96}
Z.~L. Xiao y P.~Ziemann.
\newblock Phys. Rev. B {\bf 53}, 15265  (1996).

\bibitem{Doettinger95}
S.~G. Doettinger, R.~P. Huebener y A.~K{\"u}hle.
\newblock Physica C {\bf 251}, 285  (1995).

\bibitem{Lefloch99}
F.~Lefloch, C.~Hoffmann y O.~Demolliens.
\newblock Physica C {\bf 319}, 258  (1999).
\newblock {cond-mat/9903060}.

\bibitem{Xiao98}
Z.~L. Xiao, P.~V. de~Haan, G.~Jakob y H.~Adrian.
\newblock Phys. Rev. B {\bf 57}, R736 (1998).

\bibitem{Pauly00}
M.~Pauly, R.~Ballou, G.~Fillion y J.-C. Vill{\'e}gier.
\newblock Physica B {\bf 284-288}, 721 (2000).

\bibitem{Kamm00}
F.-M. Kamm, A.~Plettl y P.~Ziemann.
\newblock Physica B {\bf 284-288}, 561 (2000).

\bibitem{Onnes11}
H.~K. Onnes.
\newblock Leiden Commun. {\bf 120b, 122b, 124c} (1911).

\bibitem{London35}
F.~London y H.~London.
\newblock Proc. Roy. Soc. (London) {\bf A155}, 71 (1935).

\bibitem{Meissner33}
W.~Meissner y R.~Ochsenfeld.
\newblock Naturwissenschaften {\bf 21}, 787 (1933).

\bibitem{Ginzburg50}
V.~L. Ginzburg y L.~D. Landau.
\newblock Zh. Eksp. Teor. Fiz. {\bf 20}, 1064 (1950).

\bibitem{Bardeen57}
J.~Bardeen, L.~N. Cooper y J.~R. Schrieffer.
\newblock Phys. Rev. {\bf 108}, 1175 (1957).

\bibitem{Cooper56}
L.~N. Cooper.
\newblock Phys. Rev {\bf 104}, 1189 (1956).

\bibitem{Abrikosov57}
A.~A. Abrikosov.
\newblock Sov. Phys. JETP {\bf 5}, 1174 (1957).

\bibitem{Rose-Innes78}
A.~C. Rose-Innes y E.~H. Rhoderick.
\newblock {\em Introduction to Superconductivity\/} (Pergamon Press, Oxford,
  1978), 2\raise1ex\hbox{\scriptsize a} ed.

\bibitem{Blatter94}
G.~Blatter, M.~V. Feigel'man, V.~B. Geshkenbein, A.~I. Larkin y V.~M. Vinokur.
\newblock Rev. Mod. Phys. {\bf 66}, 1125  (1994).

\bibitem{Kunchur93}
M.~N. Kunchur y D.~K. Christen.
\newblock Phys. Rev. Lett. {\bf 70}, 998 (1993).

\bibitem{Tinkham96}
M.~Tinkham.
\newblock {\em Introduction to Superconductivity\/} (McGraw-Hill, New York,
  1996), 2\raise1ex\hbox{\scriptsize a} ed.

\bibitem{Samoilov95}
A.~V. Samoilov, M.~Konczykowski, N.-C. Yeh, S.~Berry y C.~C. Tsuei.
\newblock Phys. Rev. Lett. {\bf 75}, 4118  (1995).

\bibitem{Ruck97}
B.~J. Ruck, J.~C. Abele, H.~J. Trodahl, S.~A. Brown y P.~Lynam.
\newblock Phys. Rev. Lett. {\bf 78}, 3378 (1997).

\bibitem{Montgomery71}
H.~C. Montgomery.
\newblock J. Appl. Phys. {\bf 42}, 2971 (1971).

\bibitem{Salgueiro96}
J.~R. Salgueiro~Pi{\~n}eiro.
\newblock {\em Escritura l\'aser directa en la fabricaci\'on de m\'ascaras para
  \'optica integrada\/}.
\newblock Tesina de Graduaci{\'o}n, \'Area de \'Optica, Dpto. de F\'{\i}sica
  Aplicada. Universidade de Santiago de Compostela. (1996).

\bibitem{Moreau89}
W.~M. Moreau.
\newblock {\em Semiconductor Lithography. Principles, Practices and
  Materials.\/} (Plenum Press, New York, 1989), 2\raise1ex\hbox{\scriptsize a}
  ed.

\bibitem{Curras95}
S.~Rodr{\'{\i}}guez~Curr{\'a}s.
\newblock {\em Crecimiento de pel\'{\i}culas delgadas de
  YBa$_2$Cu$_3$O$_{7-\delta}$ por pulverizaci\'on cat\'odica y su
  caracterizaci\'on estructural\/}.
\newblock Tesina de Graduaci{\'o}n, Facultade de F\'{\i}sica, Universidade de
  Santiago de Compostela. (1995).

\bibitem{Bhatt94}
D.~Bhatt, S.~N. Basu, A.~C. Westerheim y A.~C. Anderson.
\newblock Physica C {\bf 222}, 283 (1994).

\bibitem{Catana93}
A.~Catana, J.~G. Bednorz, C.~Gerber, J.~Mannhardt y D.~G. Schlom.
\newblock Appl. Phys. Lett. {\bf 63}, 553 (1993).

\bibitem{Loquet93}
J.~P. Loquet, Y.~Jaccard, C.~Gerber y E.~M{\"a}chler.
\newblock Appl. Phys. Lett. {\bf 63}, 10 (1993).

\bibitem{Kim92}
D.~H. Kim, D.~J. Miller, J.~C. Smith, R.~A. Holbolf, J.~H. Kang y
  J.~Talvacchio.
\newblock Phys. Rev. B {\bf 44}, 7607 (1992).

\bibitem{Wuyts92}
B.~Wuyts, Z.~X. Gao, S.~Libbrecht, M.~Maenhoudt, E.~Osquiguil y
  Y.~Bruynseraede.
\newblock Physica C {\bf 203}, 235 (1992).

\bibitem{Gavaler91}
J.~R. Gavaler, J.~Talvacchio, T.~T. Braggins, M.~G. Forrester y J.~Greggi.
\newblock J. Appl. Phys. {\bf 70}, 4383 (1991).

\bibitem{Habermeier91}
H.-U. Habermeier, G.~Beddies, B.~Leibold, G.~Lu y G.~Wagner.
\newblock Physica C {\bf 180}, 17 (1991).

\bibitem{Li95}
X.~Y. Li, N.~J. Wu, K.~X. abd J.~S.~Liu, H.~Lin, T.~Q. Huang y A.~Ignatiev.
\newblock Physica C {\bf 248}, 281 (1995).

\bibitem{Curras02}
S.~R. Curr{\'a}s, J.~Vi{\~n}a, M.~Ruibal, M.~T. Gonz{\'a}lez, M.~R. Osorio,
  J.~Maza, J.~A. Veira y F.~Vidal.
\newblock Physica C {\bf 372-376}, 1095 (2002).

\bibitem{Nahum91}
M.~Nahum, S.~Verghese, P.~L. Richards y K.~Char.
\newblock Appl. Phys. Lett. {\bf 59}, 2034 (1991).

\bibitem{Mosqueira93}
J.~Mosqueira, O.~Cabeza, M.~Fran{\c c}ois, C.~Torr{\'o}n y F.~Vidal.
\newblock Supercond. Sci. Technol. {\bf 6}, 584  (1993).

\bibitem{Mosqueira93b}
J.~Mosqueira, O.~Cabeza, M.~Fran{\c c}ois y F.~Vidal.
\newblock En {\em {ICMAS}-93 Superconducting Materials\/} (J. Etourneau, J. B. Torrance and H. Yamauchi, eds.)
\newblock pp. 285 -- 290 
\newblock (Institute for Industrial Technology Transfer, Gournay-sur-Marne, 1993).

\bibitem{Mosqueira94}
J.~Mosqueira, O.~Cabeza, F.~Migu{\'e}lez, M.~Fran{\c c}ois y F.~Vidal.
\newblock Physica C {\bf 235-240}, 2105  (1994).

\bibitem{NIan007}
{National Instruments}.
\newblock {\em Data Acquisition {(DAQ)} Fundamentals\/} (AN007).
\newblock Application Note 007, http://www.ni.com.

\bibitem{Wordenweber91}
R.~W{\"o}rdenweber, M.~O. Abd-El-Hamed, J.~Schneider y O.~Laborde.
\newblock J. Appl. Phys. {\bf 70}, 2230 (1991).

\bibitem{Wen01}
H.~H. Wen, S.~L. Li, G.~H. Chen y X.~S. Ling.
\newblock Phys. Rev. B {\bf 64}, 054507 (2001).

\bibitem{Hettinger89}
J.~D. Hettinger, A.~G. Swanso, W.~J. Skocpol, J.~S. Brooks, J.~M. G. P.~M.
  Mankiewich, R.~E. Howard, B.~L. Straughn y E.~G. Burkhardt.
\newblock Phys. Rev. Lett. {\bf 62}, 2044 (1989).

\bibitem{Zeldov90}
E.~Zeldov, N.~M. Amer, G.~Koren, A.~Gupta y M.~W. McElfresh.
\newblock App. Phys. Lett. {\bf 56}, 680  (1990).

\bibitem{Bernstein90}
P.~Bernstein, S.~Lamarti y J.~Bok.
\newblock Solid State Comm. {\bf 75}, 587 (1990).

\bibitem{Wordenweber90}
R.~W{\"o}rdenweber, M.~O. Abd-El-Hamed y J.~Scheneider.
\newblock Physica C {\bf 171}, 1 (1990).

\bibitem{Tinkham91}
M.~Tinkham.
\newblock Physica B {\bf 169}, 66 (1991).

\bibitem{Gupta93}
S.~K. Gupta, P.~Berdahl, R.~E. Russo, G.~Brice{\~n}o y A.~Zettl.
\newblock Physica C {\bf 206}, 335  (1993).

\bibitem{Roberts94}
J.~M. Roberts, B.~Brown, B.~A. Hermann y J.~Tate.
\newblock Phys. Rev. B {\bf 49}, 6890 (1994).

\bibitem{Woltgens95}
P.~J.~M. W{\"o}ltgens, C.~Dekker, R.~H. Koch, B.~W. Hussey y A.~Gupta.
\newblock Phys. Rev. B {\bf 52}, 4536 (1995).

\bibitem{Kilic98}
A.~Kili{\c c}, K.~Kili{\c c}, S.~Senoussi y K.~Demir.
\newblock Physica C {\bf 294}, 203 (1998).

\bibitem{Prester98}
M.~Prester.
\newblock Supercond. Sci. Technol. {\bf 11}, 333  (1998).

\bibitem{Wang00}
Z.~H. Wang, H.~Zhang y X.~W. Cao.
\newblock Physica C {\bf 337}, 62 (2000).

\bibitem{Landau00}
I.~L. Landau y H.~R. Ott.
\newblock Physica C {\bf 331}, 1 (2000).

\bibitem{Landau01}
I.~L. Landau y H.~R. Ott.
\newblock Phys. Rev. B {\bf 63}, 184516 (2001).

\bibitem{Landau02}
I.~L. Landau y H.~R. Ott.
\newblock Phys. Rev. B {\bf 65}, 064511 (2002).

\bibitem{Xiao98b}
Z.~L. Xiao, E.~Y. Andrei y P.~Ziemann.
\newblock Phys. Rev. B {\bf 58}, 11185  (1998).

\bibitem{Skocpol74}
W.~J. Skocpol, M.~R. Beasley y M.~Tinkham.
\newblock J. Appl. Phys. {\bf 45}, 4054  (1974).

\bibitem{Skokov93}
V.~N. Skokov y V.~P. Koverda.
\newblock Cryogenics {\bf 33}, 1072  (1993).

\bibitem{Jelila98}
F.~S. Jelila, J.-P. Maneval, F.-R. Landan, F.~Chibane, A.~Marie-de Ficquelmont,
  L.~M{\'e}chin, J.-C. Vill{\'e}gier, M.~Aprili y J.~Lesueur.
\newblock Phys. Rev. Lett. {\bf 81}, 1933  (1998).

\bibitem{Maneval01}
J.-P. Maneval, F.~Boyer, K.~Harrabi y F.-R. Ladan.
\newblock Journal of Superconductivity {\bf 14}, 347  (2001).

\bibitem{Reymond02}
S.~Reymond, L.~Antognazza, M.~Decroux, E.~Koller, P.~Reinert y {\O}.~Fischer.
\newblock Phys. Rev. B {\bf 66}, 014522 (2002).

\bibitem{Yang99}
C.~Yang, O.~Miura, D.~Ito, M.~Morita y Tokunaga.
\newblock IEEE Trans. Appl. Supercond. {\bf 9}, 1339  (1999).

\bibitem{Elschner99}
S.~Elschner, J.~Bock, G.~Brommer y L.~Cowey.
\newblock Appl. Supercond. {\bf 167}, 1029  (1999).

\bibitem{Tournier00}
R.~Tournier, E.~Beaugnon, O.~Belmont, X.~Chaud, D.~Bourgault, D.~Isfort,
  L.~Porcar y P.~Tixador.
\newblock Supercond. Sci. Technol. {\bf 13}, 886  (2000).

\bibitem{Larkin76}
A.~I. Larkin y Y.~N. Ovchinnikov.
\newblock Sov. Phys. JETP {\bf 41}, 960  (1976).

\bibitem{Bezuglyj92}
A.~I. Bezuglyj y V.~A. Shklovskij.
\newblock Physica C {\bf 202}, 234  (1992).

\bibitem{Doettinger97}
S.~G. Doettinger, S.~Kittelberger, R.~P. Huebener y C.~C. Tsuei.
\newblock Phys. Rev. B {\bf 56}, 14157 (1997).

\bibitem{Pla91}
O.~Pla y F.~Nori.
\newblock Phys. Rev. Lett. {\bf 67}, 919 (1991).

\bibitem{Xiao97}
Z.~L. Xiao y P.~Ziemann.
\newblock Physica C {\bf 282-287}, 2363  (1997).

\bibitem{Bassler98}
K.~E. Bassler y M.~Paczuski.
\newblock Phys. Rev. Lett. {\bf 81}, 3761 (1998).

\bibitem{Altshuler02}
E.~Altshuler, T.~H. Johansen, Y.~Paltiel, P.~Jing, K.~E. Bassler, O.~Ramos,
  G.~F. Reiter, E.~Zeldov y C.~W. Chu.
\newblock {\em Vortex avalanches and self organized criticallity in
  superconducting niobium\/} (2002).
\newblock {cond-mat/0208266 v2}.

\bibitem{Kleiner92}
R.~Kleiner, F.~Steinmeyer, G.~Kunkel y P.~M{\"u}ller.
\newblock Phys. Rev. Lett. {\bf 68}, 2394 (1992).

\bibitem{Kleiner94}
R.~Kleiner y P.~M{\"u}ller.
\newblock Phys. Rev. B {\bf 49}, 1327 (1994).

\bibitem{Kleiner94b}
R.~Kleiner, P.~M{\"u}ller, H.~Kohlstedt, N.~F. Pedersen y S.~Sakai.
\newblock Phys. Rev. B {\bf 50}, 3942 (1994).

\bibitem{Tanabe96}
K.~Tanabe, Y.~Hidaka, S.~Karimoto y M.~Suzuki.
\newblock Phys. Rev. B {\bf 53}, 9348 (1996).

\bibitem{Koshelev94}
A.~E. Koshelev y V.~M. Vinokur.
\newblock Phys. Rev. Lett. {\bf 73}, 3580 (1994).

\bibitem{Sabouret02}
G.~Sabouret, C.~Williams y R.~Sobolewski.
\newblock Phys. Rev. B {\bf 66}, 132501 (2002).

\bibitem{Fuchs98}
D.~T. Fuchs, E.~Zeldov, T.~Tamegi, S.~Ooi, M.~Rappaport y H.~Shtrikman.
\newblock Phys. Rev. Lett. {\bf 80}, 4971  (1998).

\bibitem{Fuchs98B}
D.~T. Fuchs, R.~A. Doyle, E.~Zeldov, S.~F. W.~R. Rycroft, T.~Tamegai, S.~Ooi,
  M.~L. Rappaport y Y.~Myasoedov.
\newblock Phys. Rev. Lett. {\bf 81}, 3944  (1998).

\bibitem{Zhang01}
Y.~H. Zhang, H.~Luo, X.~F. Wu y S.~Y. Ding.
\newblock Supercond. Sci. Technol. {\bf 14}, 346  (2001).

\bibitem{Liu02}
Y.~Liu, H.~Luo, T.~Y.~X. Leng, X.~F. Wu, J.~W. Lin, Z.~H. Wang, H.~M. Luo y
  S.~Y. Ding.
\newblock Supercond. Sci. Technol. {\bf 15}, 373  (2002).

\bibitem{Gurevich87}
A.~V. Gurevich y R.~G. Mints.
\newblock Rev. Mod. Phys. {\bf 59}, 941  (1987).

\bibitem{Kahan93}
A.~Kahan.
\newblock Supercond. Sci. Techol. {\bf 6}, 476 (1993).

\bibitem{Xu94}
X.~Xu, J.~Fang, X.~Cao, K.~Li, W.~Yao y Z.~Qi.
\newblock Solid State Comm. {\bf 92}, 501 (1994).

\bibitem{Cao97}
X.~W. Cao, Z.~H. Wang, J.~Fang, X.~J. Xu y K.~B. Li.
\newblock J. Appl. Phys. {\bf 81}, 7392 (1997).

\bibitem{Duran68}
E.~Duran.
\newblock {\em Magn\'etostatique\/} (Masson et Cie, Paris, 1968).

\bibitem{Flik92}
M.~I. Flik, Z.~M. Zhang, K.~E. Goodson, M.~P. Siegal y J.~M. Phillips.
\newblock Phys. Rev. B {\bf 46}, 5606 (1992).

\bibitem{Bonn93}
D.~A. Bonn, R.~Liang, T.~M. Riseman, D.~J. Baar, D.~C. Morgan, K.~Zhang,
  P.~Dosanjh, T.~L. Duty, A.~MacFarlane, G.~D. Morris, J.~H. Brewer, W.~N.
  Hardy, C.~Kallin y A.~J. Berlinsky.
\newblock Phys. Rev. B {\bf 47}, 11314 (1993).

\bibitem{Gao93}
F.~Gao, J.~W. Kruse, C.~E. Platt, M.~Feng y M.~V. Klein.
\newblock Appl. Phys. Lett. {\bf 63}, 2274 (1993).

\bibitem{Chiaverini01}
J.~Chiaverini, J.~N. Eckstein, I.~Bozovic, S.~Doniach y A.~Kapitulnik.
\newblock {\em Nonlinear I-V, high flux-flow velocity instability and evidence
  for quantum fluctuations in {BSCCO} superconducting films\/} (2000).
\newblock {cond-mat/0007479}.


\bibitem{Xu90}
M.~Xu, D.~Shi y R.~F. Fox.
\newblock Phys. Rev. B {\bf 42}, 10773  (1990).

\bibitem{Bean62}
C.~P. Bean.
\newblock Phys. Rev. Lett. {\bf 8}, 250  (1962).

\bibitem{Anderson62}
P.~W. Anderson.
\newblock Phys. Rev. Lett. {\bf 9}, 309  (1962).

\bibitem{Anderson64}
P.~W. Anderson y Y.~B. Kim.
\newblock Rev. Mod. Phys. {\bf 36}, 39  (1964).

\bibitem{Jakob00}
G.~Jakob, P.~Voss-de Haan, M.~Wagner, Z.~Xiao y H.~Adrian.
\newblock Physica B {\bf 284-288}, 897 (2000).

\bibitem{Chapman84}
A.~J. Chapman.
\newblock {\em Heat Transfer\/} (Macmillan, New York, 1984),
  4\raise1ex\hbox{\scriptsize a} ed.

\bibitem{Marshall93}
C.~D. Marshall, A.~Tokmakoff, I.~M. Fishman, C.~B. Eom, J.~M. Phillips y M.~D.
  Fayer.
\newblock J. Appl. Phys. {\bf 73}, 850 (1993).

\bibitem{Touloukian70}
En Y.~S. Touloukian (ed.), {\em Thermophysical Properties of Matter\/}, tomos 2,
  5 (IFI/Plenum, New York, 1970).

\end{thebibliography}

\clearemptydoublepage

% emplea "tablas" en vez de "cuadros".
\def\listtablename{\'Indice de tablas}

\phantomsection \addcontentsline{toc}{chapter}{\'Indice de tablas}
 \listoftables
 \clearemptydoublepage

\phantomsection \addcontentsline{toc}{chapter}{\'Indice de
figuras}
 \listoffigures

\clearemptydoublepage

\setlength{\parskip} {1ex plus 0.5ex minus 0.2ex}
\setlength{\parskip} {1ex plus 0.5ex minus 0.2ex}

%%%%%%%%%%%%%%%%%%%%%%%%%%%%%%%%%%%%%%%%%%%%%%%%%%%%%%%%%%%%%%%%%%

\selectlanguage{spanish}

\chapter*{Agradecimientos}
\chaptermark{Agradecimientos}
% \addtocontents{toc}{\protect\vspace{0.2cm}}
\addcontentsline{toc}{chapter}{Agradecimientos}

% Esta versi\'on, ligeramente corregida para arXiv, me permite
% desquitarme parcialmente del mal cuerpo que me dejaron los
% agradecimientos de la tesis original.

El autor agradece a los directores de este trabajo, los profesores
F\'elix Vidal y Jos\'e Antonio Veira, las orientaciones que
determinaron el transcurrir de esta investigaci\'on, as\'{\i} como
los m\'ultiples recursos materiales y humanos puestos a su
disposici\'on. En especial, agradece al Dr. Veira el trabajo
cotidiano en el laboratorio, en el que tantas cosas ha aprendido
en su compa\~n\'{\i}a.

La mayor parte de esta investigaci\'on se llev\'o a cabo en
estrecha colaboraci\'on con el Dr. Severiano R. Curr\'as y con la
Dra. M. Teresa Gonz\'alez. Con Seve aprend\'{\i} a hacer filmes, y
de \'el hered\'e sus temas de investigaci\'on. El trabajo previo
de Teresa fue una base fundamental para mis experimentos, y llevar
nuestras tesis paralelas result\'o muy enriquecedor. De la
reflexi\'on conjunta sobre nuestros resultados sali\'o, por
ejemplo, la idea del modelo de calentamiento. Tambi\'en
contribuy\'o en las medidas de los filmes Mauricio Ruibal. A
ellos, muchas gracias por sus valiosas e indispensables
aportaciones. En la parte de las fluctuaciones, el autor
colabor\'o con los doctores Manuel V. Ramallo y Carlos
Carballeira, junto con la Dra. Carolina Torr\'on y el Dr. Jes\'us
Mosqueira. A todos agradezco lo que me aportaron, en el trabajo y
en lo personal.

Gracias en general a los miembros del LBTS, y en concreto al
profesor Jes\'us Maza, entre otras cosas por las apreciadas
charlas cient\'{\i}ficas que hemos mantenido, y de las que tantas
ideas, buenas y malas, han salido. Manuel R. Osorio, compa\~nero
desde hace a\~nos, es parte indispensable de estas reuniones.
Otros \emph{bolseiros} con los que comparto fatigas, que ya se han
ido o que llegan con renovadas ilusiones, son Silvia Vidal,
Luc\'{\i}a Cabo, Gonzalo Ferro y F\'elix Soto.

El apoyo del m\'ultiple personal t\'ecnico que pasa por nuestro
departamento ha sido imprescindible en diversos momentos de este
trabajo. De manera significativa, y en representaci\'on de los
restantes, menciono a Juan Ponte, Ra\'ul Santos, Fernando Naya,
Laura M. Pouso, Marcos L. Presas, Mar\'{\i}a Jes\'us Cantelar,
Juan Turnes y Beatriz Barral.

Quedo muy agradecido a las personas que, tan generosamente, han
hecho una lectura total o parcial de este trabajo, aportando
siempre provechosas correcciones, tanto cient\'{\i}ficas como
relativas al estilo o al idioma; entre otras, los profesores
Jos\'e Mar\'{\i}a Vi\~na L., Emilia Rebolledo, Pedro S.
Rodr\'{\i}guez H., Luis Vi\~na y Luisa Baus\'a. El Dr. Calum Byron
soport\'o con estoicismo la lectura de algunos textos ingleses. La
Dra. Irene Sendi\~na contribuy\'o amablemente a hacer la edici\'on
de esta memoria una tarea mucho m\'as f\'acil.

El agradecimiento m\'as especial, cari\~noso y sincero a mis
familiares y amigos, entre los que se encuentran varios de los ya
mencionados. Han sido fundamentales soportes personales durante
todo este largo tiempo como \emph{bolseiro} de investigaci\'on, y
tambi\'en durante toda la vida, no tan larga y demasiado fugaz.

Durante estos a\~nos de doctorado el autor recibi\'o
financiaci\'on de la Secretar\'{\i}a Xeral de Investigaci\'on e
Desenvolvemento de la Xunta de Galicia, como \emph{bolseiro} de
tercer ciclo (conjuntamente con la Direcci\'on Xeral de
Universidades) y \emph{bolseiro} predoctoral; tambi\'en de la
Comisi\'on Interministerial de Ciencia y Tecnolog\'{\i}a como
becario de cooperaci\'on asociado a un proyecto (CICYT
MAT2001-3053).

Este trabajo, junto con otros del LBTS, recibi\'o diferentes
ayudas de investigaci\'on e infraestructura de la Uni\'on Europea
(SCNET-2), la CICYT (MAT2001-3272 y MAT2001-3053), la Xunta de
Galicia (PGIDIT 02PXIC20609PN) y Uni\'on Fenosa (proyectos 0666-98
y 0666-02).

%En particular, estas financiaciones nos han permitido disponer de
%la infraestructura necesaria para la realizaci\'on de los
%experimentos descritos en esta memoria.

\clearemptydoublepage

% \emph{ PGIDIT Plan Galego de Investigaci\'on, Desenvolvemento e Innovaci\'on Tecnol\'oxica} ;

\clearemptydoublepage

\setlength{\parskip}{0ex plus 0.5ex}
%%%%%%%%%%%%%%%%%%%%%%%%%%%%%%%%%%%%%%%%%%%%%%%%%%%%%%%%%%%%%%%%%%

\selectlanguage{spanish}

\chapter*{Lista de publicaciones}
\chaptermark{Lista de publicaciones}
\addcontentsline{toc}{chapter}{Lista de publicaciones}
%\addtocontents{toc}{\protect\vspace{0.2cm}}

Los art\'{\i}culos directamente vinculados con diversos aspectos
de esta Tesis Doctoral, y publicados o enviados a publicar hasta
la fecha, se relacionan a continuaci\'on.

De la primera parte de esta memoria:

\begin{itemize}

\item \emph{Transition to the normal state of superconducting \YBCOf\ thin
films induced by high current densities}. S.~R. Curr\'as, P. Wagner,
M. Ruibal, J. Vi\~na, M.~R. Osorio, M.~T. Gonz\'alez, J.~A. Veira, J.
Maza, F. Vidal. Supercond. Sci. Technol. {\bf 14}, 748 (2001).

\item \emph{Normal-state resistivity versus critical current in
\YBCOf\ thin films at high current densities}. S.R. Curr\'as, J.
Vi\~na, M. Ruibal, M.~T. Gonz\'alez, M.~R. Osorio, J. Maza, J.~A.
Veira, F. Vidal. Physica C, {\bf 372-376} 1095 (2002).

\item \emph{Electric field versus current density curves in melt-textured
samples of \YBCOf\ under currents well above the critical
current}. M.~T. Gonz\'alez, S. Vidal, J. Vi\~na, M.~R. Osorio, J.
Maza, F. Vidal. Physica C, {\bf 372-376} 1852 (2002).

\item \emph{Normal-superconducting transition induced by high current
densities in \YBCOf\ melt-textured samples and thin films:
Similarities and differences}. M.~T. Gonz\'alez, J. Vi\~na, S.~R.
Curr\'as, J. Maza, F. Vidal. (2002). En tr\'amites de
publicaci\'on. cond-mat/0305020.

\end{itemize}

\newpage

De la segunda parte de esta memoria: \label{listads}

\begin{itemize}

\item \emph{Relaxation time of the Cooper pairs near \Tc\ in cuprate
superconductors}. M.~V. Ramallo, C. Carballeira, J. Vi\~na, J.~A.
Veira, T. Mishonov, D. Pavuna and F. Vidal. Europhys. Lett. {\bf
48}, 79-85 (1999).

\item \emph{Some experimental aspects of the thermal fluctuations around \Tc\ in
cuprate superconductors: Application to the``vortex matter''}. F.
Vidal, C. Torr\'on, J. Vi\~na and J. Mosqueira. Physica C, {\bf 332},
166-172 (2000).

\item \emph{Paraconductivity at high-reduced-temperatures in \YBCOf}. C.
Carballeira, S.~R. Curr\'as, J. Vi\~na, M.~V. Ramallo, J.~A. Veira and
F. Vidal. Phys. Rev. B. {\bf 63}, 144515 (2001).

\item \emph{Universal behaviour of the in-plane paraconductivity of cuprate
superconductors in the short-wavelength fluctuation regime}. J.
Vi\~na, J.~A. Camp\'a, C. Carballeira, S.~R. Curr\'as, A. Maignan, M.~V.
Ramallo, I. Rasines, J.~A. Veira, P. Wagner, F. Vidal. Phys. Rev.
B, {\bf 65} 212509 (2002).

\item \emph{On the consequences of the uncertainty principle on the superconducting
fluctuations well inside the normal state}. F. Vidal, C.
Carballeira, S.~R. Curr\'as, J. Mosqueira, M.~V. Ramallo, J.~A.
Veira, J. Vi\~na. Europhys. Lett., {\bf 59} 754 (2002).

\item \emph{The in-plane paraconductivity in {La$_{2-x}$Sr$_x$CuO$_4$} thin film
superconductors at high reduced-temperatures: Independence of the
normal-state pseudogap}. S.~R. Curr\'as, M.~T. Gonz\'alez, M.~V.
Ramallo, M. Ruibal, J.~A. Veira, J. Vi\~na, P. Wagner, F. Vidal.
(2002). En tr\'amites de publicaci\'on.

\end{itemize}

\clearemptydoublepage \setlength{\parskip}{1ex plus 0.5ex minus
0.2ex}

\phantomsection  \addcontentsline{toc}{chapter}{\'Indice
alfab\'etico} \printindex \clearemptydoublepage

%\be   % uno u otro,
%\[     % depende de si se numera o no
%{{a\;\tan (x) \over {delta}}} = \int {a\;dx}
%\]
%\ee

\end{document}